%% file: main.tex
\documentclass[a4paper,11pt]{article}

\usepackage{geometry}
\title{Fixed Budget vs. Covering Target:\\
The Partial Set Cover Boundary for Bounded VC-Dimension}

\author{
Madhumita Kundu\textsuperscript{1} \quad
Souvik Saha\textsuperscript{2} \quad
Saket Saurabh\textsuperscript{1,2} \quad
Anannya Upasana\textsuperscript{2}\\[0.5em]
\small
\textsuperscript{1} University of Bergen\\
\small
\textsuperscript{2}The Institute of Mathematical Sciences, HBNI
}

\usepackage{amsmath}
\usepackage{amssymb}
\usepackage{amsthm}
\usepackage[boxed,lined,linesnumbered, ruled]{algorithm2e}
\SetKwInput{KwInput}{Input}
\SetKwInput{KwOutput}{Output}
\SetKwInput{KwInvariant}{Invariant}
\DontPrintSemicolon

\SetKwIF{If}{ElseIf}{Else}{if}{}{else if}{else}{}

\SetKwFor{For}{for}{}{}
\SetKwFor{ForEach}{for each}{}{}
\SetKwFor{While}{while}{}{}
\SetKwFor{Repeat}{repeat}{until}

\SetKwFunction{Return}{return}
\SetKw{KwAnd}{and}
\SetKw{KwOr}{or}
\usepackage{diagbox}
\usepackage{booktabs}
\usepackage{array}
\usepackage{pdflscape}
\usepackage{makecell} 
\usepackage{adjustbox} 
\usepackage[dvipsnames]{xcolor}
\usepackage{colortbl}
\definecolor{headergray}{RGB}{230,230,230}
\definecolor{firstcolgray}{RGB}{240,240,240}
\definecolor{lightgray}{RGB}{248,248,248}

\colorlet{mix}{red!50!black}
\definecolor{bblue}{rgb}{0.1,0.4,0.6} 
\usepackage[xetex, unicode, plainpages = false, pdfpagelabels, 
bookmarks=false,
bookmarksopen = true,
bookmarksnumbered = true,
breaklinks = true,
linktocpage,
 pagebackref,
colorlinks = true,  
linkcolor = mix,
urlcolor  = blue,
filecolor=magenta,
menucolor=blue,
citecolor = blue,
anchorcolor = green,
runcolor=red,
hyperindex = true,
hyperfigures
]{hyperref}

\usepackage{enumerate}

\usepackage{cleveref}
\usepackage[many]{tcolorbox}

\tcbset {
  base/.style={
    arc=0mm, 
    bottomtitle=0.5mm,
    boxrule=0mm,
    colbacktitle=black!10!white, 
    coltitle=black, 
    fonttitle=\bfseries, 
    left=2.5mm,
    leftrule=1mm,
    right=3.5mm,
    title={#1},
    toptitle=0.75mm, 
  }
}

\definecolor{brandblue}{rgb}{0.34, 0.7, 1}
\newtcolorbox{mainbox}[1]{
  colframe=brandblue, 
  base={#1}
}

\newtcolorbox{subbox}[1]{
  colframe=black!30!white,
  base={#1}
}

 \tcolorboxenvironment{highlight}{
    colframe=yellow, sharp corners=west, 
    colback=yellow!10,
	rightrule=0pt,
	toprule=0pt,
	bottomrule=0pt,
	top=0pt,
	right=0pt,
	bottom=0pt,
 }
 \tcolorboxenvironment{highlight*}{
    colframe=yellow, sharp corners=west, 
    colback=yellow!10,
	rightrule=0pt,
	toprule=0pt,
	bottomrule=0pt,
	top=0pt,
	right=0pt,
	bottom=0pt,
 }

\newtheorem*{highlight*}{}

\newtheorem{claim}{Claim}[section]
\newtheorem{property}{Property}
\usepackage{enumitem}
\usepackage{etoolbox}

\newtoggle{includeFootnote}
\toggletrue{includeFootnote}

\newtheorem{theorem}{Theorem}[section]

\newtheorem{lemma}[theorem]{Lemma}

\newtheorem{definition}[theorem]{Definition}
\newtheorem{observation}[theorem]{Observation}
\newtheorem{proposition}[theorem]{Proposition}
\newtheorem{redrule}[theorem]{Reduction Rule}

\usepackage{nicefrac}
\usepackage{graphicx}
\usepackage{thmtools,thm-restate}
\usepackage{mathtools}
\usepackage[framemethod=tikz]{mdframed}
\mdfsetup{frametitlealignment=\centering}

\usepackage{subcaption} 
\usepackage{tikz}
\usetikzlibrary{shapes.geometric}
\usetikzlibrary{arrows.meta}
\usetikzlibrary{positioning}
\usetikzlibrary{shapes.multipart}
\usetikzlibrary{backgrounds,fit,calc,decorations.pathreplacing}
\usepackage[multiple]{footmisc}

\newcommand{\fpt}{\textsf{FPT}\xspace}

\newcommand{\cS}{\ensuremath{\mathcal{S}}\xspace}

\newcommand{\bN}{\ensuremath{\mathbb{N}}\xspace}

\newcommand{\prob}[1]{\ensuremath{\text{\bf Pr}\left [#1\right]}\xspace}

\newcommand{\iletter}{\alpha}
\newif\ifinmainbody
\date{}
\input{macros.tex}

\hypersetup{
    colorlinks=true, 
    linktoc=all,     
    linkcolor=blue,
    citecolor=BrickRed,  
}
\begin{document}
\maketitle
\begin{abstract}
\textsc{Maximum Coverage} and \textsc{Partial Set Cover} are two fundamental
parameterized covering problems. The former is the fixed-budget version: choose
at most \(k\) sets so as to maximize the number of covered elements. The latter
is the target-covering version: cover a prescribed target using as few sets as
possible. Prior work of Badanidiyuru, Kleinberg, and Lee [BKL, SoCG'12] gives an
EPAS for \textsc{Maximum Coverage} on set systems of bounded VC-dimension, while
a separate line of work [JKPSSSU, SODA'23] establishes an additive approximation for
\textsc{Partial Set Cover} on set systems whose incidence graph is \(K_{d,d}\)-free:
whenever \(k\) sets meet the target, one can find \(k+1\) sets that also meet it. This is a strictly more restrictive class
than bounded VC-dimension. We ask whether the same target-covering guarantee
extends to all bounded VC-dimension set systems.

Our first result gives a negative answer. Unless \(\mathrm{FPT}=\mathrm{W[1]}\),
\textsc{Partial Set Cover} admits no parameterized \((2-\delta)\)-approximation,
even on instances of VC-dimension seven. Moreover, under ETH, it admits neither
a parameterized approximation scheme on such instances nor a
\(2^{o(d)}\)-approximation on instances of VC-dimension \(d\).

On the positive side, we show that bounded semi-ladder index restores the
target-covering guarantee. This condition is stronger than bounded
VC-dimension, but it strictly generalizes the \(K_{d,d}\)-free setting. We prove
the following analogue for \textsc{Weighted Partial Set Cover}: if \(k\) sets
cover elements of total weight at least \(W\), then our algorithm finds \(k+1\)
sets covering elements of total weight at least \(W\) in time
\(2^{O(\Gamma k\log k)}N\), where \(\Gamma\) is the downward intersection
complexity of the incidence graph and $N$ is the input size.

The framework also supports per-class coverage targets together with a matroid
independence constraint on the selected sets. We complement these extensions
with applications to partial dominating set, geometric partial covering, and
covering bounded-size sets by bounded-size sets.

Finally, we give a deterministic FPT reduction from \textsc{Weighted CC-MaxSAT}
to a bounded family of \textsc{Weighted Maximum Coverage} instances. The reduction
preserves the relevant incidence-graph structure and, with a constant-factor
loss in the accuracy parameter, preserves approximation schemes. As a
consequence, we obtain an EPAS for \textsc{Weighted CC-MaxSAT} on instances of
bounded semi-ladder index. We also give a more efficient deterministic
implementation of the bounded-VC-dimension EPAS of Badanidiyuru, Kleinberg, and
Lee [BKL, SoCG'12]. Combining this implementation with our reduction yields an
EPAS for bounded-VC-dimension instances of \textsc{Weighted CC-MaxSAT} with
running time $
2^{\widetilde{O}(kd/\varepsilon)}\,N^{O(1)} $.
\footnote{ \(\widetilde{O}\) suppresses factors polynomial in logarithms of the parameters.}
\end{abstract}

\newpage 

\tableofcontents

\newpage 
\pagenumbering{arabic}
\input{introduction}

\input{technical_overview}

\input{vc_dimension_lower_bounds}

\input{prelims.tex}

\input{randomized_max_coverage.tex}

\input{weighted_max_coverage.tex}

\input{partial_set_cover.tex}

\input{deterministic_algorithms.tex}

\input{bkl_trace_enumeration}

\input{maxsat_reduction}

\input{applications}

\input{matroid_partition_constraints}

\input{constrained_maxsat}

\input{conclusion.tex}

\bibliographystyle{alpha}
\bibliography{references}

\appendix 
\input{appendix}

\end{document}

%% file: macros.tex
\usepackage{mathrsfs}
\usepackage{halloweenmath}
\usepackage{tabularx}
\usepackage{latexsym}
\usepackage{amsmath,amsfonts}
\usepackage{amssymb}
\usepackage{amsthm}
\usepackage{epsfig}
\usepackage{xfrac}
\usepackage{xspace}
\usepackage{float}
\usepackage{thm-restate}
\usepackage[framemethod=tikz]{mdframed}
\usepackage{tcolorbox}
\usepackage{tikz}

\definecolor{mycolor}{rgb}{0.122, 0.435, 0.698}
\newmdenv[innerlinewidth=0.02pt, roundcorner=4pt,linecolor=mycolor,innerleftmargin=6pt,
innerrightmargin=6pt,innertopmargin=6pt,innerbottommargin=6pt]{mybox}
\newmdenv[innerlinewidth=0.5pt, roundcorner=4pt,linecolor=black,innerleftmargin=6pt,
innerrightmargin=6pt,innertopmargin=6pt,innerbottommargin=6pt]{myboxblack}
\newmdenv[innerlinewidth=0.5pt, roundcorner=4pt,linecolor=mycolor,innerleftmargin=6pt,
innerrightmargin=6pt,innertopmargin=6pt,innerbottommargin=6pt]{myboxthick}

\newcommand{\cR}{\mathcal{R}}
\newcommand{\cE}{\mathcal{E}}

\newcommand{\hidestuff}[1]{}

\newcommand{\cI}{\mathcal{I}}
\newcommand{\cP}{\mathcal{P}}
\newcommand{\cA}{\mathcal{A}}
\newcommand{\cQ}{\mathcal{Q}}
\newcommand{\cB}{\mathcal{B}}
\newcommand{\bO}{\mathbb{O}}
\newcommand{\bo}{\mathbb{O}}
\newcommand{\bQ}{\mathbb{Q}}
\newcommand{\cC}{\mathcal{C}}

\newcommand{\cD}{\mathcal{D}}
\newcommand{\cU}{\mathcal{U}}
\newcommand{\cH}{\mathcal{H}}

\newcommand{\cT}{\mathcal{T}}

\newcommand{\cX}{\mathcal{X}}
\newcommand{\cO}{\mathcal{O}}
\newcommand{\cV}{\mathcal{V}}
\newcommand{\cM}{\mathcal{M}}
\newcommand{\cF}{\mathcal{F}}
\newcommand{\cJ}{\mathcal{J}}
\newcommand{\cL}{\mathcal{L}}

\newcommand{\npc}{ {\sf NP-Complete}\xspace}

\newcommand{\no}{{\sc No}\xspace}
\newcommand{\yes}{{\sc Yes}\xspace}
\newcommand{\opt}{{\sf OPT}\xspace}

\newcommand{\whard}{\textsc{W[1]-hard}\xspace}

\usepackage{wrapfig}
\usepackage{bm}

\newcommand{\opti}{${\sf OPT}_k(\cI)$ \xspace}

\newcommand{\cov}{{\sf cov}\xspace}
\newcommand{\pot}{{\sf Psf}\xspace}
\newcommand{\fcov}{{\sf FreeCov}\xspace}
\newcommand{\free}{{\sf Free}\xspace}
\newcommand{\nfree}{{\sf Newly Free}\xspace}
\newcommand{\mcon}{{\sf MustInclude}\xspace}
\newcommand{\mi}{\cM\cI}

\newcommand{\cons}[1]{{\sf ConsTuple}(#1)}

\newcommand{\lhs}{\textsf{LHS}\xspace}
\newcommand{\rhs}{\textsf{RHS}\xspace}

\newcommand{\OO}{{\mathcal O}}

\newcommand{\wtilde}[1]{\widetilde{#1}}

\newtheorem{reduction rule}{Reduction Rule}
\newtheorem*{reduction rule*}{Reduction Rule}

\newtheorem*{thm*}{Theorem}

\newcommand{\defparopt}[4]{
\begin{tcolorbox}[colback=gray!5!white,colframe=gray!75!black]
  \vspace{-1mm}
  \begin{tabular*}{\textwidth}{@{\extracolsep{\fill}}lr} #1  & {\bf{Parameter:}} #3 \\ \end{tabular*}
  {\bf{Input:}} #2  \\
  {\bf{Output:}} #4
  \vspace{-1mm}
\end{tcolorbox}
}

\newcommand{\hit}{\textnormal{\textsf{Intersect}}\xspace}
\newcommand{\pas}{\textsf{PAS}\xspace}
\newcommand{\epas}{\textsf{EPAS}\xspace}
\newcommand{\smlindex}{\textnormal{\textsf{semi-ladder index}}\xspace}

\newcommand{\psc}{\textsc{Partial Set Cover}\xspace}
\newcommand{\wpsc}{\textsc{Weighted Partial Set Cover}\xspace}

\newcommand{\pplc}{\textsc{Partial Point Line Cover}\xspace}
\newcommand{\pvc}{\textsc{Partial Vertex Cover}\xspace}
\newcommand{\plpc}{\textsc{Partial Line Point Cover}\xspace}

\newcommand{\ccL}{\mathcal{L}}

\newcommand{\wpheds}{\textsc{W-PHEDS}\xspace}
\newcommand{\decwpheds}{\textsc{Dec-W-PHEDS}\xspace}

\newcommand{\wphedsfull}{\textsc{Weighted Partial $d$-Hyperedge Dominating Set }\xspace}
\newcommand{\decwphedsfull}{{\sf Decision Weighted Partial $d$-Hyperedge Dominating Set }\xspace}
\newcommand{\slf}{\textnormal{\textsf{semi-ladder-free}}\xspace}
\newcommand{\sml}{\textnormal{\textsf{semi-ladder}}\xspace}

\newcommand{\nbrst}{{\sf N}^{\star} }
\newcommand{\nbr}{{\sf N} }
\newcommand{\prbdsfull}{\textsc{Partial Red Blue Dominating Set}\xspace}
\newcommand{\prbds}{\textsc{Partial RBDS}\xspace}
\newcommand{\mrbds}{\textsc{Max RBDS}\xspace}
\newcommand{\prbdsi}{$(G=(\cR \cup \cB, E), k)$\xspace}

\newcommand{\algprbds}{\textnormal{\texttt{Alg-MaxRBDS}}\xspace}

\newcommand{\wmrbds}{\textsc{Weighted Max RBDS}\xspace}

\newcommand{\wmrbdsfull}{\textsc{Weighted Maximum Red Blue Dominating Set}\xspace}
\newcommand{\wprbdsi}{$(\graph, \w, k)$\xspace}
\newcommand{\algwprbds}{\textnormal{\texttt{Alg-Wt-MaxRBDS}}\xspace}
\newcommand{\wopt}{{\sf WOPT} \xspace}

\newcommand{\prds}{\textsc{Partial-$r$-Dominating Set}\xspace}

\newcommand{\wms}{\textsc{Weighted CC-MaxSAT}\xspace}
\newcommand{\wmsfull}{\textsc{Weighted Max SAT with Cardinality Constraint}\xspace}
\newcommand{\algwsat}{\textnormal{\texttt{Reduction-Wt-CCMaxSAT}}\xspace}
\newcommand{\wmc}{\textsc{Weighted Max Coverage}\xspace}

\newcommand{\cnf}{{\sf CNF-SAT}\xspace}
\newcommand{\wnd}{\textnormal{\texttt{wnndeg}}\xspace}

\newcommand{\sat}[2]{{\sf sat}_{#1}(#2)}
\newcommand{\n}[1]{{\sf neg}(#1)}
\newcommand{\wt}[2]{\textnormal{\texttt{weight}}_{#1}(#2)\xspace}
\newcommand{\woptc}{{\sf WOPT_k^{\text{COV}}} \xspace}
\newcommand{\wopts}{{\sf WOPT_k^{\text{SAT}}} \xspace}
\newcommand{\itm}{i_{\text{term}}\xspace}
\usepackage{pifont}

\newcommand{\algaddwprbds}{\textnormal{\texttt{Alg-Additive-Wt-PartialRBDS}}\xspace}
\newcommand{\decwprbdsfull}{\textsc{Decision-Weighted-Partial Red Blue Dominating Set}\xspace}
\newcommand{\decwprbds}{\textsc{Dec-Wt-Partial RBDS}\xspace}

\newcommand{\optds}{{\sf OPT_{BDS}}\xspace}

\newcommand{\pic}
{
    \resizebox{0.015\textwidth}{!}{ 
        \begin{tikzpicture}
            \def\d{5} 

        \foreach \i in {1,2,3,4,5} {
            \node[circle,draw,minimum size=10pt] (r\i) at (0,-\i) {};
            \node[circle,draw,minimum size=10pt] (b\i) at (2,-\i) {};
            \draw[dashed] (r\i) -- (b\i); 
        }
        
        \foreach \i in {2,3,4,5} {
            \foreach \j in {1,2,3,4} {
                \ifnum \i>\j
                    \draw (r\i) -- (b\j);
                \fi
            }
        }
        \end{tikzpicture}
    }
\xspace 
    }

\newcommand{\dintercomp}{\textit{downward intersection complexity}\xspace}
\newcommand{\dintcmp}{{\mathbb{D}\mathbb{I}}\xspace}

\newcommand{\realizable}{{realizable}\xspace}
\newcommand{\realizing}{{realizing}\xspace}
\newcommand{\realext}{{\textit{intersection extension}}\xspace}
\newcommand{\intcmp}{{\Gamma}\xspace}
\newcommand{\graph}{{G=(\cR \cup \cB, E)}\xspace}

\newcommand{\intbip}{{ intersection bipartite}\xspace}

\newcommand{\incld}{\textnormal{\textsf{Include}}\xspace}
\newcommand{\reex}{\textnormal{\textsf{IntExt}}\xspace}

 \newcommand{\wdeltanet}{\textnormal{{weighted}-$\delta$-{net}}\xspace}

\newcommand{\w}{\textnormal{\texttt{w}}\xspace}

\newcommand{\plc}{\textsc{Point Line Cover}\xspace}
 
\newcommand{\dimension}{\textnormal{\texttt{dim}}}

\newcommand{\cmaxsat}{\textsc{CC-MaxSAT}\xspace}
\newcommand{\maxcov}{\textsc{Maximum Coverage}\xspace}
\newcommand{\dmaxcov}{\textsc{Dec-Maximum Coverage}\xspace}

\newcommand{\cover}{{\sf cover}\xspace}
\newcommand{\pardimsc}{{\sc Partial-Dim-Set-Cover}\xspace}
\newcommand{\set}{{\sf set}\xspace}
\newcommand{\decpardimsc}{{\sc Dec-Partial-Dim-Set-Cover}\xspace}
\newcommand{\vpn}{$\delta$-net\xspace}
\newcommand{\wvpn}{weighted $\delta$-net\xspace}

\newcommand{\sizeg}{{\lVert G \rVert}\xspace}
\newcommand{\sizephi}{{\lVert \Phi \rVert}\xspace}
\newcommand{\dist}{{\textsf{dist}}\xspace}
\newcommand{\sizesetsys}{{\lVert (\cU, \cF) \rVert}\xspace}

\newcommand{\rrac}{\textnormal{\texttt{RecursiveRandomApproxCover}}\xspace}

\newcommand{\col}{{f}\xspace}
\newcommand{\matfull}{\mathcal{M}=(\mathcal{E},\mathcal{I})\xspace}
\newcommand{\mat}{\mathcal{M}\xspace}
\newcommand{\wconprbds}{\textsc{Weighted Constrained Max RBDS}\xspace}
\newcommand{\wconprbdsi}{(\graph, \w, \col,t, k,\mat,\varepsilon)\xspace}
\newcommand{\buck}{\textnormal{bucketing}\xspace}
\newcommand{\bag}{\textnormal{{bag}}\xspace}
\newcommand{\bucket}{\textnormal{\texttt{bucket}}\xspace}

\newcommand{\algwconprbds}{\textnormal{\texttt{Alg-Con-MRBDS}}\xspace}

\newcommand{\wmsc}{\textsc{Weighted Constrained CC-MaxSAT}\xspace}
\newcommand{\wmscfull}{\textsc{Weighted CC-MaxSAT with Matroid and Partition Constraints}\xspace}
\newcommand{\wmcc}{\textsc{Weighted Constrained Max Coverage}\xspace}
\newcommand{\wmccfull}{\textsc{Weighted Max Coverage with Matroid and Partition Constraints}\xspace}
\newcommand{\wndc}{\textnormal{\texttt{wcnndeg}}\xspace}
\newcommand{\algwcsat}{\textnormal{\texttt{Reduction:Weighted-Constrained-CC-MaxSAT}}\xspace}
\newcommand{\sizeinst}{{\lVert \cI \rVert}\xspace}

\newcommand{\opts}{{\sf OPT_k^{\text{SAT}}} \xspace}
\newcommand{\optc}{{\sf OPT_k^{\text{COV}}} \xspace}

\renewcommand{\footnotesize}{\scriptsize}

\usepackage{booktabs,tabularx,array,makecell}
\newcolumntype{L}[1]{>{\raggedright\arraybackslash}p{#1}}

\usetikzlibrary{decorations.text}

%% file: introduction.tex
\section{Introduction}\label{section: intro}
\textsc{Maximum Coverage} and \textsc{Partial Set Cover} represent two complementary formulations of the
same covering problem.  In the fixed-budget version, given a budget of \(k\) sets,
the goal is to maximize the number of elements covered by at
most \(k\) sets.  In the target-covering version, the objective is to minimize the number of sets needed to cover a given target $t$. Classically, these two formulations are closely related, however, our results show that structural assumptions can lead to a sharp separation between them from the perspective of parameterized approximation.

Without any structural restriction, parameterizing by the budget \(k\) alone
does not bypass the classical approximation barrier for \maxcov.  The
classical greedy algorithm gives a \((1-1/e)\)-approximation, and this ratio is
optimal in polynomial time by Feige's hardness
result~\cite{DBLP:journals/jacm/Feige98}.  Manurangsi strengthened this
barrier in the parameterized setting: assuming Gap-ETH, for every fixed
\(\varepsilon>0\), no algorithm running in time \(f(k,\varepsilon)N^{o(k)}\),
where \(N\) is the input size, can achieve a
\((1-1/e+\varepsilon)\)-approximation for \maxcov~\cite{Manurangsi20}.
This is why the positive results in the literature are framed by additional structure,
such as bounded VC-dimension or $K_{d,d}$-freeness.

For the fixed-budget formulation, the relevant benchmark is due to
Badanidiyuru, Kleinberg, and Lee~\cite{DBLP:conf/compgeom/BadanidiyuruKL12}, and it applies to the broadest structural class currently known for this problem:
set systems of bounded VC-dimension admit an efficient parameterized
approximation scheme for \textsc{Maximum Coverage}.
It is natural to ask whether the same
bounded VC-dimension machinery can be pushed to the target-covering problem
\textsc{Partial Set Cover}.

\medskip
\noindent\textbf{Bounded VC-dimension does not suffice for target covering.}
We answer this question in the negative: the structure that makes fixed-budget
coverage tractable does not carry over to target covering.

We first fix the terminology used to state the bounds.  A subset
$Z\subseteq\cU$ is \emph{shattered} by a set system $(\cU,\cF)$ if
$\cF|_Z:=\{F\cap Z:F\in\cF\}=2^Z$; the \emph{VC-dimension} of
$(\cU,\cF)$, introduced by Vapnik and Chervonenkis~\cite{doi:10.1137/1116025}, is
the largest cardinality of a shattered subset of $\cU$.  A \emph{parameterized approximation scheme}~(\pas) is a family of algorithms, one for
each $\varepsilon>0$, that on inputs of size $N$ returns a
$(1\pm\varepsilon)$-approximate solution in time $f(k,\varepsilon)\,N^{g(\varepsilon)}$,
for some computable functions $f$ and $g$; it is \emph{efficient}~(\epas) when the
exponent of $N$ is a constant independent of $\varepsilon$.

Our lower bounds make this precise, as summarized below.

\begin{tcolorbox}[
  enhanced,
  frame hidden,
  colback=teal!3!white,
  arc=1.8mm,
  left=2mm,right=2mm,top=1.2mm,bottom=1.2mm,
  borderline west={1.4pt}{0pt}{teal!45!black}
]
\textbf{This paper (informal).}
Bounded VC-dimension makes fixed-budget \maxcov\ tractable, but does not help the
target-covering problem \psc:
\begin{itemize}[leftmargin=1.4em,itemsep=2pt,topsep=3pt,parsep=0pt]
\item \emph{No approximation below factor two.} Assuming $\mathrm{FPT}\ne\mathrm{W[1]}$,
  \psc\ has no \fpt\ approximation of factor below $2$, even at VC-dimension~$7$.
\item \emph{No approximation scheme.} Assuming ETH, \psc\ has no \pas\ at
  VC-dimension~$7$.
\item \emph{Hardness scales with the dimension.} Assuming ETH, no \fpt\ algorithm
  approximates \psc\ within factor $2^{o(d)}$ on instances of VC-dimension~$d$.
\end{itemize}
\end{tcolorbox}

These bounds are stated formally and proved in \Cref{sec:vc-boundary}.  Thus the algorithmic
advantage of bounded VC-dimension is specific to the fixed-budget formulation: for
target covering it disappears already at VC-dimension seven.

\medskip
\noindent\textbf{A positive precedent under stronger structure.}
Bounded VC-dimension is therefore too weak an assumption for target covering:
unlike fixed-budget coverage, this objective becomes tractable only under a stronger
structural restriction. We therefore ask: what restriction is sufficient?  The positive precedent is due to Jain et
al.~\cite{DBLP:conf/soda/0001KPSS0U23}.  They work with
\emph{biclique-free} set systems, those whose incidence graph excludes a fixed
complete bipartite subgraph $K_{d,d}$, and show that on this class the fixed-budget
coverage guarantee lifts to the target-covering objective: if $k$ sets meet the
target, then one can find $k+1$ sets that meet the same target.  Biclique-freeness
is a strict strengthening of bounded VC-dimension, so this result sits safely on
the tractable side of the boundary above.  We note that this biclique-free line of
work appears to have been developed independently of the earlier bounded-VC-dimension
scheme of Badanidiyuru, Kleinberg, and Lee~\cite{DBLP:conf/compgeom/BadanidiyuruKL12},
whose fixed-budget \epas already implies the coverage guarantee on the narrower
biclique-free class; the genuinely new contribution of that work is instead the target-covering
guarantee described above, which we take as the starting point for our results.  It is natural to ask how far
this coexistence extends: is biclique-freeness necessary, or does a weaker structural
assumption already suffice?

\medskip
\noindent\textbf{Our answer: bounded semi-ladder index.}
Our positive results do not leave the bounded-VC-dimension world; they refine it.  We
isolate a single further combinatorial restriction, \emph{bounded semi-ladder index},
that a set system may satisfy on top of bounded VC-dimension, and show that this is
exactly what the target-covering side needs.  The restriction is a property of the
incidence bipartite graph $\graph$ of the set system (a red vertex in $\cR$ for each
set, a blue vertex in $\cB$ for each element, and an edge for each membership), and it
has appeared before in algorithms for \textsc{Dominating Set} and \textsc{Set
Cover}~\cite{DBLP:conf/stacs/FabianskiPST19,DBLP:journals/toct/G25}. A closely related metric notion, the \(\varepsilon\)-scatter dimension, was introduced
independently by Abbasi et al.~\cite{DBLP:conf/focs/AbbasiBBCGKMSS23} for parameterized
clustering; Bourneuf and Pilipczuk~\cite{DBLP:conf/soda/BourneufP25} later made the
connection to semi-ladders explicit, proving bounds for metrics induced by proper
minor-closed graph classes.

\begin{definition}[$d$-semi-ladder,~\cite{DBLP:conf/stacs/FabianskiPST19}]
A bipartite graph $G=(\cR,\cB,E)$ contains a \emph{semi-ladder of order $d$} if there are vertices $r_1,\ldots,r_d\in \cR$ and $b_1,\ldots,b_d\in \cB$ such that $(r_i,b_j)\in E$ whenever $i>j$, and $(r_i,b_i)\notin E$ for every $i\in [d]$. The \emph{semi-ladder index} of $G$ is the maximum $d$ for which $G$ contains a semi-ladder of order $d$.
\end{definition}

\begin{figure}[t]
\centering
\includegraphics[scale=0.25]{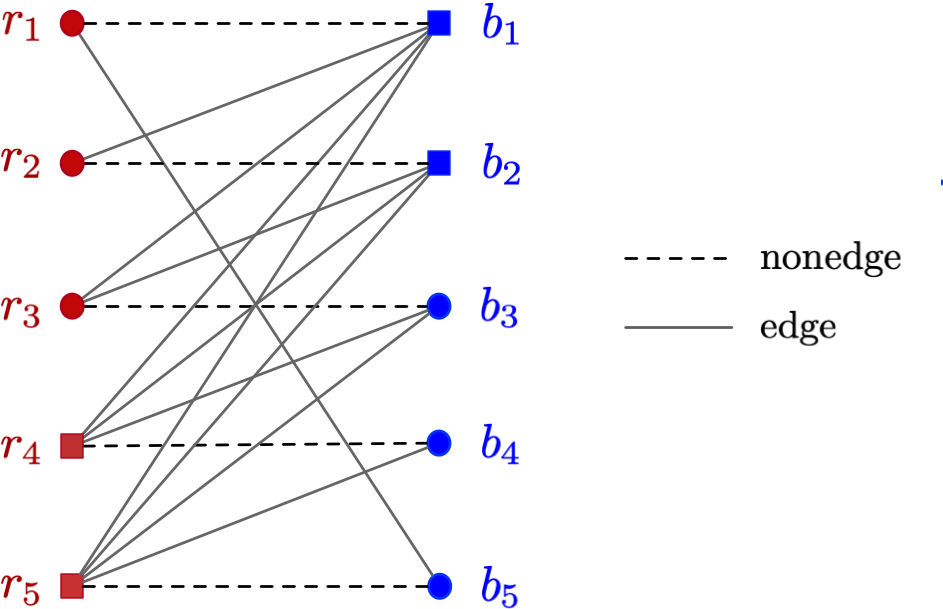}
\caption{A semi-ladder of order $5$.}
\label{fig:intro-semiladder}
\end{figure}

 What matters is where this class sits: strictly between $K_{d,d}$-freeness and bounded
VC-dimension, with a separate short argument on each side.  For the left boundary,
observe that \emph{a semi-ladder of order $2d$ contains a copy of $K_{d,d}$}: in a
semi-ladder $r_1,\dots,r_{2d},b_1,\dots,b_{2d}$ the $d$ red vertices
$r_{d+1},\dots,r_{2d}$ and the $d$ blue vertices $b_1,\dots,b_d$ satisfy $i>j$ for every
pair, so all $d^2$ edges are present; hence a $K_{d,d}$-free graph has no semi-ladder of
order $2d$, that is, its semi-ladder index is below $2d$.  For the right boundary, no
external tool is needed: if $Z=\{b_1,\dots,b_{d+1}\}$ is shattered, then for each
$i\in[d+1]$ some set has trace exactly $\{b_1,\dots,b_{i-1}\}$ on $Z$; letting $r_i$ be
the corresponding red vertex, we get $(r_i,b_j)\in E$ for all $j<i$ and
$(r_i,b_i)\notin E$, so $r_1,\dots,r_{d+1}$ and $b_1,\dots,b_{d+1}$ form a semi-ladder of
order $d+1$.  Hence the VC-dimension never exceeds the semi-ladder index, and semi-ladder
index at most $d$ forces VC-dimension at most $d$.  Together,
\[
K_{d,d}\text{-free} \;\Longrightarrow\; \text{semi-ladder index} < 2d
\;\Longrightarrow\; \text{VC-dimension} < 2d,
\]
and both implications are strict as inclusions of classes.  A $K_{d,d}$-free graph has
only $\cO(n^{2-1/d})$ edges (Kővári--Sós--Turán), whereas a complete biclique has
semi-ladder index $0$ and yet $\Theta(n^2)$ edges; and a half-graph has VC-dimension $1$
but unbounded semi-ladder index, so bounded VC-dimension alone puts no bound on the
semi-ladder index.  The second implication is the one that anchors our results: bounded
semi-ladder index keeps the VC-dimension bounded, so we stay inside the regime where the
fixed-budget \epas\ of Badanidiyuru, Kleinberg, and
Lee~\cite{DBLP:conf/compgeom/BadanidiyuruKL12} applies, and only add the structure that
makes target covering tractable; the first says the class is strictly larger than the
$K_{d,d}$-free set systems of prior work.  Beyond biclique-freeness, bounded
semi-ladder index also covers many structured graph and geometric classes, such as
radius-neighborhood incidence graphs of map graphs and powers of nowhere-dense
classes~\cite{DBLP:journals/corr/abs-1811-06799} and geometric set systems induced by
hyperplanes or bounded-degree algebraic varieties~\cite{DBLP:journals/dcg/LangermanM05}.
Further examples and consequences are developed in~\Cref{sec:applications}.

\begin{figure}[t!]
\centering
\includegraphics[scale=0.2]{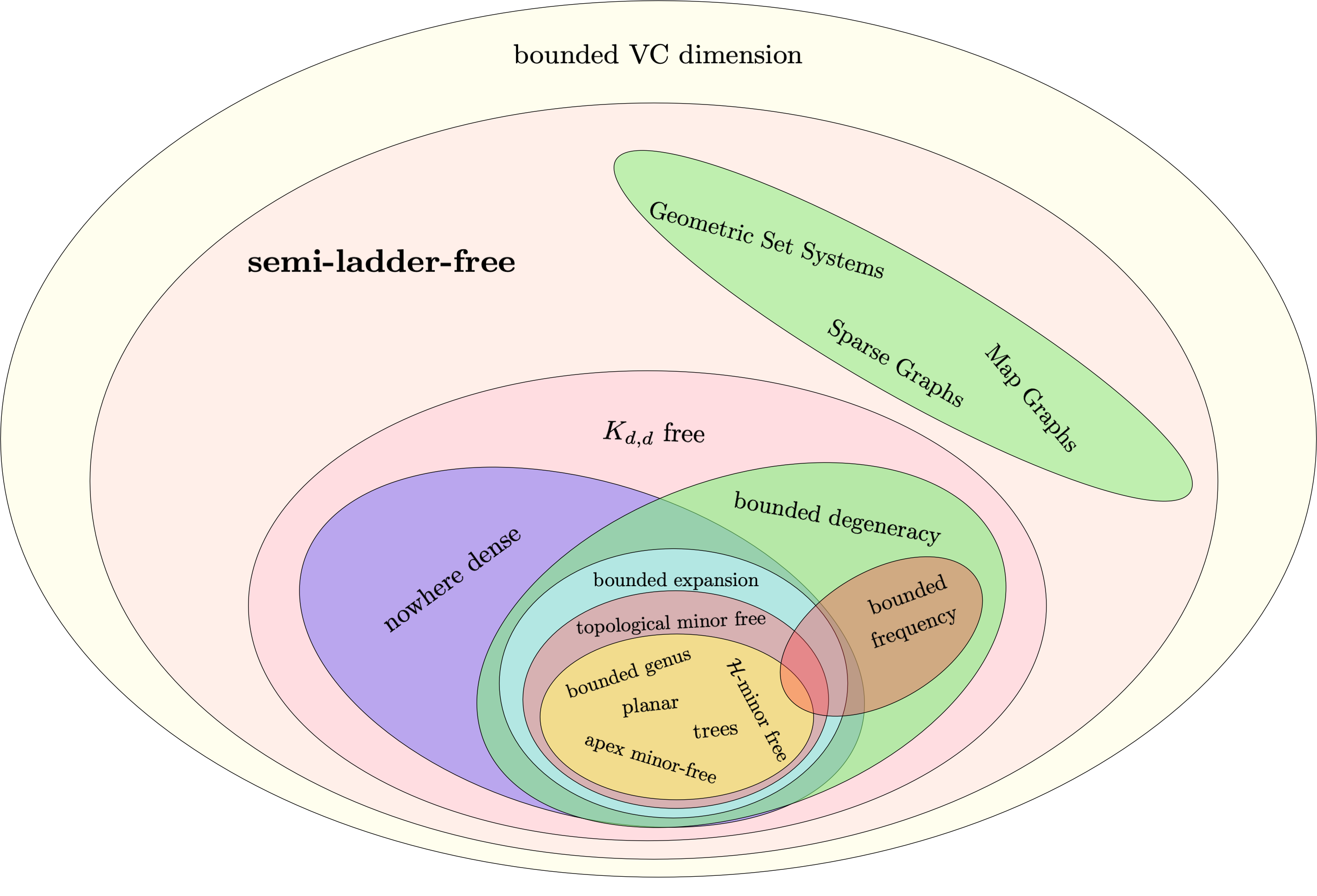}
\caption{Structural hierarchy of set systems via their incidence bipartite graph.}
\label{fig:overviewIntro}
\end{figure}

Our running times are stated in terms of the closely related \emph{downward
intersection complexity} $\Gamma=\dintcmp(G)$, which differs from the semi-ladder index
by at most an additive constant of $1$ (see \Cref{sec: sml_comp}); all results below
hold verbatim with the semi-ladder index in place of $\Gamma$.  As
\Cref{section:techidea} explains, $\Gamma$ is exactly the measure that caps how long
our randomized branching can defer committing to a set.

\noindent\textbf{Main algorithmic results.}
We phrase our algorithms in the equivalent red--blue domination formulation on the
incidence graph: a solution selects at most $k$ red vertices, and its value is the
total weight of the blue vertices they cover, where each blue vertex $b\in\cB$ carries
a weight $\w(b)\in\bQ^+$ (an unweighted instance takes all weights equal to $1$).  We
measure an instance by $\sizeg=|\cR|+|\cB|+|E(G)|+\sum_{b\in\cB}\ln\w(b)$ and write
$\widetilde{\cO}(\cdot)$ for factors polynomial in the logarithms of the displayed
parameters.

\defparopt{\wmrbds}{A bipartite graph $\graph$, a weight function $\w: \cB \to \bQ^+$ and $k\in \mathbb{N}$.}{$k$}{A set $\cR'\subseteq \cR$ such that $|\cR'|\leq k$ and the sum of weights of the vertices in $\cB$ that are adjacent to $\cR'$ is maximized.}
Given an instance $\cI $ = \wprbdsi of the \wmrbds problem, the optimal value $\wopt_k(\cI)$ is defined as
\[
\wopt_k(\cI) = \max_{\cS \subseteq \cR, |\cS| \leq k}\w(\nbr(\cS)),
\]
and we omit the subscript $k$ and the instance $\cI$ when they are clear from the
context.  Our target is the covering-target objective of \Cref{thm:addrandom}.  The
branching that gets us there is run on the fixed-budget problem, and so it first delivers,
as an intermediate step, a randomized \epas for \wmrbds.  We state it separately because
the rest of the paper calls it as a subroutine.

\begin{restatable}{theorem}{randprbdsthm}
\label{thm:mainWeighted}
Given an instance $\cI=(\graph,\w,k)$ of \wmrbds and a parameter $\varepsilon>0$, there exists a randomized algorithm running in time
\[
\left(\frac{1}{\varepsilon}\right)^{\cO(\Gamma k\log k)}\sizeg
\]
that, with probability at least $1-\frac{1}{e}$\setcounter{footnote}{0}
\footnote{Furthermore, by running \( \cO\left(\log \frac{1}{\delta}\right) \) independent iterations, we can increase the success probability to \( 1 - \delta \), where \( \delta \) is user-defined.}, outputs a subset $\cS\subseteq \cR$ of size at most $k$ such that
\[
\w(\nbr(\cS)) \geq (1-\varepsilon)\wopt_k(\cI),
\]
where $\Gamma = \dintcmp(G)$ is the downward intersection complexity of the input graph.
\end{restatable}

We do not claim \Cref{thm:mainWeighted} as an advance on the fixed-budget side.  Bounded
semi-ladder index implies bounded VC-dimension, so an \epas for \maxcov on this class
already follows from Badanidiyuru, Kleinberg, and
Lee~\cite{DBLP:conf/compgeom/BadanidiyuruKL12}.  What \Cref{thm:mainWeighted} offers is a
different kind of proof: it is purely combinatorial, falling out of the branching rule with
no appeal to $\varepsilon$-approximations or geometric sampling, it handles arbitrary
rational weights directly, and its dependence on the encoded input size $\sizeg$ is linear.
These are exactly the properties that let us reuse it inside the additive argument, where a
black-box coverage scheme would not suffice.  Relative to the biclique-free coverage results
of Jain et al.~\cite{DBLP:conf/soda/0001KPSS0U23} and Inamdar et
al.~\cite{DBLP:conf/icalp/00020LS0U24}, whose analyses do not apply to $d$-\slf instances, it
does cover a strictly larger class and does handle weights.

Our main algorithmic result addresses the target-covering objective that bounded
VC-dimension provably cannot support, and for which bounded semi-ladder index yields an
\emph{additive} approximation.

\begin{restatable}{theorem}{randadditiveprbdsthm}
\label{thm:addrandom}
Given an instance $\cI=(\graph,\w,k,W)$ of \decwprbds, there exists a randomized algorithm running in time
\[
2^{\cO(\Gamma k\log k)}\sizeg
\]
such that, if $\cI$ is a \yes-instance, then with probability at least $1-1/e$, the algorithm outputs a subset $\cS\subseteq \cR$ of size at most $k+1$ satisfying $\w(\nbr(\cS)) \geq W$.
However, if $\cI$ is a \no-instance, then the algorithm either correctly reports that the instance is a \no-instance, or outputs a subset $\cS \subseteq \cR$ of size at most $k+1$ satisfying $\w(\nbr(\cS)) \geq W$. Here $\Gamma = \dintcmp(G)$.
\end{restatable}

\noindent Our additive algorithm nonetheless provides a \pas for \decwprbds:
for $\varepsilon>\tfrac1k$ it gives a $(1+\varepsilon)$-approximation of the
solution size, while for $\varepsilon\le\tfrac1k$ one can simply guess the
solution within the allowed running time $n^{\cO(1/\varepsilon)}$.

\medskip
\noindent\textbf{Reduction to satisfiability.}
We next study cardinality-constrained maximum satisfiability. In \cmaxsat, we are given a CNF formula $\Phi=(\cV,\cC)$, a weight function on the clauses, and an integer $k$, and the task is to find an assignment of Hamming weight at most $k$ maximizing the total weight of satisfied clauses. We give a deterministic lossy Turing \fpt reduction from \cmaxsat to \maxcov.

\begin{theorem}[Informal, Reduction]
\label{thm:reduction-cmaxsat-to-maxcov}
There is a deterministic lossy Turing \fpt reduction from \cmaxsat to \maxcov. Given an instance $(\Phi=(\cV,\cC),k,t)$ of \cmaxsat and an accuracy parameter $\varepsilon>0$, the reduction runs in time $\bigl(\frac{k\log k}{\varepsilon}\bigr)^{\cO(k)}\sizeinst$ and outputs a family of \maxcov instances such that at least one of them preserves the target up to a loss of a factor $(1-\varepsilon)$. Moreover, the reduction preserves the structural parameters relevant to this paper, including bounded VC-dimension and bounded semi-ladder index.
\end{theorem}

Combining \Cref{thm:reduction-cmaxsat-to-maxcov} with \Cref{thm:mainWeighted}
yields an \epas for \cmaxsat on instances of bounded semi-ladder index.  Combining
it instead with the bounded-VC-dimension \epas of Badanidiyuru, Kleinberg, and
Lee~\cite{DBLP:conf/compgeom/BadanidiyuruKL12} yields, to the best of our knowledge,
the first \epas for \cmaxsat on instances of bounded VC-dimension.

\begin{restatable}{theorem}{vcdimsatepas}
\label{thm: vcdim_satepas}
For any \( \varepsilon > 0 \), there exists a deterministic algorithm that takes as input an instance \( \mathcal{I} = (\Phi, k) \) of \cmaxsat, and in time
\[
\left(\frac{k \log k}{\varepsilon}\right)^{\mathcal{O}(k)} \cdot
2^{\widetilde{\mathcal{O}}(kd/\varepsilon)}\,
\sizeinst^{\mathcal{O}(1)}
=
2^{\widetilde{\mathcal{O}}(kd/\varepsilon)}\,
\sizeinst^{\mathcal{O}(1)}
\]
returns an assignment \( \psi \) of weight at most \( k \) such that the total number of clauses satisfied by \( \psi \) is at least \( (1 - \varepsilon) \cdot \opts(\mathcal{I}) \). Here, \( d \) is the VC-dimension of the input formula \( \Phi \), and \( \opts(\mathcal{I}) \) denotes the maximum number of clauses that can be satisfied by any assignment of weight at most \( k \).
\end{restatable}

\medskip
\noindent\textbf{A faster deterministic bounded-VC scheme.}
The running time in \Cref{thm: vcdim_satepas} is faster than a black-box appeal to
Badanidiyuru, Kleinberg, and Lee would give, and tracing where the saving comes from
led to a second contribution on the fixed-budget side.  To run our reduction on
bounded-VC formulas we must feed it a concrete bounded-VC \maxcov algorithm, and the
only candidate is the scheme of~\cite{DBLP:conf/compgeom/BadanidiyuruKL12}.  Working
through that scheme closely enough to pin down the running time our \cmaxsat result
would inherit, we found that its deterministic implementation does more work than
needed: it enumerates \emph{arbitrary} subsets of the samples drawn in its
nondeterministic preprocessing, whereas the only guesses that can ever matter are the
\emph{traces} of the input sets on those samples.  Enumerating just these realized
traces, whose count the Sauer--Shelah lemma~\cite{DBLP:journals/jct/Sauer72,Shelah1972}
keeps small, sharpens the deterministic running time of their \maxcov \epas
from $2^{\widetilde{\cO}(k^2 d/\varepsilon^5)}$ to $2^{\widetilde{\cO}(kd/\varepsilon)}$
without touching the approximation guarantee (\Cref{sec:bkl-trace-enumeration},
\Cref{thm:bkl-trace-enumerated-runtime}).  This sharper scheme is of independent
interest, and it is exactly the implementation that \Cref{thm: vcdim_satepas} plugs
into.

\begin{table*}[t]
\centering
\small
\renewcommand{\arraystretch}{1.2}
\setlength{\tabcolsep}{6pt}
\begin{adjustbox}{max width=\textwidth}
\begin{tabular}{@{}lllll@{}}
\toprule
\textbf{Problem} & \textbf{Class} & \textbf{Weighted} & \textbf{Running time} & \textbf{Reference} \\
\midrule

\maxcov
  & VC-dim $d$ (larger)
  & Yes
  & $2^{\tilde{O}(kd/\varepsilon)}\cdot\sizeg^{O(1)}$
  & BKL~\cite{DBLP:conf/compgeom/BadanidiyuruKL12} + \Cref{sec:bkl-trace-enumeration} \\

\maxcov, \psc
  & $K_{d,d}$-free (smaller)
  & No
  & $(kd/\varepsilon)^{O(kd)}\cdot\sizeg^{O(1)}$
  & Jain et al.~\cite{DBLP:conf/soda/0001KPSS0U23} \\

\cmaxsat
  & $K_{d,d}$-free (smaller)
  & No
  & $(kd/\varepsilon)^{O(kd)}\cdot\sizeg^{O(1)}$
  & Manurangsi~\cite{DBLP:journals/tcs/Manurangsi25} \\

\maxcov, \cmaxsat
  & $K_{d,d}$-free (smaller)
  & No
  & $(kd/\varepsilon)^{O(kd)}\cdot\sizeg^{O(1)}$
  & Inamdar et al.~\cite{DBLP:conf/icalp/00020LS0U24} \\

\midrule

\wmc
  & $d$-\slf
  & \textbf{Yes}
  & $\mathbf{(1/\varepsilon)^{O(kd\log k)}\cdot\sizeg}$
  & \textbf{This paper} (\Cref{thm:mainWeighted}) \\

\wpsc\ (additive, $k+1$ sets)
  & $d$-\slf
  & \textbf{Yes}
  & $\mathbf{2^{O(kd\log k)}\cdot\sizeg}$
  & \textbf{This paper} (\Cref{thm:addrandom}) \\

\cmaxsat
  & VC-dim $d$
  & \textbf{Yes}
  & $2^{\tilde{O}(kd/\varepsilon)}\cdot\sizeg^{O(1)}$
  & \textbf{This paper} \\

\bottomrule
\end{tabular}
\end{adjustbox}
\caption{Comparison of FPT-approximation results. In the ``This paper'' rows the class parameter is written \(d\) (the \(d\)-\slf class); the algorithms in fact depend on the downward-intersection complexity \(\Gamma\le d+1\), and all our running times hold verbatim with \(\Gamma\) in place of \(d\). The \wmc\ row is not new as an existence statement, since \(d\)-\slf\ instances have bounded VC-dimension; it is listed because the additive \wpsc\ row is obtained from it.}
\label{tab:comparison}
\end{table*}

\medskip
\noindent\textbf{Constrained coverage.}
The branching framework behind our algorithm extends to a constrained version of the fixed-budget
problem. Suppose the elements are colored into $r$ classes with per-class targets
$t_1,\dots,t_r$ (a partition constraint), and the chosen sets must be independent in a
matroid $\mat$ (a matroid constraint); the task is to pick at most $k$ sets,
independent in $\mat$, covering weight at least $t_i$ in each class~$i$. On
bounded-semi-ladder instances this still admits a randomized
$(1-\varepsilon)$-approximation, as formalized below and proved in
\Cref{section:matroidpartition}.

\begin{restatable}{theorem}{mainWeightedcon}
\label{thm:mainWeightedcon}
Let $\varepsilon\in(0,1)$. Given a feasible instance
$\cI=(\graph,\w,\col,t,k,\mat)$ of \wconprbds, there is a randomized algorithm
running in time
\[
\left(\frac{kr\ln k}{\varepsilon}\right)^{\cO(r\Gamma k)}\sizeg
\]
that, with probability at least $1-1/e$, outputs an independent set
$\cS\subseteq\cR$ of size at most $k$ satisfying
\[
\w\bigl(\nbr(\cS)\cap\col^{-1}(i)\bigr)\ge (1-\varepsilon)t_i
\qquad\text{for every }i\in[r],
\]
where $\Gamma=\dintcmp(G)$.
\end{restatable}

The same extension can be combined with the constrained version of the
cardinality-satisfiability reduction introduced above; see
\Cref{sec: weightedconstrainedsat}.

\begin{restatable}{theorem}{applicationwtconccsat}
\label{thm:application_wtconccsat}
Let $\varepsilon\in(0,1)$. Given a feasible instance
$\cI=(\Phi,\mat,r,f,\w,k,t)$ of \wmsc, there is a randomized algorithm running in
time
\[
\left(\frac{rk^2}{\varepsilon}\right)^{\cO(k)}
\left(\frac{2kr\ln k}{\varepsilon}\right)^{\cO(r\Gamma k)}\sizeinst
=\left(\frac{rk}{\varepsilon}\right)^{\cO(\Gamma rk)}\sizeinst
\]
that, with probability at least $1-1/e$, returns an assignment $\psi$ of weight
at most $k$ such that $\psi^{-1}(1)$ is independent in $\mat$ and, for every
color class $j\in[r]$, the clauses of color $j$ satisfied by $\psi$ have total
weight at least $(1-\varepsilon)t_j$. Here
$\Gamma=\dintcmp(G_{\Phi})$.
\end{restatable}

\medskip
\noindent\textbf{Technical novelty and comparison with previous work.}
\Cref{tab:comparison} places our guarantees alongside the prior bounded-VC-dimension
and biclique-free results; we now explain the technical relationship.

\emph{Lower bounds.}
The point of \Cref{sec:vc-boundary} is not that \psc\ is hard, but that it is hard
\emph{at bounded VC-dimension}, the very regime in which the fixed-budget problem is
easy.  Our reductions are accordingly designed to output set systems whose
VC-dimension is small and controlled.  For the purely parameterized bound we reduce
from Lin's constant-factor inapproximability of
\(k\)-\textsc{Clique}~\cite{DBLP:conf/stoc/Lin21} and already rule out
\((2-\delta)\)-approximation at VC-dimension \(7\); for the ETH bounds we reduce from
the regular, low-soundness 2-CSP hardness of Bafna, Karthik, and
Minzer~\cite{BafnaKarthikMinzer2025}, where a label-selection gadget keeps the
VC-dimension at \(7\) for a single gap and a list construction raises it only
logarithmically as the ruled-out factor grows.  The technical core, described in
\Cref{sec:overview-lower-bounds}, is a \emph{local} representation of the constructed set
systems, in which each element is active in only a few sets, so that the VC-dimension is
bounded independently of the instance size.

\emph{The algorithm.}
Our positive framework builds on the \fpt\ machinery for \textsc{Set Cover} and
\textsc{Dominating Set} on \(d\)-\slf\ instances of
Guillemot~\cite{DBLP:journals/toct/G25} and Fabiański et
al.~\cite{DBLP:conf/stacs/FabianskiPST19}, but coverage is genuinely harder: in full
covering one always makes progress by branching on an element that \emph{must} be
covered, whereas in maximum or partial coverage no such forcing element need exist.
The difficulty is thus not the branching rule itself (its inclusion branch is standard
and its exclusion branch has a natural greedy reading), but showing that exclusion
cannot be repeated indefinitely without committing to a set.  Our main algorithmic
contribution is this analysis: we maintain a growing set of {\sf MustInclude}
constraints and use the downward intersection complexity \(\Gamma\) to bound the
recursion depth (\Cref{section:techidea}).  This is exactly the handle that the
biclique-free algorithms of Jain et al.~\cite{DBLP:conf/soda/0001KPSS0U23} and Inamdar
et al.~\cite{DBLP:conf/icalp/00020LS0U24} lack: they bound, for each \(v\in\cR\), the
number of vertices overlapping an \(\varepsilon/k\)-fraction of \(v\)'s neighborhood, a
quantity that is controlled on \(K_{d,d}\)-free graphs but unbounded on \(d\)-\slf\
graphs, which may contain large bicliques.

\emph{Comparison with prior algorithms.}
Against the bounded-VC-dimension \epas\ of Badanidiyuru, Kleinberg, and
Lee~\cite{DBLP:conf/compgeom/BadanidiyuruKL12}, which runs in time
\(2^{\widetilde{\cO}(k^2 d/\varepsilon^5)}\,\sizeg^{\cO(1)}\), our semi-ladder results
are a recovery theorem under stronger structure: on this smaller class we recover the
additive \wpsc\ guarantee that bounded VC-dimension provably cannot support.  Weights
add no real difficulty here.  They cannot be encoded by simply duplicating elements,
since they may be exponentially large, but a rounding step (\Cref{section:techidea})
still reduces any weighted instance to a polynomially larger unweighted one of the same
semi-ladder index, up to a \((1-\varepsilon)\) factor; our algorithm in fact handles
weights directly, keeping the dependence on the encoded input size \(\sizeg\) linear.
The tools also
differ: theirs use \(\varepsilon\)-approximations and geometric sampling, ours a
combinatorial branching controlled by \(\Gamma\).  Our results are incomparable to the
lossy kernel of Manurangsi~\cite{DBLP:journals/tcs/Manurangsi25} for \cmaxsat\ on
\(K_{d,d}\)-free formulas: we produce no kernel, but our reduction from \wms\ to
\wmc\ handles weighted clauses and needs no subgraph-closed family.  Finally, all
randomized results derandomize via weighted \(\delta\)-nets: the relevant set family
has VC-dimension at most \(\Gamma\), so it admits a net of size
\(\cO(\Gamma\log(1/\delta)/\delta)\) computable through known \fpt\ algorithms for
\textsc{Dominating Set} on \(d\)-\slf\ instances~\cite{DBLP:journals/toct/G25}
(\Cref{section: weighteddetfptas}).

\medskip
\noindent\textbf{Organization.}
\Cref{section:techidea} is a technical overview of the whole paper: its first half
(\Cref{sec:overview-lower-bounds}) explains the row-and-blocking-list syntax shared by our
three lower-bound reductions and why that syntax alone caps the VC-dimension, and its second
half explains the randomized branching idea behind the positive results.  The lower bounds
themselves are then proved in~\Cref{sec:vc-boundary}, showing that the fixed-budget guarantee
of Badanidiyuru--Kleinberg--Lee does not extend to the target-covering setting.  We then
develop the positive semi-ladder-free framework, beginning with
\Cref{section: prelims}, which collects the preliminaries on set systems, incidence
graphs, semi-ladder index, and downward intersection complexity. The main
randomized branching algorithms are developed in~\Cref{section: implement},
with the weighted fixed-budget version in~\Cref{subsection:wprbds}. Since the weighted version differs from the unweighted one only in minor technical details, we include the unweighted presentation for clarity. Readers interested in the main ideas may focus on the unweighted case. The additive
approximation for the partial version is proved in~\Cref{section: weighted_additive},
and deterministic counterparts are obtained in~\Cref{section: weighteddetfptas}.
We then return to the bounded-VC comparison in~\Cref{sec:bkl-trace-enumeration},
where we give the improved trace-enumerated implementation of the
Badanidiyuru--Kleinberg--Lee scheme. The reduction to satisfiability appears
in~\Cref{sec: weightedsat}, followed by applications in~\Cref{sec:applications},
the matroid/partition-constrained extension in~\Cref{section:matroidpartition},
and its constrained-satisfiability consequence in
\Cref{sec: weightedconstrainedsat}. We conclude in~\Cref{section: conclusion}.
An extended discussion of related work can be found in~\Cref{sec: related_work}.
Readers mainly interested in the core ideas may focus first on
\Cref{section:techidea,section: implement,section: weighted_additive,sec: weightedsat}.

%% file: technical_overview.tex
\section{Technical Overview}
\label{section:techidea}
In this section, we present a high-level overview of both halves of the paper.  We begin
with the lower bounds of \Cref{sec:vc-boundary}, which show that bounded VC-dimension does
not suffice for the target-covering objective, and then turn to the randomized branching
that underlies our positive results.

\subsection{Overview of the Lower Bounds}
\label{sec:overview-lower-bounds}
The point of \Cref{sec:vc-boundary} is not that \psc\ is hard, but that it is hard
\emph{at bounded VC-dimension}, the very regime in which the fixed-budget problem is easy.
A reduction that merely produces hard \psc\ instances is therefore useless to us: it must
produce instances whose VC-dimension is small and, moreover, small for a reason we can
verify from the syntax of the construction.  All three of our reductions share a single
such syntax, which we isolate once and reuse.

\paragraph{Rows and blocking lists.}
Every set system we construct has its family partitioned into \(k\) \emph{rows}
\(\cF_i=\{S_{i,a}:a\in\Sigma\}\), one row per position of the object we are encoding (a
clique position, or a CSP variable).  Selecting \(S_{i,a}\) means assigning label \(a\) to
position \(i\), and a dedicated \emph{row element} \(r_i\), contained in exactly the sets
of \(\cF_i\), forces every cover to select at least one set from each row.  Every other
universe element \(u\) is described by a set \(I(u)\subseteq[k]\) of \emph{active rows} and,
for each active row, a \emph{blocking list} \(A_i(u)\subseteq\Sigma\), with the single
membership rule
\[
  u\in S_{i,a}\iff a\notin A_i(u),
\]
and \(u\) in no set of an inactive row.  Thus \(u\) is uncovered precisely when every active
row selects only labels from its blocking list.  The construction is \emph{\(q\)-local} if
\(|I(u)|\le2\) and \(|A_i(u)|\le q\) throughout: each element is sensitive to at most two
positions, and each sensitivity is witnessed by at most \(q\) forbidden labels.

\paragraph{Why locality caps the VC-dimension.}
Locality alone bounds the number of traces, and hence the VC-dimension, independently of
\(|\Sigma|\) and of the instance size.  Fix \(Z\subseteq\cU\) with \(|Z|=s\).  Only rows
active for some element of \(Z\) can produce a nonempty trace, and there are at most \(2s\)
of them.  Within one such row \(i\), let \(B_i\) be the union of the blocking lists of the
elements of \(Z\) active in \(i\), so \(|B_i|\le qs\).  Every label \(a\notin B_i\) gives the
\emph{same} trace, namely the set of elements of \(Z\) active in row \(i\); only the \(|B_i|\)
labels inside \(B_i\) can deviate.  So each row contributes at most \(qs+1\) traces, and
\[
  |\cF|_Z|\le 2s(qs+1)+1.
\]
Shattering demands \(2^s\le 2s(qs+1)+1\), which forces \(s=\cO(\log(q+1))\), and for \(q=1\)
already fails at \(s=8\) since \(256>145\).  This is
\Cref{lem:vc-local-list-traces}; each reduction below only has to exhibit its active rows
and blocking lists, after which the VC-dimension bound is automatic.

\paragraph{The three reductions.}
The purely parameterized bound reduces from Lin's constant-factor inapproximability of
\(k\)-\textsc{Clique}~\cite{DBLP:conf/stoc/Lin21}.  Rows are the \(k\) clique positions and
\(\Sigma=V(H)\).  For positions \(i<j\) and a non-adjacent or equal pair \((u,v)\) we add an
element \(e_{i,j,u,v}\) with active rows \(\{i,j\}\) and blocking lists \(\{u\}\) and
\(\{v\}\): it is uncovered exactly when position \(i\) takes only \(u\) and position \(j\)
takes only \(v\).  The construction is \(1\)-local, so its VC-dimension is at most \(7\).  A
cover of size \((2-\delta)k\) can have at most \((1-\delta)k\) rows selecting two or more
labels, leaving \(\delta k\) rows with a unique label; those labels must be pairwise
adjacent, so they form a clique.  This is the whole soundness argument, and it yields
\Cref{thm:vc-no-fpt-below-two}: no \((2-\delta)\)-approximation in \fpt\ time, at
VC-dimension \(7\).

For the ETH bounds we reduce from the regular, low-soundness \(2\)-CSP hardness of Bafna,
Karthik, and Minzer~\cite{BafnaKarthikMinzer2025}.  In the \emph{label-selection}
construction the rows are the CSP variables, and each rejecting pair \((\lambda,\mu)\) of a
constraint \(e=x_ix_j\) contributes an element with active rows \(\{i,j\}\) and blocking
lists \(\{\lambda\}\), \(\{\mu\}\).  Again \(1\)-local, so VC-dimension at most \(7\).  A
cover of size \((1+\alpha)k\) leaves all but \(\alpha k\) rows with a unique label; the
regularity of the constraint graph turns those few ambiguous rows into at most
\(D\alpha k=2\alpha m\) violated constraints, which contradicts soundness once
\(\alpha<\eta/2\).  This gives a constant gap, hence no \pas\ (\Cref{thm:vc-no-aptas}).

The constant is all the label-selection gadget can give: a cover that spends many labels in
many rows defeats the decoding.  The \emph{list} construction removes this by blocking not
single labels but whole lists.  For each constraint and each pair of nonempty
\(A,B\subseteq\Sigma\) with \(|A|,|B|\le q\) and \((A\times B)\cap R_e=\emptyset\), it adds an
element with blocking lists \(A\) and \(B\).  Now a cover of size \(\rho k\) has at most
\((\rho-1)k/q\) rows selecting more than \(q\) labels, and on every constraint between two
non-heavy rows the two selected lists must contain an accepting pair; picking a label
uniformly from each list satisfies that constraint with probability \(1/q^2\), so
\(\operatorname{val}(\Lambda)>1/(2q^2)\), contradicting soundness \(1/(4q^2)\).  The price is
that blocking lists now have size \(q\), so the construction is only \(q\)-local and the
trace bound degrades to \(\cO(\log(q+1))\).  Since \(q=\Theta(\rho)\), this is exactly the
trade-off recorded in \Cref{thm:vc-any-constant} and, read in the other direction,
in \Cref{thm:vc-exp-d}: the VC-dimension grows like \(\log\rho\), so the forbidden
approximation ratio grows like \(2^{\Omega(d)}\).

\subsection{Overview of the Algorithms}
We now turn to the positive side, and outline the core ideas behind the algorithm used in the proof of \Cref{thm:mainWeighted}. Let \( \cI = (G, \w, k) \) be an instance of \wmrbds and let \( \varepsilon > 0 \) be a given approximation parameter. For clarity of exposition, we assume that \( \cI = (\graph, k) \) is an instance of \mrbds (i.e., the unweighted version).

\subsubsection{High-Level Overview of the Algorithm for \mrbds}
\label{sec:overviewprbds}
Suppose we are designing a recursive algorithm for \mrbds that incrementally builds a
solution set \( X \subseteq \cR \), initialized as the empty set.  Write
\[
W = \cB \setminus \nbr_G(X)
\]
for the set of blue vertices still uncovered, and let
\[
\opt(X) \;=\; \max_{X \subseteq \cS \subseteq \cR,\ |\cS| \leq k} |\nbr_G(\cS) \cap W|
\]
denote the best residual coverage achievable by an extension of \( X \) to at most \( k \)
vertices.  We call an extension \( \cS \supseteq X \) with \( |\cS| \le k \) a
\emph{\((1-\varepsilon)\)-approximate extension of \( X \)} if
\( |\nbr_G(\cS) \cap W| \geq (1-\varepsilon)\opt(X) \).

Let \( v \) be a vertex in \( \cR \setminus X \) that has the largest number of neighbors in
\( W \), and write \( \ell = |\nbr_G(v) \cap W| \).  We consider a standard branching
strategy: either some \((1-\varepsilon)\)-approximate extension of \( X \) includes \( v \),
or none does.  In the first case, we include \( v \) in \( X \), which reduces the number of
vertices still to be chosen and thus ensures progress.  However, no such guarantee exists in
the branch that excludes \( v \).

Our first key technical insight is that if no \((1-\varepsilon)\)-approximate extension of
\( X \) includes \( v \), then \emph{every} optimal extension of \( X \) must already cover a
noticeable fraction of \( v \)'s uncovered neighborhood.  Crucially, the fraction is
\( \varepsilon \) itself: it does not degrade with the recursion depth.  We record this as
the following lemma, stated in contrapositive form.

\begin{lemma}\label{lemma:overparloss}
Let \( X \subseteq \cR \) with \( |X| < k \), let \( W = \cB \setminus \nbr_G(X) \), and let
\( v \in \cR \setminus X \) be a vertex maximizing \( |\nbr_G(v) \cap W| =: \ell \).  Let
\( \bO \) be an extension of \( X \) attaining \( \opt(X) \).  If
\[
|\nbr_G(\bO) \cap \nbr_G(v) \cap W| \;<\; \varepsilon \cdot \ell,
\]
then there is a \((1-\varepsilon)\)-approximate extension of \( X \) that contains \( v \).
Equivalently, if no \((1-\varepsilon)\)-approximate extension of \( X \) contains \( v \),
then every extension \( \bO \) attaining \( \opt(X) \) satisfies
\[
|\nbr_G(\bO) \cap \nbr_G(v) \cap W| \;\geq\; \varepsilon \cdot \ell.
\]
\end{lemma}

\begin{proof}
Since \( |X| < k \) and \( |\bO| \le k \), we may pick \( o \in \bO \setminus X \) minimizing
\( |\nbr_G(o) \cap W| \).  Set \( \cS = (\bO \setminus \{o\}) \cup \{v\} \).  Then
\( X \subseteq \cS \), \( |\cS| \le k \), and \( v \in \cS \).  Since
\( \nbr_G(\cS) \supseteq (\nbr_G(\bO) \setminus \nbr_G(o)) \cup \nbr_G(v) \), we obtain
\begin{align*}
|\nbr_G(\cS) \cap W|
&\geq |\nbr_G(\bO) \cap W| - |\nbr_G(o) \cap W| + \bigl|(\nbr_G(v) \cap W) \setminus \nbr_G(\bO)\bigr| \\
&=    \opt(X) - |\nbr_G(o) \cap W| + \ell - |\nbr_G(\bO) \cap \nbr_G(v) \cap W| \\
&>    \opt(X) - |\nbr_G(o) \cap W| + \ell - \varepsilon\ell \\
&\geq \opt(X) - \varepsilon \ell \\
&\geq (1-\varepsilon)\opt(X).
\end{align*}
The third line uses the hypothesis, the fourth uses \( |\nbr_G(o) \cap W| \le \ell \) (as
\( o \in \cR \setminus X \) and \( v \) maximizes this quantity), and the last uses
\( \ell \le \opt(X) \), which holds because \( X \cup \{v\} \) is itself a feasible extension
of \( X \).  Hence \( \cS \) is a \((1-\varepsilon)\)-approximate extension of \( X \)
containing \( v \).
\end{proof}

\noindent
Observe that \Cref{lemma:overparloss} loses nothing across recursive calls: the same
\( \varepsilon \) appears in the hypothesis and in the conclusion, so the invariant
``the current \( X \) extends to a \((1-\varepsilon)\)-approximate solution'' is maintained
verbatim at every level, rather than being weakened as the depth grows.

\Cref{lemma:overparloss} allows us to make the following dichotomy at every recursive step:
\begin{enumerate}
    \item Either \( v \) lies in a \( (1 - \varepsilon) \)-approximate extension of \( X \), or
    \item every optimal extension \( \cS \) of \( X \) covers an \( \varepsilon \)-fraction of
    the uncovered neighbors of \( v \), that is,
    \[
    |(\nbr_G(v) \setminus \nbr_G(X)) \cap \nbr_G(\cS)| \geq \varepsilon \cdot | \nbr_G(v) \setminus \nbr_G(X) |.
    \]
\end{enumerate}

In the second case, a uniformly random choice from \( \nbr_G(v) \setminus \nbr_G(X) \) yields
an element \( w \in \cB \setminus \nbr_G(X) \) that is covered by \( \cS \) with probability
at least \( \varepsilon \).  These ideas immediately lead to the algorithm described in
Algorithm~\ref{alg:random-approx-cover}.  Let \( \Delta(\cB) \) denote the maximum degree of a
vertex in \( \cB \).  In each recursive step, we select the vertex \( v \) with probability
\( \tfrac12 \), and with the remaining probability \( \tfrac12 \) we first sample a vertex
\( w \in \nbr_G(v) \setminus \nbr_G(X) \) uniformly at random and then choose a vertex
\( u \in \cR_w \) uniformly at random.  Since \( |\cR_w| \leq \Delta(\cB) \), the correct
vertex is selected with probability at least
\( \min\bigl\{\tfrac12, \tfrac{\varepsilon}{2\Delta(\cB)}\bigr\}
   = \tfrac{\varepsilon}{2\Delta(\cB)} \)
at each of the \( k \) steps.  Thus, we get a polynomial time algorithm with success
probability at least
\( \left(\frac{\varepsilon}{2\Delta(\cB)} \right)^{k} \), and we obtain a randomized
\textsf{EPAS} by independently repeating the algorithm
\( \left(\frac{2\Delta(\cB)}{\varepsilon}\right)^{k} \) times.  This repetition ensures that,
with high probability, we recover a solution whose coverage is at least
\( (1 - \varepsilon) \cdot \opt_k(\cI) \).

This algorithm already yields the desired \textsf{EPAS} for various natural classes of set systems. The case where \( \Delta(\cB) \) is bounded corresponds to set systems of bounded frequency, that is, when each element belongs to only a bounded number of sets. By interchanging the roles of \( \cR \) and \( \cB \), the same algorithm also applies to the case where the size of each set in the family is bounded. In particular, this directly yields an \textsf{EPAS} for problems such as \textsc{Partial Vertex Cover} and \textsc{Partial $d$-Hitting Set}, where the underlying set system naturally satisfies bounded set size constraints.

\begin{algorithm}[t]
\caption{$\rrac(G, \cR, \cB, k, \varepsilon, X)$}
\label{alg:random-approx-cover}

\KwIn{Graph $G = (\cR \cup \cB, E)$, integer $k$, parameter $\varepsilon > 0$, partial solution $X \subseteq \cR$}
\KwOut{A set $\cS \subseteq \cR$, $|\cS| \leq k$, approximating maximum coverage}

\If{$|X| = k$}{
    \Return{$X$}
}

$W \gets \cB \setminus \nbr_G(X)$\;

Choose $v \in \cR \setminus X$ with maximum $|\nbr_G(v) \cap W|$\;

Sample vertex $u$ as follows:\;
\Begin{
    With probability $\frac{1}{2}$, set $u \gets v$\;
    Otherwise, sample $w \in \nbr_G(v) \cap W$ uniformly at random,\;
    set $\cR_w \gets \{ x \in \cR \setminus X \mid w \in \nbr_G(x) \}$,\;
    and sample $u$ uniformly from $\cR_w$\;
}

\Return{$\rrac(G, \cR, \cB, k, \varepsilon, X \cup \{u\})$}
\end{algorithm}

\paragraph{Our Strategy.} The set systems we are working with may have unbounded \( \Delta(\cB) \), that is, the maximum degree of a vertex in \( \cB \) could be arbitrarily large. Therefore, we cannot directly use the algorithm described in Algorithm~\ref{alg:random-approx-cover} for our purposes, as the success probability of sampling from \( \cR_w \) becomes too small.

Instead, we follow the algorithm up to the point where we uniformly sample a vertex \( w \in \nbr_G(v) \setminus \nbr_G(X) \). However, rather than immediately sampling from the set
\[
\cR_w \gets \{ u \in \cR \setminus X \mid w \in \nbr_G(u) \},
\]
we simply record that we intend to select a vertex from \( \cR_w \) at a later stage. We achieve this as follows.

The algorithm maintains a collection of $k$ sets at any point during its execution, which serve as \emph{skeletons} for the $k$ solution vertices the algorithm aims to construct. These skeletons are implemented using sets denoted by $\mcon_{\iletter}$ for each $\iletter \in [k]$. Specifically, the algorithm maintains a tuple
\[
\mi = \langle \mcon_1, \dots, \mcon_k \rangle,
\]
where each $\mcon_{\iletter} \subseteq \cB$ represents the $\iletter^{\textnormal{th}}$ coordinate of the skeleton for the input instance $\cI$. The final solution will be a tuple $\cS = \langle s_1, \dots, s_k \rangle$ with $s_\iletter \in \cR$, such that for all $\iletter \in [k]$, we have $\mcon_{\iletter} \subseteq \nbr_G(s_\iletter)$. That is, each $s_\iletter$ must cover the corresponding skeleton $\mcon_\iletter$.

Since $\mcon_{\iletter} \subseteq \nbr_G(s_\iletter)$, the only valid candidates for $s_\iletter$ are vertices in $\cR$ whose neighborhoods contain $\mcon_{\iletter}$. We define the corresponding set of potential candidates as
\[
\pot_{\iletter} \coloneqq \{ v \in \cR \mid \mcon_{\iletter} \subseteq \nbr_G(v) \}.
\]
Let \( \pot \coloneqq \bigcup_{\alpha \in [k]} \pot_{\alpha} \) denote the union over all candidate sets.

Moreover, consider any $\pot_{\iletter}$ for $\iletter \in [k]$. If a vertex in $\cB$ is in the neighborhood of \emph{every} vertex in $\pot_{\iletter}$, then that vertex is guaranteed to be covered by \emph{any} choice of $s_\iletter$ from $\pot_\iletter$. This set of definitely covered vertices is defined as
\[
\fcov_{\iletter} \coloneqq \bigcap_{v \in \pot_{\iletter}} \nbr_G(v).
\]
Let \( \fcov \coloneqq \bigcup_{\alpha \in [k]} \fcov_{\alpha} \) be the union over all such definitely covered vertices. These definitions imply that once the skeletons \( \mcon_1, \dots, \mcon_k \) are fixed—
and updated as the algorithm progresses—many vertices in \( \cB \) become 
\emph{freely covered}, as they are guaranteed to be covered by 
\emph{every} completion of the skeleton into a full solution.

Let \( W = \cB \setminus \fcov \) denote the set of blue vertices that are not freely covered. A tuple \( \cS = \langle s_1, \dots, s_k \rangle \), where each \( s_\iletter \in \pot_\iletter \) and \( \mcon_\iletter \subseteq \nbr_G(s_\iletter) \) for all \( \iletter \in [k] \), is called a \emph{consistent} tuple.

We now prove a version of Lemma~\ref{lemma:overparloss} adapted to this setting. Let \( v \in \pot \) be a vertex that maximizes \( |\nbr_G(v) \cap W| \). Then one of the following holds:

\begin{enumerate}
    \item The vertex \( v \) belongs to a \( (1 - \varepsilon) \)-approximate consistent solution \( \cS = \langle s_1, \dots, s_k \rangle \), or
    \item There exists a \( (1 - \varepsilon) \)-approximate consistent solution \( \cS = \langle s_1, \dots, s_k \rangle \) such that
    \[
    |(\nbr_G(v) \cap W) \cap \nbr_G(\cS)| \geq \varepsilon \cdot |\nbr_G(v) \cap W|.
    \]
\end{enumerate}

In the second case, we sample a vertex
\[
w \in \nbr_G(v) \cap W,
\]
which by \Cref{lemma:overparloss} is covered by \( \cS \) with probability at least
\( \varepsilon \).  We then guess an index \( \iletter \in [k] \) uniformly at random and
include \( w \) into the corresponding skeleton set \( \mcon_\iletter \).  Combining the
coin that chooses between the two branches, the sample of \( w \), and the guess of
\( \iletter \), a single recursive step preserves the invariant with probability at least
\( \frac{\varepsilon}{2k} \), and it makes progress by increasing the size of one of the
skeletons \( \mcon_\iletter \).

The remaining question is how many times this can happen before some \( \mcon_\iletter \)
determines its solution vertex.  Since \( \mcon_\iletter \subseteq \nbr_G(s_\iletter) \),
each skeleton can grow at most \( \Delta(\cR) \) times, where \( \Delta(\cR) \) is the
maximum degree of a vertex in \( \cR \), that is, the maximum set size in \( \cF \).  This
bound alone gives a recursion depth of \( k\Delta(\cR) \), which is useless in general.  The
whole point of the structural parameter is to replace \( \Delta(\cR) \) by a quantity that
stays bounded even when sets are large.

In order to bound the running time of the algorithm as a function of the structural parameter \( \Gamma = \dintcmp(G) \), we maintain the skeleton sets \( \mcon_{\iletter} \) for each \( \iletter \in [k] \) in such a way that we can extract a \sml of size \( |\mcon_{\iletter}| \). 
In particular, the following lemma plays a key role in bounding the running time of the algorithm.

\begin{lemma}
\label{lem:overviewSML}
Let \( \cI = (G, k) \) be an instance of \prbds, and let
\[
\mi = \langle \mcon_1, \dots, \mcon_{k} \rangle
\]
be the tuple maintained by the algorithm, where each \( \mcon_\iletter \) is initialized to \( \emptyset \) at the beginning of the algorithm. Then, for every set \( \mcon_\iletter = \{b_1, \dots, b_\ell\} \), there exists a corresponding set \( \{r_1, \dots, r_\ell\} \subseteq \cR \) such that the sequences \( r_1, \dots, r_\ell \) and \( b_1, \dots, b_\ell \) form a \sml in the graph, of size \( \ell = |\mcon_\iletter| \).
\end{lemma}

\begin{proof}
We prove this by induction on the quantity \( \zeta = \sum_{\iletter \in [k]} |\mcon_{\iletter}| \). 

\paragraph{Base case:} When \( \zeta = 0 \), all sets \( \mcon_\iletter \) are empty. Hence, the claim vacuously holds.

\paragraph{Inductive step:} Assume that the lemma holds for all values \( \zeta' < \zeta \), and we now show it holds for \( \zeta \).

Let \( w \) be a vertex in \( W = \cB \setminus \fcov \), and let \( \iletter \in [k] \) be any index. We consider the effect of extending \( \mcon_\iletter \) by adding \( w \), that is, analyzing the set \( \mcon_\iletter \cup \{w\} \). Let \( b_1, \dots, b_\ell \) denote the order in which the elements of \( \mcon_\iletter \) were added, so that \( \ell = |\mcon_\iletter| \). We argue that regardless of the choice of \( \iletter \), this extension results in a semiladder of order \( \ell + 1 \).

By the induction hypothesis, there exists a sequence \( r_1, \dots, r_\ell \in \cR \) such that the sequences \( r_1, \dots, r_\ell \) and \( b_1, \dots, b_\ell \) form a \sml of order \( \ell \). That is:
\begin{itemize}
    \item \( (r_i, b_i) \notin E(G) \) for all \( i \in [\ell] \), and
    \item \( (r_i, b_h) \in E(G) \) for all \(  h <i \).
\end{itemize}

Recall that \( W = \cB \setminus \fcov \). Now, consider the addition of vertex \( w \). Since \( w \in \nbr_G(v) \cap W \), there exists a vertex \( r \in \pot_\iletter \) such that \( w \notin \nbr_G(r) \). Observe that \( \{b_1, \dots, b_\ell\} \subseteq \nbr_G(r) \), because \( r \in \pot_\iletter \) and \( \mcon_\iletter = \{b_1, \dots, b_\ell\} \subseteq \nbr_G(r) \). Also, \( (r, w) \notin E(G) \).

Define \( r_{\ell+1} = r \) and \( b_{\ell+1} = w \). Then, the extended sequences \( r_1, \dots, r_\ell, r_{\ell+1} \) and \( b_1, \dots, b_\ell, b_{\ell+1} \) satisfy:
\begin{itemize}
    \item \( (r_i, b_i) \notin E(G) \) for all \( i \in [\ell+1] \), and
    \item \( (r_i, b_h) \in E(G) \) for all \(  h <i \).
\end{itemize}
This forms a \sml of order \( \ell + 1 = |\mcon_\iletter \cup \{w\}| \), completing the inductive step.

Hence, by induction, the lemma holds for all \( \zeta \). This conludes the proof. 
\end{proof}

\Cref{lem:overviewSML} supplies exactly this replacement.  Every set \( \mcon_\iletter \)
maintained by the algorithm certifies a \sml of order \( |\mcon_\iletter| \), so
\( |\mcon_\iletter| \) never exceeds the \smlindex of \( G \).  Since the \smlindex is at
most \( \Gamma = \dintcmp(G) \) (see \Cref{sec: sml_comp}), the algorithm stops after at
most \( k\Gamma \) steps.  Together with the per-step success probability
\( \frac{\varepsilon}{2k} \) established above, this yields a polynomial-time randomized
algorithm with success probability at least
\( \left(\frac{\varepsilon}{2k}\right)^{k\Gamma} \).  By independently running the algorithm
\( \left(\frac{2k}{\varepsilon}\right)^{\cO(\Gamma k)}
   = \left(\frac{1}{\varepsilon}\right)^{\cO(\Gamma k \log k)} \)
times, we boost the success probability to \( 1 - \frac{1}{e} \), which is precisely the
running time claimed in \Cref{thm:mainWeighted}.  Furthermore, by running an additional
\( \cO\left(\log \frac{1}{\delta}\right) \) independent iterations, we can increase the
success probability to \( 1 - \delta \), where \( \delta \) is user-defined. This concludes the technical overview of \prbds. All our randomized algorithms build on the structural insights formalized in \Cref{lemma:overparloss} and \Cref{lem:overviewSML}, and follow the general framework presented in this section.

\subsubsection{\texorpdfstring{\wmrbds: \Cref{thm:mainWeighted}}{Weighted Max Red-Blue Dominating Set}}
In this section, we first show how to reduce the weighted problem to its
unweighted counterpart while preserving the approximation guarantee. This
reduction, however, incurs a quadratic blow-up in the size of the instance. We
therefore present a direct modification of the unweighted algorithm that handles
weights explicitly and avoids this quadratic overhead.
\input{weighted_reduction_overview}

\paragraph{A Direct Modification.}
For the \wmrbds problem, we follow the same high-level strategy outlined in \Cref{sec:overviewprbds} for \mrbds. The key difference is that all arguments must now be phrased in terms of the \emph{weighted degree} of a vertex, defined as the sum of the weights of its neighbors. 
Specifically, for a fixed \( \varepsilon > 0 \), we show that if there is no \( (1 - \varepsilon) \)-approximate consistent solution that includes a vertex \( h^\star \in \cR \) (chosen as the vertex with maximum weighted degree in \( W = \cB \setminus \fcov \)), then there exists a consistent approximate solution whose neighborhood intersects the set \( \nbr(h^\star) \cap W = \{u_1, \dots, u_\ell\} \) in a significant way. More precisely, the total weight of the intersection must be at least an \( \varepsilon \)-fraction of the total weight of \( \{u_1, \dots, u_\ell\} \), i.e.,
\[
\w(\nbr(\cS) \cap \{u_1, \dots, u_\ell\}) \geq \varepsilon \cdot \w(\{u_1, \dots, u_\ell\}).
\]
This step is implemented by randomly selecting an item \(\clubsuit\) from the set \(\{ h^\star, u_1, \dots, u_\ell \}\) with a probability distribution  
\[
    \mathbb{P}[\clubsuit] = 
    \begin{cases} 
    \frac{1}{2} & \text{if } \clubsuit = h^\star \\
    \frac{\w(u_\iletter)}{2 \omega} & \text{if } \clubsuit \in \{u_\iletter \mid \iletter \in [k]\}
    \end{cases}
    \]
where $\omega = \sum_{\iletter \in [k]} \w(u_\iletter)$. That is, we select the vertex \(h^\star\) with probability \(\frac{1}{2}\), or a vertex, $u$, in \(\{u_\iletter \mid \iletter \in [k]\}\) with probability \(\frac{\w(u_\iletter)}{2 \omega}\). This leads to the desired algorithm mentioned in  \Cref{thm:mainWeighted}.

\subsubsection{\texorpdfstring{\decwprbds: \Cref{thm:addrandom}}{Weighted Partial Red-Blue Dominating Set}}
For an additive approximation algorithm for the \decwprbds problem, we use the algorithm for \wmrbds. Suppose we are given an instance 
\[ \cI = (\graph, \w, k, W) \] such that there exists a tuple (set) \( \mathbb{O} = \langle o_1, \dots, o_k \rangle \) satisfying \( \w(\nbr(\mathbb{O})) \geq W \).

The algorithm first computes a set \( S' \subseteq \cR \) such that \( \w(\nbr(S')) \geq \left(1 - \frac{1}{2k} \right) W \), using \Cref{thm:mainWeighted}. Since we are allowed to include one additional vertex in the solution, we check whether there exists a vertex \( v \in \cR \setminus S' \) such that 
\[
\w(\nbr(S' \cup \{v\})) \geq W.
\]
If such a vertex exists, then we return \( S' \cup \{v\} \) as the desired approximate solution. Assume now that no such vertex \( v \) exists. This implies that for all \( o_\alpha \in \mathbb{O} \), we have
\[
\w(\nbr_G(o_\alpha) \setminus \nbr_G(S')) < \frac{W}{2k}.
\]
Summing over all \( o_\alpha \in \mathbb{O} \), it follows that
\[
\w(\nbr_G(\mathbb{O}) \setminus \nbr_G(S')) < \frac{W}{2}.
\]
In particular, this means that at least half of the total weight of \( \nbr_G(\mathbb{O}) \) is already captured by \( \nbr_G(S') \). This is formalized in the next lemma.

\begin{lemma}\label{lemma:overviewaddstructure}
    Given a bipartite graph $\graph$ and $\cS \subseteq \cR$ with $|\cS|\leq k$, if
    $\w(\nbr(\cS))\geq \left(1-\frac{1}{2k}\right)W$, then at least one of the following two
    cases holds.
    \begin{description}
        \item[1. \label{ccaseone}] there exists $h \in \cR \setminus \cS$ such that $\w(\nbr(\{h\} \cup \cS))\geq W$;
        \item[2. \label{ccasetwo}] $\w(\nbr(\bO) \cap \nbr(\cS))\geq \frac{1}{2}\w(\nbr(\cS))$.
    \end{description}
Here $\bO$ is a $k$-tuple in $\cR$ such that $\w(\nbr (\bO))\geq W$.
\end{lemma}
\begin{proof}
    We may assume $\w(\nbr(\cS)) < W$; otherwise \hyperref[ccaseone]{Case 1} holds trivially
    with any $h$.  If \hyperref[ccaseone]{Case 1} of \Cref{lemma:overviewaddstructure} holds,
    then the statement of the lemma is true.  Suppose Case $1$ does not hold, i.e., for any
    $h \in \cR \setminus \cS$ we have $\w(\nbr(\{h\} \cup \cS))< W$.  Since
    $\w(\nbr(\cS))\geq (1-\frac{1}{2k})W$, this implies that
    $\w(\nbr(h)\setminus \nbr(\cS)) < \frac{W}{2k}$ for every $h \in \cR \setminus \cS$.
    In particular, for every $o_\iletter$, $\iletter \in [k]$, we have
    $\w(\nbr(o_\iletter) \setminus \nbr(\cS))<\frac{W}{2k}$, which by summing over the $k$
    coordinates gives $\w(\nbr(\bO) \setminus \nbr(\cS))<\frac{W}{2}$.  Since
    $\w(\nbr(\bO))\geq W$, this yields
    $$\w(\nbr(\bO) \cap  \nbr(\cS))\geq \frac{W}{2} \geq \frac{1}{2}\w(\nbr(\cS)),$$
    where the last inequality uses $\w(\nbr(\cS)) < W$.  Hence \hyperref[ccasetwo]{Case 2}
    of \Cref{lemma:overviewaddstructure} holds, proving the lemma.
\end{proof}

Lemma~\ref{lemma:overviewaddstructure} immediately implies that if we cannot extend \( S' \) by adding a vertex \( h \in \cR \setminus S' \) such that \( \w(\nbr(h \cup S')) \geq W \), then we can proceed by sampling a vertex from \( \nbr(S') \) according to the probability distribution \( \left\{ \frac{\w(v)}{\w(\nbr(S'))} \right\} \), i.e., proportional to its weight.

With probability at least \( \frac{1}{2} \), the sampled vertex \( v \in \nbr(S') \) also belongs to \( \nbr(\mathbb{O}) \), where \( \w(\nbr(\mathbb{O})) \geq W \). Hence, with probability \( \frac{1}{2} \), we successfully identify a vertex that must be covered. From this point, we can follow a strategy similar to the one described in Section~\ref{sec:overviewprbds} to complete the algorithm and prove \Cref{thm:addrandom}.

\subsubsection{Our Result for Weighted SAT: \Cref{thm:reduction-cmaxsat-to-maxcov}}\label{sec:satresult}

\wms is \npc and admits a \( \left(1 - \frac{1}{e} \right) \)-approximation algorithm in polynomial time, which is optimal unless \( \text{P} = \text{NP} \)~\cite{Sviridenko01}. All known intractability results for \wmc also extend to \wms, since if the input formula contains only positive variables, the problem reduces to {\sc Maximum Coverage}.

Jain et al.~\cite{DBLP:conf/soda/0001KPSS0U23} provided an \epas for the unweighted version of \wms when the variable-clause incidence graph is biclique-free. This result was independently improved by two subsequent works. Manurangsi~\cite{DBLP:journals/tcs/Manurangsi25} presented a lossy kernel for the problem parameterized by \( k \), which led to an \epas with improved running time. In a separate development, Inamdar et al.~\cite{DBLP:conf/icalp/00020LS0U24} gave an \fpt-time randomized reduction from \cmaxsat to \maxcov with matching runtime. Their reduction preserves both the approximation guarantee and the structure of the incidence bipartite graph, provided the graph belongs to a subgraph-closed family. 

However, if the clause-variable incidence graph has bounded \smlindex, this reduction does not directly apply, as bounded-\smlindex graphs are not subgraph-closed. Moreover, none of the above works consider the weighted variant of the problem, where the goal is to satisfy clauses such that the total weight of satisfied clauses is maximized.

We provide a more robust and deterministic reduction from \wms to a family of \wmc instances that runs in \fpt time and handles weighted inputs. This reduction preserves both the approximation factor (up to a factor of \( (1 - \varepsilon) \)) and the structure of the clause-variable incidence bipartite graph, without requiring it to be from a subgraph-closed family. As a consequence, our \epas for \textsc{Weighted Max Coverage} on set systems with bounded \smlindex immediately implies an \epas for \textsc{Weighted Max-SAT} when the clause-variable incidence graph has bounded \smlindex.

Formally, given an instance \( \mathcal{I} = (\Phi, \w, k) \) of \wms and any \( \varepsilon > 0 \), the reduction constructs in deterministic \fpt time (i.e., time \( f(k, \varepsilon) \cdot (|\mathcal{V}| + |\mathcal{C}|)^{\mathcal{O}(1)} \) for some computable function \( f \)) a bounded-size family of \wmc instances. For at least one of these instances, a \( (1 - \varepsilon/2) \)-approximate solution yields a \( (1 - \varepsilon) \)-approximate solution for the original \wms instance. This establishes a formal equivalence between \wms and \wmc in the context of parameterized approximation, thereby allowing us to lift our algorithmic results for \wmc to \wms. The starting point of the reduction in Theorem~\ref{thm:reduction-cmaxsat-to-maxcov} is inspired by the work of Manurangsi~\cite{DBLP:journals/tcs/Manurangsi25}, who designed a lossy kernel for the unweighted version of \wms.

We view a \cnf formula $\Phi = (\cV, \cC)$ as a set system, say $(\cU_\Phi, \cF_\Phi)$, where the ground set is the set of clauses $\cC$ and for each variable $v \in \cV$, there is a set $C_v$ in the set family, where $C_v$ contains all the clauses containing $v$. We show that VC dimension of the set system of any reduced \maxcov instance obtained by our reduction algorithm is bounded by that of the input \cmaxsat instance. Hence using the result of \cite{DBLP:conf/compgeom/BadanidiyuruKL12} we obtain Theorem~\ref{thm: vcdim_satepas}.

\subsubsection{Deterministic Algorithms}
We demonstrate that our algorithms can be derandomized using a \wvpn. The key combinatorial ingredient enabling this derandomization is captured in the following definition.

\begin{definition}[Weighted $\delta$-net and Weighted Heavy graph]
    Let $G= (\cR, \cB)$ be a bipartite graph and $\delta > 0$. Further let $\w: \cB \to \bQ^+$ be a weight function defined on the blue vertices of $G$.  A set $X \subseteq \cB$ is called $\wdeltanet(G)$ if for all $r\in \cR$, if $\w(\nbr(r)) \geq \delta \cdot \w(\cB)$, then $\nbr(r)\cap X\neq \emptyset$. 
    \noindent 
    Furthermore, $G$ is called weighted $\delta$-heavy graph if for all $r\in \cR$, $\w(\nbr(r)) \geq \delta \cdot \w(\cB)$. Here, $\nbr(r)$ denotes the set of neighbors of $r$ and $\w(X)=\sum_{x\in X} \w(x)$ for any subset $X\subseteq \cB$. 
\end{definition}
The definition of \wvpn in the unweighted setting is known as \vpn \footnote{This is commonly referred to as an $\varepsilon$-net in the literature, but we use the term $\delta$-net to avoid confusion, as $\varepsilon$ is already used to denote approximation ratios in this paper.}. It has been widely studied and applied in computational geometry. For foundational results on \wvpn, \vpn, and their constructions, we refer the reader to~\cite{DBLP:conf/stoc/BlumerEHW86,DBLP:conf/compgeom/Matousek90,DBLP:journals/dcg/BronnimannG95,DBLP:journals/siamcomp/BronnimannCM99}. See also the relevant book chapters in~\cite{10.5555/2031416}.

It is known that small \wvpn exist for set systems of bounded VC-dimension (see Section~\ref{section: weighteddetfptas} for the definition). Specifically, any set system with VC-dimension \( d \) admits a \wvpn of size \( \mathcal{O}\left(\frac{d}{\delta} \log{\frac{1}{\delta}}\right) \) for any \( \delta \in (0,1) \)~\cite{DBLP:conf/compgeom/HausslerW86,DBLP:conf/stoc/BlumerEHW86}. In our setting, we show that the relevant set family used to design deterministic algorithms has VC-dimension bounded by \( \Gamma = \dintcmp(G) \); recall that the VC-dimension is at most the \smlindex, which in turn is at most \( \Gamma \). Hence, we can leverage these known bounds and existing deterministic constructions to compute small \wvpn suited to our needs.

The running time obtained by directly applying the known algorithm is
\[
    d^{\cO(\log n + \log m)} \cdot \left(\frac{k}{\varepsilon}\right)^{(\log n + \log m)} \cdot \log^{(\log n + \log m)}\left(\frac{dk}{\varepsilon}\right) \cdot m.
\]
This runtime is too prohibitive for our purposes. Even when \( d \) is constant, the term \( \left(\frac{k}{\varepsilon}\right)^{(\log n + \log m)} \) already yields a bound of \( (n + m)^{\mathcal{O}(\log k)} \), which is too large to be useful for designing \epas or \pas algorithms.

Thus, we take an alternate route by computing a dominating set in an appropriate bipartite graph. For a given weighted \( \delta \)-heavy bipartite graph \( G = (\cR, \cB) \), let \( \optds(G) \) denote the minimum size of a set \( X \subseteq \cB \) such that \( X \) forms a \( \wdeltanet(G) \), formally defined as:
\[
    \optds(G) = \min_{X \subseteq \cB,\ \nbr(X) = \cR} |X|.
\]
The next lemma serves as the main technical tool for constructing \( \wdeltanet(G) \). It provides an upper bound on \( \optds(G) \) in a weighted \( \delta \)-heavy graph.

\begin{lemma}
\label{lem:UpperBoundBDSINTRO}
Let $\delta >0$, let $\graph$ with weight function $\w: \cB \to \bQ^+$ be a weighted $\delta$-heavy graph, and let $\Gamma = \dintcmp(G)$. Then, 
   \[ \optds(G) \leq  \cO\left(\frac{\Gamma}{\delta}\log{\frac{1}{\delta}}\right).\]
\end{lemma}

Observe that the bounds in \Cref{lem:UpperBoundBDSINTRO} are independent of the size of the graph \( G \). By replacing the probabilistic sampling step with the deterministic computation of a \( \delta \)-covering set, i.e., a \( \wdeltanet(G) \), we obtain a deterministic algorithm. This construction relies on known polynomial-time approximation or parameterized algorithms for {\sc Dominating Set}~\cite{DBLP:books/daglib/0004338,DBLP:journals/toct/G25} on set families of bounded \smlindex.

\subsection{Matroid and partition constraints}
The positive framework is robust under two simultaneous restrictions: the
selected red vertices must be independent in a matroid, and the blue vertices
are partitioned into \(r\) color classes with a separate coverage target for
each class. The central additional tool is a representative family. After
bucketing candidate vertices by their weighted degree in every color class, we
replace each bag by a \((k-1)\)-representative subfamily of size at most \(k\).
Thus, if a bag contains the next member of a feasible solution, it also contains
a representative that can extend the remaining independent set.

The recursive step either selects such a representative and contracts it in
the matroid, or samples a still-uncovered neighbor that reveals further
structure about the unknown solution. Bucketing contributes
\(\bigl(\cO(\log_{1+\varepsilon}k)\bigr)^r\) choices per step, while the same
downward-intersection measure as in the unconstrained algorithm bounds the
number of structural updates. This yields the running time in
\Cref{thm:mainWeightedcon}; the complete algorithm and analysis appear in
\Cref{section:matroidpartition}. Combining this algorithm with the
structure-preserving satisfiability reduction gives
\Cref{thm:application_wtconccsat}, proved in
\Cref{sec: weightedconstrainedsat}.

%% file: weighted_reduction_overview.tex
\subsubsection{Relating Weighted and Unweighted Instances}

In this part, we focus on the relationship between weighted and unweighted instances of \mrbds. Recall that in \wmrbds we are given a bipartite graph $G = (\cR \cup \cB, E)$, a weight function $\w: \cB \to \mathbb{Q}^+$ and an integer $k$. The objective is to find a set $\cR'\subseteq \cR$ such that $|\cR'|\leq k$ and the sum of weights of the vertices in $\cB$ that are adjacent to $\cR'$ is maximized. Let $\opt$ denote the weight of the optimal solution. In the unweighted version (\mrbds), our objective is to maximize the cardinality of the dominated vertices in the blue vertex set.

Because the weights $\w(v)$ can be exponentially large with respect to the input size, we cannot directly transform the weights into vertex multiplicities. Instead, we show that for any $\varepsilon > 0$, we can construct a polynomially-sized unweighted \mrbds instance such that an optimal solution in the reduced instance corresponds to a $(1-\varepsilon)$-approximate solution to the original \wmrbds instance. Furthermore, if the input graph has \smlindex $d$, the reduced graph has \smlindex at most $d$.

\paragraph{Scaling and Construction :}
Let $n = |\cB|$ be the number of blue vertices, and let $W_{max} = \max_{v \in \cB} \w(v)$. Since a single optimal red vertex can dominate the heaviest blue vertex, it is guaranteed that $\opt \geq W_{max}$.

We define a scaling factor $K = \frac{\varepsilon W_{max}}{n}$ and assign a new scaled weight $\w'(v)$ to every $v \in \cB$:
\[ \w'(v) = \left\lfloor \frac{\w(v)}{K} \right\rfloor \]
Notice that for all $v \in \cB$, the scaled weight is bounded: $\w'(v) \leq \frac{W_{max}}{K} = \frac{n}{\varepsilon}$.

We now construct an unweighted \mrbds instance defined on a new bipartite graph $G' = (\cR \cup \cB', E')$:
\begin{itemize}
    \item \textbf{Red Vertices:} The set of red vertices remains unchanged ($\cR' = \cR$).
    \item \textbf{Blue Vertices:} For every $v \in \cB$, we create a set $B_v$ consisting of exactly $\w'(v)$ independent vertices. The new blue vertex set is $\cB' = \bigcup_{v \in \cB} B_v$. Since $\w'(v) \leq n/\varepsilon$, the total size of $\cB'$ is bounded by $n^2/\varepsilon$, ensuring the construction takes polynomial time.
    \item \textbf{Edges:} For every edge $(u, v) \in E$, we add an edge between $u$ and every vertex in the set $B_v$.
\end{itemize}

\paragraph{Equivalence of Scaled and Unweighted Instances :}
For any candidate solution $S \subseteq \cR$, the construction of $E'$ ensures that the neighborhood of $S$ in $G'$ is the disjoint union of the sets corresponding to its neighborhood in $G$: $\nbr_{G'}(S) = \bigcup_{v \in \nbr_G(S)} B_v$. Because the sets $B_v$ are mutually disjoint for distinct $v \in \cB$ and each has size $|B_v| = \w'(v)$, the total number of blue vertices dominated by $S$ in $G'$ is exactly the scaled weight dominated in $G$:
$$|\nbr_{G'}(S)| = \sum_{v \in \nbr_G(S)} |B_v| = \sum_{v \in \nbr_G(S)} \w'(v)$$
Thus, maximizing the unweighted cardinality $|\nbr_{G'}(S)|$ in $G'$ is strictly equivalent to maximizing the total scaled weight of dominated vertices in $G$.

\paragraph{Approximation Ratio :}
Let $S^* \subseteq \cR$ be an optimal solution in $G$ with weight $\w(\nbr_G(S^*)) = \opt$, and let $S \subseteq \cR$ be an optimal solution to the unweighted instance $G'$. Because both sets are valid solutions of size at most $k$, and $S$ maximizes the unweighted cardinality of dominated vertices in $G'$, it must hold that $|\nbr_{G'}(S)| \geq |\nbr_{G'}(S^*)|$. 

Applying our established equivalence, this purely unweighted comparison directly translates to the scaled weights in $G$:
\[ \sum_{v \in \nbr_G(S)} \w'(v) \geq \sum_{v \in \nbr_G(S^*)} \w'(v) \]

To relate these scaled weights back to the original weights, we rely on the fundamental properties of the floor function. For any real number $x$, it holds that $x - 1 < \lfloor x \rfloor \leq x$. Substituting our definition $\w'(v) = \lfloor \w(v)/K \rfloor$ yields the strict double inequality:
\[ \frac{\w(v)}{K} - 1 < \w'(v) \leq \frac{\w(v)}{K} \]

Multiplying all terms by the strictly positive scaling factor $K$ provides the two crucial bounds for our approximation:
\begin{align}
    K \cdot \w'(v) &\leq \w(v) \label{eq:upperbound} \\
    \w(v) - K &< K \cdot \w'(v) \label{eq:lowerbound}
\end{align}

Inequality (\ref{eq:upperbound}) ensures that scaling the weights back up never overestimates the true weight, while Inequality (\ref{eq:lowerbound}) bounds the maximum weight lost due to rounding down. Using these bounds, we lower-bound the true weight of the neighborhood of our solution $S$ in $G$:
\begin{align*}
\w(\nbr_G(S)) = \sum_{v \in \nbr_G(S)} \w(v) &\geq \sum_{v \in \nbr_G(S)} K \cdot \w'(v) \quad &&\text{(Using Inequality \ref{eq:upperbound})} \\
&\geq \sum_{v \in \nbr_G(S^*)} K \cdot \w'(v) \quad &&\text{(Since $|\nbr_{G'}(S)| \geq |\nbr_{G'}(S^*)|$)} \\
&> \sum_{v \in \nbr_G(S^*)} (\w(v) - K) \quad &&\text{(Using Inequality \ref{eq:lowerbound})} \\
&= \sum_{v \in \nbr_G(S^*)} \w(v) - \sum_{v \in \nbr_G(S^*)} K \\
&= \opt - K|\nbr_G(S^*)|
\end{align*}

Because the optimal solution $S^*$ can dominate at most $n$ blue vertices, $|\nbr_G(S^*)| \leq n$. Substituting our scaling factor $K = \frac{\varepsilon W_{max}}{n}$ yields $K|\nbr_G(S^*)| \leq \varepsilon W_{max}$. Furthermore, since a single red vertex can always dominate the single heaviest blue vertex, $W_{max} \leq \opt$. Thus:
\begin{align*}
\w(\nbr_G(S)) &> \opt - \varepsilon W_{max} \\
&\geq \opt - \varepsilon \opt = (1-\varepsilon)\opt
\end{align*}

This proves that solving the constructed polynomially-sized unweighted instance $G'$ yields a $(1-\varepsilon)$-approximate solution to the original weighted instance $G$.

\paragraph{Bounding the \smlindex :}
Finally, we prove that the reduced unweighted graph $G'$ does not increase the \smlindex of the original graph $G$. Let $\smlindex(G) = d$. 

Suppose for the sake of contradiction that $G'$ contains a \sml of order $d+1$. Let the sequences forming this \sml be $r'_1, \dots, r'_{d+1} \in \cR$ and $b'_1, \dots, b'_{d+1} \in \cB'$. By the definition of a \sml, for any two indices $x$ and $y$ such that $1 \leq x < y \leq d+1$, it must hold that $(r'_y, b'_x) \in E'$ and $(r'_y, b'_y) \notin E'$.

Recall that every vertex in $\cB'$ is a copy of some original vertex in $\cB$. Suppose there exist indices $x < y$ such that $b'_x$ and $b'_y$ are copies of the exact same original vertex $b \in \cB$. By construction, $b'_x, b'_y \in B_b$ and therefore share identical open neighborhoods in $\cR$. However, the \sml definition requires the red vertex $r'_y$ to distinguish them (adjacent to $b'_x$ but not $b'_y$). This is a contradiction.

Therefore, the sequence $b'_1, \dots, b'_{d+1}$ must correspond to strictly distinct original vertices $b_1, \dots, b_{d+1} \in \cB$. Because adjacencies between $\cR$ and $\cB'$ precisely mirror those between $\cR$ and $\cB$, the original vertices $r'_1, \dots, r'_{d+1}$ and $b_1, \dots, b_{d+1}$ would form a $(d+1)$-\sml in $G$. This contradicts our premise that $\smlindex(G) = d$. Thus, $G'$ cannot contain a \sml of order $d+1$, meaning $\smlindex(G') \leq d$.

%% file: vc_dimension_lower_bounds.tex
\section{The Bounded VC-Dimension Boundary for Partial Set Cover}
\label{sec:vc-boundary}

The theorem of Badanidiyuru, Kleinberg, and Lee~\cite{DBLP:conf/compgeom/BadanidiyuruKL12}
shows that bounded VC-dimension is enough to give an \epas for \maxcov with a fixed budget (a $(1-\varepsilon)$-approximation to the best coverage by $k$ sets).
This section establishes the complementary message for the covering target
objective: bounded VC-dimension alone does not yield a \pas for \psc.

For a set system \((\cU,\cF)\) and target \(t\), write
\[
  \opt(\cU,\cF,t)
  =
  \min\bigl\{|\cA|:\cA\subseteq \cF,\
       |\bigcup_{A\in\cA}A|\ge t\bigr\}.
\]
All reductions in this section set \(t=|\cU|\).  Thus, the lower bounds
already hold for ordinary \textsc{Set Cover}, which is a special case of
\psc.
Throughout the section, we measure the size of a produced set system by its
incidence-list encoding:
\[
  N=|\cU|+|\cF|+\sum_{S\in\cF}|S|.
\]

The section is organized in two layers.  We first prove a purely
parameterized lower bound under the assumption
\(\mathrm{FPT}\ne\mathrm{W[1]}\): using Lin's constant-factor
inapproximability of \(k\)-\textsc{Clique}, we give a clique-to-\(\psc\)
reduction that rules out \((2-\delta)\)-approximations in
\(f(k)N^{\cO(1)}\) time even for set systems of VC-dimension at most \(7\).
The ETH part uses Bafna, Karthik, and
Minzer~\cite[Theorem~1.1]{BafnaKarthikMinzer2025}: for every constant
\(\zeta>0\), satisfiable 2-CSP instances are ETH-hard to distinguish from
instances in which every assignment satisfies at most a \(\zeta\)-fraction
of the constraints, even when each variable appears in exactly \(D_\zeta\)
constraints, where \(D_\zeta\) is a constant depending only on
\(\zeta\).  The label-selection reduction gives a \((1+\alpha)\)-gap at
VC-dimension \(7\), while the list construction extends the lower bound to
every fixed approximation factor, with VC-dimension growing only
logarithmically in that factor.

All reductions below use the same row structure.  We record it once, before
introducing the particular reductions.

\begin{definition}[Rows and blocking lists]
\label{def:vc-row-blocking}
Fix an alphabet \(\Sigma\) and a positive integer \(k\).  A set system
\((\cU,\cF)\) has \emph{rows and blocking lists} if its set family is
partitioned into \(k\) rows
\[
  \cF=\bigcup_{i=1}^k\cF_i,
  \qquad
  \cF_i=\{S_{i,a}:a\in\Sigma\}.
\]
For a selected subfamily \(\cA\subseteq\cF\), its list of selected labels in
row \(i\) is
\[
  L_i(\cA)=\{a\in\Sigma:S_{i,a}\in\cA\}.
\]

For each universe element \(u\in\cU\), the representation specifies a set
\(I(u)\subseteq[k]\) of \emph{active rows}.  For every \(i\in I(u)\), it
also specifies a \emph{blocking list} \(A_i(u)\subseteq\Sigma\).  No set
from an inactive row contains \(u\).  For every active row \(i\in I(u)\),
membership is governed by
\[
  u\in S_{i,a}
  \quad\Longleftrightarrow\quad
  a\notin A_i(u).
\]
We call such a set system \(q\)-local if
\(|I(u)|\le2\) and \(|A_i(u)|\le q\) for every element \(u\) and active row
\(i\in I(u)\).
\end{definition}

For a selected subfamily \(\cA\subseteq\cF\), an element \(u\) is uncovered
exactly when
\[
  L_i(\cA)\subseteq A_i(u)
  \qquad\text{for every }i\in I(u).
\]
Indeed, sets from inactive rows do not contain \(u\), while an active row
\(i\) covers \(u\) precisely when it selects some label outside
\(A_i(u)\).

Recall that for \(Z\subseteq\cU\), the trace family of \(\cF\) on \(Z\) is
\[
  \cF|_Z=\{S\cap Z:S\in\cF\}.
\]
Each set \(S\cap Z\) is called the \emph{trace of \(S\) on \(Z\)}.  Thus,
the number of traces on \(Z\) is the number of distinct subsets of \(Z\)
realized by sets in \(\cF\).  As in the introduction, \(Z\) is
\emph{shattered} if
\(\cF|_Z=2^Z\), and the VC-dimension of \((\cU,\cF)\) is the maximum
cardinality of a shattered subset of \(\cU\).

The next lemma shows that
locality and bounded blocking-list size alone imply bounded VC-dimension.
Later we will only need to identify the active rows and blocking lists in
each construction.

\begin{lemma}[Trace bound for rows and blocking lists]
\label{lem:vc-local-list-traces}
Every \(q\)-local set system with rows and blocking lists has VC-dimension
\(\cO(\log(q+1))\).  In particular, when \(q=1\), its VC-dimension is at
most \(7\).
\end{lemma}

\begin{proof}
Fix \(Z\subseteq\cU\) with \(|Z|=s\), and write
\(\cT_Z=\cF|_Z\) for the trace family on \(Z\).  Define
\[
  J(Z)=\bigcup_{u\in Z} I(u).
\]
Thus, \(J(Z)\) is exactly the set of rows that are active for at least one
element of \(Z\).  Since the set system is \(q\)-local,
\[
  |J(Z)|
  \le \sum_{u\in Z}|I(u)|
  \le 2s.
\]
If \(i\notin J(Z)\), then \(i\notin I(u)\) for every \(u\in Z\).
By the inactive-row condition, \(u\notin S_{i,a}\) for every \(u\in Z\)
and every \(a\in\Sigma\).  Consequently, \(S_{i,a}\cap Z=\emptyset\).
Therefore, all rows outside \(J(Z)\) together contribute only the empty
trace.

Now fix \(i\in J(Z)\), and define
\[
  Z_i=\{u\in Z:i\in I(u)\}
  \qquad\text{and}\qquad
  B_i=\bigcup_{u\in Z_i}A_i(u).
\]
Because every blocking list has size at most \(q\),
\[
  |B_i|
  \le \sum_{u\in Z_i}|A_i(u)|
  \le q|Z_i|
  \le qs.
\]
Consider a label \(a\in\Sigma\setminus B_i\).  For every \(u\in Z_i\), we
have \(a\notin A_i(u)\), and hence \(u\in S_{i,a}\).  For every
\(u\in Z\setminus Z_i\), row \(i\) is inactive for \(u\), so the
inactive-row condition gives \(u\notin S_{i,a}\).  It follows that
\[
  S_{i,a}\cap Z=Z_i
  \qquad\text{for every }a\in\Sigma\setminus B_i.
\]
Thus, all labels outside \(B_i\) produce the same trace.  The labels in
\(B_i\) produce at most \(|B_i|\) additional traces.  Hence row \(i\)
produces at most
\[
  |B_i|+1\le qs+1
\]
distinct traces on \(Z\).

Summing over the rows in \(J(Z)\), and adding the empty trace contributed
by all remaining rows, gives
\[
  |\cT_Z|
  \le 1+\sum_{i\in J(Z)}(|B_i|+1)
  \le 2s(qs+1)+1.
\]
If \(Z\) is shattered, then \(|\cT_Z|=2^s\), so
\[
  2^s\le 2s(qs+1)+1.
\]

To obtain the asymptotic bound, let \(L=\log_2(q+1)\ge1\).  Suppose, for
contradiction, that \(Z\) is shattered and \(s\ge16L\).  Then \(s\ge16\),
and the elementary inequality \(s^2\le2^{s/2}\) gives
\[
\begin{aligned}
  2s(qs+1)+1
  &\le 4qs^2+1\\
  &\le 2^{2+L+s/2}+1\\
  &\le 2^{2+9s/16}+1
  <2^s,
\end{aligned}
\]
contradicting the necessary inequality \(2^s\le2s(qs+1)+1\).  Hence every
shattered set satisfies
\[
  s<16\log_2(q+1),
\]
which proves that the VC-dimension is \(\cO(\log(q+1))\).

For the sharper bound when \(q=1\), return to the exact trace estimate.  If
a set \(Z\) of size \(8\) were shattered, then
\[
  256
  =2^8
  =|\cT_Z|
  \le 2\cdot8(8+1)+1
  =145,
\]
which is impossible.  Moreover, every subset of a shattered set is also
shattered.  Therefore no set of size at least \(8\) is shattered, and the
VC-dimension is at most \(7\).
\end{proof}

\subsection{Hardness under assumption \(\mathrm{FPT}\ne\mathrm{W[1]}\)}

\begin{proposition}[Parameterized clique inapproximability~{\cite[Theorem~1.2]{DBLP:conf/stoc/Lin21}}]
\label{lem:vc-lin-clique-source}
Assume \(\mathrm{FPT}\ne\mathrm{W[1]}\).  For every fixed
\(\delta\in(0,1)\), for every computable function \(F\) and every constant
\(a\), no algorithm running in time \(F(k)|V(H)|^a\) distinguishes graphs
containing a \(k\)-clique from graphs \(H\) satisfying
\(\omega(H)<\delta k\).
\end{proposition}

This is the \(c=1/\delta\) case of Lin's theorem, which excludes an
\(F(k)|V(H)|^{\cO(1)}\)-time algorithm distinguishing \(\omega(H)\ge k\) from
\(\omega(H)<k/c\), for every constant \(c\ge1\).

\paragraph{Clique construction.}
Let \((H,k)\) be a graph instance, and write \(n_0=|V(H)|\).  We construct
a \psc instance \((\cU,\cF,t)\).

The universe contains one row element \(r_i\) for every \(i\in[k]\).
Additionally, for every pair of positions \(i<j\) and every ordered pair
\((u,v)\in V(H)^2\) satisfying \(u=v\) or \(uv\notin E(H)\), introduce an
invalid-pair element \(e_{i,j,u,v}\).  These are all the elements of
\(\cU\).

For every row \(\ell\in[k]\) and vertex \(x\in V(H)\), define a set
\(S_{\ell,x}\subseteq\cU\) as follows.  It contains \(r_\ell\) and no other
row element.  For every invalid-pair element \(e_{i,j,u,v}\), define
\[
  e_{i,j,u,v}\in S_{\ell,x}
  \quad\Longleftrightarrow\quad
  (\ell=i\text{ and }x\ne u)
  \ \text{or}\
  (\ell=j\text{ and }x\ne v).
\]
In other words, \(e_{i,j,u,v}\) belongs to every set in row \(i\) except
\(S_{i,u}\), to every set in row \(j\) except \(S_{j,v}\), and to no set
from any other row.
Let \(\cF_i=\{S_{i,x}:x\in V(H)\}\) and
\(\cF=\bigcup_{i=1}^k\cF_i\).  Set \(t=|\cU|\), so every feasible solution
must cover the entire universe.

\begin{figure}[ht!]
  \centering
  \resizebox{0.95\textwidth}{!}{\input{figures/clique_construction.tex}}
  \caption{The clique construction. Rows represent the \(k\) clique positions
  and columns represent vertices of \(H\).  The row element \(r_i\) forces a
  choice in row \(i\).  For positions \(i<j\) and an invalid pair \((u,v)\),
  the element \(e_{i,j,u,v}\) is contained in all sets of rows \(i\) and \(j\)
  except \(S_{i,u}\) and \(S_{j,v}\), blocking that simultaneous assignment.}
  \label{fig:clique-construction}
\end{figure}
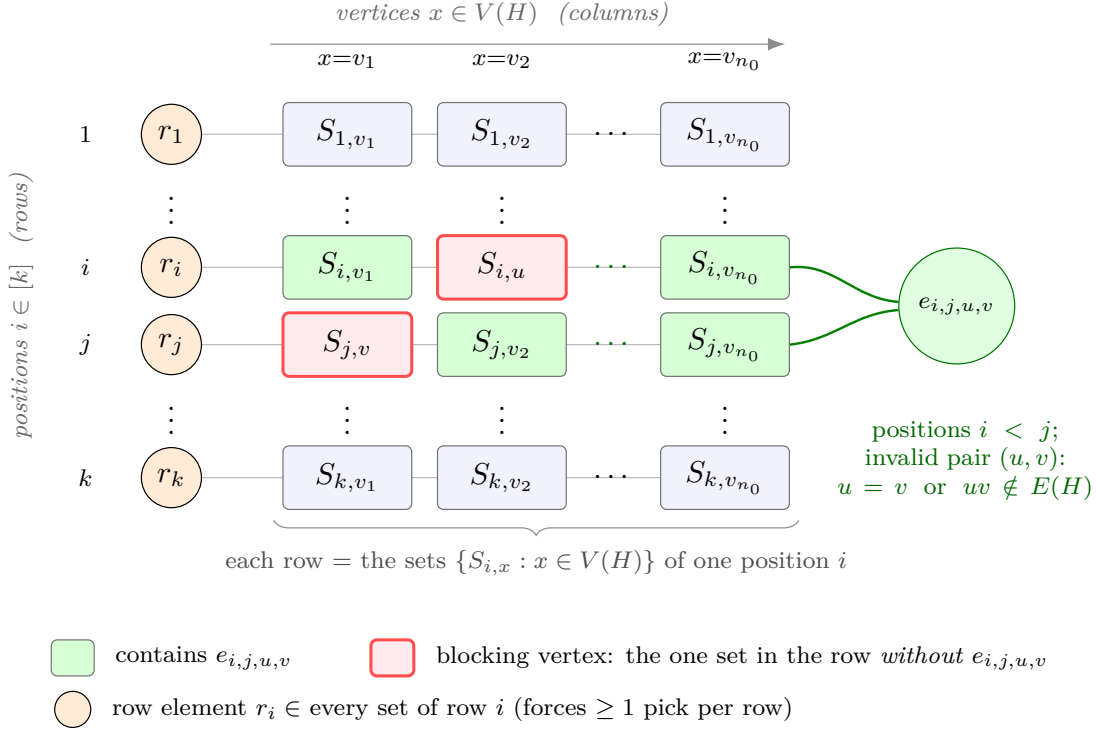

Since \(r_i\) belongs precisely to the sets of \(\cF_i\), every cover must
select at least one set from each row (see \Cref{fig:clique-construction}).
Suppose rows \(i<j\) each contribute exactly one selected set, say
\(S_{i,u}\) and \(S_{j,v}\).  If \((u,v)\) is invalid, then
\(e_{i,j,u,v}\) is uncovered.  Consequently, a selection containing
exactly one set from every row covers \(\cU\) only if the selected vertices
are distinct and pairwise adjacent.  Thus, row \(i\) represents the
\(i\)-th position of a prospective \(k\)-clique, and selecting \(S_{i,x}\)
assigns vertex \(x\) to that position.

The construction has
\[
  |\cF|=kn_0
  \qquad\text{and}\qquad
  |\cU|\le k+\binom{k}{2}n_0^2.
\]
The row elements contribute \(kn_0\) to \(\sum_{S\in\cF}|S|\).  Each
invalid-pair element belongs to exactly \(2(n_0-1)\) sets, and hence
\[
  \sum_{S\in\cF}|S|
  \le
  kn_0+2\binom{k}{2}n_0^2(n_0-1).
\]
Therefore,
\[
  N
  =
  |\cU|+|\cF|+\sum_{S\in\cF}|S|
  =
  \cO(k^2n_0^3).
\]
In particular, \(N\le(kn_0)^{c_{\mathrm{clq}}}\) for some absolute
constant \(c_{\mathrm{clq}}>0\).

\begin{lemma}[Correctness of the clique construction]
\label{lem:vc-clique-correctness}
Let \((\cU,\cF,t)\) be the instance obtained from \((H,k)\) by the clique
construction.  If \(H\) contains a \(k\)-clique, then
\(\opt(\cU,\cF,t)\le k\).  If \(\omega(H)<\delta k\), then every cover has
size greater than \((2-\delta)k\).
\end{lemma}

\begin{proof}
Suppose first that \(H\) contains a \(k\)-clique with distinct vertices
\(v_1,\ldots,v_k\).  Select
\[
  \cA=\{S_{i,v_i}:i\in[k]\}.
\]
For every \(i\in[k]\), the set \(S_{i,v_i}\) contains \(r_i\), so
\(\cA\) covers all row elements.  Consider an invalid-pair element
\(e_{i,j,u,v}\).  Since \(v_i\) and \(v_j\) are distinct and adjacent, the
pair \((v_i,v_j)\) is valid, whereas \((u,v)\) is invalid.  Hence
\((v_i,v_j)\ne(u,v)\), so either \(v_i\ne u\) or \(v_j\ne v\).  In the
first case \(e_{i,j,u,v}\in S_{i,v_i}\), and in the second case
\(e_{i,j,u,v}\in S_{j,v_j}\).  Thus \(\cA\) covers \(\cU\), and
\(\opt(\cU,\cF,t)\le k\).

For soundness, suppose that \(\omega(H)<\delta k\), and assume for
contradiction that a subfamily \(\cA\subseteq\cF\) covers \(\cU\) with
\[
  |\cA|\le(2-\delta)k.
\]
For each \(i\in[k]\), recall the selected-label list
\[
  L_i(\cA)=\{v\in V(H):S_{i,v}\in\cA\}.
\]
Since \(r_i\) belongs only to sets of row \(i\), every \(L_i(\cA)\) is
nonempty.  Moreover, since the rows partition \(\cF\),
\[
  |\cA|=\sum_{i=1}^k|L_i(\cA)|.
\]
Let
\[
  M=\{i\in[k]:|L_i(\cA)|\ge2\}.
\]
Then
\[
  |M|
  \le\sum_{i=1}^k\bigl(|L_i(\cA)|-1\bigr)
  =|\cA|-k
  \le(1-\delta)k.
\]
Therefore \(I=[k]\setminus M\) satisfies \(|I|\ge\delta k\).  For every
\(i\in I\), the list \(L_i(\cA)\) is nonempty and has size less than \(2\);
hence \(L_i(\cA)=\{v_i\}\) for a unique vertex \(v_i\in V(H)\).

We claim that the vertices \(\{v_i:i\in I\}\) form a clique.  Consider
\(i,j\in I\) with \(i<j\).  If \(v_i=v_j\) or \(v_iv_j\notin E(H)\), then
the construction contains the invalid-pair element
\(e_{i,j,v_i,v_j}\).  The only selected sets in rows \(i\) and \(j\) are
\(S_{i,v_i}\) and \(S_{j,v_j}\), and neither contains
\(e_{i,j,v_i,v_j}\).  No set from another row contains this element.
Therefore it is uncovered, contradicting that \(\cA\) covers \(\cU\).

Thus the vertices \(v_i\), \(i\in I\), are distinct and pairwise adjacent.
Consequently,
\[
  \omega(H)\ge |I|\ge\delta k,
\]
contradicting \(\omega(H)<\delta k\).  Hence every cover has size greater
than \((2-\delta)k\).
\end{proof}

\begin{lemma}[VC-dimension of the clique construction]
\label{lem:vc-clique-dimension}
The set system produced by the clique construction has VC-dimension at
most \(7\).
\end{lemma}

\begin{proof}
We use \(\Sigma=V(H)\) in \Cref{def:vc-row-blocking}.

For every row element \(r_i\), the incidence rule gives
\[
  r_i\in S_{\ell,x}
  \quad\Longleftrightarrow\quad
  \ell=i.
\]
Hence
\[
  I(r_i)=\{i\},
  \qquad
  A_i(r_i)=\emptyset.
\]

For every invalid-pair element \(e_{i,j,u,v}\), we have
\[
  e_{i,j,u,v}\in S_{\ell,x}
  \quad\Longleftrightarrow\quad
  (\ell=i\text{ and }x\ne u)
  \ \text{or}\
  (\ell=j\text{ and }x\ne v).
\]
Therefore,
\[
  I(e_{i,j,u,v})=\{i,j\},
  \qquad
  A_i(e_{i,j,u,v})=\{u\},
  \qquad
  A_j(e_{i,j,u,v})=\{v\}.
\]
Thus every element has at most two active rows, and every blocking list has
size at most \(1\).  Hence the set system is \(1\)-local, and
\Cref{lem:vc-local-list-traces} gives VC-dimension at most \(7\).
\end{proof}

\begin{restatable}[Parameterized hardness below factor two]{theorem}{vcbelowtwohardthm}
\label{thm:vc-no-fpt-below-two}
Assume \(\mathrm{FPT}\ne\mathrm{W[1]}\).  For every fixed
\(\delta\in(0,1)\), for every computable function \(f\) and every constant
\(a\), no algorithm running in time \(f(k)N^a\) can distinguish between
\[
  \opt(\cU,\cF,t)\le k
  \qquad\text{and}\qquad
  \opt(\cU,\cF,t)>(2-\delta)k,
\]
even when \(t=|\cU|\) and \((\cU,\cF)\) has VC-dimension at most \(7\).
Consequently, no \((2-\delta)\)-approximation for this class runs in time
\(f(k)N^{\cO(1)}\), for any computable function \(f\).
\end{restatable}

\begin{proof}
Fix \(\delta\in(0,1)\), and apply the clique construction to \((H,k)\).
By \Cref{lem:vc-clique-correctness}, a \(k\)-clique in \(H\) produces an
instance satisfying \(\opt(\cU,\cF,t)\le k\), whereas
\(\omega(H)<\delta k\) produces an instance satisfying
\(\opt(\cU,\cF,t)>(2-\delta)k\).  By
\Cref{lem:vc-clique-dimension}, the set system has VC-dimension at most
\(7\).  The construction preserves \(k\), has polynomial running time in
the output size, and satisfies \(N\le(kn_0)^{c_{\mathrm{clq}}}\), where
\(n_0=|V(H)|\).

Suppose, for contradiction, that a target algorithm distinguishes the two
cases in time \(f(k)N^a\).  Let \(b\) be a constant such that the clique
construction runs in time at most \(N^b\).  The reduction followed by the
target algorithm solves the clique gap in time at most
\[
  f_1(k)N^{\max\{a,b\}}
  \le
  f_2(k)n_0^{c_{\mathrm{clq}}\max\{a,b\}},
\]
where \(f_1\) and \(f_2\) absorb factors depending only on \(k\).  This is
an \(f_2(k)n_0^{\cO(1)}\)-time algorithm for the source gap, contradicting
\Cref{lem:vc-lin-clique-source}.

Finally, a \((2-\delta)\)-approximation would distinguish the two cases:
in the YES case it returns a cover of size at most \((2-\delta)k\), while
in the NO case every feasible cover has size greater than
\((2-\delta)k\).  Therefore such an approximation cannot run in time
\(f(k)N^{\cO(1)}\).
\end{proof}

\subsection{Hardness under ETH}

We now prove near-optimal running-time lower bounds under ETH.  The source
problem is regular 2-CSP.  Recall that a 2-CSP consists of a variable set
\(X\), an alphabet \(\Sigma\), a constraint graph \(G=(X,E)\), and, for
every edge \(e=x_ix_j\in E\), an accepting relation
\(R_e\subseteq\Sigma^2\).  Its value is
\[
  \operatorname{val}(\Lambda)
  =
  \max_{\sigma:X\to\Sigma}
  \frac{1}{|E|}
  \left|
    \left\{
      e=x_ix_j\in E:
      (\sigma(x_i),\sigma(x_j))\in R_e
    \right\}
  \right|,
\]
the maximum fraction of constraints satisfied by an assignment.

\begin{proposition}[ETH hardness source for regular 2-CSP~{\cite[Theorem~1.1]{BafnaKarthikMinzer2025}}]
\label{lem:vc-csp-eth-source}
Assume ETH.  For every constant \(\zeta>0\), there are constants
\(C_\zeta>0\) and \(D_\zeta\in\mathbb{N}\) such that, for every computable
function \(F\), no algorithm running in time
\[
  F(k)|\Sigma|^{k/\log^{C_\zeta}k}
\]
distinguishes a satisfiable 2-CSP on \(k\) variables over alphabet
\(\Sigma\) from one of value at most \(\zeta\), even when the constraint
graph is a \(D_\zeta\)-regular bipartite graph.
\end{proposition}

\paragraph{Label selection construction.}
We begin with a direct reduction from 2-CSP.  Let
\(\Lambda=(X,\Sigma,E,\{R_e\}_{e\in E})\) be a 2-CSP, where
\(X=\{x_1,\ldots,x_k\}\) and \(G=(X,E)\) is its constraint graph.  We
construct a \psc instance \((\cU,\cF,t)\).  For every constraint \(e\in E\),
write \(e=x_ix_j\), fixing this order so that the first and second
coordinates of \(R_e\) correspond to \(x_i\) and \(x_j\), respectively.

The universe is the disjoint union
\[
  \cU=\cU_{\mathrm{row}}\mathbin{\dot\cup}\cU_{\mathrm{con}},
\]
where
\[
  \cU_{\mathrm{row}}=\{r_i:i\in[k]\}
\]
contains one row element for each variable, and
\[
  \cU_{\mathrm{con}}
  =
  \{u_{e,\lambda,\mu}:e=x_ix_j\in E,\
    (\lambda,\mu)\in\Sigma^2\setminus R_e\}
\]
contains one constraint element for every rejecting label pair of every
constraint.

For every variable \(x_i\) and label \(a\in\Sigma\), introduce a set
\(S_{i,a}\).  Define
\[
  \cF_i=\{S_{i,a}:a\in\Sigma\}
  \qquad\text{and}\qquad
  \cF=\bigcup_{i=1}^k\cF_i.
\]
The incidences are defined as follows.  Each row element satisfies
\[
  r_h\in S_{i,a}
  \quad\Longleftrightarrow\quad
  h=i.
\]
For a constraint \(e=x_ix_j\) and a rejecting pair
\((\lambda,\mu)\notin R_e\), define
\[
  u_{e,\lambda,\mu}\in S_{\ell,\gamma}
  \quad\Longleftrightarrow\quad
  (\ell=i\text{ and }\gamma\ne\lambda)
  \ \text{or}\
  (\ell=j\text{ and }\gamma\ne\mu).
\]
These are all the incidences.  Finally, set \(t=|\cU|\), so a feasible
solution must cover every universe element.

Writing \(s=|\Sigma|\), the construction has \(ks\) sets and at most
\[
  k+|E|s^2
\]
universe elements.

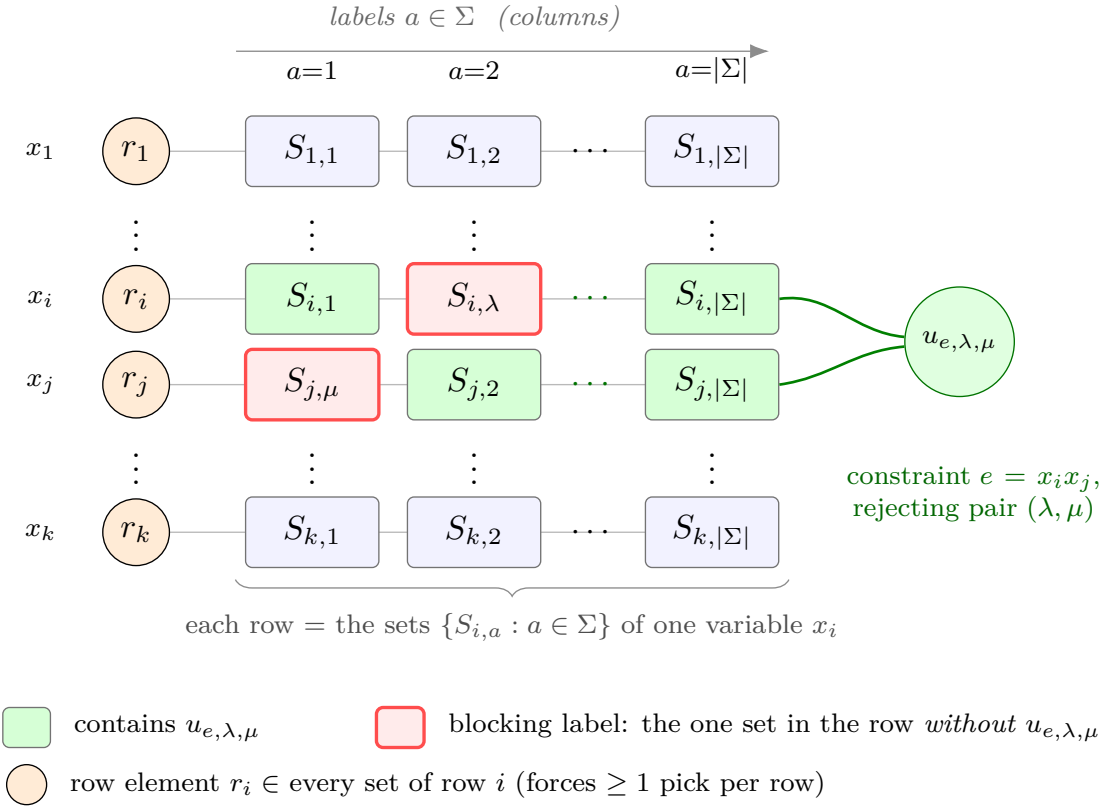
\begin{figure}[ht!]
  \centering
  \resizebox{0.95\textwidth}{!}{\input{figures/label_selection_construction.tex}}
  \caption{The label selection construction. Rows correspond to variables and
  columns to labels; selecting \(S_{i,a}\) assigns label \(a\) to \(x_i\).
  The row element \(r_i\) forces a selection from row \(i\).  For a constraint
  \(e=x_ix_j\) and rejecting pair \((\lambda,\mu)\), the element
  \(u_{e,\lambda,\mu}\) is contained in all sets of rows \(i\) and \(j\)
  except \(S_{i,\lambda}\) and \(S_{j,\mu}\), so it is uncovered exactly when
  those two labels are selected.}
  \label{fig:label-selection-construction}
\end{figure}

In words (illustrated in \Cref{fig:label-selection-construction}), selecting \(S_{i,a}\) represents assigning label \(a\) to
\(x_i\), and the row element \(r_i\) forces every cover to select at least
one set from row \(\cF_i\).  The constraint element
\(u_{e,\lambda,\mu}\) represents the forbidden simultaneous choice
\(x_i=\lambda\) and \(x_j=\mu\).  It belongs to every set in row \(i\)
except \(S_{i,\lambda}\), and to every set in row \(j\) except
\(S_{j,\mu}\); no set from any other row contains it.  Thus, selecting a
label different from \(\lambda\) for \(x_i\), or a label different from
\(\mu\) for \(x_j\), covers this element.  If exactly one set is selected
from each endpoint row, \(u_{e,\lambda,\mu}\) remains uncovered precisely
when the rejecting pair \((\lambda,\mu)\) is selected.  Hence, a solution
that selects exactly one set from each row covers all constraint elements
exactly when the corresponding assignment satisfies every constraint.

\begin{lemma}[Correctness of the label selection construction]
\label{lem:vc-label-correctness}
Let \((\cU,\cF,t)\) be the instance obtained from \(\Lambda\) by the label
selection construction.  If \(\Lambda\) is satisfiable, then
\(\opt(\cU,\cF,t)\le k\).  If the constraint graph of \(\Lambda\) is
\(D\)-regular and \(\operatorname{val}(\Lambda)\le1-\eta\), then, for
every \(0<\alpha<\eta/2\),
\[
  \opt(\cU,\cF,t)>(1+\alpha)k.
\]
\end{lemma}

\begin{proof}
Let \(\sigma:X\to\Sigma\) be a satisfying assignment, and select
\[
  \cA_\sigma=\{S_{i,\sigma(x_i)}:i\in[k]\}.
\]
The set \(S_{i,\sigma(x_i)}\) covers \(r_i\), so \(\cA_\sigma\) covers
every row element.  Consider a constraint element \(u_{e,\lambda,\mu}\),
where \(e=x_ix_j\).  Since \(\sigma\) satisfies \(e\),
\[
  (\sigma(x_i),\sigma(x_j))\in R_e,
\]
whereas \((\lambda,\mu)\notin R_e\).  Hence either
\(\sigma(x_i)\ne\lambda\) or \(\sigma(x_j)\ne\mu\).  In the first case
\(S_{i,\sigma(x_i)}\) covers \(u_{e,\lambda,\mu}\), and in the second case
\(S_{j,\sigma(x_j)}\) covers it.  Thus \(\cA_\sigma\) covers \(\cU\), and
\(\opt(\cU,\cF,t)\le k\).

Now suppose that \(\operatorname{val}(\Lambda)\le1-\eta\).  Assume, for
contradiction, that a subfamily \(\cA\subseteq\cF\) covers \(\cU\) and
satisfies
\[
  |\cA|\le(1+\alpha)k.
\]
For each row \(i\), use the previously defined list \(L_i(\cA)\).  Because
\(r_i\) is covered only by sets from row \(i\), every \(L_i(\cA)\) is
nonempty.  Since the rows partition \(\cF\),
\[
  |\cA|=\sum_{i=1}^k|L_i(\cA)|.
\]
Let
\[
  M=\{i\in[k]:|L_i(\cA)|\ge2\}.
\]
For every \(i\in M\), we have \(|L_i(\cA)|-1\ge1\).  Therefore,
\[
  |M|
  \le\sum_{i=1}^k\bigl(|L_i(\cA)|-1\bigr)
  =|\cA|-k
  \le\alpha k.
\]

Construct an assignment \(\tau:X\to\Sigma\) as follows.  If \(i\notin M\),
then \(1\le|L_i(\cA)|<2\), so \(L_i(\cA)=\{a_i\}\) for a unique label
\(a_i\); set \(\tau(x_i)=a_i\).  If \(i\in M\), assign an arbitrary label
by choosing any \(a_i\in L_i(\cA)\subseteq\Sigma\) and setting
\(\tau(x_i)=a_i\).

Let \(V_\tau\subseteq E\) be the set of constraints violated by \(\tau\).
We claim that every \(e=x_ix_j\in V_\tau\) satisfies
\(\{i,j\}\cap M\ne\emptyset\).  Suppose instead that
\(e=x_ix_j\in V_\tau\) with \(i,j\notin M\).  Then
\[
  (a_i,a_j)=(\tau(x_i),\tau(x_j))\notin R_e,
\]
so the universe contains the constraint element \(u_{e,a_i,a_j}\).  The
only selected sets in rows \(i\) and \(j\) are \(S_{i,a_i}\) and
\(S_{j,a_j}\), neither of which contains \(u_{e,a_i,a_j}\).  No set from
another row contains this element.  Hence \(u_{e,a_i,a_j}\) is uncovered,
contradicting that \(\cA\) covers \(\cU\).

The number of constraints having an endpoint indexed by \(M\) is at most
\(D|M|\).  Writing \(m=|E|=Dk/2\), we therefore obtain
\[
  |V_\tau|
  \le D|M|
  \le D\alpha k
  \le 2\alpha m.
\]
Since \(\alpha<\eta/2\), we have \(2\alpha<\eta\), and hence
\[
  |V_\tau|<\eta m.
\]
Thus the number of constraints satisfied by \(\tau\) is
\[
  m-|V_\tau|>(1-\eta)m.
\]
Consequently, \(\operatorname{val}(\Lambda)>1-\eta\), contradicting the
assumption \(\operatorname{val}(\Lambda)\le1-\eta\).  Therefore every cover
has size greater than \((1+\alpha)k\).
\end{proof}

\begin{lemma}[VC-dimension of the label selection construction]
\label{lem:vc-label-gadget}
The set system \((\cU,\cF)\) produced by the label selection construction
has VC-dimension at most \(7\).
\end{lemma}

\begin{proof}
Every element of \(\cU\) is either a row element or a constraint element.

Consider first a row element \(r_i\).  By the definition of the label
selection construction, \(r_i\) belongs to every set \(S_{i,a}\) in row
\(i\), and it belongs to no set in any other row.  Hence we may take
\[
  I(r_i)=\{i\}
  \qquad\text{and}\qquad
  A_i(r_i)=\emptyset.
\]
Thus \(r_i\) has one active row and its only blocking list has size \(0\).

Now consider a constraint element \(u_{e,\lambda,\mu}\), where
\(e=x_ix_j\).  The incidence rule gives
\[
  u_{e,\lambda,\mu}\in S_{i,\gamma}
  \quad\Longleftrightarrow\quad
  \gamma\ne\lambda
\]
and
\[
  u_{e,\lambda,\mu}\in S_{j,\gamma}
  \quad\Longleftrightarrow\quad
  \gamma\ne\mu.
\]
No set from any other row contains this element.  Therefore we may take
\[
  I(u_{e,\lambda,\mu})=\{i,j\},
  \qquad
  A_i(u_{e,\lambda,\mu})=\{\lambda\},
  \qquad
  A_j(u_{e,\lambda,\mu})=\{\mu\}.
\]
Thus \(u_{e,\lambda,\mu}\) has two active rows, and each of its blocking
lists has size \(1\).

Consequently, every universe element has at most two active rows and every
blocking list has size at most \(1\).  The set system is therefore
\(1\)-local.  Applying \Cref{lem:vc-local-list-traces} with \(q=1\) shows
that its VC-dimension is at most \(7\).
\end{proof}

\begin{restatable}[Near-optimal time lower bound for VC-dimension at most seven]{theorem}{vcsevendimeththm}
\label{thm:vc-no-aptas}
Assume ETH.  There exist constants \(\alpha,C>0\) such that, for every
computable function \(F\), no algorithm running in time
\[
  F(k)N^{k/\log^C k}
\]
distinguishes between
\[
  \opt(\cU,\cF,t)\le k
  \qquad\text{and}\qquad
  \opt(\cU,\cF,t)>(1+\alpha)k,
\]
even when \(t=|\cU|\) and the set system \((\cU,\cF)\) has VC-dimension at
most \(7\).  Consequently, \psc parameterized by \(k\) admits no \pas on
set systems of VC-dimension at most \(7\), and hence no \epas.
\end{restatable}

\begin{proof}
Fix a constant \(\zeta\in(0,1)\).  Define \(\eta=1-\zeta\), and choose
\(\alpha=\eta/4\), so \(0<\alpha<\eta/2\).  Let \(C_\zeta\) and
\(D_\zeta\) be the constants supplied by
\Cref{lem:vc-csp-eth-source}.  We reduce from the corresponding
\(D_\zeta\)-regular source gap: distinguish a satisfiable 2-CSP
\(\Lambda\) on \(k\) variables from one satisfying
\[
  \operatorname{val}(\Lambda)\le\zeta=1-\eta.
\]
Apply the label selection construction to \(\Lambda\), and denote its output
by \((\cU,\cF,t)\).  Write \(s=|\Sigma|\).  We have \(|\cF|=ks\).  The
source constraint graph has \(|E|=D_\zeta k/2\), and hence
\[
  |\cU|\le k+\frac{D_\zeta k}{2}s^2.
\]
The row elements contribute \(ks\) to \(\sum_{S\in\cF}|S|\), while the
constraint elements contribute at most
\[
  D_\zeta ks^2(s-1).
\]
Since \(D_\zeta\) is fixed once \(\zeta\) is fixed, it follows that
\(N=\cO(ks^3)\).  In particular, there is a constant
\(c_{\mathrm{lab}}>0\) such that
\[
  N\le(ks)^{c_{\mathrm{lab}}}.
\]
The construction preserves the parameter \(k\).

If \(\Lambda\) is satisfiable, then
\Cref{lem:vc-label-correctness} gives
\[
  \opt(\cU,\cF,t)\le k.
\]
If \(\operatorname{val}(\Lambda)\le1-\eta\), then the same lemma, together
with \(\alpha<\eta/2\), gives
\[
  \opt(\cU,\cF,t)>(1+\alpha)k.
\]
Moreover, \Cref{lem:vc-label-gadget} shows that the output set system has
VC-dimension at most \(7\).

By \Cref{lem:vc-csp-eth-source}, for every computable function \(G\),
no algorithm running in time
\[
  G(k)s^{k/\log^{C_\zeta}k}
\]
distinguishes between satisfiable source instances and source instances
satisfying \(\operatorname{val}(\Lambda)\le\zeta\).  Assume, for
contradiction, that
for some computable function \(F\), there is an algorithm running in time
\(F(k)N^{k/\log^{C_\zeta+1}k}\) that distinguishes the two target cases on
the constructed instances.  Let \(d\) be a constant such that the
construction time is at most \(N^d\).  Choose \(k_\star\) so that, for
every \(k\ge k_\star\),
\[
  \frac{k}{\log^{C_\zeta+1}k}\ge d
  \qquad\text{and}\qquad
  \log k\ge c_{\mathrm{lab}}.
\]
For such \(k\), the construction time is absorbed by the target running
time.  Using \(N\le(ks)^{c_{\mathrm{lab}}}\), the composed algorithm for
the source instance runs in time at most
\[
  F_1(k)N^{k/\log^{C_\zeta+1}k}
  \le
  F_2(k)s^{c_{\mathrm{lab}}k/\log^{C_\zeta+1}k}
  \le
  F_2(k)s^{k/\log^{C_\zeta}k},
\]
where \(F_1\) and \(F_2\) are computable functions absorbing factors that
depend only on \(k\).  This contradicts
\Cref{lem:vc-csp-eth-source}, and proves the theorem with
\(C=C_\zeta+1\).

Now consider a \((1+\alpha/2)\)-approximation, and let \(\cA\) be the
feasible cover it returns.  In a YES instance,
\[
  |\cA|
  \le\left(1+\frac{\alpha}{2}\right)\opt(\cU,\cF,t)
  \le\left(1+\frac{\alpha}{2}\right)k.
\]
In a NO instance, every feasible cover has size greater than
\[
  (1+\alpha)k
  >
  \left(1+\frac{\alpha}{2}\right)k.
\]
Thus comparing \(|\cA|\) with \((1+\alpha/2)k\) distinguishes the two
cases, so such an approximation cannot have running time
\(F(k)N^{k/\log^C k}\).

Finally, suppose that \psc admitted a \pas on this class.  Running it with
the fixed accuracy \(\varepsilon_0=\alpha/2\) would give a
\((1+\alpha/2)\)-approximation in time
\[
  f(k,\varepsilon_0)N^{g(\varepsilon_0)}.
\]
Since \(g(\varepsilon_0)\) is constant, it is at most
\(k/\log^C k\) for every \(k\ge k_1\), for some \(k_1\).  On all instances
with \(k\ge k_1\), this contradicts the preceding lower bound.  Since every
\epas is also a \pas, the \epas exclusion follows.
\end{proof}

\paragraph{List construction with parameter \(q\).}
The label-selection construction proves only a constant gap: if a cover
selects more than one label in many rows, the decoding argument loses those
rows.  The list construction below removes this limitation by adding, for
each constraint, elements that rule out every pair of selected label lists
of size at most \(q\) that contains no accepting pair.  This allows the
soundness proof to decode a satisfying assignment on all non-heavy rows,
where a row is heavy only if it selects more than \(q\) labels.

Fix \(\rho>1\), and choose an integer \(q>4(\rho-1)\).  Set
\(\zeta=1/(4q^2)\), and let \(D_q:=D_\zeta\) be the degree guaranteed by
\Cref{lem:vc-csp-eth-source}.  The reduction takes as input a 2-CSP
instance \(\Lambda=(X,\Sigma,E,\{R_e\}_{e\in E})\) on \(k\) variables whose
constraint graph is \(D_q\)-regular.  Let \(s=|\Sigma|\).  Since the
constraint graph is \(D_q\)-regular, \(m:=|E|=D_qk/2\).

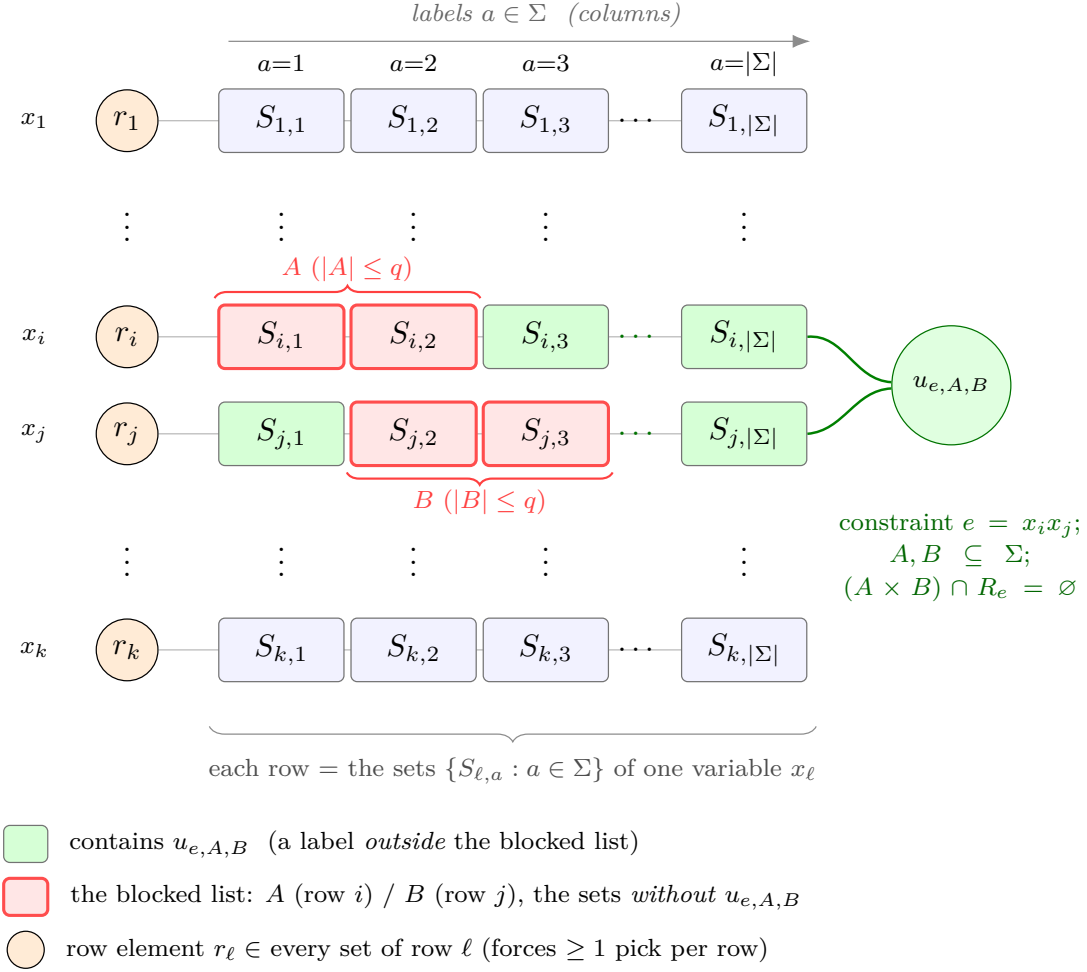
\begin{figure}[ht!]
  \centering
  \resizebox{0.95\textwidth}{!}{\input{figures/list_construction.tex}}
  \caption{The list construction with parameter \(q\).  For a constraint
  \(e=x_ix_j\), each element \(u_{e,A,B}\) corresponds to two lists
  \(A,B\subseteq\Sigma\) of size at most \(q\) with
  \((A\times B)\cap R_e=\emptyset\).  It is contained in the sets of rows
  \(i\) and \(j\) whose labels lie outside \(A\) and \(B\), respectively, and
  therefore blocks the simultaneous selection of labels contained in these
  two lists.}
  \label{fig:list-construction}
\end{figure}

The universe contains one row element \(r_i\) for every \(i\in[k]\).  For
each constraint, write \(e=x_ix_j\), fixing this order so that the two
coordinates of \(R_e\) correspond to \(x_i\) and \(x_j\), respectively, and
for every pair of nonempty sets
\(A,B\subseteq\Sigma\) satisfying
\[
  |A|\le q,
  \qquad
  |B|\le q,
  \qquad
  (A\times B)\cap R_e=\emptyset,
\]
introduce an element \(u_{e,A,B}\).  These are all the elements of \(\cU\) (illustrated in \Cref{fig:list-construction}).

For every row \(\ell\in[k]\) and label \(a\in\Sigma\), define
\(S_{\ell,a}\subseteq\cU\) as follows.  The set \(S_{\ell,a}\) contains
\(r_\ell\) and no other row element.  For a constraint element \(u_{e,A,B}\),
where \(e=x_ix_j\), define
\[
  u_{e,A,B}\in S_{\ell,a}
  \quad\Longleftrightarrow\quad
  (\ell=i\text{ and }a\notin A)
  \ \text{or}\
  (\ell=j\text{ and }a\notin B).
\]
Thus \(u_{e,A,B}\) belongs to every set of row \(i\) whose label is outside
\(A\), to every set of row \(j\) whose label is outside \(B\), and to no
set from another row.

Let \(\cF_i=\{S_{i,a}:a\in\Sigma\}\) and
\(\cF=\bigcup_{i=1}^k\cF_i\).  Finally, set \(t=|\cU|\).  For a selected
subfamily \(\cA\subseteq\cF\) and a constraint element \(u_{e,A,B}\), where
\(e=x_ix_j\), the element \(u_{e,A,B}\) is uncovered exactly when
\[
  L_i(\cA)\subseteq A
  \qquad\text{and}\qquad
  L_j(\cA)\subseteq B.
\]

It remains to bound the encoding size.  Recall that \(s=|\Sigma|\).  The
number of nonempty subsets of \(\Sigma\) having size at most \(q\) is
\[
  B_q(s)
  :=
  \sum_{\ell=1}^{\min\{q,s\}}\binom{s}{\ell}
  \le q s^q.
\]
Therefore each constraint contributes at most \(B_q(s)^2\) elements.  Each
such element belongs to exactly \((s-|A|)+(s-|B|)\le2s\) sets.
Consequently,
\[
\begin{aligned}
  |\cF|&=ks,\\
  |\cU|&\le k+mB_q(s)^2,\\
  \sum_{S\in\cF}|S|&\le ks+2smB_q(s)^2.
\end{aligned}
\]
Using \(m=D_qk/2\) and \(B_q(s)\le qs^q\), we obtain
\[
\begin{aligned}
  N
  &=|\cU|+|\cF|+\sum_{S\in\cF}|S|\\
  &\le k+2ks+(1+2s)mB_q(s)^2\\
  &\le k+2ks+\frac{D_qk}{2}(1+2s)q^2s^{2q}.
\end{aligned}
\]
For fixed \(q\), the degree \(D_q\) is constant, and hence
\(N=\cO(ks^{2q+1})\), where the hidden constant depends only on \(q\).
Thus there exists a constant \(c_q>0\), depending only on \(q\), such that
\[
  N\le(k|\Sigma|)^{c_q}.
\]
The construction is polynomial for fixed \(q\) and preserves the parameter
\(k\).

\begin{lemma}[Correctness of the list construction with parameter \(q\)]
\label{lem:vc-q-list-correctness}
Let \((\cU,\cF,t)\) be the instance obtained from \(\Lambda\) by the list
construction with parameter \(q\).  If \(\Lambda\) is satisfiable, then
\(\opt(\cU,\cF,t)\le k\).  If
\(\operatorname{val}(\Lambda)\le1/(4q^2)\) and \(q>4(\rho-1)\), then every
cover has size greater than \(\rho k\).
\end{lemma}

\begin{proof}
For completeness, let \(\sigma\) satisfy every constraint and select
\(S_{i,\sigma(x_i)}\) from each row.  These sets cover all row elements.  If
some \(u_{e,A,B}\) were uncovered, then
\((\sigma(x_i),\sigma(x_j))\in A\times B\).  This is impossible because
that pair is accepting, whereas \(A\times B\) contains no accepting pair.

For soundness, suppose that a cover \(\cA\) has size at most \(\rho k\), and
use the lists \(L_i(\cA)\) from \Cref{def:vc-row-blocking}.  Every
\(L_i(\cA)\) is nonempty, because \(r_i\) is covered only by sets from row
\(i\).  Moreover, the rows \(\cF_1,\ldots,\cF_k\) partition \(\cF\), so
each selected set in \(\cA\) is counted in exactly one list \(L_i(\cA)\).
Hence
\[
  \sum_{i=1}^k|L_i(\cA)|=|\cA|\le\rho k.
\]
Call a variable \(x_i\) heavy if
\(|L_i(\cA)|>q\).  A heavy variable uses at least \(q\) labels beyond its
mandatory first label, so the number \(h\) of heavy variables satisfies
\[
  h\le\frac{(\rho-1)k}{q}.
\]
At most \(D_qh\) constraints are incident with a heavy variable.  Since
\(m=D_qk/2\), their number is at most
\[
  \frac{2(\rho-1)}{q}m<\frac m2.
\]
Thus more than \(m/2\) constraints have two endpoints that are not heavy.

Consider a constraint \(e=x_ix_j\) whose endpoints are not heavy.  Then
\(|L_i(\cA)|\le q\) and \(|L_j(\cA)|\le q\).  Suppose that
\((L_i(\cA)\times L_j(\cA))\cap R_e=\emptyset\).  Then the construction
introduces the element \(u_{e,L_i(\cA),L_j(\cA)}\).  Since every selected
set in row \(i\) has its label in \(L_i(\cA)\), none of those sets contains
this element; similarly, no selected set from row \(j\) contains it.  Sets
from rows other than \(i\) and \(j\) also do not contain it.  Thus
\(u_{e,L_i(\cA),L_j(\cA)}\) is uncovered, contradicting that \(\cA\) is a
cover.  Hence \((L_i(\cA)\times L_j(\cA))\cap R_e\ne\emptyset\) for every
constraint whose endpoints are not heavy.

Now form a random assignment as follows.  For each non-heavy variable
\(x_i\), choose \(\tau(x_i)\) independently and uniformly from
\(L_i(\cA)\).  For each heavy variable \(x_i\), choose an arbitrary label
from the nonempty set \(L_i(\cA)\).  If \(e=x_ix_j\) has two non-heavy
endpoints, then \(L_i(\cA)\times L_j(\cA)\) contains some pair in \(R_e\).
Since both lists have size at most \(q\), this particular pair is chosen
with probability at least \(1/q^2\), and then \(e\) is satisfied.  More
than \(m/2\) constraints have two non-heavy endpoints, so the expected
number of satisfied constraints is greater than \(m/(2q^2)\).  Therefore
some deterministic assignment satisfies more than \(m/(2q^2)\) constraints,
giving value greater than \(1/(2q^2)\).  This contradicts
\(\operatorname{val}(\Lambda)\le1/(4q^2)\).
\end{proof}

\begin{lemma}[VC-dimension of the list construction]
\label{lem:vc-q-list-dimension}
The set system produced by the list construction with parameter \(q\) has
VC-dimension \(\cO(\log(q+1))\).
\end{lemma}

\begin{proof}
For a row element \(r_i\), set \(I(r_i)=\{i\}\) and
\(A_i(r_i)=\emptyset\).  The row element \(r_i\) belongs to every set
\(S_{i,a}\) in row \(i\), because \(a\notin\emptyset\) for every
\(a\in\Sigma\), and it belongs to no set from any row \(\ell\ne i\).
Thus the active-row description agrees with the row-element incidences.

Now consider a constraint element \(u_{e,A,B}\), where \(e=x_ix_j\).  Set
\(I(u_{e,A,B})=\{i,j\}\), \(A_i(u_{e,A,B})=A\), and
\(A_j(u_{e,A,B})=B\).  By the definition of the list construction, for
every label \(a\in\Sigma\),
\[
  u_{e,A,B}\in S_{i,a}
  \quad\Longleftrightarrow\quad
  a\notin A,
\]
and
\[
  u_{e,A,B}\in S_{j,a}
  \quad\Longleftrightarrow\quad
  a\notin B.
\]
Moreover, no set from a row \(\ell\notin\{i,j\}\) contains
\(u_{e,A,B}\).  Thus \(u_{e,A,B}\) has active rows exactly \(i\) and \(j\),
with blocking lists \(A\) and \(B\), as required by
\Cref{def:vc-row-blocking}.

Every universe element has at most two active rows.  Row elements have
blocking lists of size \(0\), while every constraint element has blocking
lists \(A\) and \(B\), both of size at most \(q\).  Therefore the set system
is \(q\)-local.  Applying \Cref{lem:vc-local-list-traces} gives
VC-dimension \(\cO(\log(q+1))\).
\end{proof}

\begin{restatable}[Near-optimal time lower bound for every fixed factor]{theorem}{vcfixedfactorhardthm}
\label{thm:vc-any-constant}
Assume ETH.  For every fixed \(\rho>1\), there are constants
\(d_\rho=\cO(\log(\rho+1))\) and \(C_\rho>0\) such that, for every
computable function \(F\), no algorithm can, on set systems of
VC-dimension at most \(d_\rho\), distinguish
\[
  \opt(\cU,\cF,|\cU|)\le k
  \qquad\text{from}\qquad
  \opt(\cU,\cF,|\cU|)>\rho k
\]
in time
\[
  F(k)N^{k/\log^{C_\rho}k}.
\]
Consequently, no \(\rho\)-approximation for this class can run in time
\(F(k)N^{k/\log^{C_\rho}k}\) for every computable function \(F\).  In
particular, for every fixed \(\rho>1\), \(\psc\) on set systems of
VC-dimension at most \(d_\rho\) admits no \(\rho\)-approximation running in
time \(f(k)N^{\cO(1)}\) for any computable function \(f\).
\end{restatable}

\begin{proof}
Let \(q\) be the smallest integer strictly larger than \(4(\rho-1)\).  Then
\(q=\cO(\rho)\), and \(q\) is fixed throughout the proof.  Set
\(\zeta=1/(4q^2)\), and let \(C_0=C_\zeta\) be the exponent supplied by
\Cref{lem:vc-csp-eth-source}.  Thus, writing \(s=|\Sigma|\), no
algorithm running in time \(F(k)s^{k/\log^{C_0}k}\), for any computable
function \(F\), distinguishes satisfiable source instances from instances
with value at most \(\zeta\).

Apply the list construction with parameter \(q\) to a source instance
\(\Lambda\).  By \Cref{lem:vc-q-list-correctness}, satisfiable instances
produce set cover instances with optimum at most \(k\), while instances with
\(\operatorname{val}(\Lambda)\le\zeta\) produce set cover instances with
optimum greater than \(\rho k\).  By \Cref{lem:vc-q-list-dimension}, the
VC-dimension is at most
\(d_\rho=\cO(\log(q+1))=\cO(\log(\rho+1))\).  The construction gives
\(N\le(k|\Sigma|)^{c_q}\) for a constant \(c_q\) depending only on \(q\).

Suppose, for contradiction, that a target algorithm distinguishes the two
cases in time \(F(k)N^{k/\log^{C_0+1}k}\).  Since \(q\) is fixed, the
construction is polynomial in \(N\); let its running time be at most
\(N^d\).  Choose \(k_\star\) so that \(k/\log^{C_0+1}k\ge d\) and
\(\log k\ge c_q\) for every \(k\ge k_\star\).  For such \(k\), the
construction time is absorbed by the target running time, and the composed
algorithm for the source instance runs in time at most
\[
  F_1(k)N^{k/\log^{C_0+1}k}
  \le
  F_2(k)s^{c_qk/\log^{C_0+1}k}
  \le
  F_2(k)s^{k/\log^{C_0}k},
\]
where \(F_1\) and \(F_2\) absorb factors depending only on \(k\).  This
contradicts the source lower bound.  Hence the claimed gap lower bound
holds with \(C_\rho=C_0+1\).

Finally, a \(\rho\)-approximation would distinguish the two cases: in the
YES case it returns a cover of size at most \(\rho k\), while in the NO case
every feasible cover has size greater than \(\rho k\).  Therefore such an
approximation cannot run in time \(F(k)N^{k/\log^{C_\rho}k}\) for every
computable function \(F\).  In particular, if a \(\rho\)-approximation ran
in time \(f(k)N^a\) for some constant \(a\), then
there exists \(k_1\) such that \(a\le k/\log^{C_\rho}k\) for every
\(k\ge k_1\), again contradicting the gap lower bound.
\end{proof}

The exponent \(C_\rho\) in \Cref{thm:vc-any-constant} is not uniform in
\(\rho\): in the proof, \(C_\rho=C_{1/(4q^2)}+1\), where
\(q=\Theta(\rho)\), and \Cref{lem:vc-csp-eth-source} allows
\(C_\zeta\) to depend on the soundness parameter \(\zeta\).  Thus
\Cref{thm:vc-any-constant} gives a separate near-optimal lower bound for
each fixed approximation factor \(\rho\).  The next result records a
different consequence in which the polynomial exponent in \(N\) is uniform,
while the forbidden approximation ratio grows with the VC-dimension.

\begin{restatable}[Exponential dependence on VC-dimension]{theorem}{vcexponentialdthm}
\label{thm:vc-exp-d}
Assume ETH.  There is a constant \(\gamma>0\) such that, for infinitely many
integers \(d\), for every computable function \(f\) and every constant
\(a\), no algorithm running in time \(f(k,d)N^a\) can distinguish, on set
systems of VC-dimension at most \(d\), between
\[
  \opt(\cU,\cF,|\cU|)\le k
  \qquad\text{and}\qquad
  \opt(\cU,\cF,|\cU|)>2^{\gamma d}k.
\]
Consequently, \(\psc\) on set systems of VC-dimension \(d\), parameterized
by \(k+d\), has no approximation algorithm with ratio \(2^{o(d)}\) and
running time \(f(k,d)N^{\cO(1)}\), for any computable function \(f\).
\end{restatable}

\begin{proof}
Let \(C>0\) be an absolute constant such that the VC-dimension in
\Cref{lem:vc-q-list-dimension} is at most \(C\log_2(q+1)\) for the list
construction with parameter \(q\).  We associate one VC-dimension bound with
each positive integer \(q\) by setting \(d_q=\lceil C\log_2(q+1)\rceil\).

Next set \(\rho_q=1+q/8\).  This choice ensures
\(q>4(\rho_q-1)\), which is the condition required in
\Cref{lem:vc-q-list-correctness}.  Apply
\Cref{lem:vc-csp-eth-source} with soundness parameter
\(\zeta=1/(4q^2)\).  The source problem is to distinguish satisfiable
2-CSP instances from instances of value at most \(\zeta\).  Apply the list
construction with parameter \(q\) to such a source instance \(\Lambda\).  By
\Cref{lem:vc-q-list-correctness}, the resulting set systems distinguish
\(\opt(\cU,\cF,|\cU|)\le k\) from
\(\opt(\cU,\cF,|\cU|)>\rho_q k\).  By
\Cref{lem:vc-q-list-dimension}, their VC-dimension is at most \(d_q\).

Choose \(\gamma>0\) small enough that \(\gamma C\le1/4\).  Since
\(d_q\le C\log_2(q+1)+1\), we have
\(2^{\gamma d_q}\le 2^\gamma(q+1)^{1/4}\).  Choose \(q_0\) such that
\(2^\gamma(q+1)^{1/4}\le 1+q/8\) for every \(q\ge q_0\).  Then
\(\rho_q\ge2^{\gamma d_q}\) for every \(q\ge q_0\).  The sequence
\((d_q)_{q\ge q_0}\) is unbounded, so after discarding repetitions we obtain
infinitely many values of \(d\).  For each such \(d=d_q\), the constructed
instances distinguish optimum at most \(k\) from optimum greater than
\(2^{\gamma d}k\).

It remains to justify the running-time lower bound for these values of
\(d\).  Fix \(q\ge q_0\).  The source hardness
\Cref{lem:vc-csp-eth-source}, with soundness parameter \(1/(4q^2)\),
gives a constant \(C_q>0\) such that no algorithm solves the corresponding
source gap in time \(F(k)|\Sigma|^{k/\log^{C_q}k}\), for any computable
function \(F\).  The list construction has output size
\(N\le(k|\Sigma|)^{c_q}\), where \(c_q\) depends only on \(q\).  Let
\(d_{\mathrm{red}}\) be a constant such that the construction time is at
most \(N^{d_{\mathrm{red}}}\).  Since \(q\) is fixed, the quantities
\(d_q,c_q,a\), and \(d_{\mathrm{red}}\) are constants.  Choose \(k_\star\)
so that \(c_q\max\{a,d_{\mathrm{red}}\}\le k/\log^{C_q}k\) for every
\(k\ge k_\star\).  Then the reduction together with any target algorithm
running in time \(f(k,d_q)N^a\) gives a source algorithm running in time
\(F(k)|\Sigma|^{k/\log^{C_q}k}\), after absorbing factors depending only on
\(k\) into \(F\).  This contradicts the source hardness.  Hence the claimed
lower bound holds for every \(d=d_q\) in the infinite subsequence.

Finally, suppose, toward contradiction, that there is an approximation
algorithm with ratio \(g(d)=2^{o(d)}\) and running time \(f(k,d)N^a\), for
some computable function \(f\) and constant \(a\).  Since
\(g(d)=2^{o(d)}\), there exists \(d_1\) such that
\(g(d)<2^{\gamma d}\) for every \(d\ge d_1\).  Choose \(q\) from the
infinite subsequence with \(d_q\ge d_1\).  On the corresponding instances,
the algorithm returns a cover of size at most
\(g(d_q)k<2^{\gamma d_q}k\) in the YES case, while every feasible cover has
size greater than \(2^{\gamma d_q}k\) in the NO case.  Thus it distinguishes
the two cases in time \(f(k,d_q)N^a\), contradicting the lower bound just
proved.
\end{proof}

%% file: figures/clique_construction.tex
\begin{tikzpicture}[
  font=\small,
  base/.style ={draw=black!55, rounded corners=2pt, minimum width=1.5cm, minimum height=0.74cm, fill=blue!5},
  ucell/.style={draw=black!55, rounded corners=2pt, minimum width=1.5cm, minimum height=0.74cm, fill=green!16},
  hole/.style ={draw=red!70, line width=1pt, rounded corners=2pt, minimum width=1.5cm, minimum height=0.74cm, fill=red!8},
  reln/.style ={draw, circle, minimum size=0.7cm, inner sep=0pt, fill=orange!16},
  uelt/.style ={draw=green!45!black, circle, minimum size=1.35cm, inner sep=0pt, fill=green!12, align=center, font=\footnotesize},
  rin/.style  ={gray!55, thin},
  uin/.style  ={green!50!black, thick},
]

\def\xa{3.4}\def\xb{5.2}\def\xdots{6.5}\def\xc{7.8}
\def\yone{4.5}\def\ydotsA{3.72}\def\yi{2.95}\def\yj{2.05}\def\ydotsB{1.27}\def\yk{0.5}
\def\xr{1.35}\def\xlab{0.35}

\node[font=\footnotesize] at (\xa,5.35) {$x{=}v_1$};
\node[font=\footnotesize] at (\xb,5.35) {$x{=}v_2$};
\node[font=\footnotesize] at (\xc,5.35) {$x{=}v_{n_0}$};
\node[font=\footnotesize\itshape, text=black!60] at (\xb,5.9) {vertices $x\in V(H)$ \ (columns)};
\draw[-{Latex[length=2mm]}, black!45] (2.5,5.55) -- (8.5,5.55);
\node[rotate=90, font=\footnotesize\itshape, text=black!60] at (-0.4,2.5) {positions $i\in[k]$ \ (rows)};

\node[reln] (r1) at (\xr,\yone) {$r_1$};
\node[base] (a1) at (\xa,\yone) {$S_{1,v_1}$};
\node[base] (b1) at (\xb,\yone) {$S_{1,v_2}$};
\node[base] (c1) at (\xc,\yone) {$S_{1,v_{n_0}}$};
\node at (\xdots,\yone) {$\cdots$};
\node[font=\footnotesize] at (\xlab,\yone) {$1$};

\node at (\xr,\ydotsA) {$\vdots$};
\node at (\xa,\ydotsA) {$\vdots$}; \node at (\xb,\ydotsA) {$\vdots$}; \node at (\xc,\ydotsA) {$\vdots$};

\node[reln] (ri) at (\xr,\yi) {$r_i$};
\node[ucell] (ai) at (\xa,\yi) {$S_{i,v_1}$};
\node[hole]  (bi) at (\xb,\yi) {$S_{i,u}$};
\node[ucell] (ci) at (\xc,\yi) {$S_{i,v_{n_0}}$};
\node[text=green!45!black] at (\xdots,\yi) {$\cdots$};
\node[font=\footnotesize] at (\xlab,\yi) {$i$};

\node[reln] (rj) at (\xr,\yj) {$r_j$};
\node[hole]  (aj) at (\xa,\yj) {$S_{j,v}$};
\node[ucell] (bj) at (\xb,\yj) {$S_{j,v_2}$};
\node[ucell] (cj) at (\xc,\yj) {$S_{j,v_{n_0}}$};
\node[text=green!45!black] at (\xdots,\yj) {$\cdots$};
\node[font=\footnotesize] at (\xlab,\yj) {$j$};

\node at (\xr,\ydotsB) {$\vdots$};
\node at (\xa,\ydotsB) {$\vdots$}; \node at (\xb,\ydotsB) {$\vdots$}; \node at (\xc,\ydotsB) {$\vdots$};

\node[reln] (rk) at (\xr,\yk) {$r_k$};
\node[base] (ak) at (\xa,\yk) {$S_{k,v_1}$};
\node[base] (bk) at (\xb,\yk) {$S_{k,v_2}$};
\node[base] (ck) at (\xc,\yk) {$S_{k,v_{n_0}}$};
\node at (\xdots,\yk) {$\cdots$};
\node[font=\footnotesize] at (\xlab,\yk) {$k$};

\begin{scope}[on background layer]
  \draw[rin] (r1.east) -- (c1.center);
  \draw[rin] (ri.east) -- (ci.center);
  \draw[rin] (rj.east) -- (cj.center);
  \draw[rin] (rk.east) -- (ck.center);
\end{scope}

\node[uelt] (u) at (10.5,2.5) {$e_{i,j,u,v}$};
\draw[uin] (u) to[out=176,in=6] (ci.east);
\draw[uin] (u) to[out=184,in=6] (cj.east);
\node[font=\footnotesize, text=green!40!black, align=center, text width=3.6cm]
   at (10.6,0.7) {positions $i<j$;\\ invalid pair $(u,v)$:\\ $u=v$ \ or \ $uv\notin E(H)$};

\draw[decorate,decoration={brace,amplitude=5pt,mirror},black!45]
   ($(ak.south west)+(-0.1,-0.12)$) -- ($(ck.south east)+(0.1,-0.12)$)
   node[midway,below=6pt,font=\footnotesize,text=black!70]
   {each row $=$ the sets $\{S_{i,x}:x\in V(H)\}$ of one position $i$};

\begin{scope}[shift={(0.2,-1.6)}, font=\footnotesize]
  \node[ucell, minimum width=0.5cm, minimum height=0.42cm] (lg) {};
  \node[right=3pt of lg, align=left] {contains $e_{i,j,u,v}$};
  \node[hole, minimum width=0.5cm, minimum height=0.42cm, right=3.2cm of lg] (lh) {};
  \node[right=3pt of lh, align=left] {blocking vertex: the one set in the row \emph{without} $e_{i,j,u,v}$};
\end{scope}
\begin{scope}[shift={(0.2,-2.2)}, font=\footnotesize]
  \node[reln, minimum size=0.42cm] (lr) {};
  \node[right=3pt of lr, align=left]
    {row element $r_i\in$ every set of row $i$ (forces $\ge 1$ pick per row)};
\end{scope}

\end{tikzpicture}

%% file: figures/label_selection_construction.tex
\begin{tikzpicture}[
  font=\small,
  base/.style ={draw=black!55, rounded corners=2pt, minimum width=1.4cm, minimum height=0.74cm, fill=blue!5},
  ucell/.style={draw=black!55, rounded corners=2pt, minimum width=1.4cm, minimum height=0.74cm, fill=green!16},
  hole/.style ={draw=red!70, line width=1pt, rounded corners=2pt, minimum width=1.4cm, minimum height=0.74cm, fill=red!8},
  reln/.style ={draw, circle, minimum size=0.7cm, inner sep=0pt, fill=orange!16},
  uelt/.style ={draw=green!45!black, circle, minimum size=1.15cm, inner sep=0pt, fill=green!12, align=center, font=\footnotesize},
  rin/.style  ={gray!55, thin},
  uin/.style  ={green!50!black, thick},
]

\def\xa{3.2}\def\xb{4.9}\def\xdots{6.15}\def\xc{7.4}
\def\yone{4.5}\def\ydotsA{3.72}\def\yi{2.95}\def\yj{2.05}\def\ydotsB{1.27}\def\yk{0.5}
\def\xr{1.35}\def\xlab{0.35}

\node[font=\footnotesize] at (\xa,5.35) {$a{=}1$};
\node[font=\footnotesize] at (\xb,5.35) {$a{=}2$};
\node[font=\footnotesize] at (\xc,5.35) {$a{=}|\Sigma|$};
\node[font=\footnotesize\itshape, text=black!60] at (\xb,5.9) {labels $a\in\Sigma$ \ (columns)};
\draw[-{Latex[length=2mm]}, black!45] (2.4,5.55) -- (8.0,5.55);

\node[reln] (r1) at (\xr,\yone) {$r_1$};
\node[base] (a1) at (\xa,\yone) {$S_{1,1}$};
\node[base] (b1) at (\xb,\yone) {$S_{1,2}$};
\node[base] (c1) at (\xc,\yone) {$S_{1,|\Sigma|}$};
\node at (\xdots,\yone) {$\cdots$};
\node[font=\footnotesize] at (\xlab,\yone) {$x_1$};

\node at (\xr,\ydotsA) {$\vdots$};
\node at (\xa,\ydotsA) {$\vdots$}; \node at (\xb,\ydotsA) {$\vdots$}; \node at (\xc,\ydotsA) {$\vdots$};

\node[reln] (ri) at (\xr,\yi) {$r_i$};
\node[ucell] (ai) at (\xa,\yi) {$S_{i,1}$};
\node[hole]  (bi) at (\xb,\yi) {$S_{i,\lambda}$};
\node[ucell] (ci) at (\xc,\yi) {$S_{i,|\Sigma|}$};
\node[text=green!45!black] at (\xdots,\yi) {$\cdots$};
\node[font=\footnotesize] at (\xlab,\yi) {$x_i$};

\node[reln] (rj) at (\xr,\yj) {$r_j$};
\node[hole]  (aj) at (\xa,\yj) {$S_{j,\mu}$};
\node[ucell] (bj) at (\xb,\yj) {$S_{j,2}$};
\node[ucell] (cj) at (\xc,\yj) {$S_{j,|\Sigma|}$};
\node[text=green!45!black] at (\xdots,\yj) {$\cdots$};
\node[font=\footnotesize] at (\xlab,\yj) {$x_j$};

\node at (\xr,\ydotsB) {$\vdots$};
\node at (\xa,\ydotsB) {$\vdots$}; \node at (\xb,\ydotsB) {$\vdots$}; \node at (\xc,\ydotsB) {$\vdots$};

\node[reln] (rk) at (\xr,\yk) {$r_k$};
\node[base] (ak) at (\xa,\yk) {$S_{k,1}$};
\node[base] (bk) at (\xb,\yk) {$S_{k,2}$};
\node[base] (ck) at (\xc,\yk) {$S_{k,|\Sigma|}$};
\node at (\xdots,\yk) {$\cdots$};
\node[font=\footnotesize] at (\xlab,\yk) {$x_k$};
\begin{scope}[on background layer]
  \draw[rin] (r1.east) -- (c1.center);
  \draw[rin] (ri.east) -- (ci.center);
  \draw[rin] (rj.east) -- (cj.center);
  \draw[rin] (rk.east) -- (ck.center);
\end{scope}

\node[uelt] (u) at (10.0,2.5) {$u_{e,\lambda,\mu}$};
\draw[uin] (u) to[out=175,in=8] (ci.east);
\draw[uin] (u) to[out=185,in=8] (cj.east);
\node[font=\footnotesize, text=green!40!black, align=center, text width=3.1cm]
   at (10.15,0.9) {constraint $e=x_ix_j$,\\ rejecting pair $(\lambda,\mu)$};

\draw[decorate,decoration={brace,amplitude=5pt,mirror},black!45]
   ($(ak.south west)+(-0.1,-0.12)$) -- ($(ck.south east)+(0.1,-0.12)$)
   node[midway,below=6pt,font=\footnotesize,text=black!70]
   {each row $=$ the sets $\{S_{i,a}:a\in\Sigma\}$ of one variable $x_i$};

\begin{scope}[shift={(0.2,-1.55)}, font=\footnotesize]
  \node[ucell, minimum width=0.5cm, minimum height=0.42cm] (lg) {};
  \node[right=3pt of lg, align=left] {contains $u_{e,\lambda,\mu}$};
  \node[hole, minimum width=0.5cm, minimum height=0.42cm, right=3.4cm of lg] (lh) {};
  \node[right=3pt of lh, align=left] {blocking label: the one set in the row \emph{without} $u_{e,\lambda,\mu}$};
\end{scope}
\begin{scope}[shift={(0.2,-2.15)}, font=\footnotesize]
  \node[reln, minimum size=0.42cm] (lr) {};
  \node[right=3pt of lr, align=left]
    {row element $r_i\in$ every set of row $i$ (forces $\ge 1$ pick per row)};
\end{scope}

\end{tikzpicture}

%% file: figures/list_construction.tex
\begin{tikzpicture}[
  font=\small,
  base/.style ={draw=black!55, rounded corners=2pt, minimum width=1.42cm, minimum height=0.72cm, fill=blue!5},
  ucell/.style={draw=black!55, rounded corners=2pt, minimum width=1.42cm, minimum height=0.72cm, fill=green!16},
  hole/.style ={draw=red!70, line width=1pt, rounded corners=2pt, minimum width=1.42cm, minimum height=0.72cm, fill=red!10},
  reln/.style ={draw, circle, minimum size=0.7cm, inner sep=0pt, fill=orange!16},
  uelt/.style ={draw=green!45!black, circle, minimum size=1.35cm, inner sep=0pt, fill=green!12, align=center, font=\footnotesize},
  rin/.style  ={gray!55, thin},
  uin/.style  ={green!50!black, thick},
  lbrace/.style={decorate, decoration={brace,amplitude=4pt}, red!75, line width=0.7pt},
]

\def\xa{3.1}\def\xb{4.6}\def\xc{6.1}\def\xdots{7.15}\def\xd{8.35}
\def\yone{5.5}\def\ydotsA{4.4}\def\yi{3.05}\def\yj{1.95}\def\ydotsB{0.6}\def\yk{-0.5}
\def\xr{1.35}\def\xlab{0.3}

\node[font=\footnotesize] at (\xa,6.15) {$a{=}1$};
\node[font=\footnotesize] at (\xb,6.15) {$a{=}2$};
\node[font=\footnotesize] at (\xc,6.15) {$a{=}3$};
\node[font=\footnotesize] at (\xd,6.15) {$a{=}|\Sigma|$};
\node[font=\footnotesize\itshape, text=black!60] at (\xc,6.7) {labels $a\in\Sigma$ \ (columns)};
\draw[-{Latex[length=2mm]}, black!45] (2.5,6.4) -- (9.1,6.4);

\node[reln] (r1) at (\xr,\yone) {$r_1$};
\node[base] (a1) at (\xa,\yone) {$S_{1,1}$};
\node[base] (b1) at (\xb,\yone) {$S_{1,2}$};
\node[base] (c1) at (\xc,\yone) {$S_{1,3}$};
\node[base] (d1) at (\xd,\yone) {$S_{1,|\Sigma|}$};
\node at (\xdots,\yone) {$\cdots$};
\node[font=\footnotesize] at (\xlab,\yone) {$x_1$};

\foreach \x in {\xr,\xa,\xb,\xc,\xd} \node at (\x,\ydotsA) {$\vdots$};

\node[reln] (ri) at (\xr,\yi) {$r_i$};
\node[hole]  (ia) at (\xa,\yi) {$S_{i,1}$};
\node[hole]  (ib) at (\xb,\yi) {$S_{i,2}$};
\node[ucell] (ic) at (\xc,\yi) {$S_{i,3}$};
\node[ucell] (id) at (\xd,\yi) {$S_{i,|\Sigma|}$};
\node[text=green!45!black] at (\xdots,\yi) {$\cdots$};
\node[font=\footnotesize] at (\xlab,\yi) {$x_i$};
\draw[lbrace] ($(ia.north west)+(-0.03,0.06)$) -- ($(ib.north east)+(0.03,0.06)$)
   node[midway, above=1.5pt, font=\scriptsize, text=red!75] {$A\ (|A|\le q)$};

\node[reln] (rj) at (\xr,\yj) {$r_j$};
\node[ucell] (ja) at (\xa,\yj) {$S_{j,1}$};
\node[hole]  (jb) at (\xb,\yj) {$S_{j,2}$};
\node[hole]  (jc) at (\xc,\yj) {$S_{j,3}$};
\node[ucell] (jd) at (\xd,\yj) {$S_{j,|\Sigma|}$};
\node[text=green!45!black] at (\xdots,\yj) {$\cdots$};
\node[font=\footnotesize] at (\xlab,\yj) {$x_j$};
\draw[lbrace, decoration={mirror}] ($(jb.south west)+(-0.03,-0.06)$) -- ($(jc.south east)+(0.03,-0.06)$)
   node[midway, below=1.5pt, font=\scriptsize, text=red!75] {$B\ (|B|\le q)$};

\foreach \x in {\xr,\xa,\xb,\xc,\xd} \node at (\x,\ydotsB) {$\vdots$};

\node[reln] (rk) at (\xr,\yk) {$r_k$};
\node[base] (ak) at (\xa,\yk) {$S_{k,1}$};
\node[base] (bk) at (\xb,\yk) {$S_{k,2}$};
\node[base] (ck) at (\xc,\yk) {$S_{k,3}$};
\node[base] (dk) at (\xd,\yk) {$S_{k,|\Sigma|}$};
\node at (\xdots,\yk) {$\cdots$};
\node[font=\footnotesize] at (\xlab,\yk) {$x_k$};

\begin{scope}[on background layer]
  \draw[rin] (r1.east) -- (d1.center);
  \draw[rin] (ri.east) -- (id.center);
  \draw[rin] (rj.east) -- (jd.center);
  \draw[rin] (rk.east) -- (dk.center);
\end{scope}

\node[uelt] (u) at (10.7,2.5) {$u_{e,A,B}$};
\draw[uin] (u) to[out=177,in=5] (id.east);
\draw[uin] (u) to[out=183,in=5] (jd.east);
\node[font=\footnotesize, text=green!40!black, align=center, text width=3.7cm]
   at (10.8,0.55) {constraint $e=x_ix_j$;\\[1pt] $A,B\subseteq\Sigma$;\\[1pt] $(A\times B)\cap R_e=\varnothing$};

\draw[decorate,decoration={brace,amplitude=5pt,mirror},black!45]
   ($(ak.south west)+(-0.1,-0.5)$) -- ($(dk.south east)+(0.1,-0.5)$)
   node[midway,below=6pt,font=\footnotesize,text=black!70]
   {each row $=$ the sets $\{S_{\ell,a}:a\in\Sigma\}$ of one variable $x_\ell$};

\begin{scope}[shift={(0.2,-2.7)}, font=\footnotesize]
  \node[ucell, minimum width=0.5cm, minimum height=0.42cm] (lg) {};
  \node[right=3pt of lg, align=left] {contains $u_{e,A,B}$ \ (a label \emph{outside} the blocked list)};
\end{scope}
\begin{scope}[shift={(0.2,-3.3)}, font=\footnotesize]
  \node[hole, minimum width=0.5cm, minimum height=0.42cm] (lh) {};
  \node[right=3pt of lh, align=left] {the blocked list: $A$ (row $i$) / $B$ (row $j$), the sets \emph{without} $u_{e,A,B}$};
\end{scope}
\begin{scope}[shift={(0.2,-3.9)}, font=\footnotesize]
  \node[reln, minimum size=0.42cm] (lr) {};
  \node[right=3pt of lr, align=left]
    {row element $r_\ell\in$ every set of row $\ell$ (forces $\ge 1$ pick per row)};
\end{scope}

\end{tikzpicture}

%% file: prelims.tex
\section{Notations and Preliminaries}
\label{section: prelims}

In this section, we introduce essential notation related to graphs, set systems, graph structures we use, satisfiability, and other symbols commonly used throughout the paper.

\subsection{Notation for Graphs}
Given a graph \( G \), we denote its vertex set by \( V(G) \) and its edge set by \( E(G) \). For an edge \(e \in E(G)\) joining vertices \(u\) and \(v\), we denote it by \((u,v)\).  
A \emph{bipartite graph} \( G \) is denoted by \(\graph \), where the vertex set \( V(G) \) is partitioned into two disjoint subsets: the red set \( {\color{red}\cR} \) and the blue set \( {\color{blue}\cB} \). We typically denote the edge set by \( E(G) \), though we may omit it for clarity when it is clear from context. Here, we refer to $n$ as the number of vertices in $\cR$ and $\cB$, and we use $m$ to refer to the number of edges in $E(G)$ throughout the paper.

For a graph \( G \) and a vertex \( v \in V(G) \), the \emph{neighborhood} \( \nbr_G(v) \) is defined as the set of vertices adjacent to \( v \). More generally, for a subset \( X \subseteq V(G) \), we define its neighborhood as $\nbr_G(X) = \left(\bigcup_{v \in X} \nbr_G(v)\right) \setminus X$.
When the context clearly specifies the graph, we omit the subscript and simply write \( \nbr(v) \) or \( \nbr(X) \). For a pair of vertices $u, v \in V(G)$, we define $\dist_G(u,v)$ to be the shortest distance between $u$ and $v$ in $G$. Further for a given weight function \( \w: \cB \to \bQ^+ \) and a set \( S \subseteq \cB \), we define \( \w(S) \) as  
\[
\w(S) = \sum_{b \in S} \w(b).
\]

\subsection{Sets, Set Systems, Tuples,  and Operations}
We use \( \mathbb{N} \) and \( \mathbb{Q} \) to denote the sets of natural numbers and rational numbers, respectively. Additionally, we define \( \mathbb{Q}^+ \) to represent the set of positive rational numbers, that is $\mathbb{Q}^+ = \{ q \mid q \in \mathbb{Q}, q > 0 \}$. For a number $\iletter \in \bN$, we use $[\iletter]$, to denote the set $\{1,\dots, \alpha\}$. 
Similarly, for $a,b \in \bN, b \geq a$, we use $[a,b]$ to denote the set $\{a, a+1, \dots, b\}$. For a set $B$, the power set is the set of all possible subsets of $B$, including $\emptyset$ and $B$ itself. We denote the power set of $B$ as $2^B$.

\vspace{0.3 em}
\noindent
\textbf{Set System:} A \emph{set system} \((\cU, \cF)\) consists of a universe \(\cU \coloneqq \{u_1,  \dots, u_n\}\) of \( n \) elements and a family \(\cF \subseteq 2^{\cU}\) containing \( m \) subsets of \(\cU\). For a collection of subsets \(\cS \subseteq \cF\), we define $\cov(\cS) = \bigcup_{F \in \cS} F$.
as the set of elements appearing in at least one set from \(\cS\). If the family \(\cS\) consists of a single subset \( f \), we simplify notation and write \(\cov(f)\) instead of \(\cov(\{f\})\).

\vspace{0.3 em}
\noindent
\textbf{Tuple:} For a set system, we define a \( t \)-tuple \(\mathcal{X} = \langle X_1, \dots, X_t \rangle\), where \( X_i \subseteq \cU \), for all \( i \in [t] \). For an integer \( i \in [t] \), we define \(\mathcal{X}[\downarrow i]\) to be the tuple obtained from \(\mathcal{X}\) by removing the set \( X_i \). Formally, 
\[
\mathcal{X}[\downarrow i] = \langle X_1, \dots, X_{i-1}, X_{i+1}, \dots, X_t \rangle.
\]
Similarly, given a \( t \)-tuple \(\mathcal{X} = \langle X_1, \dots, X_t \rangle\) of subsets of \( \cU \), a subset \( Y \subseteq \cU \), and an integer \( i \in [t+1] \), we define \(\mathcal{X}[\uparrow i,Y]\) as the tuple obtained by inserting the set \( Y \) into \(\mathcal{X}\) at the \( i^{\text{th}} \) position. Formally,
\[
\mathcal{X}[\uparrow i,Y] = \langle X_1, \dots, X_{i-1}, Y, X_{i}, \dots, X_t \rangle.
\]
Note that, for an integer $j\in [t]$,  $\cX[j]=X_j$ be the $j^{\text{th}}$ set of the tuple. 

We similarly define a $t$-tuple in a graph $G$ as $\cS = \langle s_1, \dots, s_t \rangle$, where each $s_i \in V(G)$ for all $i \in [t]$. To define the neighborhood of such a tuple, we treat it as a set and consider the union of the neighborhoods of its elements. More specifically, the neighborhood of a $t$-tuple $\cS = \langle s_1, \dots, s_t\rangle$ in $G$ is defined as
\[
\nbr_G(\cS) = \nbr_G(\{s_1, \dots, s_t\}) =  \left ( \bigcup_{i \in [t]}\nbr_G(s_i) \right ) \setminus \left (  \bigcup_{i \in [t]}s_i\right )
\]

We abuse the notation by using $\nbr_G(\cS)$ to refer to the neighborhood of $\cS$, whether $\cS$ is a set or a tuple. When the context clearly specifies the graph, we omit the subscript and simply write $\nbr(\cS)$.

\subsection{CNF-Satisfiability (CNF-SAT)}
Let \(\Phi\) be a CNF formula on \(n\) variables and \(m\) clauses. We denote the set of variables and clauses in \(\Phi\) by \(\cV_{\Phi}\) and \(\cC_{\Phi}\), respectively; subscripts may be omitted when the context is clear. A variable is said to be \emph{positive} if it does not occur as a negative literal anywhere in the formula; otherwise, it is called a \emph{negative} variable. We use \(\cV_{neg} \subseteq \cV_{\Phi}\) to denote the set of negative variables. Let \(\cC_{neg} \subseteq \cC_{\Phi}\) denote the set of clauses containing at least one negative literal, and define \(\cC_{pos} = \cC_{\Phi} \setminus \cC_{neg}\) as the set of clauses without any negative literals.

For a clause \(c \in \cC_{\Phi}\), let \(\n{c}\) denote the set of variables that appear as negative literals in \(c\). For a set of clauses \(C \subseteq \cC_{\Phi}\), we define
\[
\n{C} = \bigcup_{c \in C} \n{c}.
\]

Given a weight function \(\w: \cC_{\Phi} \to \bQ^+\), the weight of a set of clauses \(C \subseteq \cC_{\Phi}\) is defined as
\[
\w(C) = \sum_{c \in C} \w(c).
\]

An assignment \(\psi\) is a binary function on \(\cV_{\Phi}\), i.e., \(\psi: \cV_{\Phi} \rightarrow \{0, 1\}\). The \emph{weight} of an assignment \(\psi\), denoted by \(\wt{\Phi}{\psi}\), is the number of variables set to \(1\) by \(\psi\). For an assignment \(\psi\), we denote by \(\sat{\Phi}{\psi}\) the set of clauses of \(\Phi\) satisfied by \(\psi\). The incidence graph of $\Phi$, denoted by $G_{\Phi} = (P, Q)$, is the bipartite graph that has a vertex in $P$ for each variable in $\cV_{\Phi}$, a vertex in $Q$ for each clause in $\cC_{\Phi}$ and an edge $(u, v)$ if the variable corresponding to $u \in P$ occurs as a positive or negative literal in the clause corresponding to $v \in Q$.

\subsection{Matroids and representative families}\label{subsec:matroidprelim}
A matroid $\matfull$ consists of a finite ground set $\cE$ and a nonempty
family $\cI\subseteq 2^{\cE}$ such that (i) every subset of a member of $\cI$
belongs to $\cI$, and (ii) if $A,B\in\cI$ and $|A|<|B|$, then some
$b\in B\setminus A$ satisfies $A\cup\{b\}\in\cI$. Members of $\cI$ are
called \emph{independent}. A maximal independent set is a \emph{basis}, and
the rank of $\mat$ is the common size of its bases. Throughout, a matroid is
given by an independence oracle.

A family whose members all have size $p$ is called a $p$-family. Let
$\cA$ be a $p$-family of independent sets in $\mat$. A subfamily
$\cA'\subseteq\cA$ \emph{$q$-represents} $\cA$ if, whenever $B\subseteq\cE$
has size $q$ and $A\cup B$ is independent for some $A\in\cA$, there is an
$A'\in\cA'$ for which $A'\cup B$ is independent. We write
$\cA'\subseteq^q_{rep,\mat}\cA$, omitting $\mat$ when it is clear.

\begin{lemma}[{\cite[Lemma~9]{DBLP:conf/icalp/00020LS0U24}}]
\label{lem:oraclerepset}
Let $\matfull$ be a rank-$k$ matroid given by an independence oracle, and let
$\cA$ be a $1$-family of independent subsets of $\cE$. In polynomial time one
can compute $\cA'\subseteq^{k-1}_{rep,\mat}\cA$ with $|\cA'|\le k$.
\end{lemma}

We next recall the notion of \emph{contraction} in matroids. 
Let $\matfull$ be a matroid, and let $Z \subseteq \cE$ be a subset with $X$ denoting a basis of $Z$. 
We define the family $\cI_Z$ as follows:
\begin{align}\label{eqn:constructedfamily}
    \cI_Z \coloneqq \{ I \subseteq \cE \setminus Z \mid I \cup X \in \cI \}.
\end{align}
Let $\mat / Z = (\cE \setminus Z, \cI_Z)$ denote the set system thus obtained. 
It is well known that $\mat / Z$ is itself a matroid.

\begin{proposition}[\cite{schrijver2003combinatorial}]
\label{prop:contrmatroid}
    $\mat / Z = (\cE \setminus Z, \cI_Z)$ is a matroid.
\end{proposition}

Furthermore, when $Z$ is a singleton set, i.e., $Z = \{v\}$, we write 
$\mat / v = (\cE \setminus \{v\}, \cI_v)$ instead of $\mat / \{v\} = (\cE \setminus \{v\}, \cI_{\{v\}})$.

\subsection{Parameterized Approximation Schemes}
\paragraph{Parameterized Approximation Scheme (\pas).}
A \emph{Parameterized Approximation Scheme (\pas)} is an algorithm that,
given an instance with parameter \(k\) and an accuracy \(\varepsilon>0\),
returns a solution whose value is at least \((1-\varepsilon)\opt\) for a
maximization problem, or at most \((1+\varepsilon)\opt\) for a minimization
problem.  Its running time is
\[
  f(k,\varepsilon)\,n^{g(\varepsilon)},
\]
where \(n\) is the input size and \(f\) and \(g\) are computable functions.

\paragraph{Efficient Parameterized Approximation Scheme (\epas).}
An \emph{Efficient Parameterized Approximation Scheme (\epas)} is a \pas
whose running time is
\[
  f(k,\varepsilon)\,n^{\mathcal{O}(1)},
\]
where the exponent of \(n\) is independent of \(k\) and \(\varepsilon\).
For further details and a comprehensive discussion of these parameterized
approximation classes, we refer the reader to the survey by Feldmann et
al.~\cite{DBLP:journals/algorithms/FeldmannSLM20} and the paper by
Lokshtanov et al.~\cite{DBLP:conf/stoc/LokshtanovPRS17}.

\section{Semi-ladder and Time Complexity Measure}
\label{sec: sml_comp}
An instance of the \prbdsfull problem consists of a bipartite graph \(\graph\). The key definition that allows us to establish upper bounds on the running times of our algorithms is a quantitative measure called a \( d \)-\sml, which captures essential structural properties of the input bipartite graph \( G \). The formal definition of a \( d \)-\sml is presented below.

\begin{figure}[ht!]
    \centering
    \includegraphics[scale=0.33]{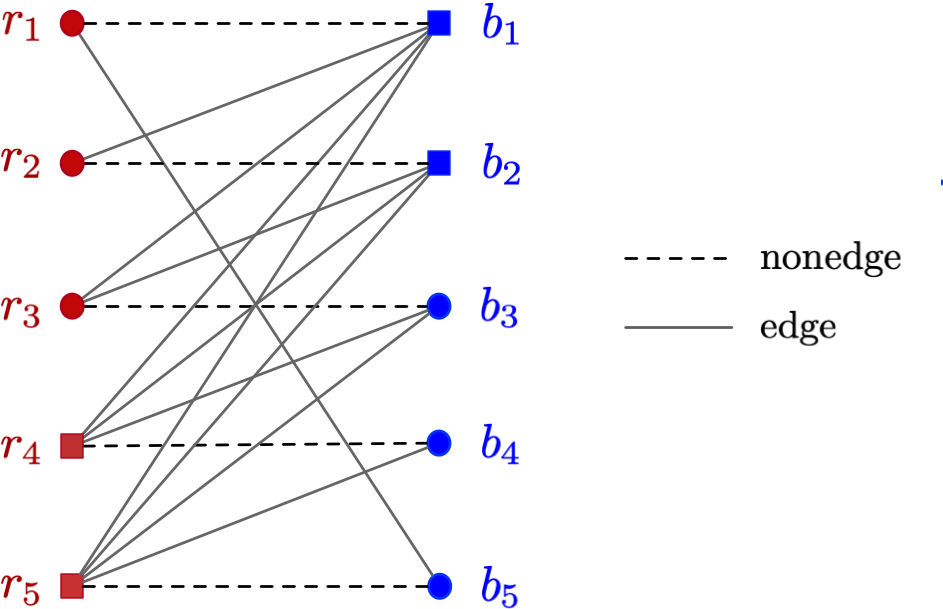}
    \caption[scale=0.40]{A \sml of order $5$. Square vertices induce a $K_{2,2}$}
    \label{fig:semiladder}
\end{figure}

\begin{definition}[$d$-\sml, \cite{DBLP:conf/stacs/FabianskiPST19}]\label{defn:semiladder}
    Given a bipartite graph $\graph$, two sequences, $r_1,\dots,r_d \in \cR$ and $b_1,\dots,b_d \in \cB$ form a \sml of order $d$ in $G$ if $(r_i,b_j) \in E(G)$ for all $i,j \in [d]$ with $i>j$, and $(r_i,b_i)\notin E(G)$ for all $i \in [d]$. See~\Cref{fig:semiladder}. 
    
The \emph{\smlindex} of a bipartite graph is the largest order of any \sml it contains. A class of graphs, $\cC$, is said to have \emph{bounded \smlindex} if there exists a finite upper bound on the \smlindex of all graphs in the class.

\end{definition}

To formally define the complexity measure used to upper bound the running time of our algorithms, we first introduce a few additional definitions. These are inspired by the notion of \emph{$e$-chains} in intersection-closed hypergraphs, as introduced by Guillemot~\cite{DBLP:journals/toct/G25}.

\begin{definition}
     Given a bipartite graph \( \graph \) and a subset \( X \subseteq \mathcal{B} \) we define, $$\incld(G,X)=\{y \in \cR \mid X \subseteq \nbr_G(y)\}.$$ 
\end{definition}
\noindent 
     In other words, $\incld(G,X)$ includes all the vertices in $\cR$ whose neighborhood contains $X$. 

\begin{definition}
     Given a bipartite graph \( \graph \) and a subset \( X \subseteq \mathcal{B} \), the \realext of $X$, denoted by $\reex(G,X)$ is defined as 
     \[
     \reex(G,X)= \bigcap_{y \in \incld(G,X)}\nbr_G(y)
     \]
\end{definition}
     \noindent 
   In other words, for \( X \subseteq \cB \), the set \( \reex(G, X) \) is a subset of \( \cB \) that contains \( X \). In fact, for any such \( X \), the sets \( \incld(G, X) \) and \( \reex(G, X) \) are uniquely defined. Moreover, observe that the neighborhood of every vertex in \( \incld(G, X) \) contains the set \( \reex(G, X) \). Note that, when $\incld(G,X) = \emptyset$, then we define $\reex(G, X) =  \emptyset$. This leads to the following definition.

\begin{definition}\label{defn:realizable}
   Given a bipartite graph \(\graph \), a subset \( X \subseteq \mathcal{B} \) is called {\color{blue} \realizable} if 
   \[X= \reex(G,X) \]
   and the corresponding unique set $\incld(G,X)$ is called as the {\color{red} \realizing} set.
\end{definition}

Note that a \realizable set \( X \subseteq \cB \) is non-empty iff its corresponding \realizing set \( \incld(G, X) \) is also non-empty.

\begin{figure}[ht!]
    \centering
    \includegraphics[scale=0.35]{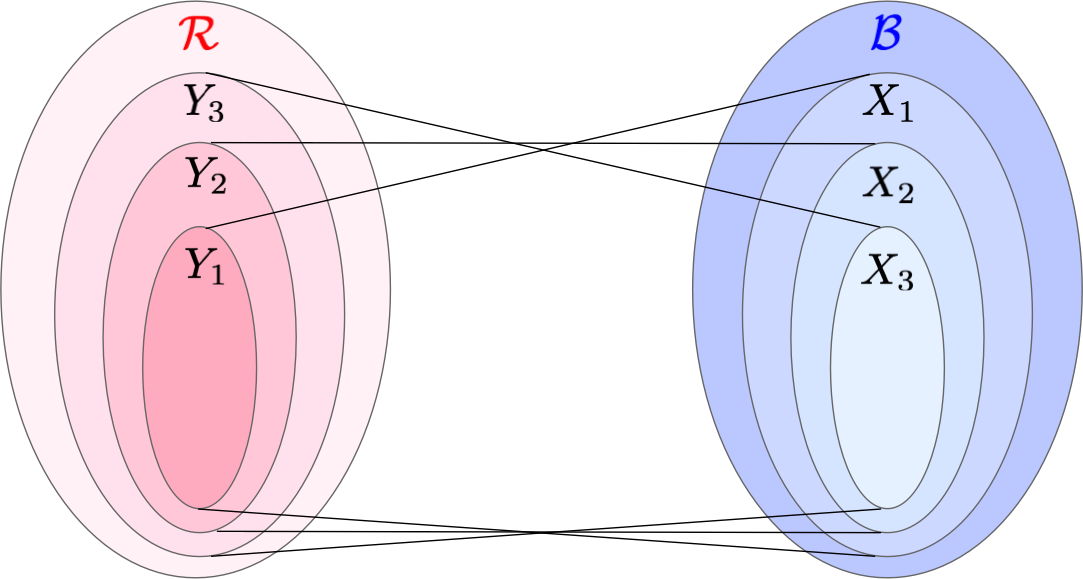}
    \caption[scale=0.4]{\realizable sets $X_1\supsetneq X_2\supsetneq X_3$ and corresponding \realizing sets $Y_1\subsetneq Y_2\subsetneq Y_3$ }
    \label{fig:realizable}
\end{figure}

\begin{definition}[Downward Intersection Complexity  Set]\label{defn:dic}
Let \( G = (\mathcal{R}, \mathcal{B}) \) be a bipartite graph and \( X \subseteq \mathcal{B} \) be a \realizable set. The \emph{downward intersection complexity} of \( X \), denoted by \( \dintcmp(G, X) \), is defined as the largest integer \( \lambda \) such that there exists a sequence of \( \lambda \) \realizable sets starting from \( X \), satisfying
\[
X = X_1 \supsetneq X_2 \supsetneq \dots \supsetneq X_\lambda.
\]
See~\Cref{fig:realizable}. We extend the definition of downward intersection complexity to non-realizable sets \( Y \subseteq \mathcal{B} \) by defining
\[
\dintcmp(G, Y) = \dintcmp(G, \reex(G, Y)).
\]
Finally, the \emph{downward intersection complexity} of the graph \( G \), denoted by \( \dintcmp(G) \), is defined as
\[
\dintcmp(G) = \max_{X \subseteq \cB} \dintcmp(G, X).
\]
We view the downward intersection complexity as a function  
\[
\dintcmp : 2^{\cB} \to \mathbb{N},
\]  
which maps any subset \( X^\star \in 2^{\cB} \) to the complexity value \( \dintcmp(G,  X^\star) \).

\end{definition}

\begin{lemma}\label{lemma:realizablerealizingreln} Given a bipartite graph $\graph$, let $X_1,X_2$ be two \realizable sets and $Y_1,Y_2$ be the corresponding $\realizing$ sets. Then \[ X_1 \subsetneq X_2 \iff Y_1\supsetneq Y_2\]
\end{lemma}

\Cref{fig:realizable} visually represents the appearance of the realizable and realizing sets.
\begin{proof}
Let \( X_1, X_2 \subseteq \cB \) be two \realizable sets, and let \( Y_1 = \incld(G, X_1) \), \( Y_2 = \incld(G, X_2) \) be their corresponding \realizing sets. By definition, we have:
\[
X_1 = \reex(G, X_1) = \bigcap_{y \in Y_1} \nbr_G(y),
\quad \text{and} \quad
X_2 = \reex(G, X_2) = \bigcap_{y \in Y_2} \nbr_G(y).
\]
By assumption, \( X_1 \subsetneq X_2 \), which implies:
\[
Y_2 = \incld(G, X_2) \subseteq \incld(G, X_1) = Y_1.
\]
We now show that this inclusion is strict. Let \( x_2 \in X_2 \setminus X_1 \). Since \( x_2 \notin X_1 = \bigcap_{y \in Y_1} \nbr_G(y) \), there exists \( y_1 \in Y_1 \) such that \( x_2 \notin \nbr_G(y_1) \). Therefore, \( y_1 \notin Y_2 \), implying that \( Y_2 \subsetneq Y_1 \).

\paragraph{Reverse Direction.}
The proof is analogous. Suppose \( Y_1 \supsetneq Y_2 \). Then,
\[
X_1 = \bigcap_{y \in Y_1} \nbr_G(y), \quad 
X_2 = \bigcap_{y \in Y_2} \nbr_G(y),
\]
and it follows that \( X_1 \subseteq X_2 \). Let \( y_1 \in Y_1 \setminus Y_2 \). Since \( y_1 \notin Y_2 \), there exists \( x_2 \in X_2 \) such that \( x_2 \notin \nbr_G(y_1) \). Therefore, \( x_2 \notin X_1 \), and hence \( X_1 \subsetneq X_2 \).
\end{proof}

We now establish a connection between the notions of \sml and downward intersection complexity. Specifically, we show that they are bounded by each other, implying that the downward intersection complexity provides an alternative characterization of the \sml.

\medskip

\begin{lemma}\label{lemma:boundingsmlindex}
For any bipartite graph $\graph$, if $\dintcmp(G)=\lambda$, then the $\smlindex$ of $G$ is at least $\lambda-1$.
\end{lemma}

\begin{figure}[ht!]
    \centering
    \includegraphics[scale=0.35]{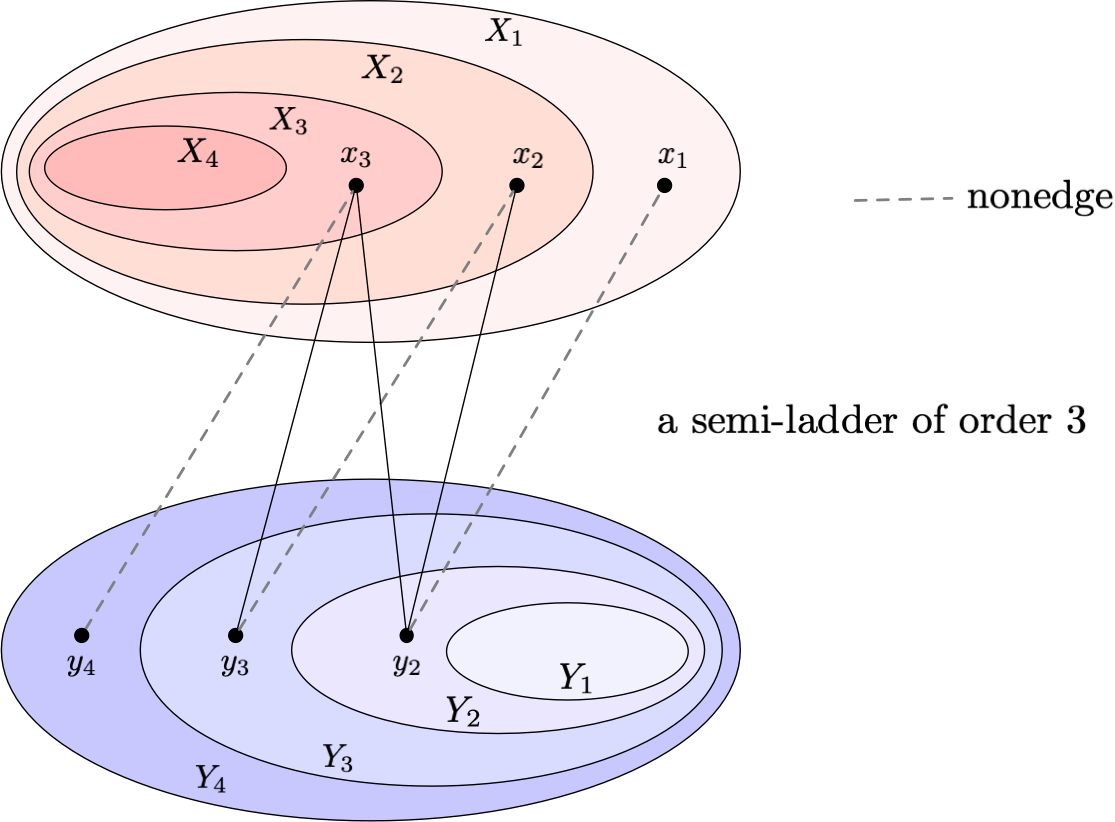}
    \caption[scale=0.4]{$3$ sized corresponding semi ladder of $\lambda = 4$ realizable sets}
    \label{fig:semiladderindex}
\end{figure}

\Cref{fig:semiladderindex} illustrates an example corresponding to \Cref{lemma:boundingsmlindex} when $\lambda = 4$.
\begin{proof}
Since \( \dintcmp(G) = \lambda \), there exists a sequence of \( \lambda \) \realizable sets  
\[
X_1 \supsetneq X_2 \supsetneq \dots \supsetneq X_\lambda.
\]
If \( \lambda = 1 \), the statement holds trivially, as the \smlindex of \( G \) is at least \( \lambda - 1 \geq 0 \). Thus, we assume \( \lambda \geq 2 \). Now consider two cases depending on whether \( X_\lambda = \emptyset \) or \( X_\lambda \neq \emptyset \).

In the first case, we construct a \sml of size \( \lambda - 1 \) using the \( \lambda - 1 \) non-empty \realizable sets \( X_1, \dots, X_{\lambda - 1} \). Observe that in this scenario, there is no vertex \( b \in \cB \) such that \( \nbr(b) = \cR \), because otherwise \( X_\lambda = \emptyset \) would not be \realizable, contradicting the definition.

In the second case, where \( X_\lambda \neq \emptyset \), we can similarly construct a \sml of size \( \lambda - 1 \) using all \( \lambda \) sets in the sequence. 
Since the construction in the latter case is analogous to the former, we omit its details and focus only on the first case.
Let \( X_1, \dots, X_{\lambda} \) be a sequence of \( \lambda  \)  \realizable sets (except $X_\lambda$, all sets are non-empty) such that
\[
X_1 \supsetneq X_2 \supsetneq \dots \supsetneq X_{\lambda }.
\]
Let \( Y_1, \dots, Y_{\lambda } \) be the corresponding \realizing sets, i.e., \( Y_i = \incld(G, X_i) \) for each \( i \in [\lambda ] \). By~\Cref{lemma:realizablerealizingreln}, we have:
\[
Y_1 \subsetneq Y_2 \subsetneq \dots \subsetneq Y_{\lambda }.
\]
We now construct a semi-ladder of order \( \lambda - 1 \). Let \( i \in [\lambda - 1] \). Consider the pair $\left( X_i, Y_{i+1} \right)$. We observe the following:

\begin{enumerate}
\setlength{\itemsep}{-2pt}
    \item \( Y_i \subsetneq Y_{i+1} \) and \( X_i \supsetneq X_{i+1} \),
    \item \( Y_{i+1} \) realizes \( X_{i+1} \) but does not realize \( X_i \),
    \item \( Y_i \) realizes \( X_i \) but does not realize \( X_{i+1} \).
\end{enumerate}
Since \( Y_{i+1} \) realizes \( X_{i+1} \) but not \( X_i \), there exists an element \( x_i \in X_i \) such that
\[
x_i \notin \bigcap_{y \in Y_{i+1}} \nbr_G(y).
\]
This implies that there exists \( y_{i+1} \in Y_{i+1} \) such that \( x_i \notin \nbr_G(y_{i+1}) \).
\noindent
Now consider the two sequences:
\[
x_1, \dots, x_{\lambda - 1} \in \cB \quad \text{and} \quad y_2, \dots, y_{\lambda} \in \cR.
\]
We claim these form a semi-ladder of order \( \lambda - 1 \). To this end, we have the following observations:

\begin{itemize}
\setlength{\itemsep}{-2pt}
    \item By construction, we have \((x_i, y_{i+1} ) \notin E(G) \) for all \( i \in [\lambda - 1] \),
    \item Furthermore, for all \(  i\leq j\) and \(i \in [2, \lambda], j\in [\lambda-1] \), since \( X_j \subseteq X_i = \bigcap_{y \in Y_i} \nbr_G(y) \), it follows that \( x_j \in \nbr_G(y_i) \), i.e., \( (x_j, y_i) \in E(G) \).
\end{itemize}

Thus it satisfies the definition of a semi-ladder of order \( \lambda - 1 \), which implies that \smlindex of $G$  is $\geq \lambda - 1$, concluding the proof.
\end{proof}

\begin{lemma}\label{lemma:boundingdicuic}
  Let $\lambda$ be the \smlindex of $\graph$. Then 
  \[ \dintcmp(G)\geq \lambda  \] 
\end{lemma}
\begin{proof} 
Let \( \lambda \) be the \smlindex of \( G \). Then, by~\Cref{defn:semiladder}, there exist two sequences 
\[
r_1, \dots, r_\lambda \in \cR \quad \text{and} \quad b_1, \dots, b_\lambda \in \cB
\]
such that \( (r_i, b_j) \in E(G) \) for all \( i, j \in [\lambda] \) with \( i > j \), and \( (r_i, b_i) \notin E(G) \) for all \( i \in [\lambda] \).

Note that \( \dintcmp(G) \geq 1 \), since the downward intersection complexity is defined as a function \( \dintcmp : 2^\cB \to \bN \). Therefore, if \( \lambda = 1 \), the claim holds trivially. So, we assume \( \lambda \geq 2 \).

For each \( i \in [\lambda] \), let \( B_i = \{b_1, \dots, b_i\} \). Given a bipartite graph \( G \) and a subset \( X \subseteq \cB \), recall that
\[
\incld(G, X) = \{ y \in \cR \mid X \subseteq \nbr_G(y) \}.
\]
Hence, we obtain the chain of inclusions:
\[
\incld(G, B_\lambda) \subseteq \dots \subseteq \incld(G, B_1).
\]

We now argue that this inclusion is strict. Specifically, observe the following:
\begin{itemize}
    \item For all \( i \in \{2, \dots, \lambda\} \), we have \( r_i \in \incld(G, B_{i-1}) \) since \( (r_i, b_j) \in E(G) \) for all \( j < i \).
    \item For all \( i \in [\lambda] \), we have \( r_i \notin \incld(G, B_i) \) because \( (r_i, b_i) \notin E(G) \).
\end{itemize}

It follows that
\[
\incld(G, B_\lambda) \subsetneq \dots \subsetneq \incld(G, B_1),
\]
which forms a strictly increasing sequence of \( \lambda \) \realizing sets. Then, by~\Cref{lemma:realizablerealizingreln}, we obtain
\[
\reex(G, B_\lambda) \supsetneq \dots \supsetneq \reex(G, B_1),
\]
which forms a strictly decreasing sequence of \( \lambda \) \realizable sets.

Thus, by~\Cref{defn:dic}, we conclude that
\[
\dintcmp(G) \geq \dintcmp(G, \reex(G, B_\lambda)) \geq \lambda.
\]
This concludes the proof.
\end{proof}

Lemmata~\ref{lemma:boundingsmlindex} and \ref{lemma:boundingdicuic} imply the following theorem. 

\begin{theorem}
\label{thm: smli_dic_rel}
    For any bipartite graph $\graph$, let $\lambda$ be the \smlindex then we have that 
\[ \lambda \leq \dintcmp(G) \leq \lambda+1 \]
    
\end{theorem}

%% file: randomized_max_coverage.tex
\section{Probabilistic Branching Algorithms for \mrbds}
\label{section: implement}
 In this section, we present a generic algorithm for the \mrbds problem and analyze its running time as a function of the input parameters \( k \), \( \varepsilon \), and the downward intersection complexity of the input bipartite graph. We begin by describing the algorithm for the unweighted version of the problem, followed by its extension to the weighted setting. This separation is intentional: working with the unweighted case first allows us to convey the key ideas of the algorithm without the additional notational overhead and technical subtleties introduced by weights.

\subsection{Unweighted \mrbds}
\label{section: unweighted_main}
Given an instance $\cI $ = \prbdsi of the \mrbds problem, the optimal value \opti is defined as  
\[
{\sf OPT}_k(\cI) = \max_{\cR' \subseteq \cR, |\cR'| \leq k} |\nbr_G(\cR')|,
\]  
where the goal is to select at most $k$ vertices from $\cR$ to maximize the size of the union of their neighborhoods, in other words maximize the number of vertices covered in $\cB$. When the values of $k$ and the instance $\cI$ are clear from the context, we omit the subscript $k$ and the instance notation in the definition. 
We first provide a high-level description of the algorithm, followed by its pseudocode. We then present an analysis of its running time, along with guarantees on the approximation factor and the success probability.

\subsubsection{Description of the algorithm.} 
Our algorithm is recursive. However, the input bipartite graph $\graph$ remains unchanged throughout all recursive calls. Therefore, the neighborhood function does not depend on any subgraph, and hence we omit the subscript in \( \nbr(\cdot) \) throughout the section.

Given an instance $\cI$ = \prbdsi of the \mrbds problem, the algorithm maintains $k$-sets at any point of time, which can be viewed as ``skeletons'' of the $k$-vertices that the algorithm aims to output. The skeletons act as placeholders that represent partial information about the neighborhoods of the solution vertices. Skeletons are implemented using sets called $\mcon_{\iletter}$, for $\iletter \in [k]$. That is, the algorithm maintains a tuple  
\[
\mi = \langle \mcon_1, \dots, \mcon_{k} \rangle
\]  
where for all $\iletter \in [k]$, $\mcon_{\iletter} \subseteq \cB$, be the $\iletter^{\textnormal{th}}$ coordinate of $\mi$ for the input instance $\cI$. The algorithm outputs a tuple $\cS = \langle s_1, \dots, s_{k} \rangle$ such that $s_\iletter \in \cR$, and as the name suggests, $\mcon_{\iletter}$ needs to be included by a vertex in the solution, that is $\mcon_{\iletter} \subseteq \nbr(s_\iletter)$ for all $\iletter \in [k]$. 
 Moreover, for all $\iletter \in [k]$, since $\mcon_{\iletter} \subseteq \nbr(s_\iletter)$, the only possible choices for $s_\iletter$ are the vertices in $\cR$ whose neighborhood contain $\mcon_\iletter$. This is captured by the set   
\[
{\sf PotentialSolutionFamily}_\iletter = \pot_\iletter \coloneqq \{ v \in \cR \mid \mcon_\iletter \subseteq \nbr(v) \}.
\] 
We also define $\pot = \bigcup_{\alpha \in [k]} \pot_{\alpha}$.
Further, consider a $\pot_\iletter$ for $\alpha \in [k]$. If any vertex in $\cB$ belongs to the {\em neighborhood} of every vertex in $\pot_\iletter$, that vertex will {\em definitely} be covered by any solution. This is precisely captured by  
\[
\fcov_\iletter = \bigcap_{v \in \pot_\iletter} \nbr(v).
\]
We also define $\fcov = \bigcup_{\alpha \in [k]} \fcov_{\alpha}$. This implies that once the algorithm establishes skeletons for the $k$-vertices, which keep evolving throughout the execution of the algorithm, many vertices in $\cB$ are freely covered. All of these together are captured in the input to our algorithm, denoted as $\algprbds$ for the \mrbds problem. The input parameters are as follows:  
\[
\left(\graph, \mi = \langle \mcon_1, \dots, \mcon_{k} \rangle, \free,\varepsilon \right)
\]
where $\free$ is the set of vertices of $\cB$ that is already covered by the partial solution constructed so far. For ease of presentation, as noted in the beginning of this section, the input bipartite graph remains unchanged throughout all recursive calls. The parts that change are $\mi$ and $\free$. In any recursive call of the algorithm, the first step is to check whether any new set of vertices in \( \cB \) has become additionally free in the current recursive call due to the current partially constructed skeleton. This is captured by the set  
\[
\nfree  := \left( \bigcup_{\iletter \in [k]} \fcov_\iletter \right) \setminus \free = \fcov \setminus \free.
\]

In each iteration, after performing some base case sanity checks, the algorithm picks a vertex $h^\star 
\in \pot$ such that
\[
\nbr(h^\star) \setminus (\free\cup \nfree) = \{ u_1, \dots, u_\ell \},
\]  
that consists of only {\em non-free} vertices, has the largest cardinality. Note that \emph{non-free} vertices refer to those vertices in \( \cB \) that are not freely covered so far till current recursive call. In other words, non-free vertices do not belong to the set $\free \cup \nfree$. Given a vertex \( h^\star \in \pot \), the algorithm recursively solves either an instance in which \( h^\star \) is included in the solution, or an instance where the skeleton of some vertex in the solution, that is \( \mcon_{\iletter} \), for \( \iletter \in [k] \), must additionally include at least one vertex from the set \( \{ u_1, \dots, u_\ell \} \). For a fixed \( \varepsilon > 0 \), these two steps correspond to the following two cases: either \( h^\star \) is part of some \( (1 - \varepsilon) \)-approximate solution, or there does not exist any such approximate solution that includes the vertex \( h^\star \).

For a fixed $\varepsilon >0$, we show that when there is no \((1 - \varepsilon)\)-approximate solution containing \(h^\star\), then any optimal solution to our instance must contain a good fraction of vertices in \(\nbr(h^\star)\setminus (\free\cup \nfree)=\{u_1,\dots,u_\ell\}\), in particular, at least \(\varepsilon \ell\) vertices. This step is implemented by randomly selecting an item \(\clubsuit\) from the set \(\{ h^\star, u_1, \dots, u_\ell \}\) with a probability distribution  
\[
    \mathbb{P}[\clubsuit] = 
    \begin{cases} 
    \frac{1}{2} & \text{if } \clubsuit = h^\star \\
    \frac{1}{2\ell} & \text{if } \clubsuit \in \{u_1,\dots,u_\ell\}
    \end{cases}
    \]
That is, we select the vertex \(h^\star \in \pot\) with probability \(\frac{1}{2}\), or a vertex $u$ from \(\{u_1,\dots,u_\ell\} \subseteq \cB\) with probability \(\frac{1}{2\ell}\). We also guess uniformly at random an index \( \beta \in [k] \) to which we plan to add the vertex \( u \). Let $\mi$ is updated to $\mi'$ where
\[\mi' \gets \langle \mcon_1, \dots, \mcon_{\beta} \cup \{u\}, \dots, \mcon_{k} \rangle.\]
Observe that with probability at least \( \frac{\varepsilon}{2k} \), when we recursively call  
\[
\algprbds(\graph, \mi', \free \cup \nfree \cup \{ u \}, \varepsilon),
\]  
the skeletons maintain the invariant that there exists an optimal solution  
$\mathbb{O} = \langle o_1, \dots, o_k \rangle$ 
such that \( o_\iletter \in \cR \), and for every component of the tuple $\mi'$ is included in the neighborhood of the solution, that is \( \mcon_{\iletter} \subseteq \nbr(o_\iletter) \) for all \( \iletter \in [k] \).

\begin{figure}[ht!]
    \centering
    \includegraphics[scale=0.4]{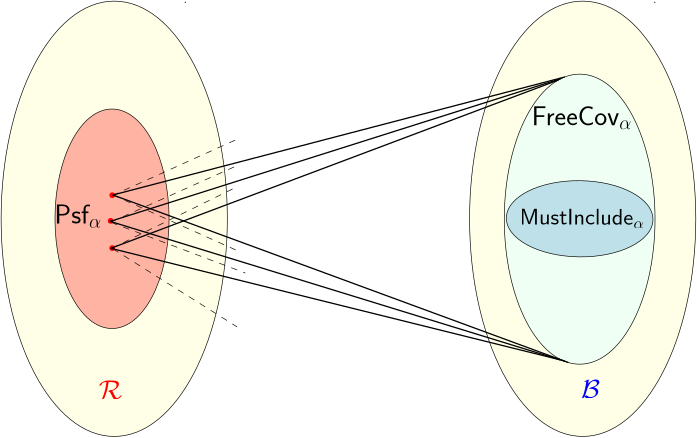}
    \caption[scale=0.4]{{\sf Potential Solution Family}, {\sf Free Coverage} and {\sf Must Include} for one coordinate}
    \label{fig:potsolfamily}
\end{figure}

We now compile all the definitions we have established so far. For ease of understanding, refer to \Cref{fig:potsolfamily}.

\begin{definition}[Potential, Freely Covered and Non-Free neighbors] 
Let 
\[
\cJ=(\graph,\mi = \langle \mcon_1, \dots, \mcon_{k} \rangle,\free, \varepsilon),
\]  
be an instance for \Cref{alg:framework-prbds}.  Then, the potential solution family, the set of freely covered, newly-free and non-free neighbors of any subset of $\cB$ are defined as follows. 
\begin{description}
    \item[1.]
    ${\sf PotentialSolutionFamily}_\iletter =\pot_\iletter= \{ v \in \cR \mid \mcon_\iletter \subseteq \nbr(v) \}$ for all $\iletter \in [k]$,  and $\pot=\cup_{\iletter \in [k]}\pot_\iletter $
    \item[2.] 
$\fcov_\iletter =\bigcap_{v \in \pot_\iletter} \nbr(v)$ for all $\iletter \in [k]$ and $\fcov = \cup_{\iletter \in [k]}\fcov_\iletter$
\item[3.] $\nfree_\iletter  =   \fcov_\iletter \setminus \free$ for all $\iletter \in [k]$ and 
 $\nfree = \fcov \setminus \free$\;
 \item[4.] $\nbrst(\cS, \cJ) \coloneq \nbr(\cS) \setminus \free$,
where $\cS \subseteq \cR $. The definition of $\nbrst(,)$ captures the idea of newly covered vertices in $\cB$.
\end{description} 
\end{definition}
Finally, we define the notion of a consistent tuple, which helps us compare objects across different recursive calls of our algorithm.

\begin{definition}[Consistent Tuple]
Given an instance  
\[
\cJ=(\graph,\mi = \langle \mcon_1, \dots, \mcon_{k} \rangle,\free, \varepsilon),
\]  
we say a tuple \(\cS=\langle s_1,\dots, s_k\rangle\) is \emph{consistent for} \(\cJ\) if  
\[
s_\iletter \in \cR, \quad \text{and} \quad \fcov_\iletter = \bigcap_{v \in \pot_\iletter} \nbr(v) \subseteq \nbr(s_\iletter), \quad \forall \iletter \in [k].
\]  
We use \(\cons{\cJ}\) to denote the set of all consistent tuples for \(\cJ\). 
\end{definition}
A complete description of the algorithm with pseudocode is provided in Algorithm~\ref{alg:framework-prbds}.  

\begin{algorithm}[ht!]
\caption{\algprbds \label{alg:framework-prbds}}
  \KwInput{A bipartite graph $\graph$, a $k$-tuple $\mi = \langle \mcon_1, \dots, \mcon_{k} \rangle $, $\free\subseteq \cB$,  and  $\varepsilon \in (0,1)$ }
  \KwOutput{A consistent $k$-tuple $\cS$  such that $|\nbrst(\cS, \cJ)|$ is at least $(1-\varepsilon)\opt_k(\cJ)$, where $\cJ=(\graph,\mi,\free,\varepsilon)$ .}
Let $\pot_\iletter = \{ v \in \cR \mid \mcon_\iletter \subseteq \nbr(v) \}$,  
$\fcov_\iletter=\bigcap_{v \in \pot_\iletter} \nbr(v)$ for all $\iletter \in [k]$ and $\nfree = \left(\bigcup_{\iletter \in [k]} \fcov_\iletter \right) \setminus \free$.\label{line:datastr}\\
\If {$\mi$ is a $0$-tuple \label{line1:datastr1}} {\Return $0$-tuple}\label{line:basecase1}
\If {there exists $\iletter \in [k]$ such that $\bigcup_{v \in \pot_\iletter }\nbr(v) \subseteq \free \cup \nfree $\label{algprbds:basetwo}} { Let $v$ be an arbitrary vertex in $\pot_\iletter$.\\
Let $\cS$ be the tuple returned by
$\algprbds (\graph,\mi[\downarrow \iletter],\free \cup \nfree  \cup \nbr(v), \varepsilon)$\\ \Return $\cS[\uparrow \iletter,v]$\label{algprbds:return_red_rule}}
Let $h^\star$ be a vertex in $\pot$ such that $\nbr(h^\star) \setminus (\free \cup \nfree)= \{u_1,\dots u_\ell \}$ is of the highest cardinality. \label{algprbds:select_high_deg}\\
Randomly select an item $\clubsuit$ from the set $\{h^\star,u_1,\dots,u_\ell\}$ with  probability distribution $(\frac{1}{2},\frac{1}{2\ell},\dots,\frac{1}{2\ell})$. That is, we select the vertex $h^\star$ with probability $\frac{1}{2}$, or a non-free neighbor of $h^\star$ with probability $\frac{1}{2\ell}$.   \label{algprbds:prob_dist} \\
\If{$\clubsuit=h^\star \in  \pot_\iletter$, for some $\iletter\in [k]$ } {Let $\cS$ be a tuple returned by $\algprbds (\graph,\mi[\downarrow \iletter],\free \cup \nfree  \cup \{u_1,\dots,u_\ell\}, \varepsilon)$\label{step:returnj'} \\ 
\Return $\cS[\uparrow \iletter,h^\star]$\label{algprbds:return_high_deg}} 
\Else {Let $\clubsuit = u \in \{u_1, \dots, u_\ell\}$. Pick an integer  $\beta\in[k]$ uniformly at random.\\
	 Assign $\mi \gets \langle \mcon_1, \dots, \mcon_{\beta} \cup \{u\}, \dots, \mcon_{k} \rangle$\\
	 Let $\cS$ be a tuple returned by  $\algprbds(\graph, \mi, \free \cup \nfree \cup \{u\}, \varepsilon)$\\ 
	\Return $\cS$\label{algprbds:return_opt_elem}} 
\end{algorithm}

\subsubsection{Algorithm Analysis}
In this section, we show that our algorithm indeed outputs a solution with the desired approximation ratio, that is an \epas. Recall that, for an instance \(\cJ\) and a consistent tuple \(\cS\), we have  
\[
\nbrst(\cS, \cJ) = \nbr(\cS) \setminus \free,
\] Further for an instance $\cJ$ of \Cref{alg:framework-prbds}, we define the value of an optimal solution as follows: 
\[
{\sf OPT}(\cJ) =\max_{\cS= \langle s_1,\dots, s_k\rangle \in \cons{\cJ} } | \nbr(\cS) \setminus \free| = \max_{\cS \in \cons{\cJ} }|\nbrst(\cS, \cJ)|.
\]  
Let \( \mathbb{O}=\langle o_1,\dots, o_k\rangle\) be a tuple such that  ${\sf OPT}(\cJ)  = |\nbrst(\mathbb{O}, \cJ)|$.  
Such a tuple will be called an \emph{optimal tuple}. In what follows, we present a sequence of lemmata that capture structural relationships between two ``similar-looking'' instances arising in consecutive recursive calls. To avoid ambiguity between identically named objects across different recursive calls, we explicitly associate each object with its corresponding instance using brackets and omit those when the context is clear. 
\begin{lemma}
\label{lem:propertyonealgoone}
 Let $\cJ$ denote the input instance 
 \[(\graph,\mi = \langle \mcon_1, \dots, \mcon_{k} \rangle,\free, \varepsilon).\] 
 Consider $v\in \pot_{\iletter}(\cJ)$ for some $\iletter \in [k]$ and let $\cJ'$ denote the instance
  \[ \left(\graph,\mi[\downarrow \iletter],\free(\cJ) \cup \nfree(\cJ) \cup \nbr(v), \varepsilon\right).\]   
Then, for any consistent tuple $\cS$ for $\cJ'$, we have the following.
\begin{description}
\setlength{\itemsep}{-2pt}
    \item[\rm \bf 1)  Consistent:]  $\cS[\uparrow \iletter, v]$ is a consistent tuple for $\cJ$. 
    \item[\rm \bf 2) Size Equality:]
    \begin{equation}\label{eqn:firstpartcovstarsum}
	|\nbrst(\cS[\uparrow \iletter,v], \cJ)|=|\nbrst(\cS,\cJ')| +|\nfree(\cJ)| + |\nbr(v) \setminus (\free(\cJ) \cup \nfree(\cJ))|. 
\end{equation}
\item [\rm \bf 3) Pairwise Disjoint:] Each of the sets $A_1=\nbrst(\cS,\cJ')$, $A_2=\nfree(\cJ)$ and $A_3=\nbr(v) \setminus (\free(\cJ) \cup \nfree(\cJ))$ are pairwise disjoint. Furthermore, $\nbrst(\cS[\uparrow \iletter,v], \cJ)=\bigcup_{i=1}^3 A_i$.
\end{description}

\end{lemma}
\begin{figure}[t!]
    \centering
    \includegraphics[scale=0.34]{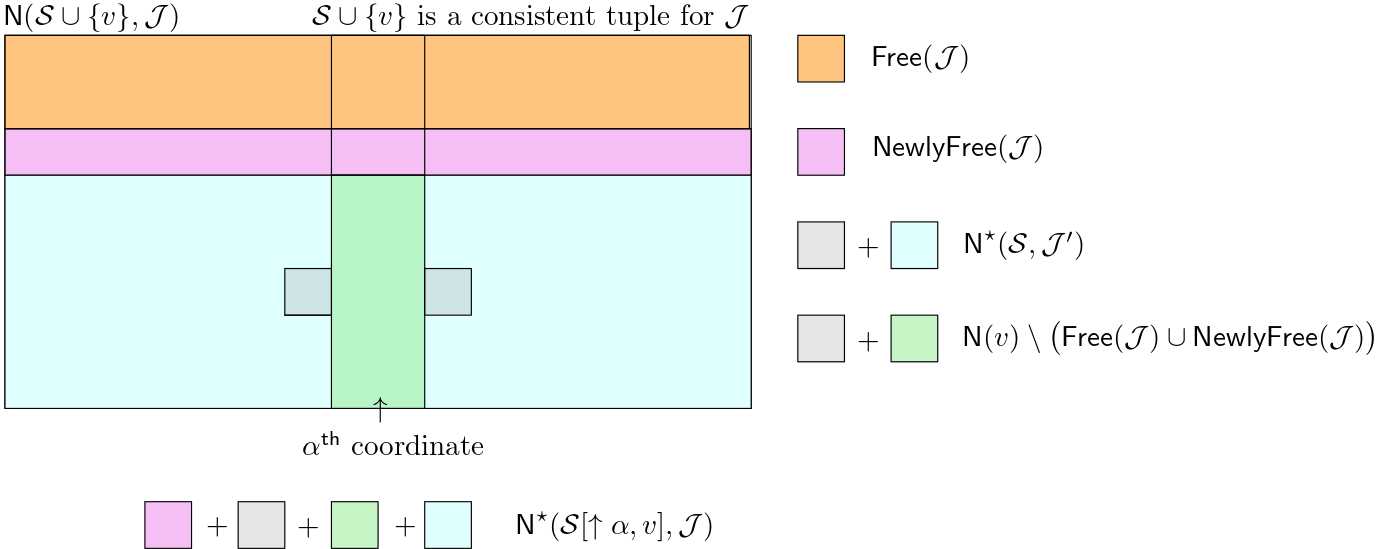}
    \caption[scale=0.4]{A visual representation of ~\Cref{lem:propertyonealgoone}}
    \label{fig:freenfreevv}
\end{figure}

For ease of understanding, refer to \Cref{fig:freenfreevv}.

\begin{proof}
We will prove each of the statements one by one. Let $\cS = \langle s_1, \dots, s_{\alpha-1}, s_{\alpha+1}, \dots, s_k \rangle$ be a consistent tuple for $\cJ'$.

\medskip
\noindent 
 {\rm \bf 1) Consistent:} We first show that \( \cS[\uparrow \iletter, v] \) is a consistent tuple for \( \cJ \). To this end, observe that since \( \cS \) is a consistent tuple for \( \cJ' \), we have
\[
\fcov_i(\cJ') \subseteq \nbr(s_i) \quad \text{for all } i \in [k] \setminus \{\iletter\}.
\]
Moreover, for all \( i \in [k] \setminus \{\iletter\} \), we have \( \mcon_i(\cJ) = \mcon_i(\cJ') \), and thus \( \fcov_i(\cJ) = \fcov_i(\cJ') \subseteq \nbr(s_i) \). Now consider index \( \iletter \). Since \( v \in \pot_\iletter(\cJ) \), we know that
\[
\mcon_\iletter(\cJ) \subseteq \fcov_\iletter(\cJ) \subseteq \nbr(v).
\]
Therefore, for all \( i \in [k] \), the neighborhood of the \( i^\text{th} \) vertex in the tuple \( \cS[\uparrow \iletter, v] \) contains \( \fcov_i(\cJ) \). This proves that \( \cS[\uparrow \iletter, v] \) is a consistent tuple for \( \cJ \).

\noindent

\smallskip

\noindent 
 {\rm \bf 2) Size Equality:} We show the size equality by showing that the left hand side (\lhs)  of \Cref{eqn:firstpartcovstarsum} is less than or equal to the right hand side (\rhs) and vice versa. 
 
 \smallskip
 
 \noindent 
 We first show that  $\lhs \leq\rhs$.   Pick a vertex \( x \in \nbrst(\cS[\uparrow \iletter, v], \cJ) \). We will show that \( x \) belongs to one of the sets on the right-hand side (\rhs). Recall that  
\[
    \nbrst(\cS[\uparrow \iletter, v], \cJ) = \nbr(\cS[\uparrow \iletter, v]) \setminus \free(\cJ).
\] 
Since $\nbr(\cS[\uparrow \iletter, v]) = \nbr(\cS) \cup \nbr(v)$, this implies that \( x \) must lie in either \( \nbr(\cS) \setminus (\free(\cJ) \cup \nbr(v)) \) or \( \nbr(v) \setminus \free(\cJ) \). 

Suppose \( x \in \nbr(\cS) \setminus (\free(\cJ) \cup \nbr(v)) \). Then, either \( x \in \nfree(\cJ) \) or \( x \notin \nfree(\cJ) \). In the former case, we already have that $x \in \nfree(\cJ)$. In the latter case, we have $x \in \nbr(\cS) \setminus (\free(\cJ) \cup \nfree(\cJ) \cup \nbr(v)) = \nbrst(\cS, \cJ')$,
and hence \( x \in \nbrst(\cS, \cJ') \). 

Now suppose \( x \in \nbr(v) \setminus \free(\cJ) \). Again, either \( x \in \nfree(\cJ) \) or \( x \notin \nfree(\cJ) \). In the former case we have $x \in \nfree(\cJ)$. In the later case, we have $x \in \nbr(v) \setminus (\free(\cJ) \cup \nfree(\cJ)).
$

Thus, we have shown that every element on the left-hand side (\lhs) appears in at least one of the sets on the right-hand side. Therefore, \( \lhs \leq \rhs \). In particular, we have
\begin{equation}
\label{eqn:SetDropLemmaEqnA}
    \nbrst(\cS[\uparrow \iletter, v], \cJ) \subseteq \bigcup_{i=1}^3 A_i.
\end{equation}

 \noindent 
 Next, we show that  $\lhs \geq\rhs$. To do this, we will show that the three sets appearing in \( \rhs \) are pairwise disjoint and that every element in these sets also appears in the set on the left-hand side. Note that, by construction, all the three sets $A_1$, $A_2$ and $A_3$ are pairwise disjoint.  
 
 Suppose $ x \in \nbrst(\cS, \cJ') = \nbr(\cS) \setminus (\free(\cJ) \cup \nfree(\cJ) \cup \nbr(v)).$ Then by definition, \( x \) belongs to \( \nbrst(\cS[\uparrow \iletter, v], \cJ) \). 

Now suppose, \( x \in \nfree(\cJ) \), then clearly $x \in \nbrst(\cS[\uparrow \iletter, v], \cJ)$, because $\cS[\uparrow \iletter, v]$ is a consistent tuple for $\cJ$, which implies that $\nfree(\cJ) \subseteq \fcov(\cJ) \subseteq \nbr(\cS[\uparrow \iletter, v], \cJ)$ (as seen in the proof for the first property).

Finally, if \( x \in \nbr(v) \setminus (\free(\cJ) \cup \nfree(\cJ)) \), then clearly $x \in \nbrst(\cS[\uparrow \iletter, v], \cJ)$,  
and does not belong to \( \nfree(\cJ) \) or \( \nbrst(\cS, \cJ') \) as \( \nbrst(\cS, \cJ') \) excludes $\nbr(v)$ by definition. 
This implies $\lhs \geq\rhs$. This concludes the proof. However, we have also shown that 
\begin{equation}
\label{eqn:SetDropLemmaEqnB}
    \nbrst(\cS[\uparrow \iletter, v], \cJ) \supseteq \bigcup_{i=1}^3 A_i.
\end{equation}

\textbf{3) Pairwise Disjoint: }Equations~\ref{eqn:SetDropLemmaEqnA} and~\ref{eqn:SetDropLemmaEqnB}, together with the fact that the sets \( A_1 \), \( A_2 \), and \( A_3 \) are pairwise disjoint, establish the third property of the lemma.
\end{proof}

\begin{lemma}
\label{lem:propertytwoalgoone}
Let \( \cJ \) be the input instance  
\[
(\graph,\mi = \langle \mcon_1, \dots, \mcon_{k} \rangle, \free, \varepsilon).
\]  
Let \( u \in \cB \setminus (\free(\cJ)\cup \nfree(\cJ)) \). For any \( \beta \in [k] \),  let $$\mi' \gets \langle \mcon_1, \dots, \mcon_{\beta} \cup \{u\}\dots, \mcon_{k} \rangle,$$ and  let \( \cJ' \) denote the instance  
\[
(\graph, \mi', \free(\cJ)\cup \nfree(\cJ) \cup \{u\}, \varepsilon).
\]
 
Then, for any consistent tuple $\cS$ for $\cJ'$, we have the following.
\begin{description}
    \item {\rm \bf 1) Consistent:\label{consistent}}  $\cS$ is a consistent tuple for $\cJ$. 
    \item {\rm \bf 2) Size Equality:}
\begin{equation}\label{eqn:sizeequal2}
	|\nbrst(\cS, \cJ)|=|\nbrst(\cS,\cJ ')|+| \nfree(\cJ)|+|\{u\}|
\end{equation}
\item {\rm \bf 3) Pairwise Disjoint:} Each of the sets $A_1=\nbrst(\cS,\cJ')$, $A_2=\nfree(\cJ)$ and $A_3=\{u\}$ are pairwise disjoint. Furthermore, $\nbrst(\cS, \cJ)=\bigcup_{i=1}^3 A_i$.
\item  {\rm \bf 4) Intersection Complexity:} $\dintcmp(G,\mcon_\beta \cup \{u\})\geq \dintcmp(G,\mcon_\beta)+1$
\end{description}
\end{lemma}

\begin{figure}[ht!]
    \centering
    \includegraphics[scale=0.34]{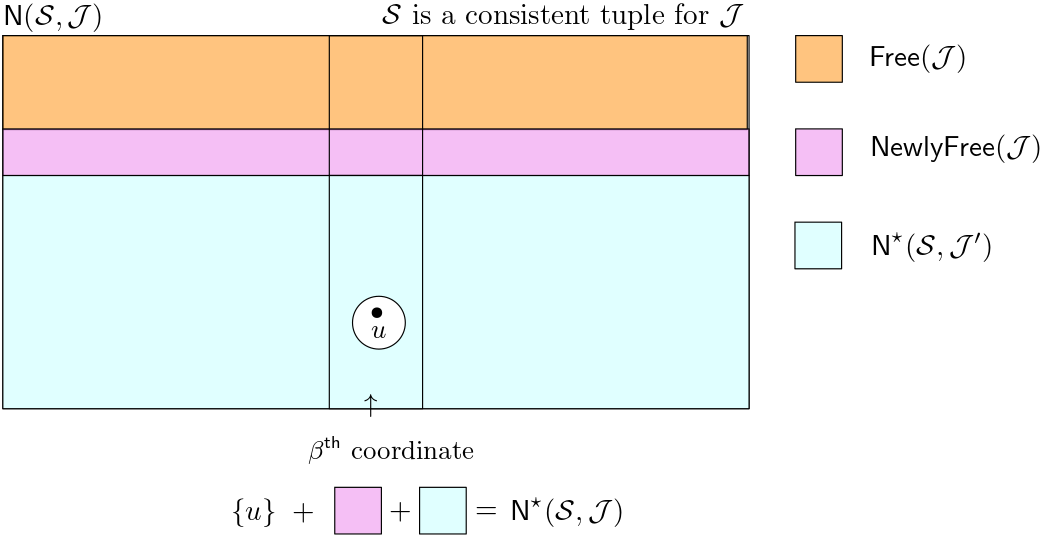}
    \caption[scale=0.4]{A visual respresentation of~\Cref{lem:propertytwoalgoone}}
    \label{fig:freenfreev}
\end{figure}

For ease of understanding, refer to \Cref{fig:freenfreev}.

\begin{proof}
Let \( \cS = \langle s_1, \dots, s_k \rangle \) be a consistent tuple for $\cJ'$.

\smallskip
\noindent 
{\rm \bf 1) Consistent:} 
We aim to show that \( \cS \) is a consistent tuple for \( \cJ \).   Since \( \cS \) is a consistent tuple for \( \cJ' \), for any \( i \in [k] \setminus \{\beta\} \), we have that $\fcov_{i}(\cJ') \subseteq \nbr(s_i)$(by definition). Further by property of $\cJ$ and $\cJ'$, for every $i \in [k] \setminus \beta$, we have that $\mcon_{i}(\cJ) = \mcon_{i}(\cJ') \subseteq \nbr(s_i)$ which implies that $\fcov_{i}(\cJ) = \fcov_{i}(\cJ') \subseteq \nbr(s_i)$.

Now, consider the coordinate \( \beta \). Since $\mcon_{\beta}(\cJ') = \mcon_{\beta}(\cJ) \cup \{u\}$, this implies that $\mcon_{\beta}(\cJ) \subseteq \mcon_{\beta}(\cJ')$.
Since \( \cS \) is a consistent tuple for \( \cJ' \), it follows that $\mcon_{\beta}(\cJ') \subseteq \fcov_{\beta}(\cJ') \subseteq \nbr(s_{\beta})$. 
Therefore,  
$\mcon_{\beta}(\cJ) \subseteq \nbr(s_{\beta})$.  
This implies that \( s_{\beta} \in \pot_{\beta}(\cJ) \). Consequently, we also have that  $\fcov_{\beta}(\cJ) \subseteq \nbr(s_{\beta})$. Thus, \( \cS \) is a consistent tuple for the instance \( \cJ \).

\smallskip
\noindent 
 {\rm \bf 2) Size Equality:} We show the size equality by showing that the left hand side (\lhs)  of \Cref{eqn:sizeequal2} is less than or equal to the right hand side (\rhs) and vice versa. 
 
 \smallskip
 
\noindent  We first show that  $\lhs \leq\rhs$.  Let $x 
 \in \nbrst(\cS, \cJ)$. Recall that $
\nbrst(\cS, \cJ) = \nbr(\cS) \setminus \free(\cJ)$. Then either $x= u$ or $x 
\in \nbr(\cS) \setminus (\free(\cJ) \cup \{u\})$. If $x = u$, then it contributes to the size of $\{u\}$ in the \rhs and we are done. 

Suppose $x \in \nbr(\cS) \setminus (\free(\cJ)\cup \{u\})$. In this case, $x$ is either in $\nfree(\cJ)$ or in $\nbr(S) \setminus (\free(\cJ) \cup \nfree(\cJ) \cup \{u\})$. If $x \in \nfree(\cJ)$, then it contributes $1$ to $|\nfree(\cJ)|$. If $x \in \nbr(\cS) \setminus (\free(\cJ) \cup \nfree(\cJ) \cup \{u\})$, then $x$ contributes $1$ to $|\nbrst(\cS, \cJ')|$ by definition. Thus, we have shown that every element occurring in the set of \lhs occurs in some set in the \rhs. That is, 
\begin{equation}
\label{eqn:SetDropLemmaEqnC}
\nbrst(\cS, \cJ)\subseteq \bigcup_{i=1}^3 A_i
\end{equation}
Hence, $\lhs \leq\rhs$. 

 \noindent 
Next, we show that \( \lhs \geq \rhs \). To do this, we will show that the three sets appearing in \( \rhs \) are pairwise disjoint and that every element in these sets also appears in the set on the left-hand side. Indeed, by definition 
$\nbrst(\cS,\cJ ') = \nbr(\cS) \setminus (\free(\cJ) \cup \nfree(\cJ) \cup \{u\})$. Thus, $\nbrst(\cS,\cJ ')$ is disjoint from  
$\nfree(\cJ)$ and $\{u\}$. Finally, since  \( u \in \cB \setminus (\free(\cJ)\cup \nfree(\cJ)) \), we have that $\{u\}$ and $\nfree(\cJ)$ are pairwise disjoint. This also implies one part of the \textbf{\rm Pairwise Disjoint} constraint stated in the lemma. As before, let \( \cS = \langle s_1, \dots, s_k \rangle \).  

 First we show that  the element $u$ appears in \lhs.  
Since,  $\cS$ is a consistent tuple for $\cJ'$, this implies that 
$u \in \fcov_{\beta}(\cJ') \subseteq \nbr(s_{\beta})$. Furthermore, by our choice, we have that \( u \in \cB \setminus (\free(\cJ)\cup \nfree(\cJ)) \), in particular $u \notin \free(\cJ)$. Hence this implies that $u\in \nbr(\cS) \setminus \free(\cJ)= \nbrst(\cS, \cJ)$. 
 
Suppose $x 
\in \nfree(\cJ)$. Since $\cS$ is a consistent tuple for $\cJ$ (proved in item one of this lemma), this implies that $\nfree(\cJ) = \fcov(\cJ) \setminus \free(\cJ)\subseteq \nbr(\cS)  \setminus \free(\cJ) = \nbrst(\cS,\cJ )$. Hence $x\in  \nbrst(\cS,\cJ )$. 

 Finally suppose $x \in \nbrst(\cS,\cJ ')$. Since $\nbrst(\cS,\cJ ') = \nbr(\cS) \setminus (\free(\cJ) \cup \nfree(\cJ) \cup \{u\})$, this implies that $x$ also is in the set $\nbr(\cS) \setminus \free(\cJ)= \nbrst(\cS, \cJ)$. Thus we have shown
 \begin{equation}
\label{eqn:SetDropLemmaEqnD}
\nbrst(\cS, \cJ)\supseteq \bigcup_{i=1}^3 A_i
\end{equation}
This implies $\lhs \geq\rhs$. Together with the fact $\lhs \leq \rhs$ we get the desired size equality. 

\smallskip
\noindent
{\rm \bf 3) Pairwise Disjoint: }Equations~\ref{eqn:SetDropLemmaEqnC} and~\ref{eqn:SetDropLemmaEqnD}, together with the fact that the sets \( A_1 \), \( A_2 \), and \( A_3 \) are pairwise disjoint, establish the Pairwise Disjoint item of the lemma.
 
\medskip
\noindent {\rm \bf 4) Intersection Complexity:}
Let $\lambda= \dintcmp(G,\mcon_\beta(\cJ))$. By definition, we have a sequence of $\lambda$ \realizable sets $$\reex(G,\mcon_\beta(\cJ))=X_1 \supsetneq X_2 \supsetneq \dots \supsetneq X_{\lambda}.$$ Let $$ \incld(G,\mcon_\beta(\cJ)) = Y_1 \subsetneq Y_2 \subsetneq \dots \subsetneq Y_{\lambda}$$ be the corresponding sequence of \realizing sets (due to~\Cref{lemma:realizablerealizingreln}). Note that in this case, $\fcov_\beta(\cJ) = \reex(G,\mcon_\beta(\cJ))$.

Since, by definition, $u \notin \fcov(\cJ)$, there exists $y \in Y_1$ such that $u \notin \nbr(y)$. Let $\widetilde{Y}_1=\{y \in Y_1 \mid u \notin \nbr(y)\}$. Clearly $\widetilde{Y}_1 \neq \emptyset$. Also observe that $Y_1 \setminus \widetilde{Y}_1$ is a \realizing set for $X_1 \cup \{u\}$. Thus,
$$\reex(G,\mcon_\beta(\cJ) \cup \{u\})=\reex(G,X_1 \cup \{u\}) \supsetneq X_1 \supsetneq X_2 \supsetneq \dots \supsetneq X_{\lambda}$$ is a sequence of $\lambda+1$ realizable sets. Thus, $\dintcmp(G,\mcon_\beta(\cJ) \cup \{u\})$ is at least $\lambda+1$ due to~\Cref{defn:dic}.
\end{proof}

Let \(\intcmp = \dintcmp(G)\). We define a measure associated with \(\cJ\), denoted by \(\mu(\cJ)\), as  
\[
\mu(\cJ) = k + k\intcmp - \sum_{\iletter \in [k]} \dintcmp( G, \mcon_{\iletter}(\cJ))
\]  
We use this measure for induction during our proof for both approximation analysis as well as probability analysis. Next, we present and prove the main technical lemma of this section.
\begin{lemma} \label{lem:framework-prbds}
For a given input instance $$\cJ=(\graph,\mi = \langle \mcon_1, \dots \mcon_{k} \rangle,\free, \varepsilon)$$ Algorithm \ref{alg:framework-prbds}~\textnormal{(\algprbds)} runs in polynomial time and outputs a consistent tuple $\cS = \langle s_1, s_2, \dots, s_k \rangle$ such that  
\[|\nbrst(\cS, \cJ)| \geq (1-\varepsilon)\opt(\cJ)\]
with probability at least $\left( \frac{\varepsilon}{2 k} \right)^{\mu(\cJ)}$.
\end{lemma}

\begin{proof}
We divide the proof into two parts. First, we establish the approximation ratio of our algorithm, and then we analyze its success probability. Our proof proceeds by induction, using the measure defined above: 
\[
\mu(\cJ) = k + k\intcmp - \sum_{\iletter \in [k]} \dintcmp( G, \mcon_{\iletter}(\cJ))
\]
In particular, we prove the desired statement of \Cref{lem:framework-prbds} by induction on \(\mu(\cJ)\).

\medskip

\noindent 
\paragraph{\bf Non-Negativity of Measure:} 
Note that \(\mu(\cJ) = k + k\intcmp - \sum_{\iletter \in [k]} \dintcmp( G, \mcon_{\iletter}(\cJ))\) is always non-negative, since for all \(\iletter \in [k]\), we have \( \dintcmp( G,\mcon_{\iletter}(\cJ)) \leq \intcmp \), which implies \( \sum_{\iletter =1}^k \dintcmp (G,\mcon_{\iletter}(\cJ)) \leq k \intcmp \). Moreover, since any subset cannot have a negative size, we know that \( k \geq 0 \). Hence, it follows that  
\[
\mu(\cJ) = k + k\intcmp - \sum_{\iletter \in [k]} \dintcmp( G, \mcon_{\iletter}(\cJ)) \geq 0.
\]

\medskip

\noindent 
\paragraph{\bf Analysis of the algorithm:} 
Consider an optimal tuple \(\bo = \langle o_1, \dots, o_k \rangle\) for the instance \(\cJ\). By definition, we have \( \opt(\cJ)  = |\nbrst(\bo, \cJ)|  \). We prove the statement of the lemma using induction on \(\mu(\cJ)\). That is, the statement of the lemma serves as our induction hypothesis.

\smallskip

\noindent 
	\textbf{Base Case:} When $\mu(\cJ) = 0$, then we have that $k=0$ since $\intcmp \geq \dintcmp(G,\mcon_{\iletter}(\cJ))$ for $\iletter \in [k]$.  This implies that $\mi$ is a $0$-tuple. Thus, $0$-tuple is the only consistent tuple in this case and \Cref{alg:framework-prbds} correctly returns a $0$-tuple with {\color{blue}\texttt{probability $1$}}.

\smallskip

\noindent 
\textbf{Induction Assumption:} Let $\mu' >0$. Assume that the statement of \Cref{lem:framework-prbds} holds for all instances, $\cJ$,  for which  $\mu(\cJ) < \mu'$. That is, for an  input instance $\cJ=(\graph,\mi = \langle \mcon_1, \dots \mcon_{k} \rangle,\free, \varepsilon)$, such that $\mu(\cJ) <\mu'$, Algorithm \ref{alg:framework-prbds}~\textnormal{(\algprbds)} outputs a consistent tuple $\cS = \langle s_1, s_2, \dots, s_k \rangle$ such that  
\[|\nbrst(\cS, \cJ)| \geq (1-\varepsilon)\opt(\cJ)\]
with probability at least $\left( \frac{\varepsilon}{2 k} \right)^{\mu(\cJ)}$.

\smallskip

\noindent 
\textbf{Inductive Step:} We prove the statement of Lemma \ref{lem:framework-prbds} for an arbitrary instance $\cJ$ for which $\mu(\cJ) = \mu'$.

\medskip

\noindent 
\underline{\textbf{Approximation Analysis:}}
 
 \begin{tcolorbox}[colback=red!5!white,colframe=gray!75!black]
 We first perform the approximation analysis \textbf{assuming} that when the algorithm recursively calls an instance \( \cJ' \) such that \( \mu(\cJ') < \mu(\cJ) \), it indeed returns a solution \( \cS' \) for \( \cJ' \) satisfying  
\[
|\nbrst(\cS', \cJ')| \geq (1 - \varepsilon) \opt(\cJ').
\]  
In reality, this outcome occurs with some probability, which we will analyze after establishing the desired approximation factor. In simple terms, we condition on the recursive subroutines ``correctly'' returning the expected approximate solution and then proceed to analyze the resulting approximation factor.
  \end{tcolorbox}

 We divide the approximation analysis in two parts. In the first part, we analyse the case when the condition of ~\Cref{algprbds:basetwo} holds. Then, we analyse the case when the algorithm executes~\Cref{algprbds:select_high_deg}. 

\smallskip
\noindent 
\textbf{Case A: Analysis for the case when condition of ~\Cref{algprbds:basetwo} holds:}
 Let $\cJ$ denote the input instance 
 \[(\graph,\mi = \langle \mcon_1, \dots \mcon_{k} \rangle,\free, \varepsilon)\] 
 and $\cJ'$ denote the instance
  \[ (\graph,\mi[\downarrow \iletter],\free \cup \nfree \cup \nbr(v), \varepsilon).\] 
Further let \( \cS \) be a consistent tuple for \( \cJ' \) returned in \Cref{algprbds:return_red_rule} of the algorithm. Then by Lemma~\ref{lem:propertyonealgoone} we know that the returned solution $\cS[\uparrow \iletter, v]$ is a consistent tuple for $\cJ$. Next, we show that it also satisfies the approximation guarantee. That is, 
	 \[|\nbrst(\cS[\uparrow \iletter,v], \cJ)| \geq (1-\varepsilon)\opt(\cJ)
	 \]
As a step towards our proof, we first establish that \( \mu(\cJ') < \mu(\cJ) \), allowing us to apply the induction hypothesis to \( \cJ' \). Observe that 
\begin{align}
	&\mu(\cJ) - \mu(\cJ') \nonumber \\ 
	 &=\big(k + k\intcmp - \sum_{\iletter \in [k]} \dintcmp( G, \mcon_{\iletter}) \big)- \big( (k-1) + (k-1)\intcmp - \sum_{\substack{\iletter\in[k]\\ \iletter\neq j}} \dintcmp(G,  \mcon_{\iletter})\big) \nonumber \\
	&=(1+\intcmp-\dintcmp(G,\mcon_{j}))\stackrel{(\P)}{>}0 \label{measure:caseA}
\end{align}
where $(\P)$ follows from the fact that $\intcmp = \dintcmp(G) \geq \dintcmp(G,\mcon_j)$. Now since $\mu(\cJ') < \mu(\cJ)$, by induction hypothesis we have 
\[|\nbrst(\cS,\cJ')| \geq (1-\varepsilon)\opt(\cJ')
\]

Recall that  $\bo$ is an optimal tuple for the instance $\cJ$. By definition, this implies that  $\opt(\cJ) = |\nbrst(\bo, \cJ)|$.  Furthermore, $\mathbb{O}[\downarrow \iletter]$  forms a consistent tuple for the instance  $\cJ'$. Therefore, it follows that  $\opt(\cJ') \geq |\nbrst(\bO[\downarrow \iletter], \cJ')|$.  
Finally, by the second statement of Lemma~\ref{lem:propertyonealgoone} (\Cref{eqn:firstpartcovstarsum}), we conclude that 
$$|\nbrst(\cS[\uparrow \iletter,v],\cJ)| =|\nbrst(\cS,\cJ ')|+| \nfree|+ |\nbr(v) \setminus (\free(\cJ) \cup \nfree(\cJ))|.$$
 Since, $\nbr(v)  \subseteq (\free(\cJ) \cup \nfree(\cJ))$, we have that $\nbr(v) \setminus (\free(\cJ) \cup \nfree(\cJ)) = \emptyset$. Thus, we get the following. 
\begin{align*}
	|\nbrst(\cS[\uparrow \iletter,v],\cJ)| &=|\nbrst(\cS,\cJ ')|+| \nfree|  & \\ 
	&\geq (1-\varepsilon)\opt(\cJ') +| \nfree| &(\textnormal{by induction hypothesis})\\
	&\geq (1-\varepsilon)|\nbrst(\bO[\downarrow \iletter], \cJ')|+| \nfree|\\
	&= (1-\varepsilon)(|\nbrst(\bO,\cJ)| -|\nfree|)+| \nfree| & \\
	&  (\textnormal{By applying \Cref{eqn:firstpartcovstarsum} on $\bO[\downarrow \iletter]$ }) & \\
	&\geq (1-\varepsilon)|\nbrst(\bO,\cJ)|\\
	&= (1-\varepsilon)\opt(\cJ)
\end{align*}

\noindent  \textbf{Case B: Analysis for the case when~\Cref{algprbds:select_high_deg} is executed:}
Let $h^\star \in  \pot_\iletter$, for some $\iletter\in [k]$, be the vertex selected in \Cref{algprbds:select_high_deg}.  We further have following two cases.

	 \medskip
	\noindent
	\textbf{Case $\textnormal{\textbf{B}}_{\textbf{1}}$:} When $|\nbr(\{o_1,\dots,o_k\}) \cap \{u_1,\dots ,u_\ell\} | < \varepsilon \ell$ 
    
    \smallskip
    For ease of understanding, refer to \Cref{fig:smallint}.
    \begin{figure}[ht!]
    \centering
    \includegraphics[scale=0.35]{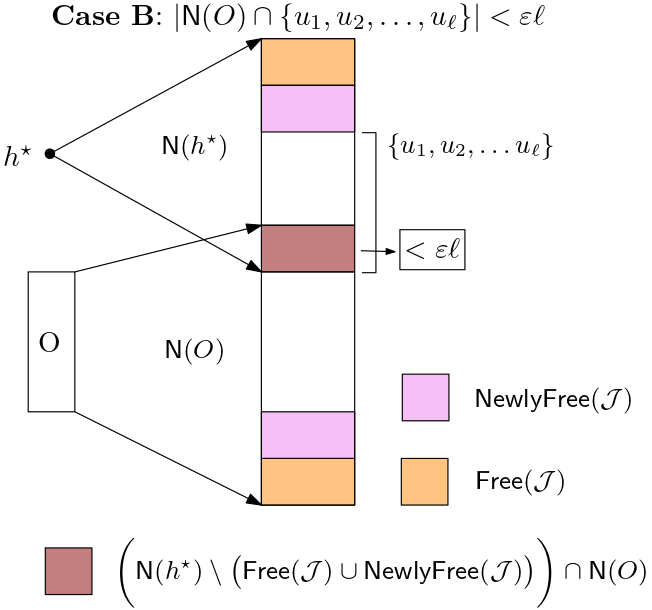}
    \caption[scale=0.4]{Representing Case B }
    \label{fig:smallint}
\end{figure}
Proof of this case is kind of similar to Case A that we considered earlier. In this case, consider the branch where the algorithm returns $\cS[\uparrow \iletter,h^\star]$ in \Cref{algprbds:return_high_deg}, where  $\cS$ is the tuple returned by \[\algprbds (\graph,\mi[\downarrow \iletter],\free\cup\nfree \cup \{u_1,\dots,u_\ell\}, \varepsilon)\] 
	 Let $\cJ$ denote the input instance 
     \[(\graph,\mi = \langle \mcon_1, \dots \mcon_{k} \rangle,\free, \varepsilon),\] 
     and $\cJ'$  denote the  instance 
     \[(\graph,\mi[\downarrow \iletter],\free(\cJ)\cup \nfree \cup  \{u_1,\dots,u_\ell\}, \varepsilon\ ),\] 
     which the algorithm recursively solves in this branch.
     
Let \( \cS \) be a consistent tuple for \( \cJ' \) returned in \Cref{algprbds:return_high_deg} of the algorithm. Then by Lemma~\ref{lem:propertyonealgoone}, we know that $\cS[\uparrow \iletter, h^\star]$ is a consistent tuple for $\cJ$. Next, we show that it also satisfies the approximation guarantee. That is, we need to show that
	 \[|\nbrst(\cS[\uparrow \iletter,h^\star], \cJ)| \geq (1-\varepsilon)\opt(\cJ). 
	 \]

As a step toward our proof, we first establish that \( \mu(\cJ') < \mu(\cJ) \), allowing us to apply the induction hypothesis on \( \cJ' \). Observe that 
\begin{align}
	&\mu(\cJ) - \mu(\cJ')\nonumber\\
	 &=\big(k + k\intcmp - \sum_{\iletter \in [k]} \dintcmp( G, \mcon_{\iletter}) \big)- \big( (k-1) + (k-1)\intcmp - \sum_{\substack{\iletter \in [k] \\ \iletter \neq j}} \dintcmp( G, \mcon_{\iletter})\big) \nonumber \\
	&=(1+\intcmp-\dintcmp(G,\mcon_{j}))\stackrel{(\P)}{>}0 \label{measure:caseB}
\end{align}
where $(\P)$ follows from the fact that $\intcmp\geq \dintcmp(G,\mcon_j)$. Now since $\mu(\cJ') < \mu(\cJ)$, by induction hypothesis, we have 
\[|\nbrst(\cS,\cJ')| \geq (1-\varepsilon)\opt(\cJ')
\]
Recall that  $\bo$ is an optimal tuple for the instance $\cJ$. By definition, this implies that  $\opt(\cJ) = |\nbrst(\bo, \cJ)|$.  Furthermore, $\mathbb{O}[\downarrow \iletter]$  forms a consistent tuple for the instance  $\cJ'$. Therefore, it follows that  $\opt(\cJ') \geq |\nbrst(\bO[\downarrow \iletter], \cJ')|$.  Now we derive a lower bound on $ \opt(\cJ')$ in terms of $\opt(\cJ) $. 
\begin{align*}
	 \opt(\cJ')& \geq |\nbrst(\mathbb{O}[\downarrow \iletter], \cJ')| \\
	 &= |\nbr(\mathbb{O}[\downarrow \iletter]) \setminus (\free(\cJ)\cup \nfree \cup \{u_1,\dots ,u_\ell\})|\\
	 &= |(\nbr(\bO [\downarrow \iletter]) \setminus (\free(\cJ)\cup \nfree)) \setminus (\{u_1,\dots ,u_\ell\})|\\
	 &= |\nbr(\bO [\downarrow \iletter])\setminus (\free(\cJ)\cup \nfree)| \\
     & \quad - |(\nbr(\bO[\downarrow \iletter])\setminus (\free(\cJ)\cup \nfree)) \cap (\{u_1,\dots ,u_\ell\})|\\
	 & \stackrel{(\bigstar)}{>} |\nbr(\bO) \setminus (\free(\cJ)\cup \nfree)| -|\nbr(o_\iletter)\setminus (\free(\cJ)\cup \nfree)|-(\varepsilon \cdot |\{u_1,\dots ,u_\ell\}|)\\
	 &\geq |\nbr(\bO) \setminus \free| - |\nfree| -|\nbr(o_\iletter)\setminus (\free(\cJ)\cup \nfree)| - \varepsilon \ell\\
     &\geq \opt(\cJ) -|\nfree| -|\nbr(o_\iletter)\setminus (\free(\cJ)\cup \nfree)| - \varepsilon\ell
\end{align*}

$(\bigstar)$ holds because we are in $\textnormal{ Case~B}_{1}$ where $|\nbr(\bO) \cap \{u_1,\dots ,u_\ell\}| < \varepsilon\ell$ which implies $|(\nbr(\bO[\downarrow \iletter])\setminus (\free(\cJ)\cup \nfree))\cap \{u_1,\dots ,u_\ell\}| \leq \varepsilon\ell $. For ease of understanding, refer \Cref{fig:largeinto}.

\begin{figure}[ht!]
    \centering
    \includegraphics[scale=0.4]{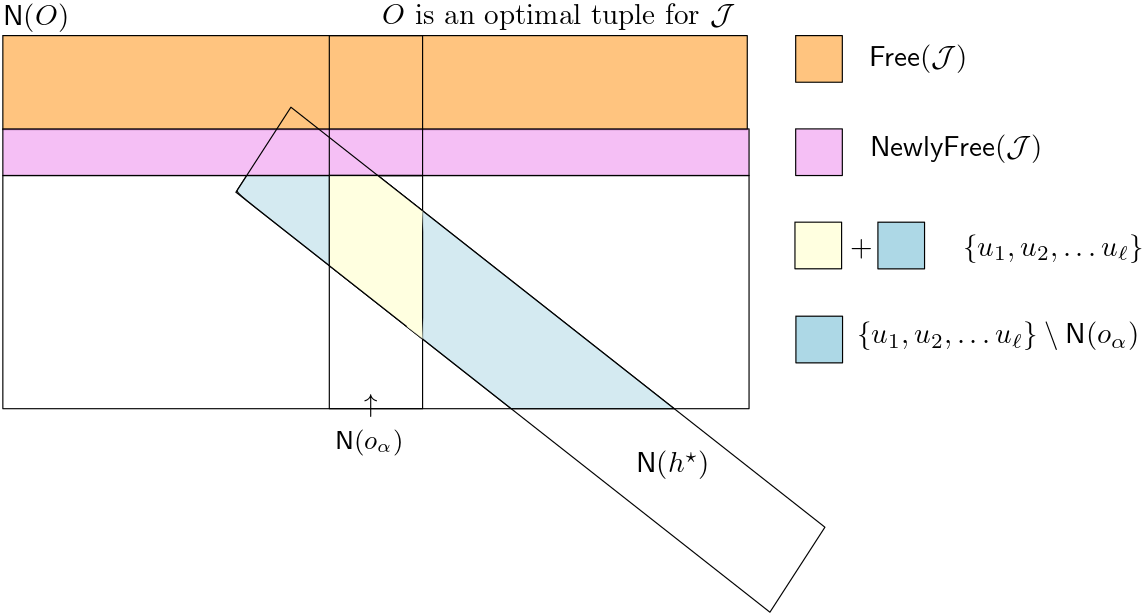}
    \caption[scale=0.4]{A diagram illustrating the interaction between $\nbr(\bO)$ and $\nbr(h^\star)$}
    \label{fig:largeinto}
\end{figure}

This gives us 
\begin{equation}\label{eqn:case1covstar}
|\nbrst(\cS,\cJ')| \geq (1-\varepsilon)\opt(\cJ')\geq (1-\varepsilon) (\opt(\cJ) -|\nbr(o_\iletter)\setminus (\free(\cJ)\cup \nfree)| - \varepsilon\ell-|\nfree|)
\end{equation}

\noindent Finally, we show the desired approximation ratio. 
\begin{align*}	&|\nbrst(\cS[\uparrow \iletter,h^\star], \cJ)|\\&=|\nbrst(\cS,\cJ')| + |\{u_1,\dots ,u_\ell\} | +|\nfree| & \\
		&  (\textnormal{By applying \Cref{eqn:firstpartcovstarsum} on $\cS$ }) & \\
	&\geq (1-\varepsilon) (\opt(\cJ) -|\nbr(o_\iletter) \setminus (\free(\cJ)\cup \nfree)| - \varepsilon\ell-|\nfree|)+ \ell +|\nfree|& \\
	& (\textnormal{from equation~\ref{eqn:case1covstar}}) & \\
	&\geq (1-\varepsilon) \opt(\cJ) + (1-\varepsilon)(\ell - |\nbr(o_\iletter)\setminus (\free(\cJ)\cup \nfree)|) +\varepsilon|\nfree|\\
	& \stackrel{(\heartsuit)}{\geq} (1-\varepsilon) \opt(\cJ)
\end{align*}

$(\heartsuit)$ follows from the fact that $\ell = |\nbr(h^\star)\setminus (\free(\cJ)\cup \nfree)| = |\{u_1, \dots, u_{\ell}\}|$ denotes the highest cardinality of the non-free neighbors of any vertex in $\pot$ and hence  $\ell \geq |\nbr(o_\iletter) \setminus (\free(\cJ)\cup \nfree)| $.\\

\noindent
\textbf{Case $\textnormal{\textbf{B}}_{\textbf{2}}$:}  When $|\nbr(\{o_1,\dots,o_k\}) \cap \{u_1,\dots ,u_\ell\} | \geq \varepsilon \ell $ 

\smallskip
For ease of understanding, refer to \Cref{fig:largeint}.
\begin{figure}[ht!]
    \centering
    \includegraphics[scale=0.35]{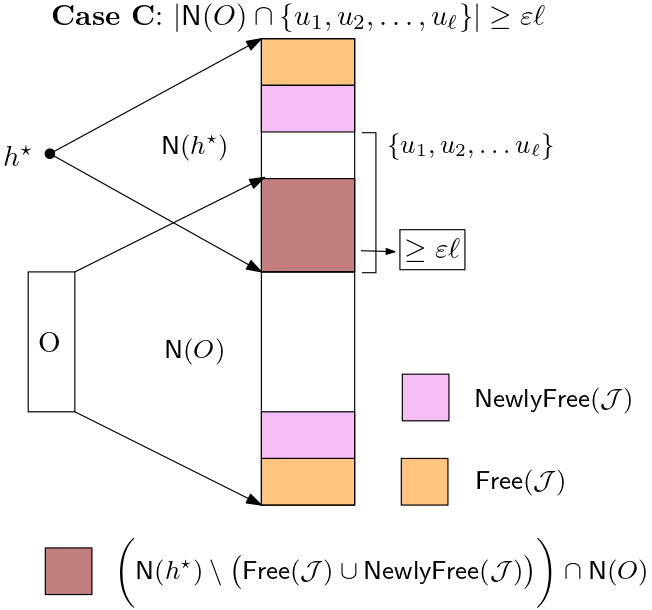}
    \caption[scale=0.4]{Representing Case C }
    \label{fig:largeint}
\end{figure}
In this case, consider the branch where the algorithm executes \Cref{algprbds:return_opt_elem} and returns  $\cS$, 
 where  $\cS$ is the tuple returned by
\[\algprbds(\graph, \mi, \free(\cJ)\cup \nfree \cup \{u\}, \varepsilon)\] 
where $\mi \gets \langle \mcon_1, \dots, \mcon_{\iletter} \cup \{u\}, \dots, \mcon_{k} \rangle$ and the selected element $u$ belongs to $\nbr(o_{\iletter})$.  Let $\cJ$ denote the input instance 
\[(\graph,\mi = \langle \mcon_1, \dots \mcon_{k} \rangle,\free, \varepsilon)\] 
and let  $\cJ'$ denote the instance 
\[(\graph,\mi',\free(\cJ)\cup \nfree \cup \{u\}, \varepsilon)\] where $\mi' = \langle  \mcon_1, \dots, \mcon_{\iletter} \cup \{u\}, \dots, \mcon_{k} \rangle$ 
 which the algorithm recursively solves in this branch. 
 
 Let $\cS$ be a consistent tuple for $\cJ'$, then due to~\Cref{lem:propertytwoalgoone}, $\cS$ is also a consistent tuple for $\cJ$.  Next, we show that it also satisfies the approximation guarantee. That is, we need to show that
\[|\nbrst(\cS,\cJ)|\geq(1-\varepsilon)\opt(\cJ) \]

As a step towards our proof, we first establish that \( \mu(\cJ') < \mu(\cJ) \), allowing us to apply the induction hypothesis on \( \cJ' \). Observe that 
$\mu(\cJ) - \mu(\cJ')$ is at least 
\begin{align}
 & \bigg(k + k\intcmp - \sum_{\iletter \in [k]} \dintcmp( G, \mcon_{\iletter}) \bigg) 
    - \bigg(k + k\intcmp - (\sum_{\iletter \in [k]} \dintcmp( G, \mcon_{\iletter})+1\bigg)\nonumber\\
        &\textnormal{ (since }\dintcmp(G,\mcon_\iletter \cup \{u\})\geq \dintcmp(G,\mcon_\iletter)+1 \textnormal{ due to~\Cref{lem:propertytwoalgoone})} \nonumber\\
	&=1 >0  \label{measure:CaseC}
\end{align}

Now since $\mu(\cJ') < \mu(\cJ)$, by induction hypothesis we have 
\[|\nbrst(\cS,\cJ')| \geq (1-\varepsilon)\opt(\cJ')
\]

Note that in this branch  $\bO= \langle o_1,\dots, o_k\rangle$ is a consistent tuple of $\cJ'$ since the selected element, $u \in \nbr(o_{\iletter})$.  Hence we can say that $\opt(\cJ') \geq  |\nbrst(\bO, \cJ')|$. Thus we have that
\begin{align*}
	|\nbrst(\cS,\cJ)| &=|\nbrst(\cS,\cJ ')|+| \nfree|+ 1  &(\textnormal{from equation~\ref{eqn:sizeequal2}})\\
	&\geq (1-\varepsilon)\opt(\cJ') +| \nfree|+1 &(\textnormal{by induction hypothesis})\\
	&= (1-\varepsilon)|\nbrst(\bO, \cJ')|+| \nfree|+1\\
	&= (1-\varepsilon)(|\nbrst(\bO,\cJ)| -|\nfree|-1)+| \nfree|+1 &(\textnormal{from equation~\ref{eqn:sizeequal2}})\\
	&\geq (1-\varepsilon)|\nbrst(\bO,\cJ)|\\
	&= (1-\varepsilon)\opt(\cJ)
\end{align*}	

This concludes the approximation analysis of the algorithm. 

\medskip

\noindent 
\underline{\textbf{Probability  Analysis:}} 

We next prove the success probability of our algorithm. In particular, we go through each case of approximation analysis and analyze the desired probability of success.  For an input instance $\cJ=(\graph,\mi = \langle \mcon_1, \dots \mcon_{k} \rangle,\free, \varepsilon)$, we say that a tuple $\cQ$ is {\em good for $\cJ$} if $|\nbrst(\cQ,\cJ)| \geq (1-\varepsilon)\opt(\cJ)$, otherwise it is {\em bad for $\cJ$}. 

\smallskip

\noindent 
\textbf{Case A: Analysis for the case when condition of ~\Cref{algprbds:basetwo} holds:}
 Let $\cJ$ denote the input instance 
 \[(\graph,\mi = \langle \mcon_1, \dots \mcon_{k} \rangle,\free, \varepsilon)\] 
 and $\cJ'$ denote the instance
  \[ (\graph,\mi[\downarrow \iletter],\free \cup \nfree \cup \nbr(v), \varepsilon).\] 
Let \( \cS \) be a consistent tuple for \( \cJ' \) returned in \Cref{algprbds:return_red_rule} of the algorithm. By \Cref{measure:caseA}, we know that $\mu(\cJ')< \mu(\cJ)$. Let  $\mathbf{P}=\prob{\cS[\uparrow \iletter,v] \mbox{ is good for }  \cJ}$.  Then,  
\begin{eqnarray*}
\mathbf{P} & = &  \prob{\cS[\uparrow \iletter,v] \mbox{ is good for }  \cJ \mid \cS \mbox{ is good for }  \cJ'}  
 \prob{\cS \mbox{ is good for }  \cJ'} \\
 && +   \prob{\cS[\uparrow \iletter,v] \mbox{ is good for }  \cJ \mid \cS \mbox{ is bad for }  \cJ'}  
 \prob{\cS \mbox{ is bad for }  \cJ'} \\
 &\geq & \prob{\cS[\uparrow \iletter,v] \mbox{ is good for }  \cJ \mid \cS \mbox{ is good for }  \cJ'}  
 \prob{\cS \mbox{ is good for }  \cJ'} \\
 & = & 1 \times  \prob{\cS \mbox{ is good for }  \cJ'} \\
& \geq  &  \left( \frac{\varepsilon}{2 k} \right)^{\mu(\cJ')} (\textnormal{by induction hypothesis})\\
& \geq &  \left( \frac{\varepsilon}{2 k} \right)^{\mu(\cJ)}
 \end{eqnarray*}

\smallskip
\noindent 
\textbf{Case B: When \Cref{algprbds:select_high_deg} is executed:}
Next we will consider the case when we select a vertex  $h^\star$  in $\pot$ such that $\nbr(h^\star) \setminus (\free(\cJ)\cup \nfree)= \{u_1,\dots u_\ell \}$ is of the highest cardinality (\Cref{algprbds:select_high_deg}). In this case the algorithm randomly selects the vertex $h^\star$ with probability $\frac{1}{2}$, or a non-free neighbor of $h^\star$ with probability $\frac{1}{2\ell}$  (\Cref{algprbds:prob_dist}).

Recall that \(\bo = \langle o_1, \dots, o_k \rangle\) is  an optimal tuple for the instance \(\cJ\). 
Now we consider two cases based on whether (a) 
$ |\nbr(\{o_1,\dots,o_k\}) \cap \{u_1,\dots ,u_\ell\} | < \varepsilon \ell $  or (b) $ |\nbr(\{o_1,\dots,o_k\}) \cap \{u_1,\dots ,u_\ell\} | \geq \varepsilon \ell $. This corresponds to execution of \Cref{algprbds:return_high_deg}, or   \Cref{algprbds:return_opt_elem}. Furthermore,  let the algorithm return a tuple $\cQ$.
\begin{eqnarray*}
 \prob{\cQ \mbox{ is good for }  \cJ} & = &  \prob{\cQ \mbox{ is good for }  \cJ \wedge |\nbr(\{o_1,\dots,o_k\}) \cap \{u_1,\dots ,u_\ell\} | < \varepsilon \ell}  
 \\
 && +   \prob{\cQ \mbox{ is good for }  \cJ \wedge |\nbr(\{o_1,\dots,o_k\}) \cap \{u_1,\dots ,u_\ell\} | \geq  \varepsilon \ell}  \\
 &\geq & \min \begin{cases}
      \prob{\cQ \mbox{ is good for }  \cJ \wedge |\nbr(\{o_1,\dots,o_k\}) \cap \{u_1,\dots ,u_\ell\} | < \varepsilon \ell}, \\
     \prob{\cQ \mbox{ is good for }  \cJ \wedge |\nbr(\{o_1,\dots,o_k\}) \cap \{u_1,\dots ,u_\ell\} | \geq  \varepsilon \ell} 
 \end{cases}
 \end{eqnarray*}
We will show that in either case  $\prob{\cQ \mbox{ is good for }  \cJ} \geq \left( \frac{\varepsilon}{2 k} \right)^{\mu(\cJ)}.$

\medskip

\noindent 
\textbf{Case $\textnormal{\textbf{B}}_{\textbf{1}}$:} When $ |\nbr(\{o_1,\dots,o_k\}) \cap \{u_1,\dots ,u_\ell\} | < \varepsilon \ell$

\smallskip
 In this case, consider the branch where the algorithm returns $\cS[\uparrow \iletter,h^\star]$ in \Cref{algprbds:return_high_deg}, where  $\cS$ is the tuple returned by \[\algprbds (\graph,\mi[\downarrow \iletter],\free\cup\nfree \cup \{u_1,\dots,u_\ell\}, \varepsilon)\] 
	 Let $\cJ$ denote the input instance 
     \[(\graph,\mi = \langle \mcon_1, \dots \mcon_{k} \rangle,\free, \varepsilon),\] 
     and $\cJ'$  denote the  instance 
     \[(\graph,\mi[\downarrow \iletter],\free(\cJ)\cup \nfree \cup  \{u_1,\dots,u_\ell\}, \varepsilon\ ),\] 
     which the algorithm recursively solves in this branch.
Let \( \cS \) be a consistent tuple for \( \cJ' \) returned in \Cref{step:returnj'} of the algorithm. By \Cref{measure:caseB} we know that 
$\mu(\cJ')< \mu(\cJ)$. Let $\mathcal{E}$ denote the event that $|\nbr(\{o_1,\dots,o_k\}) \cap \{u_1,\dots ,u_\ell\} | < \varepsilon \ell$. 
Let $\mathbf{P}=\prob{\cS[\uparrow \iletter,h^\star] \mbox{ is good for }   \cJ  \wedge \mathcal{E}}$. 
\begin{eqnarray*}
\mathbf{P} & \geq & \prob{\cS[\uparrow \iletter,h^\star]  \mbox{ is good for }  \cJ \wedge \mathcal{E} \mid \cS \mbox{ is good for }  \cJ'} \times\\ 
 & & \prob{\cS \mbox{ is good for }  \cJ'} \\
 & = & \prob{ |\nbrst(\cS[\uparrow \iletter,h^\star],\cJ)| \geq  (1-\varepsilon)\opt(\cJ) \wedge \mathcal{E} \mid \cS \mbox{ is good for }  \cJ'}\times \\ 
 & & \prob{\cS \mbox{ is good for }  \cJ'}\\
 & \geq & \prob{\clubsuit=h^\star }\times  \prob{\cS \mbox{ is good for }  \cJ'} \\
& \geq  &  \frac{1}{2}\left( \frac{\varepsilon}{2 k} \right)^{\mu(\cJ')} \quad \quad(\textnormal{by induction hypothesis})\\
& \geq &  \left( \frac{\varepsilon}{2 k} \right)^{\mu(\cJ)}
 \end{eqnarray*}

\medskip
\noindent 
\textbf{Case $\textnormal{\textbf{B}}_{\textbf{2}}$:} When $|\nbr(\{o_1,\dots,o_k\}) \cap \{u_1,\dots ,u_\ell\}| \geq \varepsilon \ell$ 

\smallskip
In this case, consider the branch where the algorithm executes \Cref{algprbds:return_opt_elem} and returns  $\cS$, 
 where  $\cS$ is the tuple returned by
\[\algprbds(\graph, \mi', \free(\cJ)\cup \nfree \cup \{u\}, \varepsilon)\] 
where 
\[\mi' \gets \langle \mcon_1, \dots, \mcon_{\iletter} \cup \{u\}, \dots, \mcon_{k} \rangle\] and the selected element, $u \in \nbr(o_{\iletter})$.  Let $\cJ$ denote the input instance 
\[(\graph,\mi = \langle \mcon_1, \dots \mcon_{k} \rangle,\free, \varepsilon)\] and let $\cJ'$ denote the instance 
\[(\graph,\mi',\free(\cJ)\cup \nfree \cup \{u\}, \varepsilon)\]
where \[\mi' = \langle \mcon_1, \dots, \mcon_{\iletter} \cup \{u\}, \dots, \mcon_{k} \rangle\]   
 which the algorithm recursively solves in this branch.

Let \( \cS \) be a consistent tuple for \( \cJ' \) returned in \Cref{algprbds:return_opt_elem} of the algorithm. By \Cref{measure:CaseC} we know that 
$\mu(\cJ')< \mu(\cJ)$. Let $\mathcal{E}$ denote the event that $|\nbr(\{o_1,\dots,o_k\}) \cap \{u_1,\dots ,u_\ell\} | \geq \varepsilon \ell$. 
 Let $\mathbf{P}=\prob{\cS \mbox{ is good for }   \cJ  \wedge \mathcal{E}}$. 
\begin{eqnarray*}
\mathbf{P} & \geq & \prob{\cS \mbox{ is good for }  \cJ \wedge \mathcal{E} \mid \cS \mbox{ is good for }  \cJ'} \times\\ 
 & & \prob{\cS \mbox{ is good for }  \cJ'} \\
 & = & \prob{ |\nbrst(\cS,\cJ)| \geq  (1-\varepsilon)\opt(\cJ) \wedge \mathcal{E} \mid \cS \mbox{ is good for }  \cJ'}\times \\ 
 & & \prob{\cS \mbox{ is good for }  \cJ'}\\
   &\geq &   \prob{\clubsuit= u \in \left ( \nbr(\{o_1,\dots,o_k\}) \cap \{u_1, \dots, u_\ell\} \right ) \wedge u\in \nbr(o_\alpha)}\times \\ & & \prob{\cS \mbox{ is good for }  \cJ'} \\
& \geq  & \prob{\clubsuit= u \in \left ( \nbr(\{o_1,\dots,o_k\}) \cap \{u_1, \dots, u_\ell\} \right )} \times \prob{u\in \nbr(o_\alpha)}\left( \frac{\varepsilon}{2 k} \right)^{\mu(\cJ')} \\ & & (\textnormal{by induction hypothesis})\\
& \geq &  \frac{\varepsilon \ell}{2 \ell} \frac{1}{k}\left( \frac{\varepsilon}{2 k} \right)^{\mu(\cJ')}\\
& \geq & \left( \frac{\varepsilon}{2 k} \right)^{\mu(\cJ)} 
 \end{eqnarray*}
Thus, we have shown that in either case 
\[
\Pr\bigl[\cQ \mbox{ is good for } \cJ\bigr] \ge \left(\frac{\varepsilon}{2k}\right)^{\mu(\cJ)}
\]
This concludes the probability analysis and also completes the proof of the lemma.
\end{proof}
\smallskip
We now state our main theorem.

\begin{theorem}
\label{thm:mainUnweighted}
Given an instance \( \cI = (\graph, k) \) of \mrbds and a parameter \( \varepsilon > 0 \), there exists an algorithm that runs in time 

\[
\left( \frac{k}{\varepsilon}  \right)^{\cO(\Gamma k)} \cdot \sizeg = \left(\frac{1}{\varepsilon}\right)^{\cO(\Gamma k \log k)} \cdot \sizeg
\]

and, with probability at least \( 1 - \frac{1}{e} \), outputs a subset \( \cS \subseteq \cR \) of size \( k \) such that 
\[
|\nbr(\cS)| \geq (1 - \varepsilon) \cdot {\sf OPT}_k(\cI),
\]
where \( \Gamma = \dintcmp(G) \).
\end{theorem}

\begin{proof}
Given an instance \( \cI \) of \mrbds, we construct an instance of \algprbds (\Cref{alg:framework-prbds}) as follows. We initialize a \( k \)-tuple \( \mi \) where each \( \mcon_{\iletter} \) is set to \( \emptyset \) for \( \iletter \in [k] \), and we set \( \free = \emptyset \). That is, we run \algprbds for the first time with the input instance $\cJ = (\graph, \mi = \langle \emptyset, \dots, \emptyset \rangle, \emptyset, \varepsilon)$.
Recall that for the initial instance $\cJ$ for \algprbds, we have
\[
{\sf OPT}(\cJ) = \max_{\cS \in \cons{\cJ}} |\nbrst(\cS, \cJ)| = \max_{\substack{\cS \subseteq \cR \\ |\cS| \leq k}} |\nbr(\cS)| = {\sf OPT}_k(\cI).
\]

Now we run \Cref{alg:framework-prbds} on \( \cJ \), and obtain a tuple \( \cS \subseteq \cR \) such that
\[
|\nbrst(\cS, \cJ)| = |\nbr(\cS) \setminus \emptyset|= |\nbr(\cS)| \geq (1 - \varepsilon) \cdot {\sf OPT}(\cJ) = (1 - \varepsilon) \cdot {\sf OPT}_k(\cI)
\]
with probability at least \( \left( \frac{\varepsilon}{2k} \right)^{\mu(\cJ)} \). We output the same set \( \cS \) for $\cI$, and therefore succeed in returning the subset \( \cS \subseteq \cR \) such that $|\nbr(\cS)| \geq (1 - \varepsilon) \cdot {\sf OPT}_k(\cI)$
with probability at least \( \left( \frac{\varepsilon}{2k} \right)^{\mu(\cJ)} \).

We have the upper bound on the measure as follows
\[
\mu(\cJ) = k + k \cdot \intcmp - \sum_{\iletter \in [k]} \dintcmp(G, \mcon_{\iletter}) \leq k + k \Gamma,
\]
where \( \Gamma = \dintcmp(G) \). Thus, the success probability is at least \( \left( \frac{\varepsilon}{2k} \right)^{k + k \Gamma} \). To boost the success probability, we run \algprbds (\Cref{alg:framework-prbds}) on \( \cJ \) independently \( \left( \frac{2k}{\varepsilon} \right)^{(\intcmp + 1)k} \) times, and return the tuple (alternately set) \( \cS \) with the maximum value of \( |\nbr(\cS)| \). The probability that all runs fail is at most
\[
\left( 1 - \left( \frac{\varepsilon}{2k} \right)^{(\intcmp + 1)k} \right)^{\left( \frac{2k}{\varepsilon} \right)^{(\intcmp + 1)k}} \leq \frac{1}{e},
\]
and hence the algorithm succeeds with probability at least \( 1 - \frac{1}{e} \).

\medskip

\noindent 
\underline{\textbf{Running time:}} One execution of \Cref{alg:framework-prbds} runs in polynomial time with respect to the input size, i.e., in \( \mathrm{poly}(\sizeg) \), and succeeds with probability at least \( \left( \frac{\varepsilon}{2k} \right)^{(\intcmp + 1)k} \). In what follows, we present an implementation where the dependence on the input size is linear. We assume that the input graph \( \graph \) is given as an adjacency list. Note that the total input size is \( \sizeg \).

\vspace{0.3 em}
\noindent 
\textbf{Data Structure maintained:} We maintain the vertices in the sets \( \mcon_\alpha \) globally using \( k \) linked lists. Additionally, we maintain an array of size \( |\cB| \) to represent the set \( \free \). These linked lists and the array are shared across all recursive calls and are implemented outside the scope of any individual call.

Now let us consider a single recursive call of \Cref{alg:framework-prbds}. In \Cref{line:datastr}, we compute a set of arrays \( \pot_\alpha \) and \( \fcov_\alpha \) for each \( \alpha \in [k] \), along with the arrays \( \pot \), \( \fcov \), and \( \nfree \), all of which are essential for the subsequent execution of the algorithm. To compute these arrays, we proceed as follows.

\begin{description}

    \item[1)] We maintain \( k \) arrays \( \pot_1, \pot_2, \dots, \pot_k \), each of size \( |\cR| \). The indices of these arrays correspond to the vertices of \( \cR \), and each entry stores a non-negative integer. All entries are initialized to $1$.

    \item[2)] Additionally, we maintain \( k \) arrays \( \fcov_1, \fcov_2, \dots, \fcov_k \), each of size \( |\cB| \). The indices of these arrays correspond to the vertices of \( \cB \), and each entry stores a non-negative integer. These arrays are also initialized to zero.

    \item[3)] Finally, we have a single array \( \nfree \) of size \( |\cB| \), where the indices correspond to the vertices of \( \cB \), and each entry stores a non-negative integer. This array is likewise initialized to zero.

\end{description}

We assume that standard set operations such as \emph{union}, \emph{intersection}, and \emph{set difference} between two arrays (possibly of different lengths) can be performed in time proportional to the sum of their sizes. These arrays effectively represent set membership: an entry set to $1$ indicates that the corresponding element is included in the set, while an entry of $0$ indicates exclusion. With this setup, we are now ready to analyze the running time of a single recursive call.

\vspace{0.3 em}
\noindent
\textbf{\underline{Consider \Cref{line1:datastr1}:}}
\begin{description}
    \item[$\blacktriangleright$ Time to compute $\pot_{\alpha}$]: Consider the linked list $\mcon_{\alpha}$, which has size at most $\Gamma$, and the array $\pot_{\alpha}$. For every vertex $v \in \mcon_{\alpha}$, we iterate over its neighbors $\nbr(v)$ using the adjacency list. For each vertex $w \in \nbr(v)$, we increment the entry indexed by $w$ in the array $\pot_{\alpha}$ by $1$, that is $\pot[w] = \pot[w] + 1$.

After processing all vertices in $\mcon_{\alpha}$, let $|\mcon_{\alpha}| = x$. We then scan the array $\pot_{\alpha}$ and set each entry with value exactly $x+1$ to $1$, and reset all other entries to $0$. This yields a $\{0,1\}$-array representing the current $\pot_{\alpha}$. The set of vertices with corresponding entries equal to $1$ forms the current $\pot_{\alpha}$. The time to process each vertex $v \in \mcon_{\alpha}$ is proportional to its degree, $\deg(v)$. Thus, the total time to compute $\pot_{\alpha}$ is 
\[
\sum_{v \in \mcon_{\alpha}} \deg(v) \leq n \Gamma.
\]
Therefore, the total time to compute $\pot = \bigcup_{\alpha \in [k]} \pot_{\alpha}$ is bounded by $k n \Gamma$.

    \item[$\blacktriangleright$ Time to compute $\fcov_{\alpha}$]:
    
   Consider the array $\pot_{\alpha}$ for every $\alpha \in [k]$. For each vertex $v \in \cR$ such that $\pot_{\alpha}[v] = 1$, we examine its neighbors $\nbr(v)$ using the adjacency list. For every vertex $w \in \nbr(v)$, we access the entry indexed by $w$ in the array $\fcov_{\alpha}$ and increment it by $1$, that is $\fcov[v] = \fcov[v] + 1$. This process is repeated for all such vertices $v \in \cR$ whose $\pot[v] = 1$. Let $|\{v \mid \pot_{\alpha}[v] = 1\}| = x$. 

Next, we scan the array $\fcov_{\alpha}$ and all the entries with value equal to $x$ are updated to $1$, and the remaining entries are set to $0$. At this stage, $\fcov_{\alpha}$ becomes a binary array with values $0$ and $1$. The set of vertices corresponding to entries with value $1$ constitutes the current set $\fcov_{\alpha}$.

For each vertex $v \in \pot_{\alpha}$, the time required to scan its adjacency list is $\deg(v)$. Thus, the total time to compute $\fcov_{\alpha}$ is $\sum_{v \in \pot_{\alpha}}\deg(v) \leq m^\star$, where $m^\star = |E(G)|$. Consequently, the total time to compute $\fcov = \bigcup_{\alpha \in [k]}\fcov_{\alpha}$ is at most $km^\star$.

    \item[$\blacktriangleright$ Time to compute $\nfree$:] 
    
    Note that we have the array $\free$ of size $|\cB|$. Additionally, we have the array $\fcov$, computed in the previous paragraph. We now assign the array $\nfree$ as follows: for every vertex $v \in \cB$, we examine the entries at index $v$ in both $\fcov$ and $\free$. If $\fcov[v] = 1$ and $\free[v] = 0$, then we set $\nfree[v] = 1$; otherwise, we set it to $0$. The time required to compute $\nfree$ is $\cO(n)$, where $|\cB| = n$.
\end{description}

\vspace{0.3 em}
\noindent
\textbf{\underline{Consider \Cref{line:basecase1}:}} The first base case takes constant time to verify.

\vspace{0.3 em}
\noindent
\textbf{\underline{Consider \Cref{algprbds:basetwo}:}} It is easy to compute the set $\free \cup \nfree$ in time $\mathcal{O}(m)$. For a fixed $\alpha \in [k]$ and a vertex $v \in \pot_{\alpha}$, checking whether $\nbr(v) \subseteq (\free \cup \nfree)$ takes $\deg(v)$ time. Therefore, verifying this condition for all vertices in $\pot_{\alpha}$ requires $\mathcal{O}(n + m^\star)$ time. To determine whether there exists an $\alpha \in [k]$ that satisfies the condition in \Cref{algprbds:basetwo}, we examine each $\alpha$ individually. This yields an overall time complexity of $\mathcal{O}(k(n + m^\star))$.

\vspace{0.3 em}
\noindent
\textbf{\underline{Consider \Cref{algprbds:select_high_deg}:}} Consider a set $\pot_{\alpha}$. For every vertex $v \in \pot_{\alpha}$, we compute the number of its neighbors that are \emph{non-free}, which takes $\mathcal{O}(\deg(v))$ time. Therefore, identifying the vertex in $\pot_{\alpha}$ with the highest number of non-free neighbors takes total time $\sum_{v \in \pot_{\alpha}} \deg(v) \leq \mathcal{O}(n + m^\star)$. Consequently, finding such a vertex across all $\pot_{\alpha}$ for $\alpha \in [k]$ takes at most $\mathcal{O}(k(n + m^\star))$ time.

Every other line of the algorithm takes no more than $\cO(k(n+m^\star))$ time. 

\vspace{0.3 em}
\noindent
\textbf{Recurrence for time complexity:}
Let 
 \[ \mu(\cJ) = k + k \cdot \intcmp - \sum_{\iletter \in [k]} \dintcmp(G, \mcon_{\iletter}) \leq k + k \Gamma.\]
 The recurrence for computing the overall running time of the algorithm is given by 
\[
T(\mu(\cJ)) \leq T(\mu(\cJ) - 1) + c^\star \cdot k(n + m^\star),
\]
where $c^\star$ is a fixed constant. The base case is $T(0) = c$ for some constant $c$. This recurrence solves to 
\[
T(\mu(\cJ)) = \mu(\cJ) \cdot c^\star \cdot k(n + m^\star).
\]
Since $\mu(\cJ) \leq k+  k \Gamma$, the overall running time of one iteration of the algorithm is 
\[
(k + k \Gamma) \cdot c^\star \cdot k(n + m^\star) = \mathcal{O}(k^2 \Gamma (n + m^\star)).
\]

To boost the success probability to a constant, we repeat the algorithm independently \( \left( \frac{2k}{\varepsilon} \right)^{(\intcmp + 1)k} \) times. Hence, the total running time of our \epas\ is
\[
\mathcal{O}\!\left( \left( \frac{2k}{\varepsilon} \right)^{(\intcmp + 1)k} \cdot \Gamma k^{2}\,(n + m^\star) \right).
\]
This completes the proof.
\end{proof}

%% file: weighted_max_coverage.tex
\subsection{\wmrbdsfull}\label{subsection:wprbds}

In this section, we present an algorithm for the \wmrbdsfull(\wmrbds) problem. The problem takes as input a bipartite graph $\graph$, a weight function $\w: \cB \rightarrow \bQ^+$ and an integer $k$ and the goal is to find a subset of $k$ vertices, $\cS \subseteq \cR$ such the sum of weight of vertices in $\nbr(\cS)$ is maximum.
For any $B\subseteq \cB$, we define
$\w(B) = \sum_{v \in B}\w(v).$

Given an instance $\cI $ = \wprbdsi of the \wmrbds problem, the optimal value $\wopt_k(\cI)$ is defined as  
\[
\wopt_k(\cI) = \max_{\cS \subseteq \cR, |\cS| \leq k}\w(\nbr(\cS)),
\]  
 When the values of $k$ and the instance $\cI$ are clear from the context, we omit the subscript $k$ and the instance notation in the definition.

For completeness, the problem is defined as follows. 

\defparopt{\wmrbds}{A bipartite graph $\graph$, a weight function $\w:\cB \rightarrow \bQ^{+}$ and an integer $k$.}{$k$}{A subset of vertices $\cS \subseteq \cR$ such that $|\cS|\leq k$ and $\w(\nbr(\cS))= \wopt_k(\cI)$.}

\subsubsection{A Brief Description of the Algorithm} 

The algorithm for \wmrbds is same as \mrbds, except for the following modification during sampling step. The algorithm picks a vertex $h^\star 
\in \pot$  such that
\[
\w(\nbr(h^\star) \setminus (\free\cup \nfree)) = \w(\{ u_1, \dots, u_\ell \}),
\]  
that consists of only {\em non-free} vertices has the largest cardinality. Note that \emph{non-free} vertices refer to those vertices in \( \cB \) that are not freely covered so far till current recursive call and hence they do not belong to $\free \cup \nfree$. Given a vertex \( h^\star \in \pot \), the algorithm recursively solves either an instance in which \( h^\star \) is included in the solution, or an instance where the skeleton of some vertex in the solution, that is \( \mcon_{\iletter} \), for \( \iletter \in [k] \), must additionally include at least one vertex from the set \( \{ u_1, \dots, u_\ell \} \). For a fixed \( \varepsilon > 0 \), these two steps correspond to the following cases: either \( h^\star \) is part of some \( (1 - \varepsilon) \)-approximate solution, or there does not exist any such approximate solution that includes the vertex \( h^\star \).

For a fixed $\varepsilon > 0$, we show that when there is no $(1 - \varepsilon)$-approximate solution containing $h^\star$, then any optimal solution to our instance must contain a good fraction of vertices in $\nbr(h^\star) \setminus (\free \cup \nfree) = \{u_1, \dots, u_\ell\}$,
such that the total weight of the intersection is a significant fraction of the total weight of $\{u_1, \dots, u_\ell\}$ in particular at least $\varepsilon \w(\{u_1, \dots, u_\ell\})$. This step is implemented by randomly selecting an item \(\clubsuit\) from the set \(\{ h^\star, u_1, \dots, u_\ell \}\) with a probability distribution  
\[
    \mathbb{P}[\clubsuit] = 
    \begin{cases} 
    \frac{1}{2} & \text{if } \clubsuit = h^\star \\
    \frac{\w(u_\iletter)}{2 \omega} & \text{if } \clubsuit \in \{u_\iletter \mid \iletter \in [k]\}
    \end{cases}
    \]
where $\omega = \sum_{\iletter \in [\ell]} \w(u_\iletter)$. That is, we select the vertex \(h^\star\) with probability \(\frac{1}{2}\), or a vertex, $u$, in \(\{u_\iletter \mid \iletter \in [\ell]\}\) with probability \(\frac{\w(u_\iletter)}{2 \omega}\).

The notion of a consistent tuple in this case is also similar to the unweighted case. A complete description of the algorithm with pseudocode is provided in Algorithm~\ref{alg:framework-wprbds}.

\medskip
\begin{algorithm}[ht!]

\caption{\algwprbds}\label{alg:framework-wprbds}
  \KwInput{A bipartite graph $\graph$, a weight function $\w:\cB \rightarrow \bQ^+$, a $k$-tuple $\mi = \langle \mcon_1, \dots, \mcon_{k} \rangle $, $\free \subseteq \cB$ and $\varepsilon \in (0,1)$. }
  \KwOutput{A consistent $k$-tuple $\cS$  such that $\w(\nbrst(S, \cJ))$ is at least $(1-\varepsilon)\wopt_k(\cJ)$, where $\cJ=(\graph, \w, \mi,\free,\varepsilon)$ .}
Let $\pot_\iletter = \{ v \in \cR \mid \mcon_\iletter \subseteq \nbr(v) \}$,  
$\fcov_\iletter=\bigcap_{v \in \pot_\iletter} \nbr(v)$ for all $\iletter \in [k]$ and $\nfree = \left(\bigcup_{\iletter \in [k]} \fcov_\iletter \right) \setminus \free$. \\
\If {$\mi$ is a $0$-tuple} {\Return $0$-tuple}
\If {there exists $\iletter \in [k]$ such that $\bigcup_{v \in \pot_\iletter }\nbr(v) \subseteq (\free\cup \nfree)$\label{algwprbds:basetwo}} { Let $v$ be an arbitrary vertex in $\pot_\iletter$. \\
Let $\cS$ be the tuple returned by
$\algwprbds (\graph, \w, \mi[\downarrow \iletter],\free\cup \nfree  \cup \nbr(v), \varepsilon)$\\ \Return $\cS[\uparrow \iletter,v]$\label{algwprbds:return_red_rule}}
Let $h^\star$ be a vertex in $\pot$ such that $\w(\nbr(h^\star) \setminus (\free\cup \nfree))$ is the largest. Let $\nbr(h^\star) \setminus (\free\cup \nfree)= \{u_1,\dots u_\ell \}$ \label{algwprbds:select_high_deg}\\
Let $\omega=\sum_{j=1}^\ell \w(u_j)$. Randomly select an item $\clubsuit$ from the set $\{h^\star,u_1,\dots,u_\ell\}$ with  probability distribution $\left(\frac{1}{2},\frac{\w(u_1)}{2\omega},\dots,\frac{\w(u_\ell)}{2\omega}\right)$. That is, we select the vertex $h^\star$ with probability $\frac{1}{2}$, or an element $u$ of $\{u_1,\dots,u_\ell\}$ with probability $\frac{\w(u)}{2\omega}$.\label{algwprbds:prob_dist} \\

\uIf{$\clubsuit=h^\star \in  \pot_\iletter$, for some $\iletter\in [k]$ } {Let $\cS$ be a tuple returned by
$\algwprbds (\graph, \w, \mi[\downarrow \iletter],\free \cup \nfree  \cup \{u_1,\dots,u_\ell\}, \varepsilon)$\\ 
\Return $\cS[\uparrow \iletter,h^\star]$\label{algwprbds:return_high_deg}}
\Else {Let $\clubsuit = u \in \{u_1, \dots, u_\ell\}$. Pick an integer  $\beta\in[k]$ uniformly at random.\\
	 Assign $\mi \gets \langle \mcon_1, \dots, \mcon_{\beta} \cup \{u\}, \dots, \mcon_{k} \rangle$\\
	 Let $\cS$ be a tuple returned by  $\algwprbds(\graph, \w, \mi, \free\cup \nfree \cup \{u\}, \varepsilon)$\\
	\Return $\cS$\label{algwprbds:return_opt_elem}} 
\end{algorithm}

\subsubsection{Algorithm Analysis}
In this section, we show that our algorithm indeed outputs a solution with the desired approximation ratio, that is \epas. We start with a  definition that we use during the analysis. For an instance \(\cJ\) and a consistent tuple \(\cS\), we define  
\[
\w(\nbrst(\cS, \cJ)) = \w(\nbr(\cS) \setminus \free),
\]  
Further for an instance $\cJ$, we define the value of an optimal solution as follows:  
\[
\wopt(\cJ) =\max_{\cS= \langle s_1,\dots, s_k\rangle \in \cons{\cJ} } \w( \nbr(\cS) \setminus \free) = \max_{\cS \in \cons{\cJ} }\w(\nbrst(\cS, \cJ)).
\]  
Let \( \mathbb{O}=\langle o_1,\dots, o_k\rangle\) be a tuple such that  $\wopt(\cJ)  = \w(\nbrst(\mathbb{O}, \cJ))$.  
Such a tuple will be called an \emph{optimal tuple}.  In the following, we present a sequence of lemmata that capture relationships between two structurally similar instances arising from consecutive recursive calls. To avoid ambiguity caused by identically named objects across different calls, we annotate each object with its corresponding instance in brackets and omit those when the context is clear.

\begin{lemma}
\label{lem:wpropertyonealgoone}
 Let $\cJ$ denote the input instance 
 \[(\graph, \w,\mi = \langle \mcon_1, \dots, \mcon_{k} \rangle,\free, \varepsilon).\] 
 Consider $v\in \pot_{\iletter}(\cJ)$, for some $\iletter \in [k]$ and let $\cJ'$ denote the instance
  \[ (\graph, \w, \mi[\downarrow \iletter],\free(\cJ) \cup \nfree(\cJ) \cup \nbr(v), \varepsilon).\]   
Then, for any consistent tuple $\cS$ for $\cJ'$, we have the following.
\begin{description}
    \item {\rm \bf 1) Consistent:}  $\cS[\uparrow \iletter, v]$ is a consistent tuple for $\cJ$. 
    \item {\rm \bf 2) Weight Equality:} We have
    \begin{equation}\label{eqn:wfirstpartcovstarsum}
	\w(\nbrst(\cS[\uparrow \iletter,v], \cJ)) = \w(\nbrst(\cS,\cJ')) +\w(\nfree(\cJ)) + \w(\nbr(v) \setminus (\free(\cJ) \cup \nfree(\cJ))). 
\end{equation}
\end{description}
\end{lemma}
\begin{proof}
 Let $\cS = \langle s_1, \dots, s_{\alpha-1}, s_{\alpha+1}, \dots, s_k \rangle$ be a consistent tuple for $\cJ'$. 
The proof of \textbf{Consistency} is identical to the one in \Cref{lem:propertyonealgoone}, and is therefore omitted.

\medskip

\noindent 
 {\rm \bf 2) Weight Equality:} Due to the  {\bf Pairwise Disjoint Property} of \Cref{lem:propertyonealgoone}, we know that in this case, the following holds:
\begin{equation}
\label{eqn: wteq}
   \nbrst(\cS[\uparrow \iletter,v], \cJ)= \nbrst(\cS,\cJ') \uplus \nfree(\cJ) \uplus (\nbr(v) \setminus (\free(\cJ) \cup \nfree(\cJ))).
\end{equation}
Hence, \Cref{eqn:wfirstpartcovstarsum} holds.
\end{proof}

\begin{lemma}
\label{lem:wpropertytwoalgoone}
Let \( \cJ \) be the input instance  
\[
(\graph, \w, \mi = \langle \mcon_1, \dots, \mcon_{k} \rangle, \free, \varepsilon).
\]  
Consider \( u \in \cB \setminus (\free(\cJ)\cup \nfree(\cJ)) \). For any \( \beta \in [k] \),  let $$\mi' \gets \langle \mcon_1, \dots, \mcon_{\beta} \cup \{u\},\dots, \mcon_{k} \rangle,$$ and  let \( \cJ' \) denote the instance  
\[
(\graph, \w, \mi', \free(\cJ)\cup \nfree(\cJ) \cup \{u\}, \varepsilon).
\]
 
Then, for any consistent tuple $\cS$ for $\cJ'$, we have the following.
\begin{description}
    \item {\rm \bf 1) Consistent:}  $\cS$ is a consistent tuple for $\cJ$. 
    \item {\rm \bf 2) Weight Equality:}
\begin{equation}\label{eqn:wtequal2}
	\w(\nbrst(\cS, \cJ))=\w(\nbrst(\cS,\cJ '))+\w( \nfree(\cJ))+\w(u)
\end{equation}
\item  {\rm \bf 3) Intersection Complexity:} $\dintcmp(G,\mcon_\beta \cup \{u\})\geq \dintcmp(G,\mcon_\beta)+1$
\end{description}

\end{lemma}
\begin{proof}
Let $\cS = \langle s_1, \dots, s_k \rangle$ be a consistent tuple for $\cJ'$. 
\noindent
The proofs for \textbf{Consistency} and \textbf{Intersection Complexity} follow identically from the corresponding arguments in \Cref{lem:propertytwoalgoone}, and are therefore omitted.

\smallskip
\noindent
{\rm \bf 2) Weight Equality:} 
Due to the \textbf{Pairwise Disjoint} property in \Cref{lem:propertytwoalgoone}, we know that in this case,
\begin{equation}
\label{eqn:wteqn2}
\nbrst(\cS, \cJ) = \nbrst(\cS,\cJ') \uplus \nfree(\cJ) \uplus \{u\}.
\end{equation}
Hence, \Cref{eqn:wtequal2} holds.
 \end{proof}

 Let \(\intcmp = \dintcmp(\graph)\). We define a measure associated with \(\cJ\), denoted by \(\mu(\cJ)\), as  
\[
\mu(\cJ) = k + k\intcmp - \sum_{\iletter \in [k]} \dintcmp( G, \mcon_{\iletter})
\]  
We use this measure for induction in our proof for both approximation as well as probability analysis. We now present the main technical lemma of this section, along with its proof.
\begin{lemma} \label{lem:framework-wprbds}
For a given input instance $$\cJ = (\graph, \w, \mi = \langle \mcon_1, \dots, \\ \mcon_{k} \rangle, \free, \varepsilon)$$ Algorithm \ref{alg:framework-wprbds}~\textnormal{(\algwprbds)} runs in time $\cO(\sizeg) $and outputs a consistent tuple $\cS = \langle s_1, s_2, \dots, s_k \rangle$ such that  
\[\w(\nbrst(\cS, \cJ)) \geq (1-\varepsilon)\wopt(\cJ)\]
with probability at least $\left( \frac{\varepsilon}{2k} \right)^{\mu(\cJ)}$.
\end{lemma}

\begin{proof}
    We divide the proof into two parts. First, we establish the approximation ratio of our algorithm, and then we analyze its success probability. Our proof proceeds by induction, using the measure defined above: 
\[
\mu(\cJ) = k + k\intcmp - \sum_{\iletter \in [k]} \dintcmp( G, \mcon_{\iletter})
\]
In particular, we prove the desired statement by induction on \(\mu(\cJ)\).

\paragraph{\bf Non-Negativity of Measure:} 
Note that \(\mu(\cJ) = k + k\intcmp - \sum_{\iletter \in [k]} \dintcmp( G, \mcon_{\iletter}(\cJ))\) is always non-negative, since for all \(\iletter \in [k]\), we have \( \dintcmp( G,\mcon_{\iletter}(\cJ)) \leq \intcmp \), which implies \( \sum_{\iletter =1}^k \dintcmp (G,\mcon_{\iletter}(\cJ)) \leq k \intcmp \). Moreover, since any subset cannot have a negative size, we know that \( k \geq 0 \). Hence, it follows that  
\[
\mu(\cJ) = k + k\intcmp - \sum_{\iletter \in [k]} \dintcmp( G, \mcon_{\iletter}) \geq 0.
\]

\paragraph{\bf Analysis of the algorithm.} 
Consider an optimal tuple \(\bo = \langle o_1, \dots, o_k \rangle\) for the instance \(\cJ\). By definition, we have \( \wopt(\cJ)  = \w(\nbrst(\bo, \cJ))  \). We prove the statement of the lemma using induction on \(\mu(\cJ)\). That is, the statement of the lemma serves as our induction hypothesis.

\smallskip

\noindent 
	\textbf{Base Case:} When $\mu(\cJ) = 0$, then we have that $k=0$ since $\intcmp \geq \dintcmp(G,\mcon_{\iletter})$ for $\iletter \in [k]$.  This implies that $\mi$ is a $0$-tuple. Thus, $0$-tuple is the only consistent tuple in this case and \Cref{alg:framework-wprbds} correctly returns a $0$-tuple with {\color{blue}\texttt{probability $1$}}.  

    \smallskip

\noindent 
\textbf{Induction Assumption:} Let $\mu' >0$. Assume that the statement of \Cref{lem:framework-wprbds} holds for all instances, $\cJ$,  for which  $\mu(\cJ) < \mu'$. That is, for an  input instance $$\cJ=(\graph, \w, \mi = \langle \mcon_1, \dots \mcon_{k} \rangle,\free, \varepsilon)$$ such that $\mu(\cJ) <\mu'$, \Cref{alg:framework-wprbds} $\textnormal{(\algwprbds)} $ outputs a consistent tuple $\cS = \langle s_1, s_2, \dots, s_k \rangle$ such that  
\[\w(\nbrst(\cS, \cJ)) \geq (1-\varepsilon)\wopt(\cJ)\]
with probability at least $\left( \frac{\varepsilon}{2k}\right)^{\mu(\cJ)}$.

\smallskip

\noindent 
\textbf{Inductive Step:} We prove the statement of Lemma \ref{lem:framework-wprbds} for an arbitrary instance, $\cJ$, for which $\mu(\cJ) = \mu'$.

\medskip

\noindent 
\underline{\textbf{Approximation Analysis.}}   We first perform the approximation analysis \textbf{assuming} that when the algorithm recursively calls an instance \( \cJ' \) such that \( \mu(\cJ') < \mu(\cJ) \), it indeed returns a solution \( \cS' \) for \( \cJ' \) satisfying  
\[
\w(\nbrst(\cS', \cJ')) \geq (1 - \varepsilon) \wopt(\cJ').
\]  
In reality, this outcome occurs with some probability, which we will analyze after establishing the desired approximation factor. In simple terms, we condition on the recursive subroutines ``correctly'' returning the expected approximate solution and then proceed to analyze the resulting approximation factor.

We divide the approximation analysis in two parts. In the first part, we analyse the case when the condition of ~\Cref{algwprbds:basetwo} holds. Then, we analyse the case when the algorithm executes~\Cref{algwprbds:select_high_deg}. 

\smallskip
\noindent 
\textbf{Case A: Analysis for the case when condition of ~\Cref{algwprbds:basetwo} holds:}
 Let $\cJ$ denote the input instance 
 \[(\graph, \w, \mi = \langle \mcon_1, \dots, \mcon_{k} \rangle,\free, \varepsilon)\] 
 and $\cJ'$ denote the instance
  \[ (\graph, \w, \mi[\downarrow \iletter],\free \cup \nfree \cup \nbr(v), \varepsilon).\] 
Let \( \cS \) be a consistent tuple for \( \cJ' \) returned in \Cref{algwprbds:return_red_rule} of the algorithm. Then, by Lemma~\ref{lem:wpropertyonealgoone}, we know that the returned solution $\cS[\uparrow \iletter, v]$ is a consistent tuple for $\cJ$. Next, we show that it also satisfies the approximation guarantee. That is, 
	 \[\w(\nbrst(\cS[\uparrow \iletter,v], \cJ)) \geq (1-\varepsilon)\wopt(\cJ)
	 \]
As a step towards our proof, we first establish that \( \mu(\cJ') < \mu(\cJ) \), allowing us to apply the induction hypothesis to \( \cJ' \). Observe that 
\begin{align}
	&\mu(\cJ) - \mu(\cJ') \nonumber \\ 
	 &=\big(k + k\intcmp - \sum_{\iletter \in [k]} \dintcmp( G, \mcon_{\iletter}) \big)- \big( (k-1) + (k-1)\intcmp - \sum_{\substack{\iletter \in [k] \\ \iletter \neq j}} \dintcmp(G,  \mcon_{\iletter})\big) \nonumber \\
	&=(1+\intcmp-\dintcmp(G,\mcon_{j})) \stackrel{(\P)}{>}0 \label{measure:wcaseA}
\end{align}
where $(\P)$ follows from the fact that $\intcmp\geq \dintcmp(G,\mcon_j)$. Now since $\mu(\cJ') < \mu(\cJ)$, by induction hypothesis we have
\[\w(\nbrst(\cS,\cJ')) \geq (1-\varepsilon)\wopt(\cJ').
\]

Recall that  $\bo$ is an optimal tuple for the instance $\cJ$. By definition, this implies that  $\wopt(\cJ) = \w(\nbrst(\bo, \cJ))$.  Furthermore, $\mathbb{O}[\downarrow \iletter]$  forms a consistent tuple for the instance  $\cJ'$. Therefore, it follows that  $\wopt(\cJ') \geq \w(\nbrst(\bO[\downarrow \iletter], \cJ'))$.  
Finally, by the second statement of Lemma~\ref{lem:wpropertyonealgoone} (\Cref{eqn:wfirstpartcovstarsum}), we conclude that 
$$\w(\nbrst(\cS[\uparrow \iletter,v],\cJ)) =\w(\nbrst(\cS,\cJ '))+\w(\nfree(\cJ))+ \w(\nbr(v) \setminus (\free(\cJ) \cup \nfree(\cJ))).$$
 Since, $\nbr(v)  \subseteq (\free(\cJ) \cup \nfree(\cJ))$, we have that $\nbr(v) \setminus (\free(\cJ) \cup \nfree(\cJ)) = \emptyset$. Thus, we get the following. 
\begin{align*}
	\w(\nbrst(\cS[\uparrow \iletter,v],\cJ)) &=\w(\nbrst(\cS,\cJ '))+ \w(\nfree(\cJ))  & \\ 
	&\geq (1-\varepsilon)\wopt(\cJ') + \w(\nfree(\cJ)) & \\ & (\textnormal{by induction hypothesis})\\
	&\geq (1-\varepsilon)\w(\nbrst(\bO[\downarrow \iletter], \cJ'))+ \w(\nfree(\cJ))\\
	&= (1-\varepsilon)(\w(\nbrst(\bO,\cJ)) -\w(\nfree(\cJ)))+ \w(\nfree(\cJ)) & \\
	&  (\textnormal{By applying \Cref{eqn:wfirstpartcovstarsum} on $\bO[\downarrow \iletter]$ for the instance $\cJ'$}) & \\
	&\geq (1-\varepsilon)\w(\nbrst(\bO,\cJ))\\
	&= (1-\varepsilon)\wopt(\cJ)
\end{align*}

\noindent  \textbf{Case B: Analysis for the case when~\Cref{algwprbds:select_high_deg} is executed:} Let $h^\star \in  \pot_\iletter$, for some $\iletter\in [k]$, be the vertex selected in \Cref{algwprbds:select_high_deg}.  We have following two cases.

\medskip
\noindent
\textbf{Case $\textnormal{\textbf{B}}_{\textbf{1}}$:} When $\w(\nbr(\{o_1,\dots,o_k\}) \cap \{u_1,\dots u_\ell\}) < \varepsilon \w(\{u_1,\dots u_\ell\}) $ \\
	Proof of this case is similar to Case A we considered earlier. In this case, consider the branch where the algorithm returns $\cS[\uparrow \iletter,h^\star]$ in \Cref{algwprbds:return_high_deg}, where  $\cS$ is the tuple returned by \[\algwprbds (\graph, \w, \mi[\downarrow \iletter],\free\cup\nfree \cup \{u_1,\dots,u_\ell\}, \varepsilon)\] 
	 Let $\cJ$ denote the input instance 
     \[(\graph, \w, \mi = \langle \mcon_1, \dots \mcon_{k} \rangle,\free, \varepsilon),\] 
     and $\cJ'$  denote the  instance 
     \[(\graph, \w, \mi[\downarrow \iletter],\free(\cJ)\cup \nfree(\cJ) \cup  \{u_1,\dots,u_\ell\}, \varepsilon\ ),\] 
     which the algorithm recursively solves in this branch.
Let \( \cS \) be a consistent tuple for \( \cJ' \) returned in \Cref{algwprbds:return_high_deg} of the algorithm. Then by Lemma~\ref{lem:wpropertyonealgoone}, we know that $\cS[\uparrow \iletter, h^\star]$ is a consistent tuple for $\cJ$. Next, we show that it also satisfies the approximation guarantee. That is, we need to show that
	 \[\w(\nbrst(\cS[\uparrow \iletter,h^\star], \cJ)) \geq (1-\varepsilon)\wopt(\cJ). 
	 \]

As a step toward our proof, we first establish that \( \mu(\cJ') < \mu(\cJ) \), allowing us to apply the induction hypothesis on \( \cJ' \). Observe that 
\begin{align}
\mu(\cJ) - \mu(\cJ')
&= \bigg(k+k\intcmp-
\sum_{\iletter\in[k]}\dintcmp(G,\mcon_{\iletter})\bigg) \nonumber\\
&\quad-\bigg((k-1)+(k-1)\intcmp-
\sum_{\substack{\iletter\in[k]\\ \iletter\ne j}}
\dintcmp(G,\mcon_{\iletter})\bigg) \nonumber \\
&= (1 + \intcmp - \dintcmp(G, \mcon_{j})) \stackrel{(\P)}{>} 0
\label{measure:wcaseB}
\end{align}
where $(\P)$ follows from the fact that $\intcmp = \dintcmp(G) \geq \dintcmp(G,\mcon_j)$. Now since $\mu(\cJ') < \mu(\cJ)$, by induction hypothesis, we have 
\[\w(\nbrst(\cS,\cJ')) \geq (1-\varepsilon)\wopt(\cJ')
\]

Recall that  $\bo$ is an optimal tuple for the instance $\cJ$. By definition, this implies that  $\wopt(\cJ) = \w(\nbrst(\bo, \cJ))$.  Furthermore, $\mathbb{O}[\downarrow \iletter]$  forms a consistent tuple for the instance  $\cJ'$. Therefore, it follows that  $\wopt(\cJ') \geq \w(\nbrst(\bO[\downarrow \iletter], \cJ'))$.  Recall that, $\omega=\sum_{j=1}^\ell \w(u_j).$

\smallskip
Now, we derive a lower bound on $ \wopt(\cJ')$ in terms of $\wopt(\cJ) $. 
\begin{align*}
	 \wopt(\cJ')& \geq \w(\nbrst(\mathbb{O}[\downarrow \iletter], \cJ')) & \\
	 &= \w\big(\nbr(\mathbb{O}[\downarrow \iletter]) \setminus (\free(\cJ)\cup \nfree(\cJ) \cup \{u_1,\dots u_\ell\})\big) \\
	 &= \w\big(\nbr(\bO [\downarrow \iletter]) \setminus (\free(\cJ)\cup \nfree(\cJ)) \setminus (\{u_1,\dots u_\ell\})\big)\\
	 &= \w\big(\nbr(\bO[\downarrow \iletter])\setminus (\free(\cJ)\cup \nfree(\cJ))\big) & \\ & \quad - \w\big(\nbr(\bO[\downarrow \iletter])\setminus ((\free(\cJ)\cup \nfree(\cJ)) \cap \{u_1,\dots u_\ell\})\big)\\
     &\geq \w\big(\nbr(\bO[\downarrow \iletter])\setminus (\free(\cJ)\cup \nfree(\cJ))\big) - \varepsilon \w(\{u_1,\dots u_\ell\})\\
     & \stackrel{(\bigstar)}{\geq} \w(\nbr(\bO) \setminus (\free(\cJ)\cup \nfree(\cJ))) -\w(\nbr(o_\iletter)\setminus (\free(\cJ)\cup \nfree(\cJ))) & \\ & \quad -\varepsilon \omega\\
	 &\geq \w(\nbr(\bO) \setminus \free) - \w(\nfree(\cJ)) -\w(\nbr(o_\iletter)\setminus (\free(\cJ)\cup \nfree(\cJ))) - \varepsilon \omega\\
     &\geq \wopt(\cJ) -\w(\nfree(\cJ)) -\w(\nbr(o_\iletter)\setminus (\free(\cJ)\cup \nfree(\cJ))) - \varepsilon \omega
\end{align*}
($\bigstar$) holds because we are in Case $\textnormal{B}_1$ where $\w(\nbr(\bO) \cap \{u_1,\dots u_\ell\}) < \varepsilon \w(\{u_1,\dots u_\ell\}) =  \varepsilon \omega$ which also implies that $\w(\nbr(\bO[\downarrow \iletter])\setminus (\free(\cJ)\cup \nfree)\cap \{u_1,\dots u_\ell\}) < \varepsilon \omega $.

This gives us 
\begin{equation}
\label{eqn:wcase1covstar}
\begin{split}
\w(\nbrst(\cS,\cJ')) &\geq (1-\varepsilon)\wopt(\cJ') \\
&\geq (1-\varepsilon) (\wopt(\cJ) -\w(\nbr(o_\iletter)\setminus (\free(\cJ)\cup \nfree(\cJ))) - \w(\nfree) - \varepsilon \omega)
\end{split}
\end{equation}

\noindent Finally, we show the desired approximation ratio. 
\begin{align*}
	&\w(\nbrst(\cS[\uparrow \iletter,h^\star], \cJ))\\&=\w(\nbrst(\cS,\cJ')) + \w(\{u_1,\dots u_\ell\}) + \w(\nfree(\cJ)) & \\
		&  (\textnormal{By applying \Cref{eqn:wfirstpartcovstarsum} on $\cS$ }) & \\
	&\geq (1-\varepsilon) (\wopt(\cJ) -\w(\nbr(o_\iletter) \setminus (\free(\cJ)\cup \nfree(\cJ))) - \varepsilon \omega- \w(\nfree(\cJ))) & \\
	& + \omega + \w(\nfree(\cJ)) ~(\textnormal{from equation~\ref{eqn:wcase1covstar}}) & \\
	&\geq (1-\varepsilon) \wopt(\cJ) + (1-\varepsilon)(\omega - \w(\nbr(o_\iletter)\setminus (\free(\cJ)\cup \nfree(\cJ)))) +\varepsilon \w(\nfree(\cJ))\\
	& \stackrel{(\heartsuit)}{\geq} (1-\varepsilon) \wopt(\cJ)
\end{align*}
$(\heartsuit)$ follows from the fact that $\omega = \w(\nbr(h^\star)\setminus (\free(\cJ)\cup \nfree)) = \w(\{u_1, \dots, u_{\ell}\})$ denotes the largest weight of the non-free neighbors of any vertex in $\pot$ and hence  $\omega \geq \w(\nbr(o_\iletter) \setminus (\free(\cJ)\cup \nfree)) $.\\

\noindent
\textbf{Case $\textnormal{\textbf{B}}_{\textbf{2}}$:} When $ \w(\nbr(\{o_1,\dots,o_k\}) \cap \{u_1,\dots u_\ell\}) \geq \varepsilon \w(\{u_1,\dots u_\ell\}) $ \\

In this case, consider the branch where the algorithm executes \Cref{algwprbds:return_opt_elem} and returns  $\cS$, 
 where  $\cS$ is the tuple returned by
\[\algwprbds(\graph, \w, \mi, \free(\cJ)\cup \nfree \cup \{u\}, \varepsilon)\] 
where 
\[\mi \gets \langle \mcon_1, \dots, \mcon_{\iletter} \cup \{u\}, \dots, \mcon_{k} \rangle\] and the selected element $u$ belongs to $\nbr(o_{\iletter})$.  Let $\cJ$ denote the input instance 
\[(\graph, \w, \mi = \langle \mcon_1, \dots \mcon_{k} \rangle,\free, \varepsilon)\] and  $\cJ'$ denote the instance 
\[(\graph, \w, \mi',\free(\cJ)\cup \nfree \cup \{u\}, \varepsilon)\] 
where $\mi' = \langle  \mcon_1, \dots, \mcon_{\iletter} \cup \{u\}, \dots, \mcon_{k} \rangle$,
 which the algorithm recursively solves in this branch.

Let $\cS$ be a consistent tuple for $\cJ'$, then due to~\Cref{lem:wpropertytwoalgoone}, $\cS$ is also a consistent tuple for $\cJ$.  Next, we show that it also satisfies the approximation guarantee. That is, we need to show that
\[\w(\nbrst(\cS,\cJ))\geq(1-\varepsilon)\wopt(\cJ) \]

As a step towards our proof, we first establish that \( \mu(\cJ') < \mu(\cJ) \), allowing us to apply the induction hypothesis on \( \cJ' \). Observe that $\mu(\cJ) - \mu(\cJ')$ is at least 
\begin{align}
	 & \left(k + k\intcmp - \sum_{\iletter \in [k]} \dintcmp(  G, \mcon_{\iletter}) \right) - \left(k + k\intcmp - \big(\sum_{\iletter \in [k]} \dintcmp(  G, \mcon_{\iletter})+1\big)\right)\nonumber\\
        & \textnormal{ (since }\dintcmp(G,\mcon_\iletter \cup \{u\})\geq \dintcmp(G, \mcon_\iletter)+1 \textnormal{ due to~\Cref{lem:wpropertytwoalgoone})} \nonumber\\
	&=1 >0  \label{measure:wCaseC}
\end{align}
Now since $\mu(\cJ') < \mu(\cJ)$, by induction hypothesis we have 
\[\w(\nbrst(\cS,\cJ')) \geq (1-\varepsilon)\wopt(\cJ')
\]

Note that in this branch,  $\bO= \langle o_1,\dots, o_k\rangle$ is a consistent tuple of $\cJ'$ since the selected element $u$ belongs to $o_{\iletter}$.  Hence we can say that $\wopt(\cJ') \geq  \w(\nbrst(\bO, \cJ'))$. Thus we have that
\begin{align*}	\w(\nbrst(\cS,\cJ)) &=\w(\nbrst(\cS,\cJ '))+ \w(\nfree(\cJ))+ \w(u)  & \\ & (\textnormal{from equation~\ref{eqn:wtequal2}})\\
	&\geq (1-\varepsilon)\wopt(\cJ') + \w( \nfree(\cJ))+\w(u) &  \\ &(\textnormal{by induction hypothesis})\\
	&\geq (1-\varepsilon)\w(\nbrst(\bO, \cJ'))+\w( \nfree(\cJ))+\w(u) &\\
	&= (1-\varepsilon)(\w(\nbrst(\bO,\cJ)) - \w(\nfree(\cJ))-\w(u))+ \w(\nfree(\cJ))+\w(u) & \\ &(\textnormal{from equation~\ref{eqn:wtequal2}})\\
	&\geq (1-\varepsilon)\w(\nbrst(\bO,\cJ))\\
	&= (1-\varepsilon)\wopt(\cJ)
\end{align*}	

This concludes the approximation analysis of the algorithm. 

\medskip

\noindent 
\underline{\textbf{Probability  Analysis:}}  We next prove the success probability of our algorithm. Just as we do in approximation analysis, we go through each case and analyze the desired probability of success.  For an input instance $\cJ=(\graph, \w, \mi = \langle \mcon_1, \dots \mcon_{k} \rangle,\free, \varepsilon)$, we say that a tuple $\cQ$ is {\em good for $\cJ$} if $\w(\nbrst(\cQ,\cJ)) \geq (1-\varepsilon)\wopt(\cJ)$, otherwise it is {\em bad for $\cJ$}.

\smallskip
\noindent 
\textbf{Case A: Analysis for the case when condition of ~\Cref{algwprbds:basetwo} holds:}
 Let $\cJ$ denote the input instance 
 \[(\graph, \w, \mi = \langle \mcon_1, \dots \mcon_{k} \rangle,\free, \varepsilon)\] 
 and $\cJ'$ denote the instance
  \[ (\graph, \w, \mi[\downarrow \iletter],\free \cup \nfree \cup \nbr(v), \varepsilon).\] 
Let \( \cS \) be a consistent tuple for \( \cJ' \) returned in \Cref{algwprbds:return_red_rule} of the algorithm. By Equation \ref{measure:wcaseA}, we know that $\mu(\cJ')< \mu(\cJ)$. Let $\mathbf{P}=\prob{\cS[\uparrow \iletter,v] \mbox{ is good for }  \cJ}$. Then
\begin{eqnarray*}
\mathbf{P} & = &  \prob{\cS[\uparrow \iletter,v] \mbox{ is good for }  \cJ \mid \cS \mbox{ is good for }  \cJ'}  
 \prob{\cS \mbox{ is good for }  \cJ'} \\
 && +   \prob{\cS[\uparrow \iletter,v] \mbox{ is good for }  \cJ \mid \cS \mbox{ is bad for }  \cJ'}  
 \prob{\cS \mbox{ is bad for }  \cJ'} \\
 &\geq & \prob{\cS[\uparrow \iletter,v] \mbox{ is good for }  \cJ \mid \cS \mbox{ is good for }  \cJ'}  
 \prob{\cS \mbox{ is good for }  \cJ'} \\
 & =& 1 \times  \prob{\cS \mbox{ is good for }  \cJ'} \\
& \geq  &  \left( \frac{\varepsilon}{2k} \right)^{\mu(\cJ')} (\textnormal{by induction hypothesis})\\
& \geq &  \left( \frac{\varepsilon}{2k} \right)^{\mu(\cJ)}
 \end{eqnarray*}

\smallskip
\noindent 
\textbf{Case B: When \Cref{algwprbds:select_high_deg} is executed:}
Next, we will consider the case when we select a vertex  $h^\star$  in $\pot$ such that $\w(\nbr(h^\star) \setminus (\free(\cJ)\cup \nfree))= \w(\{u_1,\dots u_\ell \})$ is the largest(\Cref{algwprbds:select_high_deg}). In this case the algorithm randomly selects the vertex $h^\star$ with probability $\frac{1}{2}$, or a non-free neighbor of $h^\star$ with probability $\frac{\w(u)}{2\omega}$  (\Cref{algwprbds:prob_dist}).  

Recall that \(\bo = \langle o_1, \dots, o_k \rangle\) is  an optimal tuple for the instance \(\cJ\). 
Now we consider two cases based on whether $(a)~ \w(\nbr(\bO) \cap \{u_1,\dots u_\ell\} ) < \varepsilon \w(\{u_1,\dots u_\ell\} ) $,  or $(2)~ \w(\nbr(\bO) \cap \{u_1,\dots u_\ell\} ) \geq \varepsilon \w(\{u_1,\dots u_\ell\} ) $

 This corresponds to execution of \Cref{algwprbds:return_high_deg}, or   \Cref{algwprbds:return_opt_elem}. Furthermore,  let the algorithm return a tuple $\cQ$. 
\begin{eqnarray*}
 \prob{\cQ \mbox{ is good for }  \cJ} & = &  \prob{\cQ \mbox{ is good for }  \cJ \wedge \w(\nbr(\bO) \cap \{u_1,\dots u_\ell\} ) < \varepsilon  \w(\{u_1,\dots u_\ell\} )}  
 \\
 && +   \prob{\cQ \mbox{ is good for }  \cJ \wedge \w(\nbr(\bO) \cap \{u_1,\dots u_\ell\} ) \geq \varepsilon  \w(\{u_1,\dots u_\ell\} )}  \\
 &\geq & \min \begin{cases}
      \prob{\cQ \mbox{ is good for }  \cJ \wedge \w(\nbr(\bO) \cap \{u_1,\dots u_\ell\} ) < \varepsilon  \w(\{u_1,\dots u_\ell\} )}, \\
     \prob{\cQ \mbox{ is good for }  \cJ \wedge \w(\nbr(\bO) \cap \{u_1,\dots u_\ell\} ) \geq \varepsilon  \w(\{u_1,\dots u_\ell\} )} 
 \end{cases}
 \end{eqnarray*}
We will show that in either case  $\prob{\cQ \mbox{ is good for }  \cJ} \geq \left( \frac{\varepsilon}{2k} \right)^{\mu(\cJ)}.$

\medskip

\noindent 
\textbf{Case $\textnormal{\textbf{B}}_{\textbf{1}}$:} When $ \w(\nbr(\bO) \cap \{u_1,\dots u_\ell\} ) < \varepsilon  \w(\{u_1,\dots u_\ell\} )$ \\
 In this case, consider the branch where the algorithm returns $\cS[\uparrow \iletter,h^\star]$ in \Cref{algwprbds:return_high_deg}, where  $\cS$ is the tuple returned by \[\algwprbds (\graph, \w, \mi[\downarrow \iletter],\free\cup\nfree \cup \{u_1,\dots,u_\ell\}, \varepsilon)\] 
	 Let $\cJ$ denote the input instance 
     \[(\graph, \w, \mi = \langle \mcon_1, \dots \mcon_{k} \rangle,\free, \varepsilon),\] 
     and $\cJ'$  denote the  instance 
     \[(\graph, \w, \mi[\downarrow \iletter],\free(\cJ)\cup \nfree \cup  \{u_1,\dots,u_\ell\}, \varepsilon\ ),\] 
     which the algorithm recursively solves in this branch.
Let \( \cS \) be a consistent tuple for \( \cJ' \) returned in \Cref{algwprbds:return_high_deg} of the algorithm. By Equation \ref{measure:wcaseB}, we know that 
$\mu(\cJ')< \mu(\cJ)$. Let $\mathcal{E}$ denote the event that $\w(\nbr(\bO) \cap \{u_1,\dots u_\ell\} ) < \varepsilon  \w(\{u_1,\dots u_\ell\} )$. Let $\mathbf{P}=\prob{\cS[\uparrow \iletter,h^\star] \mbox{ is good for }   \cJ  \wedge \mathcal{E}}$

\begin{eqnarray*}
 \mathbf{P} & \geq & \prob{\cS[\uparrow \iletter,h^\star]  \mbox{ is good for }  \cJ \wedge \mathcal{E} \mid \cS \mbox{ is good for }  \cJ'} \times\\ 
 & & \prob{\cS \mbox{ is good for }  \cJ'} \\
 & = & \prob{ \w(\nbrst(\cS[\uparrow \iletter,h^\star],\cJ)) \geq (1-\varepsilon)\wopt(\cJ) \wedge \mathcal{E} \mid \cS \mbox{ is good for }  \cJ'}\times \\ 
 & & \prob{\cS \mbox{ is good for }  \cJ'}\\
 & \geq & \prob{\clubsuit=h^\star }\times  \prob{\cS \mbox{ is good for }  \cJ'} \\
& \geq  &  \frac{1}{2}\left( \frac{\varepsilon}{2k} \right)^{\mu(\cJ')} (\textnormal{by induction hypothesis})\\
& \geq &  \left( \frac{\varepsilon}{2k} \right)^{\mu(\cJ)}
 \end{eqnarray*}

\medskip

\noindent 
\textbf{Case $\textnormal{\textbf{B}}_{\textbf{2}}$:} When $ \w(\nbr(\bO) \cap \{u_1,\dots u_\ell\} ) \geq \varepsilon  \w(\{u_1,\dots u_\ell\}) $

In this case, consider the branch where the algorithm executes \Cref{algwprbds:return_opt_elem} and returns  $\cS$, 
 where  $\cS$ is the tuple returned by
\[\algwprbds(\graph, \w, \mi, \free(\cJ)\cup \nfree \cup \{u\}, \varepsilon)\] 
where 
$\mi \gets \langle \mcon_1, \dots, \mcon_{\iletter} \cup \{u\}, \dots, \mcon_{k} \rangle$ and the selected element, $u$, belongs to $\nbr(o_{\iletter})$.  Let $\cJ$ denote the input instance 
\[(\graph, \w, \mi = \langle \mcon_1, \dots \mcon_{k} \rangle,\free, \varepsilon)\] and  $\cJ'$ denote the instance 
\[(\graph, \w, \mi',\free(\cJ)\cup \nfree \cup \{u\}, \varepsilon)\]  
where $\mi' = \langle  \mcon_1, \dots, \mcon_{\iletter} \cup \{u\}, \dots, \mcon_{k} \rangle$,
 which the algorithm recursively solves in this branch.

Let \( \cS \) be a consistent tuple for \( \cJ' \) returned in \Cref{algwprbds:return_opt_elem} of the algorithm. By \Cref{measure:wCaseC} we know that 
$\mu(\cJ')< \mu(\cJ)$. Let $\mathcal{E}$ denote the event that $\w(\nbr(\{o_1,\dots,o_k\}) \cap \{u_1,\dots u_\ell\} ) \geq \varepsilon  \w(\{u_1,\dots u_\ell\})$. Let $\mathbf{P}=\prob{\cS \mbox{ is good for }   \cJ  \wedge \mathcal{E}}$
\begin{eqnarray*}
\mathbf{P} & \geq & \prob{\cS \mbox{ is good for }  \cJ \wedge \mathcal{E} \mid \cS \mbox{ is good for }  \cJ'} \times\\ 
 & & \prob{\cS \mbox{ is good for }  \cJ'} \\
 & = & \prob{ \w(\nbrst(\cS,\cJ)) \geq  (1-\varepsilon)\wopt(\cJ) \wedge \mathcal{E} | \cS \mbox{ is good for }  \cJ'}\times \\ 
 & & \prob{\cS \mbox{ is good for }  \cJ'}\\
   &\geq &   \prob{\clubsuit= u \in \{u_1, \dots, u_\ell\} \cap\nbr(\bO) \wedge u\in \nbr(o_\alpha)}\times \\ & & \prob{\cS \mbox{ is good for }  \cJ'} \\
& \geq  &   \prob{\clubsuit= u \in \{u_1, \dots, u_\ell\} \cap\nbr(\bO)}\times \prob{u\in \nbr(o_\alpha)}\times \left( \frac{\varepsilon}{2k} \right)^{\mu(\cJ')} \\ & & (\textnormal{by induction hypothesis})\\
& \geq &  \frac{\varepsilon \omega}{2 \omega} \frac{1}{k}\left( \frac{\varepsilon}{2k} \right)^{\mu(\cJ')}\\
& \geq & \left( \frac{\varepsilon}{2k} \right)^{\mu(\cJ)}
 \end{eqnarray*}
 Thus, we have shown that in either case 
\[
\Pr\bigl[\cQ \mbox{ is good for } \cJ\bigr] \ge \left(\frac{\varepsilon}{2k}\right)^{\mu(\cJ)}.
\]
This concludes the probability analysis. Finally, observe that the running time is $\cO(\sizeg)$, based on the arguments regarding the linear dependence of the running time on the input size, as established in the proof of~\Cref{thm:mainUnweighted}. This completes the proof of the lemma.
\end{proof}

To boost the success probability, we run \algwprbds (\Cref{alg:framework-wprbds}) independently \( \left( \frac{2k}{\varepsilon} \right)^{\mu(\cJ)} \) times and return the tuple \( \cS \) with the maximum value of \( \w(\nbr(\cS)) \). The probability that all runs fail is at most
\[
\left(1 - \left( \frac{\varepsilon}{2k} \right)^{\mu(\cJ)} \right)^{\left( \frac{2k}{\varepsilon} \right)^{\mu(\cJ)}} \leq \frac{1}{e},
\]
so the algorithm succeeds with probability at least \( 1 - \frac{1}{e} \). This gives us the following lemma. Note that we refer to the new algorithm with boosted success probability by \algwprbds also.

\begin{lemma} \label{lem:framework-wprbdsth}
For a given input instance $$\cJ = (\graph, \w, \mi = \langle \mcon_1, \dots, \\ \mcon_{k} \rangle, \free, \varepsilon)$$ \algwprbds runs in time 
\[ \left( \frac{2k}{\varepsilon} \right)^{\mu(\cJ)}\cO(\sizeg)= \left( \frac{1}{\varepsilon} \right)^{\cO(\mu(\cJ)\log{k})}\sizeg \] 
and outputs a consistent tuple $\cS = \langle s_1, s_2, \dots, s_k \rangle$ such that  
\[\w(\nbrst(\cS, \cJ)) \geq (1-\varepsilon)\wopt(\cJ)\]
with probability at least  \( 1 - \frac{1}{e} \).
\end{lemma}

Recall that given an instance $\cI $ = \wprbdsi of the \wmrbds problem, the optimal value $\wopt_k(\cI)$ is defined as  
\[
\wopt_k(\cI) = \max_{\cS \subseteq \cR, |\cS| \leq k}\w(\nbr(\cS)),
\]  
We now state our main theorem.
\togglefalse{includeFootnote}
\randprbdsthm*
\toggletrue{includeFootnote} 
\begin{proof}
Given an instance \( \cI \) of \wmrbds, we construct an instance of 
\algwprbds (\Cref{alg:framework-wprbds}) as follows. We initialize the \( k \)-tuple \( \mi \) by setting each \( \mcon_{\iletter} = \emptyset \) for all \( \iletter \in [k] \), and we set \( \free = \emptyset \).
 That is, we run \algwprbds for the first time with the input instance $\cJ = (\graph, \w, \mi = \langle \emptyset, \dots, \emptyset \rangle, \emptyset, \varepsilon)$. For the initial instance $\cJ$ of \algwprbds, we have 

\[
\wopt(\cJ) = \max_{\cS \in \cons{\cJ}} \w(\nbrst(\cS, \cJ)) = \max_{\substack{\cS \subseteq \cR \\ |\cS| \leq k}} \w(\nbr(\cS)) = \wopt_k(\cI).
\]
 Now we run \Cref{alg:framework-wprbds} on \( \cJ \), and obtain a tuple \( \cS \subseteq \cR \) such that

\[
\w(\nbrst(\cS, \cJ)) = \w(\nbr(\cS) \setminus \emptyset) = \w(\nbr(\cS)) \geq (1 - \varepsilon) \cdot \wopt(\cJ) = (1 - \varepsilon) \cdot \wopt_k(\cI)
\]
with probability at least  \( 1 - \frac{1}{e} \). We output the same set \( \cS \) for the instance $\cI$, which satisfies $\w(\nbr(\cS)) \geq (1 - \varepsilon) \cdot \wopt_k(\cI)$ with probability at least  \( 1 - \frac{1}{e} \).

We have the upper bound on the measure as follows
\[
\mu(\cJ) = k + k \cdot \intcmp - \sum_{\iletter \in [k]} \dintcmp(G, \mcon_{\iletter}) \leq k + k \Gamma,
\]
where \( \Gamma = \dintcmp(G) \). 
Thus, from \Cref{lem:framework-wprbdsth} the running time is at least 
\[
\left( \frac{1}{\varepsilon} \right)^{\cO(\Gamma k \log{k})}\sizeg
\]

\medskip

This completes the proof.
\end{proof}

%% file: partial_set_cover.tex
\section{Additive Approximation for \decwprbdsfull}
\label{section: weighted_additive}
In this section, we present an additive approximation algorithm for the \decwprbdsfull (\decwprbds) problem. The problem takes as input a bipartite graph \( \graph \), a weight function \( \w:\cB \rightarrow \bQ^{+} \), an integer \( k \) and $W \in \bQ^+$. The goal is to select at most \( k \) vertices, say \( \cS \), from \( \cR \) such that the sum of the weights of the vertices in \( \cB \) covered by these \( k \) vertices is at least \( W \).   Recall that, the weight of a set \( B \subseteq \cB \) is defined as the sum of the weights of the vertices in $B$. More precisely,  
\[
\w(B) = \sum_{v \in B}\w(v).
\]  

The problem is formally stated as follows:  

\defparopt{\decwprbds}{A bipartite graph \( \graph \), a weight function \( \w:\cB \rightarrow \bQ^{+} \), an integer \( k \) and $W \in \bQ^+$.}{\( k \)}{A subset of vertices \( \cS \subseteq \cR \) such that \( |\cS| \leq k \) and \( \w(\nbr(\cS)) \geq W \).}

In the case of a \yes instance, we provide an algorithm that returns a set \( \cS \subseteq \cR \) of size at most \( k+1 \) such that \( \w(\nbr(\cS)) \geq W \).

\subsection{Description of Algorithm}
Suppose we are given an instance $\cI=(\graph,\w,k,W)$ such that there exists a set $\cS=\{s_1,\dots,s_k\}$ of $k$ vertices satisfying \( \w(\nbr(\cS)) \geq W \). We call such a set an optimal solution. Our algorithm maintains $k$ subsets of vertices, $\pot_\iletter \subseteq \cR$, one for each $s_\iletter$, where $\iletter \in [k]$, which can be viewed as potential candidates for $s_\iletter$. Initially, each of these sets is the entire set $\cR$. As the algorithm progresses, these candidate sets are refined based on the requirement that their neighborhood must contain a given set of vertices, denoted $\mcon_i$.

The algorithm first computes a set $S'$ such that \( \w(\nbr(S')) \geq \left(1 - \frac{1}{2k}\right)W \), using the previously described approximation algorithm. Since we are allowed to include an additional vertex in the solution, we check whether there exists a vertex $v \in \cR$ such that 
\( \w(\nbr(S' \cup \{v\})) \geq W \). If such a vertex exists, then we return $S' \cup \{v\}$ as the desired approximate solution.

If no such vertex is found we show that \( \w(\nbr(S') \cap \nbr(\cS)) \geq \frac{1}{2} \w(\nbr(S')) \). In this case, we choose a vertex $v$ from $\nbr(S')$ with probability proportional to its weight, that is, according to the distribution \( \left\{ \frac{\w(v)}{\w(\nbr(S'))} \right\} \). With probability $\frac{1}{2}$, this selected vertex $v \in \nbr(S')$ also belongs to $\nbr(\cS)$, where $\cS$ is an optimal solution. Furthermore, with probability $\frac{1}{k}$, we can identify an index $\iletter \in [k]$ such that $v \in \nbr(s_\iletter)$. In that case, we add $v$ to $\mcon_\iletter$.

Additionally, we ensure that the size of each $\pot_\iletter$ is reduced by removing any vertex that is a neighbor of all the vertices in $\pot_\iletter$, for each $\iletter \in [k]$. Thus, at any step, the algorithm either returns a solution of size $k+1$ or makes progress toward identifying an optimal solution. Since the set $\mcon_\iletter$ corresponds to a \sml{} of order $|\mcon_\iletter|$, we know that the algorithm can make progress for at most $kd$ steps, assuming that the \smlindex{} of the graph is $\Gamma$.

We start by defining the necessary objects.  The next set of definitions are similar to the ones considered in Section~\ref{section: implement}. 
\begin{definition}[Potential and Freely Covered] 
Let 
\[
\cJ=(\graph, \w, k, \mi = \langle \mcon_1, \dots, \mcon_{k} \rangle, W),
\]  
be an instance.  Then, the potential solution family and the set of freely covered neighbors are defined as follows. 
\begin{enumerate}
    \item Let 
    ${\sf PotentialSolutionFamily}_\iletter (\cJ)=\pot_\iletter (\cJ)= \{ v \in \cR \mid \mcon_\iletter \subseteq \nbr(v) \}$,  and $\pot(\cJ)=\cup_{\iletter \in [k]}\pot_\iletter(\cJ) $
    \item 
$\fcov_\iletter (\cJ) =\bigcap_{v \in \pot_\iletter} \nbr(v)$ for all $\iletter \in [k]$ and $\fcov(\cJ) = \cup_{\iletter \in [k]}\fcov_\iletter(\cJ)$
\end{enumerate}
Furthermore by,  $\mcon_{\iletter}(\cJ)$ we denote the $\iletter^{\textnormal{th}}$ coordinate of $\mcon$ for the input instance $\cJ$.  
When the context is clear, we omit the accompanying subscript and mention of the instance in the bracket. 
\end{definition}
Now, we define consistent tuple with slight modifications to the previous definition to suit our algorithm. 
\begin{definition}
    Given an instance  
\[
\cJ=(\graph, \w, k, \mi = \langle \mcon_1, \dots, \mcon_{k} \rangle, W),
\]  
we say a $(k+1)$-tuple \(\cS=\langle s_1,\dots, s_k, s_{k+1}\rangle\) is \emph{consistent for} \(\cJ\) if  
\[ 
 \forall \iletter \in [k], s_\iletter \in \cR, \quad \text{and} \quad \fcov_\iletter \subseteq \nbr(s_\iletter), \quad .
\] For consistency, we ignore the last component of the tuple.
We use \(\cons{\cJ}\) to denote the set of all consistent tuples for \(\cJ\). 
\end{definition}

A complete description of the algorithm with pseudocode is provided in \Cref{alg:add-wprbds}.

\subsection{Algorithm Analysis}
Let \( \mathbb{O}=\langle o_1,\dots, o_k\rangle\) be a tuple such that  
$\w(\nbr (\bO))\geq W$.
Such a tuple will be called an \emph{optimal tuple}. Next we present the following lemma which will be useful later.

\begin{lemma}\label{lemma:addstructure}
    Given a bipartite graph $\graph$, $\cS \subseteq \cR$  with $|\cS|\leq k$ and $B'\subseteq \cB$ if $\w(\nbr(\cS) \setminus B')\geq (1-\frac{1}{2k})W$, then  at least one of the two cases hold.
    \begin{description}
        \item[1. \label{caseone}] there exists $h \in \pot \setminus \cS$ such that $\w(\nbr(h \cup \cS) )\geq W$
        \item[2. \label{casetwo}]$\w(\nbr(\bO) \cap (\nbr(\cS)\setminus B'))\geq \frac{1}{2}\w(
        \nbr(\cS)\setminus  B')$
    \end{description}
Here  $\bO$ is a $k$-tuple (optimal tuple) in $\cR$ such that $\w(\nbr (\bO))\geq W$. 
\end{lemma}
\begin{proof}
    Given $\graph$, $B'$ and $\cS \subseteq \cR$ that satisfies the conditions of the lemma, if \hyperref[caseone]{Case 1} of \Cref{lemma:addstructure} holds, then the statement of the lemma is true. Suppose Case $1$ does not hold, i.e., for any $h \in \pot \setminus \cS$ we have $\w(\nbr(h \cup \cS))< W$. This implies that $\w(\nbr(h)\setminus (\nbr(\cS)\setminus B')) < \frac{W}{2k}$ for any $h \in \pot \setminus \cS$. Otherwise we have $$\w(\nbr(h) \cup \nbr(\cS)) \geq \w(\nbr(h) \cup  (\nbr(\cS)\setminus B'))\geq (1-\frac{1}{2k})W +\frac{W}{2k}=W.$$
    This implies that for every $o_\iletter$ where $\iletter \in [k]$, we have $\w(\nbr(o_\iletter) \setminus (\nbr(\cS)\setminus B'))<\frac{W}{2k}$ which implies $\w(\nbr(\bO) \setminus (\nbr(\cS)\setminus B'))<\frac{W}{2}$. This gives us   $$\w(\nbr(\bO) \cap( \nbr(\cS)\setminus B'))\geq \frac{W}{2} \geq \frac{1}{2}\w(\nbr(\cS)\setminus B')$$ which shows that \hyperref[casetwo]{Case 2} of \Cref{lemma:addstructure} holds and hence proves the correctness of the lemma.
\end{proof}

Now we present some lemmata that capture relationships between two structurally similar instances arising from consecutive recursive calls. To avoid ambiguity caused by identically named objects across different calls, we annotate each object with its corresponding instance in brackets and omit those when the context is clear.

\begin{lemma}\label{lemma:consistency_recursive}
    Let $\cJ$ denote the input instance 
 \[(\graph, \w, k, \mi, W).\] 
 For $u \in \cB \setminus \fcov(\cJ)=\{u_1,\dots, u_\ell \}$, let $\cJ'$ denote the instance
  \[ (\graph, \w, k, \mi', W),\]
  where $\mi' = \langle \mcon_1, \dots, \mcon_{j} \cup \{u\}, \dots, \mcon_{k} \rangle$.
Then, the following holds:
\begin{description}
    \item[Prop 1. \label{pointone}] Any optimal tuple $\bO$ for $\cJ$ is also optimal for $\cJ'$, if $u \in \nbr(o_j)$.
    \item[Prop 2. \label{pointtwo}] Any consistent tuple $H$ for $\cJ'$ is also consistent for $\cJ$.
\end{description}

\end{lemma}

\begin{proof}
\hyperref[pointone]{Prop 1.} We first show that $\bO$ is consistent for $\cJ'.$
For any $\iletter \in [k]$, $\iletter \neq j$, $$\mcon_{\iletter}(\cJ') = \mcon_{\iletter}(\cJ) \subseteq \nbr(o_{\iletter}).$$ For the coordinate $j$, $\mcon_{j}(\cJ') = \mcon_{j}(\cJ) \cup \{u\}$. Due to the fact that $u \in \nbr(o_j)$, and $\bO$ is consistent for $\cJ$ implying $\mcon_{j}(\cJ) \subseteq \nbr(o_j)$, we have that $\mcon_{j}(\cJ') \subseteq \nbr(o_j)$. This implies that $\fcov(\cJ') \subseteq \nbr(\bO)$, and thus, $\bO$ is a consistent tuple for $\cJ'$. As $\w(\nbr(\bO)) \geq W$, $\bO$ is also an optimal tuple for $\cJ'.$

\hyperref[pointtwo]{Prop 2.} We know that $\mcon_{j}(\cJ') = \mcon_{j}(\cJ) \cup \{u\}$, which implies that $\mcon_{j}(\cJ) \subseteq \mcon_{j}(\cJ')$. As $\mcon_{j}(\cJ') \subseteq \fcov_{j}(\cJ') \subseteq \nbr(h_{j})$ due to the consistency of $\cH$ for the instance $\cJ'$ , we can conclude that $\mcon_{j}(\cJ) \subseteq \nbr(h_{j})$ which also implies that $\fcov_{j}(\cJ) \subseteq \nbr(h_{j})$. Moreover, coupled with the fact that the sets $\mcon_{\iletter}(\cJ)$, for $\iletter \in [k]$, $\iletter \neq j$, are the same as $\mcon_{\iletter}(\cJ')$, we can conclude that the tuple $H$ is consistent for the instance $\cJ$.  
\end{proof}

\begin{lemma}\label{lemma:additivemiprog}
     Let $\cJ$ denote the input instance 
 \[(\graph, \w, k, \mi, W).\] 
 
 For $u \in \cB \setminus \fcov(\cJ)$, let $\cJ'$ denote the instance
  \[ (\graph, \w, k, \mi', W),\]
  where $\mi' = \langle \mcon_1, \dots, \mcon_{\beta} \cup \{u\}, \dots, \mcon_{k} \rangle$.
Then, we have
\[\dintcmp(G,\mcon_\beta \cup \{u\})\geq \dintcmp(G,\mcon_\beta)+1\]
\end{lemma}
\begin{proof}
Let $d= \dintcmp(G,\mcon_\beta)$. By definition, we have a sequence of $d$ \realizable sets $$\reex(G,\mcon_\beta)=X_1 \supsetneq X_2 \supsetneq \dots \supsetneq X_d$$  Let $$ \incld(G,\mcon_\beta) = Y_1 \subsetneq Y_2 \subsetneq \dots \subsetneq Y_{d}$$ be the corresponding sequence of \realizing sets (due to~\Cref{lemma:realizablerealizingreln}). Note that in this case, $\fcov_\beta = \reex(G, \mcon_\beta)$.

Since, by definition, $u \notin \fcov(\cJ)$, there exists $y \in Y_1$ such that $u \notin \nbr(y)$. Let $\Tilde{Y_1}=\{y \in Y_1 \mid u \notin \nbr(y)\}$. Note that $Y_1 \setminus \Tilde{Y_1}$ is a \realizing set for $X_1 \cup \{u\}$. Thus,
$$\reex(G,\mcon_\beta \cup \{u\})=\reex(G,X_1 \cup \{u\}) \supsetneq X_1 \supsetneq X_2 \supsetneq \dots \supsetneq X_d$$ is a sequence of $d+1$ realizable sets. Thus, $\dintcmp(G,\mcon_\beta \cup \{u\})$ is at least $d+1$ due to~\Cref{defn:dic}.

\end{proof}
Let \(\intcmp = \dintcmp(G) \) . We define a measure associated with \(\cJ\), denoted by \(\mu(\cJ)\), as  
\[
\mu(\cJ) =  k\intcmp - \sum_{\iletter \in [k]} \dintcmp(G,  \mcon_{\iletter})
\]  
We use this measure for induction during our proof for both approximation analysis as well as probability analysis. A complete description of the algorithm is provide in~\Cref{alg:add-wprbds}.

\begin{algorithm}[ht!]

\SetKwFunction{algo}{algo}
\caption{\algaddwprbds}\label{alg:add-wprbds}
  \KwInput{A bipartite graph $\graph$, a weight function $\w: \cB \rightarrow \bQ^+$, an integer $k$, a tuple $\mi = \langle \mcon_1, \dots, \mcon_{k} \rangle $ and $W \in \bQ^+$ }
  \KwOutput{Either a consistent $(k+1)$-tuple $\cS$  such that $\w(\nbr(\cS))\geq W$, or return $\cJ=(\graph, \w, k, \mi, W)$ is a \no instance.}
 Let $\pot_\iletter = \{ v \in \cR \mid \mcon_\iletter \subseteq \nbr(v) \}$,  
 $\fcov_\iletter=\bigcap_{v \in \pot_\iletter} \nbr(v)$ for all $\iletter \in [k]$ and $\fcov = \cup_{ \alpha \in [k]} \fcov_{\alpha}$ \; 

\If{
for all $\iletter \in [k]$, we have $ \left(\bigcup_{v \in \pot_\iletter }\nbr(v) \right) \subseteq\fcov$\label{algaddwprbds:baseone}
}
{
Let $\cS = \langle v_1,\dots,v_{\alpha}, \dots, v_k\rangle$ where, for $\iletter \in [k]$, and $v_{\iletter}$ be an arbitrary vertex in $\pot_{\iletter}$\;
\If {$\w(\nbr(\cS)) \geq W$}{ 
\Return $\cS[\uparrow k+1,\emptyset]$\;}
\Return ~\no\;
}

Let $\cS$ be a  tuple returned by  $\algwprbds((\cR, \cB), \w, \mi, \fcov,\varepsilon \gets \frac{1}{2k})$ \label{line:rem}\; 
Let $h^\star$ be a vertex in $\pot \setminus \cS$ such that $\w(\nbr(h^\star) \setminus \nbr(\cS))$ is the largest\; \label{algaddwprbds:select_high_deg}
\If {$\w(\nbr(\cS[\uparrow k+1, h^\star]))\geq W$\label{algaddwprbds:checkcasetwo}}{\Return {$\cS[\uparrow k+1, h^\star]$ \label{algaddwprbds:returnssoln}}\;}

Randomly select an item $\clubsuit$ from the set $\nbr(\cS)\setminus \fcov=\{u_1,\dots, u_\ell \}$ with  probability distribution $(\frac{\w(u_1)}{\omega},\dots,\frac{\w(u_{\ell})}{\omega})$, where $\omega = \sum_{i \in [\ell]} \w(u_i)$. That is, we select a non-free neighbor $u$ of $\cS$ with probability $\frac{\w(u)}{\omega}$\;\label{algaddwprbds:prob_dist}
Let $\clubsuit = u \in \{u_1, \dots, u_\ell\}$. Pick an integer  $\beta \in[k]$ uniformly at random \label{algaddwprbds:chooseinteger}\;
	 Assign $\mi \gets \langle \mcon_1, \dots, \mcon_{\beta} \cup \{u\}, \dots, \mcon_{k} \rangle$\;
	 Let $\cS$ be a tuple returned by  $\algaddwprbds(\graph, \w, k, \mi, W)$\label{algaddwprbds:recursive}\; 
     \If {$\w(\nbr(\cS)) \geq W$}{\Return ~$\cS$}
\Return~\no\;  
\end{algorithm}

\begin{lemma} \label{lem:addwprbds}
For a given input instance \[\cJ=(\graph, \w, k,\mi = \langle \mcon_1, \dots \mcon_{k} \rangle, W)\] Algorithm \ref{alg:add-wprbds}~\textnormal{(\algaddwprbds)} 
runs in time $$ (2k)^{2(\Gamma+1)k} \sizeg$$ and 
in case of a \yes instance outputs a consistent tuple $$\cS = \langle s_1, s_2, \dots, s_{k+1} \rangle$$ such that  
$\w(\nbr(\cS)) \geq W$
with probability at least $\left( \frac{1}{2k}(1 - \frac{1}{e}) \right)^{\mu(\cJ)}$. However in case of a \no instance the algorithm either returns \no or returns a $k+1$ size tuple $\cS$ such that $\w(\nbr(\cS)) \geq W$. 

\end{lemma}

\begin{proof}
 We prove the statement of our lemma by induction, using the measure defined above: 
\[
\mu(\cJ) =  k\intcmp - \sum_{\iletter \in [k]} \dintcmp(G,  \mcon_{\iletter})
\]
In particular, we prove the desired statement by induction on \(\mu(\cJ)\).

\medskip

\noindent 
\paragraph{\bf Non-Negativity of Measure.} 
Note that \(\mu(\cJ) =  k\intcmp - \sum_{\iletter \in [k]} \dintcmp(G,  \mcon_{\iletter})\) is always non-negative, since for all \(\iletter \in [k]\), we have \( \dintcmp(G, \mcon_{\iletter}) \leq \intcmp \), which implies \( \sum_{\iletter =1}^k \dintcmp (G,\mcon_{\iletter}) \leq k \intcmp \). Hence it follows that $\mu(\cJ)\geq 0$.

\noindent 
\paragraph{\bf Analysis of the algorithm.} 
Consider an optimal tuple \(\bo = \langle o_1, \dots, o_k \rangle\) for the instance \(\cJ\). We prove the statement of the lemma using induction on \(\mu(\cJ)\). That is, the statement of the lemma serves as our induction hypothesis.

\smallskip
\noindent 
\textbf{Base Case:} When $\mu(\cJ) = 0$, then it implies that either $k=0$ or for all $\iletter \in [k]$ we have $\dintcmp(G,\mcon_{\iletter})=\intcmp$. Now consider the case when for all $\iletter \in [k]$, $\dintcmp(G,\mcon_{\iletter})=\intcmp$. We show that any consistent tuple has the same neighborhood. For any consistent tuple $\cS$, it holds that $\fcov_{\iletter} \subseteq \nbr(s_{\iletter})$ for each $\iletter \in [k]$. If there exists $u \in \nbr(s_{\iletter})$ such that $u \notin \fcov_{\iletter}$ then we have $\dintcmp(G,\mcon_{\iletter} \cup \{u\}) \geq \intcmp +1$ due to~\Cref{lemma:additivemiprog} which will contradict the fact that \(\intcmp = \dintcmp(G)\). Thus, if for every $u \in \nbr(s_{\iletter})$ we have $u \in \fcov_{\iletter} $, then combined with the fact that $\fcov_{\iletter} \subseteq \nbr(s_{\iletter})$, we get $\fcov_{\iletter} = \nbr(s_{\iletter})$ for each $\iletter \in [k]$. Thus for any consistent tuple $\cS$ we have $\cup_{\iletter \in  [k]}\nbr(s_\iletter)=\cup_{\iletter \in  [k]}\fcov_\iletter=\fcov(\cJ)$.  
    
    Therefore, if $\w(\fcov(\cJ))\geq W$, the algorithm correctly identifies and returns a consistent tuple, with probability $1$, in~\Cref{algaddwprbds:baseone}; otherwise, it outputs \no. Now consider the case when $k=0$. In this case if $W=0$ the algorithm correctly returns $\langle \emptyset \rangle$, otherwise if $W\neq 0$ the algorithm correctly returns \no.
\smallskip

\noindent 
\textbf{Induction Assumption:} Let $\mu(\cJ) >0$. 
The statement of \Cref{lem:addwprbds} holds for all $\mu(\cJ) < \mu'$.

\smallskip

\noindent 
\textbf{Inductive Step:} We prove the statement of Lemma \ref{lem:addwprbds} for $\mu(\cJ) = \mu'$.

\medskip

\noindent 

We divide the analysis of the algorithm in two main parts as follows. The case when the condition of \Cref{algaddwprbds:baseone} is satisfied. The other one is when the algorithm executes the remaining lines, starting from \Cref{line:rem}.

\medskip
\noindent  
\textbf{Case A: Analysis when \Cref{algaddwprbds:baseone} condition holds.}

Let $\cJ=(\graph, k, \mi, W)$ be an input instance. We will first show that the union of the neighborhood of any consistent tuple for $\cJ$ is the same for this case. Consider any $\iletter \in [k]$ and any $v \in \pot_{\iletter}$. Since in this case $\nbr(v) \subseteq \cup_{\iletter \in [k]} \fcov_{\iletter}$ we have that for every $u \in \nbr(v)$ there exists $j \in [k]$ such that $u \in \fcov_j$. The fact that $u \in \fcov_j$ implies $u$ is in the neighborhood of any vertex chosen from $\pot_j$. Thus we have that for any ${\iletter} \in [k]$ and any  $v \in \pot_{\iletter}$, $\nbr(v)$ is in the neighborhood of any consistent tuple. Thus the union of the neighborhood of any consistent tuple for $\cJ$ is the same. The algorithm in this case correctly finds such a consistent tuple, \cS, and returns $\cS[\uparrow k+1,\emptyset]$ with probability $1$, if the weight of its neighborhood is at least $W$. Otherwise, it correctly returns \no, if no such consistent tuple exists. 

\medskip
\noindent  
\textbf{Case B: Analysis when the algorithm executes from \Cref{line:rem}.} In this case, we show the following claim.

\begin{claim}
   Let $\cJ = (\graph, k, \mi, W)$ be a \yes instance. Then the algorithm \algaddwprbds returns a consistent tuple $\cS$ of size at most $k+1$ with probability at least \( \left( \frac{1}{2k} \left(1 - \frac{1}{e} \right) \right)^{\mu(\cJ)} \). 
\end{claim}

The claim is proved by induction on $\mu(\cJ)$ below.  First, observe that $\cS$, which is returned by 
\[\algwprbds(\graph, \mi, \cup_{\iletter \in [k]} \fcov_{\iletter},\varepsilon \gets \frac{1}{2k})\] 
is a consistent tuple for $\cJ$. This is because for all $\iletter \in [k]$, we have $\fcov_{\iletter}\subseteq \nbr(s_{\iletter})$, since $\mi$ remains unchanged. Also $\algwprbds$ returns a set $\cS$ such that $\w(\nbr(\cS)\setminus \fcov(\cJ))\geq (1-\frac{1}{2k})W$ with probability at least  $(1-\frac{1}{e})$ in time \( \left( \frac{2k}{\varepsilon} \right)^{\intcmp(k+1)} \sizeg \) 
due to~\Cref{lem:framework-wprbdsth}. Now, assuming \algwprbds returns such a solution $\cS$, due to~\Cref{lemma:addstructure}, we know that in case it is a  \yes instance at least one of the two cases hold.
    \begin{description} 
        \item[Case 1:] there exists $h^\star \in \pot \setminus \cS$ such that $\w(\nbr(h^\star \cup \cS) )\geq W$
        \item[Case 2:] $\w(\nbr(\bO) \cap (\nbr(\cS)\setminus \fcov(\cJ))\geq \frac{1}{2}\w(
        \nbr(\cS)\setminus  \fcov(\cJ))$
    \end{description}

If Case 1 holds, the algorithm returns $\cS[\uparrow k
+1, h^\star]$ in~\Cref{algaddwprbds:returnssoln} with probability $1$. Thus, from now on, we assume that Case~$1$ does not hold. Hence, when Case~$2$ holds and we have a \yes-instance, we know that an optimal solution \( \bO \) exists and satisfies
\[
\w(\nbr(\bO) \cap (\nbr(\cS)\setminus \fcov(\cJ))) \geq \frac{1}{2} \w(\nbr(\cS) \setminus \fcov(\cJ)).
\]

In this case, the algorithm randomly selects an item $\clubsuit$ from the set \( \nbr(\cS) \setminus \fcov = \{u_1, \dots, u_\ell\} \), according to the probability distribution \( \left( \frac{\w(u_1)}{\omega}, \dots, \frac{\w(u_\ell)}{\omega} \right) \), where \( \omega = \sum_{i \in [\ell]} \w(u_i) \). That is, it selects a non-free neighbor \( u \) of \( \cS \) with probability \( \frac{\w(u)}{\omega} \). Then, it uniformly at random chooses an integer \( j \in [k] \), and recursively calls itself on the modified instance, 
\[ \cJ' = (\graph, k, \mi', W),\] 
where $\mi' = \langle \mcon_1, \dots, \mcon_j \cup \{u\}, \dots, \mcon_k \rangle$.

We consider the branch where the selected element \( u \in \nbr(o_j) \). Observe that \( \bO \) remains an optimal tuple for the instance \( \cJ' \), as guaranteed by \Cref{lemma:consistency_recursive}. Therefore, with probability $\frac{1}{2k}$, the instance \( \cJ' \) remains a \yes instance. To apply the induction hypothesis, we now show that \( \mu(\cJ') < \mu(\cJ) \).

\begin{align}
	\mu(\cJ) - \mu(\cJ') &\geq \left( k\intcmp - \sum_{\iletter \in [k]} \dintcmp(G,  \mcon_{\iletter}) \right)- \left( k\intcmp - \big(\sum_{\iletter \in [k]} \dintcmp(G,  \mcon_{\iletter})+1\big)\right)\nonumber\\
        &\textnormal{ (since }\dintcmp(G,\mcon_\iletter \cup \{u\})\geq \dintcmp(G,\mcon_\iletter)+1 \textnormal{ due to~\Cref{lemma:additivemiprog})} \nonumber\\
	&=1 >0  \label{wtmeasure:CaseC}
\end{align}

Conditioned on the event that \( \cJ' \) remains a \yes-instance, the induction hypothesis implies that the algorithm correctly returns a tuple \( \cS \) of size at most \( k+1 \), such that \( \w(\nbr(\cS)) \geq W \), with probability at least
\[
\left( \frac{1}{2k}\left(1 - \frac{1}{e} \right) \right)^{\mu(\cJ')}.
\]

Since the algorithm recursively calls itself on the instance $\cJ' = (\graph, k, \mi', W)$ where $\mi' = \langle \mcon_1, \dots, \mcon_j \cup \{u\}, \dots, \mcon_k \rangle$ and obtained a tuple \( \cS \) as described above with desired probability. It returns the same with probability at least 

\begin{align*}
\left(\frac{1}{2k}\left(1 - \frac{1}{e}\right) \right) \left( \frac{1}{2k}(1 - \frac{1}{e}) \right)^{\mu(\mathcal{J}')} \geq \left( \frac{1}{2k}\left(1 - \frac{1}{e}\right) \right)^{\mu(\mathcal{J})} \quad [\textnormal{since } \mu(\cJ')\leq \mu(\cJ)-1 ] 
\end{align*}

\noindent\textbf{Running Time:}  The running time of a single recursive call of the algorithm is governed by the running time of \algwprbds, which is \( k^{\cO(\intcmp(k))} \sizeg \). Since the measure is bounded by \( k\intcmp \) and \( \varepsilon = \frac{1}{2k} \), the total running time of the algorithm is upper bounded by \( k^{\cO(\intcmp k)} \sizeg \). This concludes the proof. 
\end{proof}

We now state our main theorem of this section.

\randadditiveprbdsthm*

\begin{proof}

Given an instance $\cI = (G, \w, k, W)$ of \decwprbds, we construct an instance of \Cref{alg:add-wprbds} as follows. We initialize a $k$-tuple $\mi$ with each $\mcon_{\iletter}$ set to $\emptyset$, for $\iletter \in [k]$. That is we run \Cref{alg:add-wprbds} for the first time with the input instance $\cJ=(\graph, \w, k, \mi = \langle \emptyset, \dots, \emptyset \rangle, W)$. A solution returned by \Cref{alg:add-wprbds} on the instance $\cJ$ is a solution for the instance $\cI$.

Further we run \Cref{alg:add-wprbds} $\left( 2k\frac{e}{e-1} \right)^{\intcmp k}$ times independently. If any of the independent runs return a tuple $\cS$ satisfying $\w(\nbr(\cS))\geq W$, then we return \yes, otherwise we return \no. 

If $\cI$ is an \yes instance then by the correctness of \Cref{lem:addwprbds}, we know that the probability that the algorithm fails in each of the independent runs is at most 
\[\left(1 - \left( \frac{1}{2k}\left(1 - \frac{1}{e}\right) \right)^{\intcmp k}\right)^{\left( 2k\frac{e}{e-1} \right)^{\intcmp k}}\leq \frac{1}{e}
\]
Thus, the algorithm succeeds with probability at least $1 - \frac{1}{e}$. In case of \no instance, the algorithm returns \no if all runs of the algorithm returns \no.  \\
\noindent \textbf{Running time:} We analyze the running time as follows. \Cref{alg:add-wprbds} runs in time  \( k^{\cO(\intcmp k)}\sizeg \)  and is repeated $\left( 2k\frac{e}{e-1} \right)^{\intcmp k}$ times. Thus, the overall running time of our \epas is $k^{\cO(\intcmp k)}\sizeg$.
\end{proof}

%% file: deterministic_algorithms.tex
\section{Weighted Deterministic \epas/\pas for Partial Coverage}
\label{section: weighteddetfptas}

In this section, we present deterministic counterparts of~\Cref{thm:mainWeighted} and~\Cref{thm:addrandom}. To this end, we first pinpoint the sources of randomness in our algorithms. 
For the purpose of this discussion, we focus on \algwprbds. Note that this algorithm involves two sources of randomness. The first, in \Cref{algaddwprbds:chooseinteger}, can be easily handled by branching over the $k$ possible choices and is thus not discussed in detail. The primary and more interesting source of randomness arises in Step~\ref{algwprbds:prob_dist}, where the algorithm samples an element. This section focuses on derandomizing that step. For ease of reference, we recollect the steps as follows.

\begin{mainbox}{Sampling Step} 
\begin{itemize}[label=$\circlearrowright$]
\item   Let $h^\star$ be a vertex in $\cR$ such that $\w(\nbr(h^\star) \setminus (\free\cup \nfree))$ is the largest. Let $\nbr(h^\star) \setminus (\free\cup \nfree)= \{u_1,\dots u_\ell \}$. Let $\mathbf{\omega}=\sum_{j=1}^\ell \w(u_j)$.
\item  Randomly selecting an item \(\clubsuit\) from the set \(\{ h^\star, u_1, \dots, u_\ell \}\) with a probability distribution  
\[
    \prob{\clubsuit} = 
    \begin{cases} 
    \frac{1}{2} & \text{if } \clubsuit = h^\star \\
    \frac{\w(u_\iletter)}{2 \omega} & \text{if } \clubsuit \in \{u_\iletter \mid \iletter \in [\ell]\}
    \end{cases}
    \]

That is, we select the vertex $h^\star$ with probability $\frac{1}{2}$, or an element $u$ of $\{u_1,\dots,u_\ell\}$ with probability $\frac{\w(u)}{2\omega}$.
\item If $\clubsuit = u$, then we add $u$ to $\mcon_\iletter$, $\iletter \in [k]$,  to signify that $u$ is covered by an optimal solution. 
\end{itemize}

\end{mainbox}
One natural way to eliminate the randomness is to replace the random sampling step with explicit branching over the set \( \{ h^\star, u_1, \dots, u_\ell \} \). However, this approach is often expensive, as the size \( \ell \) can be too large to allow for efficient branching. To better understand the role of randomness in the algorithm, we ask: what does the weighted sampling over the set \( \{ u_1, \dots, u_\ell \} \) represent in the analysis? It turns out that this sampling step captures situations where the \emph{weighted intersection} of \( \{ u_1, \dots, u_\ell \} \) with the neighborhood of the optimal solution is significant.

Formally, let \( \cI = (\graph, \w, k) \) be an instance of \wmrbds. Let \( \bO = \langle o_1, \dots, o_k \rangle \) denote an optimal solution to this instance.  Then we have:
\begin{equation}\label{eqn:weightedDet}
\w\left( \nbr(\{ o_1, \dots, o_k \}) \cap \{ u_1, \dots, u_\ell \} \right) \geq \varepsilon \cdot \w(\{ u_1, \dots, u_\ell \}).
\end{equation}

Now, if we select an element \( u \in \{ u_1, \dots, u_\ell \} \) with probability proportional to its weight — specifically, with probability \( \frac{\w(u)}{2\omega} \), where \( \omega = \w(\{ u_1, \dots, u_\ell \}) = \sum_{i \in [\ell]}\w(u_i) \) — then the probability that \( u \in \nbr(\bO) \) is at least \( \frac{\varepsilon}{2} \).

Looking again at the inequality above, we see that if the total weight of the intersection is large, then there must exist at least one \( o_\iletter \in \bO \) such that:
\begin{equation}\label{eqn:weightedDetB}
\w\left( \nbr(o_\iletter) \cap \{ u_1, \dots, u_\ell \} \right) \geq \frac{\varepsilon \cdot \w(\{ u_1, \dots, u_\ell \})}{k}.
\end{equation}
This implies that \( \nbr(o_\iletter) \cap \{ u_1, \dots, u_\ell \} \neq \emptyset \). We call this {\em special vertex $o^\star$}. 
Therefore, instead of blindly sampling or branching over all of \( \{ u_1, \dots, u_\ell \} \), we can focus on identifying a vertex \( o^\star \in \bO \) that satisfies the above inequality, and restrict our attention to elements \( u \in \nbr(o^\star) \cap \{ u_1, \dots, u_\ell \} \). 
This leads to the following definition, which plays a key role in our deterministic algorithms.

\begin{definition}[\wvpn and Weighted Heavy graph]
    Let $\graph$ be a bipartite graph and $\delta > 0$. Further let $\w: \cB \to \bQ^+$ be a weight function defined on the blue vertices of $G$.  A set $X \subseteq \cB$ is called $\wdeltanet(G)$ if for all $r\in \cR$, if $\w(\nbr(r)) \geq \delta \cdot \w(\cB)$, then $\nbr(r)\cap X\neq \emptyset$. 
    \noindent 
    Furthermore, $G$ is called weighted $\delta$-heavy graph if for all $r\in \cR$, $\w(\nbr(r)) \geq \delta \cdot \w(\cB)$. 
\end{definition}
 The definition of  \wvpn  in the unweighted case is known as \vpn.

Let \( \cI = (\graph, \w, k) \) be an instance of \wmrbds. We now construct a bipartite graph \( G^\star = (\cR^\star, \cB^\star) \), where  $\cB^\star = \{ u_1, \dots, u_\ell \} $ 
and
\[
\cR^\star = \left\{ r \in \cR \,\middle|\, \w\left( \nbr(r) \cap \{ u_1, \dots, u_\ell \} \right) \geq \frac{\varepsilon \cdot \w(\{ u_1, \dots, u_\ell \})}{k} \right\}.
\]
We define the weight function on \( \cB^\star \) to be the restriction of \( \w \) to the vertices of $\cB^\star$. Further, we set \( \delta \gets \frac{\varepsilon}{k} \).

Observe that, by~\Cref{eqn:weightedDetB}, there exists a vertex \( o^\star \in \bO \) such that \( o^\star \in \cR^\star \). Therefore, if \( X \subseteq \cB^\star \) is a \wvpn in \( G^\star \) (\( \wdeltanet(G^\star) \)), then we are guaranteed that
\[
\nbr(o^\star) \cap X \neq \emptyset.
\]

At first glance, it is not clear whether a small-sized \( \wdeltanet(G^\star) \) exists. In the following, we show that such a cover not only exists but can also be computed efficiently. It uses the classical concept of $\delta$-net for set systems with bounded VC-dimension.\footnote{We thank an anonymous reviewer for pointing us to this construction.}

\subsection{\texorpdfstring{Weighted $\delta$-Nets via Bounded VC dimension}{Weighted delta-Nets via Bounded VC dimension}}

It is known that set families of bounded VC-dimension admit small Weighted Delta Net, in short \wdeltanet. For completeness, we recall the notion of VC-dimension, introduced by Vapnik and Chervonenkis~\cite{doi:10.1137/1116025}. 

\begin{definition}[VC-dimension]\label{def:vcdimension}
    Let $(\cU,\cF)$ be a set system. Given $Y \subseteq \cU$, consider the family of sets $\cF_Y=\{Y \cap F \mid F\in\cF\}$ induced by $Y$. We say that $Y$ is shattered by $\cF$ if $\cF_Y=2^Y$. The VC-dimension of $(\cU,\cF)$ is the maximum cardinality of a shattered subset of $\cU$.
\end{definition}

The next proposition shows the existence of small \wvpn in set systems of bounded VC-dimension. 
\begin{proposition}[\cite{DBLP:conf/compgeom/HausslerW86,DBLP:conf/stoc/BlumerEHW86}]    
\label{prop:vcExist}
Set systems of VC-dimension $d$ admit \wvpn of size $\mathcal{O}(\frac{d}{\delta}\log{\frac{1}{\delta}})$ where $\delta \in (0,1)$. 
\end{proposition}

The above proposition was shown by Haussler and Welzl~\cite{DBLP:conf/compgeom/HausslerW86} with a size bound of $\mathcal{O}(\frac{d}{\delta}\log{\frac{d}{\delta}})$, and was later improved to the bound in the proposition by Blumer et al.~\cite{DBLP:conf/stoc/BlumerEHW86}. This bound was shown to be tight by Komlós et al.~\cite{komlos1992almost}. Now we have the following definition and proposition due to~\cite{DBLP:journals/dcg/BronnimannG95,DBLP:journals/siamcomp/BronnimannCM99}.

\begin{definition}
    A set system $(\cU, \cF)$ has a subsystem oracle of degree $D$ if there is an algorithm which, given a subset $Y \subseteq \cU$, returns $(Y,\cF_Y)$ (as a boolean matrix) in time $\mathcal{O}(|Y|^{D+1})$.
\end{definition}

\begin{proposition}\label{prop:epnetcom}
    If the set system has a subsystem oracle of degree $D$, then there exists a deterministic algorithm to find a \wvpn of size $\mathcal{O}(\frac{d}{\delta}\log{\frac{d}{\delta}})$ for $\delta \in (0,1)$ in set systems of VC-dimension $d$, in time 
    \[
        d^{\cO(D)} \cdot \left(\frac{1}{\delta}\right)^D \cdot \log^D\left(\frac{d}{\delta}\right) \cdot |\cU|
    \]
    in the real RAM model, for both the weighted and unweighted cases.
\end{proposition}

Now we proceed as follows. First, we show that the set system we are considering has a bounded VC-dimension. 
\begin{lemma}
   Given a graph $\graph$, if the \smlindex of $G$ is bounded by $d$, then the set system $(\cB, \{\nbr(v) \mid v \in \cR\})$ has VC-dimension bounded by $d$.
\end{lemma}
\begin{proof}
Suppose, for the sake of contradiction, that the VC-dimension of the set system \( (\cB, \{\nbr(v) \mid v \in \cR\}) \) is \( d+1 \). Let \( \cF = \{\nbr(v) \mid v \in \cR\} \). By \Cref{def:vcdimension}, there exists a set \( Y \subseteq \cB \) of size \( d+1 \) such that \( Y \) is shattered by \( \cF \) i.e, $|\cF_{Y}|=2^{Y}$. Let \( Y = \{y_1, \dots, y_{d+1} \} \). This implies that for every subset \( Y' \subseteq Y \), there exists a vertex \( v \in \cR \) such that \( \nbr(v) \cap Y = Y' \).

Hence, we can construct a set \( X \subseteq \cR \) with \( X = \{x_0, \dots, x_{d+1}\} \) such that the following conditions hold: for every \( i \in [0, d] \), we have \( \nbr(x_i) \cap Y = \{y_{i+1}, \dots, y_{d+1}\} \), and \( \nbr(x_{d+1}) \cap Y = \emptyset \).

Now, let \( G' \) be the subgraph of \( G \) induced on \( X \cup Y \). Then, we obtain two sequences \( \{x_1, \dots, x_{d+1}\} \) and \( \{y_1, \dots, y_{d+1}\} \) such that \( (x_i, y_i) \notin E(G) \) and \( (x_i, y_j) \in E(G) \) for all \( d+1 \geq j > i > 0 \). Therefore, these vertices form a semi-ladder of size \( d+1 \), which contradicts the assumption that \( G \) has \smlindex\ at most \( d \).
\end{proof}

For our set system, for any given $Y \subseteq \cB$, the running time to find the family induced by $Y$  is $\mathcal{O}(mn)=\mathcal{O}(2^{\log{n}+\log m})$. Set $D=\log n + \log m$. Hence, we have a subsystem oracle of degree $D$.  Thus due to \Cref{prop:epnetcom} we can compute a weighted-$\frac{\varepsilon}{k}$-net for our set system of size $\mathcal{O}(\frac{dk}{\varepsilon}\log{\frac{dk}{\varepsilon}})$ using ~\Cref{prop:epnetcom}. 
Thus, we get the following lemma.
\begin{lemma}
\label{lem:deltaNetVCDim}
    Given a graph $\graph$ of \smlindex $d$ we can find weighted-$\frac{\varepsilon}{k}$-net of size $\mathcal{O}(\frac{dk}{\varepsilon}\log{\frac{dk}{\varepsilon}})$  in time 
        \[
        d^{\cO(\log n + \log m)} \cdot \left(\frac{k}{\varepsilon}\right)^{(\log n + \log m)} \cdot \log^{(\log n + \log m)}\left(\frac{dk}{\varepsilon}\right) \cdot m
    \]
    
\end{lemma}

The running time required by \Cref{lem:deltaNetVCDim} to compute the desired \wvpn\ is too prohibitive for our purposes. Even for constant \( d \), the term \( \left(\frac{k}{\varepsilon}\right)^{(\log n + \log m)} \) evaluates to at least \( (n + m)^{\mathcal{O}(\log k)} \), which is too large to be useful for designing \epas or \pas algorithms.

\subsection{\texorpdfstring{Weighted $\delta$-Nets via Dominating Set}{Weighted delta-Nets via Dominating Set}}
In this section, we present an algorithmic construction of \wvpn using known algorithms for the {\sc Set Cover} problem.

Given a bipartite graph \( \graph \) and \( \delta > 0 \), let \( G_\delta \) be the bipartite graph obtained by removing all vertices \( r \in \cR \) such that \( \w(\nbr(r)) < \delta \cdot \w(\cB) \). By construction, \( G_\delta \) is a weighted $\delta$-heavy graph. The following observation establishes a connection between $\wdeltanet(G)$ and $\wdeltanet(G_\delta)$.

\begin{observation}
\label{obs:reducetoheavy}
 Let $\graph$ be a bipartite graph, $\delta > 0$ and $X\subseteq \cB$. Further let $\w: \cB \to \bQ^+$ be the weight function defined on $\cB$. Then, $X$ is a  $\wdeltanet(G)$ if and only if $X$ is a  $\wdeltanet(G_\delta)$. 
\end{observation}

\Cref{obs:reducetoheavy} implies that, for the purpose of either finding $\wdeltanet(G)$ or establishing a combinatorial upper bound on the size of $\wdeltanet(G)$, it is sufficient to focus on weighted $\delta$-heavy graphs. 
The next observation views covering sets as a \emph{blue dominating set}. More precisely, it views them as a subset \( X \) of blue vertices such that every red vertex has at least one neighbor in \( X \).

\begin{observation}
\label{obs:domsetDet}
Let $\delta >0$ and $\graph$ with weight function $\w: \cB \to \bQ^+$, be a weighted $\delta$-heavy graph. Then  $\wdeltanet(G)$, say $X \subseteq \cB$ is a blue dominating set, that is $\nbr(X) = \cR$.
\end{observation}

\noindent 
 \Cref{obs:domsetDet} and \Cref{obs:reducetoheavy} together imply that the smallest size of $\wdeltanet(G)$ equals the size of a minimum blue dominating set for \( G_{\delta} \). We now establish an upper bound on the size of such a set in a weighted $\delta$-heavy graph \( \graph \), expressed as a function of \( \delta \) and \( \dintcmp(G) \). For a given weighted $\delta$-heavy bipartite graph \( G = (\cR, \cB) \), let \( \optds(G) \) denote the size of a minimum blue dominating set \( X \subseteq \cB \), defined as follows:
 
\[ \optds(G) = \min_{X \subseteq \cB, \nbr(X) = \cR} |X|\]
The next lemma is the main technical engine for the construction of $\wdeltanet(G)$. It provides an upper bound on \( \optds(G) \) in $G$, a weighted \( \delta \)-heavy graph. 

\begin{restatable}{lemma}{optdomsetlemma}
\label{lem:UpperBoundBDS}
\texorpdfstring{%
Let $\delta > 0$, $\graph$ with weight function $\w: \cB \to \bQ^+$ be a weighted $\delta$-heavy graph and $\Gamma = \dintcmp(G)$. Then,
}{%
Let delta > 0, G with weight function w: B -> Q+ be a weighted delta-heavy graph and Gamma = dintcmp(G). Then,
}
$$
\optds(G) \leq  \mathcal{O}\!\left(\frac{\Gamma}{\delta}\log{\frac{1}{\delta}}\right)
$$
\end{restatable}

 Proof of Lemma~\ref{lem:UpperBoundBDS} follows from \Cref{prop:vcExist} and \Cref{obs:domsetDet}.

 \subsection{\texorpdfstring{Algorithms for finding Weighted $\delta$-Nets and Deterministic Algorithms}{Algorithms for finding Weighted delta-Nets and Deterministic Algorithms}}

In this section, we describe two algorithms for computing a \( \wdeltanet(G) \) in a weighted \( \delta \)-heavy bipartite graph \( \graph \). The first is a polynomial-time algorithm that achieves a \( \cO(\ln n) \)-approximation, and the second is an \fpt algorithm parameterized by $\delta$ and \( \Gamma = \dintcmp(G) \). We formalize those in the following two lemmata.

\begin{lemma}[Polynomial Time algorithm]\label{lem:polytimecover}
Given a weighted $\delta$-heavy graph $\graph$ such that $\dintcmp(G) = \Gamma$, in $\cO(\sizeg)$ time, we can compute a $\wdeltanet(G)$, say $X$,  such that $|X| \leq (\ln n +1) \cdot \cO\left(\frac{\Gamma}{\delta}\log{\frac{1}{\delta}}\right)$ where $n = |\cR|$.
\end{lemma}

Proof of \Cref{lem:polytimecover} follows from the standard application of $\cO(\ln n)$-factor greedy approximation algorithm for greedy {\sc Set Cover}~\cite{DBLP:books/daglib/0004338}. 
\begin{lemma}[FPT algorithm]\label{lem:fpttimecover}
    For a weighted $\delta$-heavy graph $\graph$ such that $\dintcmp(G) = \Gamma$, we can compute a  $\wdeltanet(G)$ of size at most  $\cO\left(\frac{\Gamma}{\delta}\log{\frac{1}{\delta}}\right)$  in time \[\left(\frac{\Gamma}{\delta}\log{\frac{1}{\delta}}\right)^{\cO\left( \frac{\Gamma^2}{\delta}\log{\frac{1}{\delta}}\right)}\sizeg.\] 
\end{lemma}

Lemma~\ref{lem:fpttimecover} follows from the parameterized algorithm for {\sc Set Cover}~\cite{DBLP:journals/toct/G25} for set systems of bounded semi-ladder index. The next lemma ties the \wvpn with our algorithms.

\begin{lemma}
\label{lem:rel_delcov_alg}
Let \( \graph \) be a weighted bipartite graph with weight function $\w : \cB \to \bQ^+$ defined on \( \cB \). Suppose there exists a subset \( \{ u_1, \dots, u_\ell \} \subseteq \cB \), and let \( \omega = \sum_{\iletter \in [\ell]} \w(u_\iletter) \), such that there exists a set \( \bO \subseteq \cR \) with \( |\bO| \leq k \) and
\[
\w\left( \nbr(\bO) \cap \{ u_1, \dots, u_\ell \} \right) \geq \varepsilon \omega.
\]
Then, we can compute a set \( Z \subseteq \{ u_1, \dots, u_\ell \} \) such that \( Z \cap \nbr(\bO) \neq \emptyset \), and one of the following holds:

\begin{enumerate}
\setlength{\itemsep}{-2pt}
    \item \( |Z| \leq (\ln n +1) \cdot \cO\left(\frac{\Gamma}{\delta}\log{\frac{1}{\delta}}\right)\), and \( Z \) can be computed in polynomial time, or
    \item \( |Z| \leq \cO\left(\frac{\Gamma}{\delta}\log{\frac{1}{\delta}}\right) \), and \( Z \) can be computed in time \( \left(\frac{\Gamma}{\delta}\log{\frac{1}{\delta}}\right)^{\cO\left( \frac{\Gamma^2}{\delta}\log{\frac{1}{\delta}}\right)}\sizeg \).
\end{enumerate}
where $\Gamma=\dintcmp(G)$. 
\end{lemma}
\begin{proof}
Let \( \widetilde{\cB} = \{ u_1, \dots, u_\ell \} \). Define 
\[
\widetilde{\cR} \coloneqq \left\{ r \in \cR \,\middle|\, \w(\nbr(r) \cap \widetilde{\cB}) \geq \frac{\varepsilon}{k} \cdot \omega \right\}.
\]
Observe that the bipartite graph \( \widetilde{G} = (\widetilde{\cR}, \widetilde{\cB}) \) is a weighted \( (\frac{\varepsilon}{k}) \)-heavy graph. The rest of the proof follows by applying either \Cref{lem:polytimecover} or \Cref{lem:fpttimecover} and \Cref{lem:UpperBoundBDS}.

Let \( Z \) be the set returned by one of these lemmas. We now argue that \( |Z \cap \nbr(\bO)| \geq 1 \). By construction, \( Z \) is a weighted-$\frac{\varepsilon}{k}$-net (or a blue dominating set for \( \widetilde{\cR} \)). 

Moreover, since \( \w(\nbr(\bO) \cap \widetilde{\cB}) \geq \varepsilon \cdot \omega \), there exists some vertex \( o \in \bO \) such that 
\[
\w(\nbr(o) \cap \widetilde{\cB}) \geq \frac{\varepsilon}{k} \cdot \omega.
\]
By the definition of \( \widetilde{\cR} \), this implies that \( o \in \widetilde{\cR} \), and therefore \( \bO \cap \widetilde{\cR} \neq \emptyset \). Since \( Z \) is a blue dominating set for \( \widetilde{\cR} \), it must contain at least one neighbor of every vertex in \( \widetilde{\cR} \), including \( o \). 

Hence, we conclude that \( Z \cap \nbr(\bO) \neq \emptyset \), i.e., \( |Z \cap \nbr(\bO)| \geq 1 \). This completes the proof.
\end{proof}

Using Lemma~\ref{lem:rel_delcov_alg} we get the following deterministic version of ~\Cref{thm:mainWeighted} and ~\Cref{thm:addrandom} as follows.

\begin{theorem}
\label{thm:detmainWeighted}
Given an instance $\cI = (\graph, \w, k)$ of \wmrbds and an $\varepsilon>0$, there exists an algorithm that runs in time \[ \left(\frac{\Gamma k}{\varepsilon}\right)^{\cO(\Gamma k)} \sizeg = \left(\frac{1}{\varepsilon}\right)^{\cO(\Gamma k  \log (\Gamma k))} \sizeg \] 
and outputs a subset $\cS \subseteq \cR$ of size $k$ such that  \[\w(\nbr(\cS)) \geq (1 - \varepsilon) \wopt_k(\cI),\] where  $\Gamma = \dintcmp(G)$.
\end{theorem}

\medskip

In \Cref{thm:detmainWeighted}, we use an \( \frac{\varepsilon}{k} \)-net to derandomize \Cref{thm:mainWeighted}. Specifically, we replace the randomized step with a branching step over an \( \frac{\varepsilon}{k} \)-net of size \( \cO\left(\frac{\Gamma}{\varepsilon} \log{\frac{1}{\varepsilon}}\right) \). Additionally, we branch over \( k \) choices of \mcon, and the measure of the algorithm is \( (\Gamma + 1)k \). These steps yield the running time stated in \Cref{thm:detmainWeighted}.

\medskip

\begin{theorem}\label{thm:detaddrandom}
Given an instance \( \mathcal{I} = (G, \w, k, W) \) of \decwprbds, there exists a deterministic algorithm with running time 
\[ \left(\Gamma k\right)^{\cO(\Gamma k)}\sizeg = 2^{\cO(\Gamma k \log (\Gamma k))} \sizeg\]
which, in the case of a \textsc{Yes}-instance, returns a subset \( \mathcal{S} \subseteq \mathcal{R} \) of size at most \( k+1 \) such that 
\[
\w(\nbr(\mathcal{S})) \geq W. 
\]
However, in the case of a \textsc{No}-instance, the algorithm either correctly reports that the instance is a \textsc{No}-instance, or returns a subset \( \mathcal{S} \subseteq \mathcal{R} \) of size at most \( k+1 \) such that
\[
\w(\nbr(\mathcal{S})) \geq W.
\]
Here, \( \Gamma = \dintcmp(G) \) denotes the downward intersection complexity of the input graph \( G \).
\end{theorem}

 In \Cref{thm:detaddrandom} we use  $\frac{1}{2} $-net to derandomize \Cref{thm:addrandom}.  
In particular, we replace the randomized step by branching over $\frac{1}{2}$-net of of size \( \cO(\Gamma) \). Furthermore, we branch over \( k \) choices of \mcon, and the measure of the algorithm is \( (\Gamma + 1)k \). These choices lead to the desired running time given in \Cref{thm:detaddrandom}.

%% file: bkl_trace_enumeration.tex
\section{A Faster Implementation of the BKL Scheme}
\label{sec:bkl-trace-enumeration}

The lower bounds in the next section are stated relative to the theorem of
Badanidiyuru, Kleinberg, and Lee~\cite{DBLP:conf/compgeom/BadanidiyuruKL12},
our positive fixed-budget benchmark: on set systems of bounded VC-dimension,
\maxcov admits an \epas.  The same theorem is also the coverage subroutine
behind our \epas for \cmaxsat on bounded-VC-dimension formulas
(\Cref{thm: vcdim_satepas}).

In this section we sharpen the \emph{deterministic} implementation of this
scheme.  The approximation guarantee is unchanged; only the running time
improves, from the deterministic bound \(2^{\widetilde{\cO}(k^2d/\varepsilon^5)}\)
of Badanidiyuru, Kleinberg, and Lee to \(2^{\widetilde{\cO}(kd/\varepsilon)}\).
The improvement rests on a single observation.  To make the preprocessing
deterministic, their implementation enumerates
\emph{all} \(2^{|A|}\) subsets \(B\subseteq A\) of each sample \(A\).  Yet the
correctness argument only ever uses guesses of the form \(S_i\cap A\) for the
sets \(S_i\) of an optimal solution, that is, \emph{traces} of input sets,
\(\{F\cap A: F\in\cF\}\).  By the Sauer--Shelah
lemma~\cite{DBLP:journals/jct/Sauer72,Shelah1972}, there are at most
\(|A|^{\cO(d)}\) such traces, vastly fewer than \(2^{|A|}\).  Enumerating only
these realized traces therefore leaves their approximation proof intact while
sharply reducing the number of branches, and it is what improves the exponent
from \(k^2d/\varepsilon^5\) to \(kd/\varepsilon\).

Throughout, \(N\) denotes the incidence-list size of \((\cU,\cF)\),
which is within a constant factor of the incidence-graph size \(\sizeg\) used
elsewhere in the paper.

\begin{proposition}[Badanidiyuru--Kleinberg--Lee~{\cite[Section~3]{DBLP:conf/compgeom/BadanidiyuruKL12}}]
\label{prop:bkl-original-vc-epas}
Let \((\cU,\cF)\) be an unweighted \maxcov instance whose set system has
VC-dimension \(d\), let \(k\) be the budget, and let
\(\varepsilon\in(0,1)\).  There is a deterministic algorithm that returns a
\((1-\varepsilon)\)-approximate \(k\)-set solution in time
\[
  2^{\widetilde{\cO}(k^2d\varepsilon^{-5})}N^{\cO(1)},
\]
where \(N\) denotes the incidence-list size of \((\cU,\cF)\).
\end{proposition}

Badanidiyuru, Kleinberg, and Lee first prove the theorem in the
subsystem-oracle model and then note that for explicitly listed set systems
the standard oracle can be implemented by scanning the input sets, increasing
the running time by only a polynomial
factor~\cite[Section~3]{DBLP:conf/compgeom/BadanidiyuruKL12}.  The rest of this
section keeps their approximation proof intact and only improves the
deterministic enumeration used in that proof.

Let \((\cU,\cF)\) be a set system of VC-dimension \(d\).  For
\(A\subseteq\cU\), write
\[
  \cF|_A=\{F\cap A:F\in\cF\}
\]
for the trace family on \(A\).  By the Sauer--Shelah
lemma~\cite{DBLP:journals/jct/Sauer72,Shelah1972},
\[
  |\cF|_A|
  \le \sum_{j=0}^{d}\binom{|A|}{j}
  \le \left(\frac{e|A|}{d}\right)^d
\]
whenever \(1\le d\le |A|\).  Thus a sample of size \(a\) has only
\(a^{\cO(d)}\) realizable traces, even though it has \(2^a\) subsets.
This is the same trace-counting principle built into their subsystem oracle:
given \(A\), the oracle lists \(\cF|_A\), and the list has size
\(\cO(|A|^d)\), by the Sauer--Shelah
lemma~\cite{DBLP:journals/jct/Sauer72,Shelah1972}; see also their subsystem-oracle formulation
\cite[Definition~3.1]{DBLP:conf/compgeom/BadanidiyuruKL12}.

\subsection{Where the deterministic enumeration loses}

We recall only the part of the algorithm of Badanidiyuru, Kleinberg, and Lee
where the deterministic running time is affected.  Their preprocessing phase
reduces the general case to the case where the optimum covers a constant
fraction of the remaining universe.  The preprocessing is described as a
nondeterministic algorithm with
\[
  p=\left\lceil \frac{6}{\varepsilon}\ln\frac{3}{\varepsilon}\right\rceil
\]
phases.  In phase \(q\), it constructs a set
\[
  A^q\subseteq \cU
\]
which is an \((\varepsilon/6p)\)-approximation for the relevant residual
set system.  The size of this sample is
\[
  |A^q|
  =
  \cO\!\left(
       k d\,\varepsilon^{-2}p^2
       \log\frac{kdp}{\varepsilon}
     \right).
\]
The nondeterministic step guesses \(k\) subsets
\[
  B^q_1,\ldots,B^q_k\subseteq A^q
\]
and then defines constrained families
\[
  \cF^q_i(B^q_i)
  =
  \{F\in\cF:F\cap A^q=B^q_i\},
  \qquad i\in[k].
\]
The phase then runs greedy for constrained maximum coverage on the residual
universe, choosing one set from each of these \(k\) constrained families.
These are exactly the choices described in the nondeterministic preprocessing
of Section~3 of Badanidiyuru, Kleinberg, and
Lee~\cite[Section~3]{DBLP:conf/compgeom/BadanidiyuruKL12}.

In the correctness proof, let \(S_1,\ldots,S_k\) be an optimum solution for
the original \maxcov instance.  The successful nondeterministic execution is
the one that guesses
\[
  B^q_i=S_i\cap A^q
  \qquad\text{for every }q\in[p]\text{ and }i\in[k].
\]
Thus the successful guesses are not arbitrary subsets of \(A^q\).  They are
traces realized by actual input sets.  However, the deterministic analysis in
Section~3 of Badanidiyuru, Kleinberg, and
Lee~\cite[Section~3]{DBLP:conf/compgeom/BadanidiyuruKL12} counts
\emph{all} subsets \(B^q_i\subseteq A^q\).  In our notation, the number of
corresponding deterministic executions is
\[
  N_{\mathrm{BKL}}(k,d,\varepsilon)
  \coloneqq
  \prod_{q=1}^{p}\prod_{i=1}^{k}2^{|A^q|}.
\]
This leads to the stated
\[
  2^{\widetilde{\cO}(k^2d\varepsilon^{-5})}
\]
parameter dependence~\cite[Section~3]{DBLP:conf/compgeom/BadanidiyuruKL12}.
Here is the calculation explicitly.  Let
\[
  a_{\max}\coloneqq \max_{q\in[p]} |A^q|.
\]
The sample-size bound from their analysis gives
\[
  a_{\max}
  =
  \cO\!\left(
       k d\,\varepsilon^{-2}p^2
       \log\frac{kdp}{\varepsilon}
     \right)
\]
and hence \(\sum_{q=1}^{p}|A^q|\le p a_{\max}\).  Since every \(B_i^q\) is
counted as an arbitrary subset of \(A^q\), one has
\[
\begin{aligned}
  N_{\mathrm{BKL}}(k,d,\varepsilon)
  &=
  \prod_{q=1}^{p}\prod_{i=1}^{k}2^{|A^q|}  \\
  &\le
  2^{k\sum_{q=1}^{p}|A^q|}
  \le
  2^{kpa_{\max}}.
\end{aligned}
\]
Therefore
\[
\begin{aligned}
  \log N_{\mathrm{BKL}}(k,d,\varepsilon)
  &\le
  k p\cdot
  \cO\!\left(
       k d\,\varepsilon^{-2}p^2
       \log\frac{kdp}{\varepsilon}
     \right)\\
  &=
  \cO\!\left(
       k^2 d\,\varepsilon^{-2}p^3
       \log\frac{kdp}{\varepsilon}
     \right).
\end{aligned}
\]
Finally, Badanidiyuru, Kleinberg, and Lee choose
\[
  p=\left\lceil \frac{6}{\varepsilon}\ln\frac{3}{\varepsilon}\right\rceil
  =
  \cO\!\left(\varepsilon^{-1}\log\frac1\varepsilon\right),
\]
and hence
\[
  \log N_{\mathrm{BKL}}(k,d,\varepsilon)
  =
  \cO\!\left(
       k^2d\,\varepsilon^{-5}
       \log^3\frac1\varepsilon
       \log\!\left(
          \frac{kd}{\varepsilon^2}\log\frac1\varepsilon
       \right)
     \right)
  =
  \widetilde{\cO}\!\left(k^2d\varepsilon^{-5}\right).
\]
The final per-execution work in their algorithm contributes only
\(\exp(\cO(kd\log(kd/\varepsilon)))\), so the subset-guessing term above
dominates the published parameter dependence
\(2^{\widetilde{\cO}(k^2d\varepsilon^{-5})}\)
\cite[Section~3]{DBLP:conf/compgeom/BadanidiyuruKL12}.

\subsection{Enumerating only realized traces}

The deterministic simulation can instead enumerate only the trace family
\(\cF|_{A^q}\).  For an explicitly listed set system this is immediate:
scan every set \(F\in\cF\), compute \(F\cap A^q\), and bucket equal
intersections together.  The number of buckets is exactly
\(|\cF|_{A^q}|\), which is at most \(|A^q|^{\cO(d)}\).  Each bucket also
stores the corresponding constrained family
\[
  \{F\in\cF:F\cap A^q=B\}.
\]
Therefore phase \(q\) has at most
\[
  |\cF|_{A^q}|^k
  \le |A^q|^{\cO(dk)}
\]
trace-tuples to try, rather than \(2^{k|A^q|}\) arbitrary subset-tuples.

\begin{lemma}[Trace-realized branch]
\label{lem:bkl-trace-realized-branch}
The trace enumeration described above contains a successful execution of
the nondeterministic preprocessing of Badanidiyuru, Kleinberg, and Lee.
\end{lemma}

\begin{proof}
Fix the optimum sets \(S_1,\ldots,S_k\) used in their analysis.  In phase
\(q\), the successful nondeterministic branch guesses
\(B^q_i=S_i\cap A^q\).  Since each \(S_i\) is an input set, \(B^q_i\) is a
member of the trace family \(\cF|_{A^q}\).  Hence the tuple
\[
  (S_1\cap A^q,\ldots,S_k\cap A^q)
\]
is explicitly enumerated by the trace-based simulation.

For this tuple, the constrained family
\[
  \cF^q_i(S_i\cap A^q)
  =
  \{F\in\cF:F\cap A^q=S_i\cap A^q\}
\]
contains \(S_i\).  Therefore the same comparison made in their proof
against the optimum tuple \(S_1,\ldots,S_k\) remains valid.  The greedy
choice for constrained maximum coverage is run on the same constrained
families as in the successful nondeterministic branch, so the rest of their
phase analysis applies without change.
\end{proof}

\begin{theorem}[Trace-enumerated BKL implementation]
\label{thm:bkl-trace-enumerated-runtime}
Let \((\cU,\cF)\) be an explicitly listed set system of VC-dimension \(d\),
let \(k\) be the budget, and measure the input size by \(N\).
For every \(\varepsilon\in(0,1)\), fixed-budget \maxcov has a deterministic
\((1-\varepsilon)\)-approximation algorithm with running time
\[
  2^{\widetilde{\cO}(kd/\varepsilon)}N^{\cO(1)}.
\]
\end{theorem}

\begin{proof}
The algorithm of Badanidiyuru, Kleinberg, and Lee and its approximation proof
are unchanged.  We only replace the deterministic simulation of the
nondeterministic guesses.

For a fixed phase \(q\), trace enumeration gives at most
\[
  |\cF|_{A^q}|^k
  \le |A^q|^{\cO(dk)}
\]
branches.  Let
\[
  a_{\max}\coloneqq \max_{q\in[p]} |A^q|.
\]
The sample-size bound from their analysis gives
\[
  a_{\max}
  =
  \cO\!\left(
       k d\,\varepsilon^{-2}p^2
       \log\frac{kdp}{\varepsilon}
     \right)
\]
and therefore, over all \(p\) phases, the number of branch sequences is
bounded by
\[
  \prod_{q=1}^{p}|A^q|^{\cO(dk)}
  \le
  a_{\max}^{\cO(pdk)}
  =
  \exp\!\left(\cO(pdk\log a_{\max})\right).
\]
Since
\[
  p=\cO\!\left(\varepsilon^{-1}\log\frac1\varepsilon\right),
\]
this is
\[
  2^{\widetilde{\cO}(kd/\varepsilon)}.
\]

For every enumerated trace tuple, the constrained families can be represented
by buckets of input sets with the specified intersections with \(A^q\), and
the constrained greedy step can be implemented by scanning the input sets.
This adds only a polynomial factor in \(N\) for explicitly listed set
systems.  Finally, by \Cref{lem:bkl-trace-realized-branch}, the deterministic
trace enumeration still contains the successful branch used in their
correctness proof.  Hence the output is a \((1-\varepsilon)\)-approximation.
\end{proof}

%% file: maxsat_reduction.tex
\section{\wmsfull}
\label{sec: weightedsat}
This section presents an \epas for \wmsfull (or \wms) for generic \cnf formulae. 
This is achieved by reducing the problem to \wmc.  We define the problem formally. 

\defparopt{\wms}{A \cnf formula $\Phi = (\cV, \cC)$ on $n$ variables and $m$ clauses, a weight function $\w: \cC \rightarrow \bQ^{+}$, and an integer $k$.}
{$k$}
{An assignment to $\cV$ of weight at most $k$ that maximizes the sum of weights of satisfied clauses.}

\medskip
\noindent 
We give a reduction from \wms to a family of \wmc instances. This reduction is deterministic and for any $\varepsilon > 0$, runs in time $f(k, \varepsilon)$ for some computable function $f$. Furthermore, the reduction is robust: it extends naturally to the weighted setting and maps a $\left(1-\frac{\varepsilon}{2}\right)$-approximate solution of one produced coverage instance to a $(1-\varepsilon)$-approximate assignment for the original instance.

This establishes the equivalence of \wms and \wmc in the context of parameterized approximation schemes, which enables us to utilize the results we obtained for \wmc to solve \wms. We recall the definition of \wmc for the reader's convenience.

\defparopt{\wmc}{A set system $(\cU, \cF)$ on $n$ elements and $m$ sets, a weight function $\w: \cU \rightarrow \bQ^{+}$, and an integer $k$.}
{$k$}
{A subfamily $\cF'$ of size at most $k$ that maximizes the sum of weights of covered elements.}

We recall our notations for {\sf SAT} here for convenience.
Let $\Phi$ be a \cnf formula on $n$ variables and $m$ clauses. Variables and clauses in $\Phi$ are denoted by $\cV_{\Phi}$ and $\cC_{\Phi}$. When the context is clear, the subscripts are omitted. A variable is said to be positive if it doesn't occur as a negative literal in the formula $\Phi$, otherwise it is a negative variable. We denote $\cV_{neg} \subseteq \cV$ as the set of negative variables in the formula. Let $\cC_{neg} \subseteq \cC$ be the set of clauses with at least one negative literal. Then, $\cC_{pos} = \cC \setminus \cC_{neg}$ is the set of clauses that do not contain any negative literal.

For a clause $c \in \cC_{\Phi}$, let $\n{c}$ denote the set of variables that appear as negative literals in $c$. For a set of clauses $C \subseteq \cC_{\Phi}$, we define
\[
\n{C} = \bigcup_{c \in C} \n{c}.
\]

Given a weight function \(\w: \cC_{\Phi} \to \bQ^+\), the weight of a set of clauses \(C \subseteq \cC_{\Phi}\) is defined as
\[
\w(C) = \sum_{c \in C} \w(c).
\]

An assignment $\psi$ is a binary function on $\cV_{\Phi}$, i.e., $\psi: \cV_{\Phi} \rightarrow \{0, 1\}$. The \emph{weight} of an assignment $\psi$, denoted by $\wt{\Phi}{\psi}$, is the number of variables set to $1$ by $\psi$. For an assignment $\psi$, we denote by $\sat{\Phi}{\psi}$ the set of clauses of $\Phi$ satisfied by $\psi$. Similary, for $S \subseteq \cV$, we have $\sat{\Phi}{S}=\sat{\Phi}{\psi}$ where the assignment $\psi$ assigns $1$ only to the variables in $S$.
The incidence graph of $\Phi$, denoted by $G_{\Phi} = (P, Q, E)$, is the bipartite graph that has a vertex in $P$ for each variable in $\cV_{\Phi}$, a vertex in $Q$ for each clause in $\cC_{\Phi}$ and an edge $(u, v)$ if the variable corresponding to $u \in P$ occurs as a positive or negative literal in the clause corresponding to $v \in Q$.
For a formula \( (\cV, \cC) \), we use \( n = |\cV| \), \( m = |\cC| \), and \( m^\star = \sum_{C \in \cC} |C| \) to denote the number of variables, the number of clauses, and the total number of variable-clause incidences, respectively. Let $\w$ be the weight function defined on the set of clauses $\cC$ and $W_{\text{rep}} = \sum_{c \in \cC} \ln(1+\w(c))$ be the number of bits required to represent the weights on the clauses. We use $\sizeinst$ to denote the number of bits required to store the instance, i.e., $m+n+m^\star+ W_{\text{rep}}$.

\subsection{A Brief description of the reduction algorithm.} 

We briefly explain our algorithm for the unweighted case. The arguments follow similarly in the presence of a weight function on the set of clauses, which we prove in detail in our analysis. 
The algorithm takes a {\sc CC-MAXSAT} instance as input with a constant $\varepsilon > 0$. It retains negative variables with {\em high} negative degree. Here, we refer to the negative degree of a variable as the number of clauses in which it is present as a negative literal, with the precise definition coming up later in the section. We show that this subset of variables is bounded by a function of $k$ and $\varepsilon$. The remaining negative variables (ones with {\em low} negative degree) are transformed into positive variables by removing clauses that contain them as negative literals. This procedure is safe because we show that any assignment of weight at most $k$ to these low negative degree variables can unsatisfy only a small fraction of these deleted clauses, in fact, even a small fraction of the clauses satisfied by any optimal assignment of weight at most $k$. To deal with the high negative degree variables, since they are of bounded size, we guess all possible assignments of weight at most $k$ to this set. For each such assignment to the high negative degree variables, we remove clauses that are satisfied by this assignment and the resulting formula is essentially a {\sc Max Coverage} instance. Since we create a {\sc Max Coverage} instance for each assignment to the high negative degree variables, we get a family of {\sc Max Coverage} instances of size bounded by $g(k, \varepsilon)$ for some computable function $g$.
Furthermore, we show that there is at least one instance in this family of reduced instances such that an approximate solution to it can be used to construct an approximate solution to our input {\sc CC-MAXSAT} instance. 

We use the concept of {\em weighted normalized negative degree} defined below in our reduction algorithm.
This has been inspired by a similar concept in \cite{DBLP:journals/tcs/Manurangsi25}

\begin{definition}[Weighted normalized negative degree]
Let $\Phi = (\cV, \cC)$ be a \cnf formula and $\wtilde{\cV}$ be a variable  subset in $\cV$. The weighted normalized negative degree of a variable $x \in \wtilde{\cV}$, denoted by $\wnd(x)$, is defined as follows.
\[
\wnd_{\Phi, \wtilde{\cV}}(x) = \sum_{\substack{c \in \cC_{neg} \\ x \in \n{c}}} \frac{\w(c)}{|\n{c} \cap \wtilde{\cV}|}
\] 
\end{definition}

We begin by applying the following reduction rule to bound the number of negative literals in a negative clause. 

\begin{redrule}
\label{rr: largeneglit}

Let $(\Phi = (\cV, \cC), \w, k)$ be an instance of \wms. If there exists a clause $c \in \cC$ such that $|\n{c}| \geq k+1$, then remove clause $c$ from $\Phi$. Return the instance $(\Phi' = (\cV, \cC'), \w', k)$, where $\cC' = \cC \setminus \{c\}$ and $\w'$ is $\w$ restricted to $\cC'$.
\end{redrule}

It is easy to see that~\Cref{rr: largeneglit} is safe. Any assignment of weight at most $k$ will set at least one variable in $\n{c}$ to $0$, effectively satisfying the clause. We apply the rule exhaustively and, in the analysis below, write $\Phi$ for the reduced formula; the removed clauses contribute the same weight to every feasible assignment, so approximation guarantees for the reduced instance transfer to the original instance.

A complete description of the algorithm with pseudocode is provided in Algorithm~\ref{alg:framework-wsat}.

\medskip
\begin{algorithm}[ht!]

\caption{\algwsat}\label{alg:framework-wsat}
\KwInput{An instance $\cI=(\Phi,\w,k)$ of \wms and
$\varepsilon\in(0,1)$.}
\KwOutput{A family $\cM$ of \wmc instances, one of whose
$(1-\varepsilon/2)$-approximate solutions lifts to a
$(1-\varepsilon)$-approximate solution to $\cI$.}
Apply~\Cref{rr: largeneglit} exhaustively.\label{redwms: rrapply}\\
Set $\tau = \frac{\varepsilon}{2k} \cdot \w(\cC_{neg})$, $Y_0 = \cV_{neg}$ and $i = 0$.\\
\While{there exists a variable $x_{i+1} \in Y_i$ such that $\wnd_{\Phi, Y_i}(x_{i+1}) > \tau$\label{redwms: greedybegin}} 
    {   $Y_{i+1} \gets Y_i \setminus \{x_{i+1}\}$\\ 
        $i \gets i+1$\\
    }\label{redwms: greedy}
    Let $\itm = i$.\\
Let $\cC' = \{c \in \cC_{neg} \mid \n{c} \cap Y_{\itm} \neq \emptyset \}$ and $Z = \cV_{neg} \setminus Y_{\itm}$. \label{redwms: z_bound}\\
Set $\cM \gets \emptyset$.\\
\For(\label{line:assnguess}){each assignment $\beta$ to $Z$ of weight at most $k$\label{redwms: guess_assgn}}
{
Let $\cC_Z = \{c \in \cC \setminus \cC' \mid c \text{ is satisfied by the partial assignment } \beta \text{ on } Z\}$.\label{redwms:findsatcl}\\
Let $\wtilde{\Phi} = (\cV_{\wtilde{\Phi}}, \cC_{\wtilde{\Phi}})$, where $\cV_{\wtilde{\Phi}} = \cV \setminus Z$ and $\cC_{\wtilde{\Phi}} = \{ c \setminus Z \mid c \in \cC \setminus (\cC' \cup \cC_Z)\}.$ Clauses in $\cC_{\wtilde{\Phi}}$ keep their original labels.\label{redwms:formmod}\\
{Create an instance of \wmc $((\cU, \cF), \wtilde{\w}, \wtilde{k})$ from $\wtilde{\Phi} = (\cV_{\wtilde{\Phi}}, \cC_{\wtilde{\Phi}})$ as follows:\label{redwms:reducetowmc}
\begin{description}
\setlength{\itemsep}{-1pt}
      \item[$\star$] For each clause $c \in \cC_{\wtilde{\Phi}}$, we have an element $e_c \in \cU$, and set $\wtilde{\w}(e_c) = \w(c),$ \\
      \item[$\star$] For each variable $x \in \cV_{\wtilde{\Phi}}$, we have a set $f_x \in \cF$, where $f_x = \{e_c \in \cU \mid \text{clause } c \text{ contains variable } x\},$\\
      \item[$\star$] Set $\wtilde{k} = k - \wt{}{\beta}.$\\
\end{description}}
$\cJ_\beta\gets((\cU,\cF),\wtilde{\w},\wtilde{k})$\\
$\cM\gets\cM\cup\{\cJ_\beta\}$\\
}
\Return $\cM$\\
\end{algorithm}

\subsection{Algorithm Analysis}

For an instance $\cI = (\Phi, \w, k)$ of \wms, $\wopts(\cI)$ denotes the largest weight of clauses that can be satisfied by an assignment of weight at most $k$. If $\beta^{\star}$ is an optimal assignment to $\cV$ of weight at most $k$, then $\w(\sat{\Phi}{\beta^{\star}}) = \wopts(\cI)$.

We next prove some claims and lemmas that will help us in proving our desired result.

\begin{claim}
\label{clm: z_bound}
    Let $Z$ be the set of negative variables in $\Phi$ in \Cref{redwms: z_bound}. Then, $|Z| \leq \mathcal{O}(\frac{k \log k}{\varepsilon})$.
\end{claim}

\begin{proof}
Let $x_i$ be the variable chosen in the $i^{\sf th}$ iteration of the while loop. For each such $x_i$, we have $\wnd_{\Phi, Y_{i-1}}(x_i) > \tau$.
Thus,
\allowdisplaybreaks
\begin{align*}
    |Z| \tau & < \sum_{i \in [|Z|]} \wnd_{\Phi, Y_{i-1}}(x_i) \\
    & = \sum_{i \in [|Z|]} \sum_{\substack{c \in \cC_{neg} \\ x_i \in \n{c}}} \frac{\w(c)}{|\n{c} \cap Y_{i-1}|} \\
    & = \sum_{c \in \cC_{neg}} \sum_{\substack{i \in [|Z|] \\ x_i \in \n{c}}} \frac{\w(c)}{|\n{c} \cap Y_{i-1}|} \\
    & = \sum_{c \in \cC_{neg}} \w(c) \sum_{\substack{i \in [|Z|] \\ x_i \in \n{c}}} \frac{1}{|\n{c} \cap Y_{i-1}|} \\
    & \stackrel{(\bigstar)}{\leq}  \sum_{c \in \cC_{neg}} \w(c) \left( \frac{1}{k} + \frac{1}{k-1} + \dots + \frac{1}{2} + 1\right) \\
    & \leq \sum_{c \in \cC_{neg}} \w(c) (\ln k + 1) \\
    & = \w(\cC_{neg}) (\ln k + 1)
\end{align*}

Notice that $(\bigstar)$ holds because of the following reason. Due to the exhaustive application of~\Cref{rr: largeneglit}, for any clause $c$, $|\n{c}| \leq k$. Moreover, for any clause $c$, during the iterations of the while loop where the variable $x_i$ being considered is in $\n{c}$, the terms $|\n{c} \cap Y_{i-1}|$ are all distinct as $x_i$ is removed from $Y_{i-1}$. Hence, we get the required upper bound.

Substituting the value of $\tau$, we get,
\begin{align*}
    |Z| < \frac{2k}{\varepsilon}(\ln k + 1)
\end{align*}

Thus, $|Z| \leq \mathcal{O} \left(\frac{k \ln k}{\varepsilon} \right)$.
\end{proof}

\Cref{clm: z_bound} tells us that the number of negative variables placed in the branching set $Z$ after the while loop ends in Line~\ref{redwms: greedy} of \Cref{alg:framework-wsat} is $\mathcal{O}\left(\frac{k \ln k}{\varepsilon} \right)$.

\begin{lemma}
\label{lem: unsat_clause_bound}
    Let $\psi$ be an assignment to $\cV$ of weight at most $k$. Let $Y_{\itm}$ be the set of variables that survive at the end of the while loop in \Cref{redwms: greedy} of \Cref{alg:framework-wsat} and let $\cC'$ be the set of clauses containing at least one variable from $Y_{\itm}$ as a negative literal. Then the total weight of clauses in $\cC'$ unsatisfied by $\psi$ is at most $\frac{\varepsilon}{2} \wopts(\cI)$.
\end{lemma}

\begin{proof}
$\cC'$ is the set of clauses that contain at least one negative literal from $Y_{\itm}$. Let $\psi$ be an assignment of weight at most $k$ to $\cV$. Any clause $c \in \cC'$ is unsatisfied by $\psi$ only if $\n{c} \subseteq \psi^{-1}(1)$. Let $\cC'_{\text{unsat}(\psi)} \subseteq \cC'$ be the subset of clauses in $\cC'$ that are unsatisfied by $\psi$. We claim that $\w(\cC'_{\text{unsat}(\psi)}) \leq \frac{\varepsilon}{2} \wopts(\cI)$,

\begin{align*}
    \w(\cC'_{\text{unsat}(\psi)}) & \leq \sum_{c \in \cC'_{\text{unsat}(\psi)}} \w(c) \\
    & \leq \sum_{c \in \cC'_{\text{unsat}(\psi)}} \sum_{x \in \left(\n{c} \cap Y_{\itm}\right)} \frac{\w(c)}{|\n{c} \cap Y_{\itm}|} \\
    & \stackrel{(\blacktriangle)}{=} \sum_{c \in \cC'_{\text{unsat}(\psi)}} \sum_{x \in \left(\n{c} \cap \psi^{-1}(1) \cap Y_{\itm}\right)} \frac{\w(c)}{|\n{c} \cap Y_{\itm}|} \\
    & \leq \sum_{x \in \left(\psi^{-1}(1) \cap Y_{\itm}\right)}  \sum_{\substack{c \in \cC'_{\text{unsat}(\psi)} \\ x \in \n{c}}} \frac{\w(c)}{|\n{c} \cap Y_{\itm}|}\\
    & \leq \sum_{x \in \left(\psi^{-1}(1) \cap Y_{\itm}\right)} \wnd_{\Phi, Y_{\itm}}(x)\\
    & \leq |\psi^{-1}(1) \cap Y_{\itm}| \tau \\
    & \leq k \tau \\
    & = \frac{\varepsilon}{2} \w(\cC_{neg})
\end{align*}

Note that $(\blacktriangle)$ holds because $\n{c} \subseteq \psi^{-1}(1)$ for any clause $c \in \cC'_{\text{unsat}(\psi)}$.
Moreover, every clause in $\cC_{neg}$ is satisfied by the all-zero assignment, and therefore $\w(\cC_{neg}) \leq \w(\sat{\Phi}{\emptyset}) \leq \wopts(\cI)$. Hence, $\w(\cC'_{\text{unsat}(\psi)}) \leq \frac{\varepsilon}{2} \wopts(\cI)$.
\end{proof}

Thus,~\Cref{lem: unsat_clause_bound} tells us that for any assignment to the variables in $Y_{\itm}$ of weight at most $k$, the total weight of unsatisfied clauses in $\cC'$ is no more than $\frac{\varepsilon}{2} \wopts(\cI)$.

\begin{lemma}
\label{lem: approx_preserve}
    Let $\cI = (\Phi, \w, k)$ be an instance of \wms and $\cM$ be the set of \wmc instances returned by \Cref{alg:framework-wsat} on input $\cI$. Then, there exists at least one \wmc instance $\cJ = ((\cU, \cF), \wtilde{\w}, \wtilde{k})$ in $\cM$ such that given a $\left( 1-\frac{\varepsilon}{2} \right)$-approximate solution to $\cJ$, we can compute a $(1-\varepsilon)$-approximate solution to $\cI$ of \wms.
\end{lemma}

\begin{proof}
Let $\beta^{\star}$ be an optimal assignment of weight at most $k$ to $\cI$. Then, $\w(\sat{\Phi}{\beta^{\star}}) = \wopts(\cI)$. 

 Let $Z$ be the set of remaining negative variables in $\Phi$ that we get in \Cref{redwms: z_bound} of the reduction algorithm.
 Clearly, $\beta^{\star}\restriction_Z$, i.e. the assignment $\beta^{\star}$ restricted to the variables in $Z$, is one of the assignments guessed in \Cref{redwms: guess_assgn} of the algorithm. Let $\cJ_{\beta^{\star}\restriction_Z} = ((\cU, \cF), \wtilde{\w}, \wtilde{k})$ be an instance of \wmc corresponding to $\beta^{\star}\restriction_Z$ that is constructed by the algorithm. Thus, we know that $\cJ_{\beta^{\star}\restriction_Z} \in \cM$. 
 
 Let $\cS$ be a $\left( 1-\frac{\varepsilon}{2} \right)$-approximate solution to $\cJ_{\beta^{\star}\restriction_Z}$. Then, $\cS$ is a subfamily of size $\wtilde{k}$ and the sum of weights of elements covered by $\cS$ is at least $\left( 1-\frac{\varepsilon}{2} \right)\woptc(\cJ_{\beta^{\star}\restriction_Z})$. Here, $\woptc(\cJ_{\beta^{\star}\restriction_Z})$ is the largest weight of elements covered by a subfamily of size at most $\wtilde{k}$ in the instance $\cJ_{\beta^{\star}\restriction_Z}$. 
 
 Let $\cV_{\cS} = \{x \in \cV \setminus Z \mid f_x \in \cF \text{ belongs to } \cS\}$ and $\sigma$ be the assignment to $\cV$ defined as follows.
\[
\sigma(x) = \begin{cases} 
      \beta^{\star}\restriction_Z(x) & \text{if } x \in Z \\
      1 & \text{if } x \in \cV \setminus Z \text{ and the corresponding set } f_x \in \cS \\
      0 & \text{otherwise}
   \end{cases}
\]
 We claim that $\w(\sat{\Phi}{\sigma}) \geq (1 - \varepsilon)\wopts(\cI)$. Clearly, $\wt{\Phi}{\sigma} \leq \wtilde{k} + \wt{}{\beta^{\star}\restriction_Z} \leq k$. 
 
We next show the approximation guarantee. Recall that $\cC'$ was the subset of clauses that contained at least one variable in $Y_{\itm}$ as a negative literal.
Let $\cC_Z^{\star}$ be the set of clauses in $\cC \setminus \cC'$ that is satisfied by the partial assignment $\beta^{\star}\restriction_Z$ to the variables in $Z$. Thus, for any assignment $\psi$ to $\cV$ that agrees with $\beta^{\star}\restriction_Z$ and is of weight at most $k$, we have 
\begin{align*}
\w(\sat{\wtilde{\Phi}}{\psi}) + \w(\cC_Z^{\star}) + \w(\cC') - \w(\cC'_{\text{unsat}(\psi)}) = \w(\sat{\Phi}{\psi}).
\end{align*}
Here, $\wtilde{\Phi} = (\cV_{\wtilde{\Phi}}, \cC_{\wtilde{\Phi}})$, where $\cV_{\wtilde{\Phi}} = \cV \setminus Z$ and $\cC_{\wtilde{\Phi}} = \{c \setminus Z \mid c \in \cC \setminus (\cC' \cup \cC_Z^{\star})\}.$

In particular, for the optimum solution $\psi^{\star} = \beta^{\star}$, which agrees with $\beta^{\star}\restriction_Z$ and has weight at most $k$, we have
\begin{align*}
    \w(\sat{\wtilde{\Phi}}{\psi^{\star}}) + \w(\cC_Z^{\star}) + \w(\cC') - \w(\cC'_{\text{unsat}(\psi^{\star})}) & = \w(\sat{\Phi}{\psi^{\star}}) \\
    & = \wopts(\cI)
\end{align*}
which implies that
\begin{align*}
    \w(\sat{\wtilde{\Phi}}{\psi^{\star}}) & = \wopts(\cI) - \w(\cC_Z^{\star}) - \w(\cC') + \w(\cC'_{\text{unsat}(\psi^{\star})}) \\
    & \geq \wopts(\cI) - \w(\cC_Z^{\star}) - \w(\cC')
\end{align*}
since $\w(\cC'_{\text{unsat}(\psi^{\star})})$ is non-negative. 

Moreover, $\woptc(\cJ_{\beta^{\star}\restriction_Z}) \geq \w(\sat{\wtilde{\Phi}}{\psi^{\star}})$ as $\cJ_{\beta^{\star}\restriction_Z}$ corresponds to the \wmc instance created from $\wtilde{\Phi}$ in \Cref{redwms:reducetowmc} of the algorithm.
Thus,
\begin{equation}
\label{eqn: bound}
\woptc(\cJ_{\beta^{\star}\restriction_Z}) \geq \wopts(\cI) - \w(\cC_Z^{\star}) - \w(\cC')
\end{equation}

Since $\cS$ is a $\left(1 - \frac{\varepsilon}{2} \right)$-approximate solution to $\cJ_{\beta^{\star}\restriction_Z}$, and the instance is created by removing the clauses in $(\cC' \cup \cC_Z^{\star})$ from $\cC$, the sum of weights of clauses in $\cC \setminus (\cC' \cup \cC_Z^{\star})$ satisfied by the assignment $\sigma$ is at least $\left(1 - \frac{\varepsilon}{2} \right)\woptc(\cJ_{\beta^{\star}\restriction_Z})$.

Thus, the sum of weights of clauses in $\cC$ satisfied by the assignment $\sigma$ is at least 
\begin{align*}
\w(\sat{\Phi}{\sigma}) & \stackrel{(\diamondsuit)}{\geq} \left(1 - \frac{\varepsilon}{2} \right)\woptc(\cJ_{\beta^{\star}\restriction_Z}) + \w(\cC') - \frac{\varepsilon}{2}\wopts(\cI) + \w(\cC_Z^{\star}) \\
& \stackrel{(\blacklozenge)}{\geq} \left(1 - \frac{\varepsilon}{2} \right)\wopts(\cI) - \frac{\varepsilon}{2}\wopts(\cI) \\
& = (1 - \varepsilon)\wopts(\cI)
\end{align*}

Note that $(\diamondsuit)$ holds because \Cref{lem: unsat_clause_bound} implies that any feasible assignment, in particular $\sigma$, can leave clauses with total weight at most $\frac{\varepsilon}{2}\wopts(\cI)$ unsatisfied in $\cC'$, and hence the weight of satisfied clauses in $\cC'$ is at least $\w(\cC') - \frac{\varepsilon}{2}\wopts(\cI)$. Moreover, as $\sigma$ agrees with $\beta^{\star}\restriction_Z$ on $Z$, it satisfies all clauses in $\cC_Z^{\star}$. Also, $(\blacklozenge)$ holds due to \Cref{eqn: bound}. This concludes our proof.
\end{proof}

Thus, we get our main result below.

\begin{theorem}
\label{thm: wtred_sat_to_cov}
Given an instance \( \mathcal{I} = (\Phi, \w, k) \) of \wms, there exists a deterministic algorithm that, in time
\[
\left(\frac{k \ln k}{\varepsilon}\right)^{\mathcal{O}(k)} \cdot \sizeinst,
\]
produces a family \( \mathcal{M} \) consisting of at most \( \left(\frac{k \ln k}{\varepsilon}\right)^{\mathcal{O}(k)} \) instances of \wmc. Moreover, there exists an instance \( \mathcal{J} \in \mathcal{M} \) such that, given a \( \left(1 - \frac{\varepsilon}{2} \right) \)-approximate solution for \( \mathcal{J} \), we can compute a \( (1 - \varepsilon) \)-approximate solution for \( \mathcal{I} \) in polynomial time.
\end{theorem}

\begin{proof}
The algorithm first applies \Cref{rr: largeneglit} to remove all clauses with at least $k+1$ negative literals, and then greedily reduces the set of negative variables in the input formula in Steps \ref{redwms: greedybegin} to \ref{redwms: greedy}. At the end of this greedy procedure, this set of negative variables $Z$ has bounded size as seen in \Cref{clm: z_bound}.  
It then constructs a family $\cM$ of \wmc instances. For each assignment $\eta$ of weight at most $k$ to $Z$, it creates a \wmc instance $\cJ_{\eta}$ as follows. It removes all clauses $\cC'$ that contain a variable from $Y_{\itm}$ as a negative literal, and also removes all clauses in $\cC \setminus \cC'$ satisfied by $\eta$. Finally, it deletes the variables in $Z$ from the remaining clauses. Every remaining negative literal would have to belong either to a variable in $Y_{\itm}$, whose clause was removed, or to a variable in $Z$, whose occurrence was deleted. Hence the resulting formula has only positive literals and is equivalent to a \wmc instance. 

By \Cref{lem: approx_preserve}, we know that there is at least one instance $\cJ \in \cM$ such that given a $\left( 1 - \frac{\varepsilon}{2}\right)$-approximate solution to $\cJ$, we can get a $(1 - \varepsilon)$-approximate solution for our input instance $\cI$. The correctness of the algorithm follows from the correctness of \Cref{rr: largeneglit} and \Cref{lem: approx_preserve}.

\medskip
\noindent 
\textbf{Running time:} Most of the steps of \Cref{alg:framework-wsat} can be executed in polynomial time and the only overhead comes from \Cref{redwms: guess_assgn} of the algorithm. In fact, we can compute the set $Z$ in time \( \cO\left(\left(\frac{k \ln k}{\varepsilon}\right)^{\cO(1)} \sizeinst \right) \) using standard data structures and bookkeeping. 
Since, by \Cref{clm: z_bound}, $|Z| \leq \mathcal{O} \left(\frac{k \ln k}{\varepsilon} \right)$, there are at most  
\[\sum_{i=0}^k  \binom{\cO\!\left(\frac{k \ln k}{\varepsilon}\right)}{i} \leq \left(\frac{k \ln k}{\varepsilon}\right)^{\mathcal{O}(k)}\]

assignments of weight at most $k$ to the variables in $Z$. For each assignment of weight at most $k$, a \wmc instance is constructed in $\cO(\sizeinst)$ time. Hence, the total running time of the algorithm is $\left(\frac{k \log k}{\varepsilon}\right)^{\mathcal{O}(k)}\sizeinst$.
\end{proof}

Before proceeding further, we state and prove a claim that helps in bounding the downward intersection complexity of the incidence graph of any reduced \wmc instance obtained by the reduction algorithm.
The incidence graph of a \wmc instance $\cJ = ((\cU, \cF), \w, k)$ is a bipartite graph $G_{\cJ} = (P, Q)$ where $P$ contains a vertex for each set in $\cF$, $Q$ contains a vertex for each element in $\cU$ and an edge joins a set vertex and an element vertex if that element is present in the set.

\begin{claim}
\label{clm: di_reduced_bound}
Let $\cI = (\Phi, \w, k)$ be an instance of \wms and $\cM$ be a family of \wmc instances that is output by \Cref{alg:framework-wsat}. Then, for any instance $\cJ$ in $\cM$, $\dintcmp(G_{\cJ}) \leq \dintcmp(G_{\cI}) + 1$.
\end{claim}

\begin{proof}
Let $\cI = (\Phi, \w, k)$ be an instance of \wms, where $G_{\cI}$ or $G_{\Phi} = (P_{\cV}, Q_{\cC})$ is the incidence graph of $\Phi$.
Let $\cJ$ in $\cM$ be a reduced \wmc instance, where $G_{\cJ}$ or $G_{(\cU, \cF)} = (P_{\cF}, Q_{\cU})$ is the incidence graph of $(\cU, \cF)$.   

Every \wmc instance $\cJ$ in $\cM$ is constructed from a residual formula $\wtilde{\Phi} = (\wtilde{\cV}, \wtilde{\cC})$, where $\wtilde{\cV} \subseteq \cV$ and each clause in $\wtilde{\cC}$ is a labeled restriction $c \setminus Z$ of some original clause $c \in \cC$. The set side of $G_{\cJ}$ corresponds to variables in $\wtilde{\cV}$ and the element side corresponds to clauses in $\wtilde{\cC}$, so $P_{\cF} = P_{\wtilde{\cV}}$ and $Q_{\cU} = Q_{\wtilde{\cC}}$.
Let the $\smlindex$ of $G_{\Phi}$ be $\lambda$ and the $\smlindex$ of $G_{\cJ}$ be $\lambda'$. 
For any remaining variable $x \in \wtilde{\cV}$ and any residual clause $c \setminus Z \in \wtilde{\cC}$, the variable $x$ occurs in $c \setminus Z$ if and only if $x$ occurs in the original clause $c$. Hence $G_{\cJ}$ is obtained from an induced subgraph of $G_{\Phi}$ by relabeling the surviving clause vertices. Therefore $G_{\cJ}$ cannot have a larger induced semi-ladder than $G_{\Phi}$, and $\lambda' \leq \lambda$.
Due to \Cref{thm: smli_dic_rel}, we know that $\lambda \leq \dintcmp(G) \leq \lambda + 1$, for any graph $G$. 
Thus, we have $\dintcmp(G_{\cJ}) \leq \lambda' + 1 \leq \lambda + 1 \leq \dintcmp(G_{\Phi}) + 1$.
\end{proof}

Equipped with \Cref{thm: wtred_sat_to_cov}, we now present our randomized and deterministic \epas result for \wms.

\begin{theorem}
\label{thm: application_randomized}
For any \( \varepsilon > 0 \), there exists a randomized algorithm that takes as input an instance \( \mathcal{I} = (\Phi, \w, k) \) of \wms, and in time
\[
\left(\frac{k \log k}{\varepsilon}\right)^{\mathcal{O}(k)} \cdot \left( \frac{4k}{\varepsilon} \right)^{\mathcal{O}(\Gamma k)} \cdot \sizeinst = \left(\frac{1}{\varepsilon}\right)^{\cO(\Gamma k \log k)} \cdot \sizeinst
\]
returns an assignment \( \psi \) of weight at most \( k \) such that the total weight of clauses satisfied by \( \psi \) is at least \( (1 - \varepsilon) \cdot \wopts(\mathcal{I}) \), with probability at least $\left(1 - \frac{1}{e}\right)$.
 
Here, \( \Gamma = \dintcmp(G_{\Phi}) \), where \( G_{\Phi} \) is the bipartite incidence graph of the formula \( \Phi \), and \( \wopts(\mathcal{I}) \) denotes the maximum total weight of clauses that can be satisfied by any assignment of weight at most \( k \).
\end{theorem}

\medskip

For an $\varepsilon > 0$, and an instance $\cI = (\Phi, \w, k)$ of \wms, we first run \Cref{alg:framework-wsat} to obtain a family of $\left(\frac{k \log k}{\varepsilon}\right)^{\mathcal{O}(k)}$ \wmc instances. The incidence graph of each instance in the family is a \wmrbds instance that is given as input to
\Cref{alg:framework-wprbds} with $\wtilde{\varepsilon}$ set to $\frac{\varepsilon}{2}$.  
By \Cref{thm:mainWeighted},
this algorithm runs in time $\left(\frac{2k}{\wtilde{\varepsilon}} \right)^{(\intcmp+1) k}\sizeg$ 
and outputs a $(1 - \wtilde{\varepsilon})$-approximate solution with probability at least $1 - \frac{1}{e}$. By \Cref{thm: wtred_sat_to_cov}, we get a $(1 - \varepsilon)$-approximate solution to $\cI$.

\medskip

\begin{theorem}
\label{thm: application_deterministic}
 For any \( \varepsilon > 0 \), there exists a deterministic algorithm that, given an instance \( \mathcal{I} = (\Phi, \w, k) \) of \wms, runs in time
\[
\left(\frac{k \log k}{\varepsilon}\right)^{\mathcal{O}(k)} \cdot \left(\frac{\Gamma k}{\varepsilon}\right)^{\cO(\Gamma k)} \sizeinst = \left ( \frac{1}{\varepsilon} \right )^{\cO(\Gamma k \log (\Gamma k))} \sizeinst,
\]
and returns an assignment \( \psi \) of weight at most \( k \) such that the total weight of clauses satisfied by \( \psi \) is at least \( (1 - \varepsilon) \cdot \wopts(\mathcal{I}) \).

Here, \( \Gamma = \dintcmp(G_{\Phi}) \), where \( G_{\Phi} \) is the bipartite incidence graph of the formula \( \Phi \), and \( \wopts(\mathcal{I}) \) denotes the maximum total weight of clauses satisfied by any assignment of weight at most \( k \).

\end{theorem}

\medskip

Just as in the case for randomized \epas, for an $\varepsilon > 0$ and a \wms instance $\cI = (\Phi, \w, k)$, we obtain $\left(\frac{k \log k}{\varepsilon}\right)^{\mathcal{O}(k)}$ \wmc instances using \Cref{alg:framework-wsat}. The incidence graph of each instance in the family is a \wmrbds instance. By \Cref{thm:detmainWeighted}, we have a deterministic algorithm that runs in time
\[\left(\frac{\Gamma k}{\wtilde{\varepsilon}}\right)^{\cO(\Gamma k)} \sizeinst\]

and outputs a $(1 - \wtilde{\varepsilon})$-approximate solution, where $\wtilde{\varepsilon}$ is set to $\frac{\varepsilon}{2}$. Finally, using \Cref{thm: wtred_sat_to_cov}, we get a $(1 - \varepsilon)$-approximate solution to $\cI$.

\subsection{\cmaxsat on formulas with bounded VC-dimension}

In this section, we give an \epas for \cmaxsat on formulas that have bounded VC-dimension. This is achieved by combining our reduction algorithm, \Cref{alg:framework-wsat}, with the \epas for \maxcov on set systems with bounded VC-dimension given by Badanidiyuru et al.~\cite{DBLP:conf/compgeom/BadanidiyuruKL12}. Since the aforementioned \epas for \maxcov works only for unweighted set systems with bounded VC-dimension, and our reduction algorithm also works for unweighted instances, we give an \epas for \cmaxsat on unweighted instances with bounded VC-dimension. 

We begin by defining what we mean by the VC-dimension of a \cnf formula $\Phi$. To do so, we view a \cnf formula $\Phi = (\cV, \cC)$ as a set system, say $(\cU_\Phi, \cF_\Phi)$, where the ground set $\cU_\Phi$ is set of all the clauses $\cC$ and for each variable $v \in \cV$, there is a subset of clauses $C_v$ in the family $\cF_\Phi$, where $C_v$ contains all the clauses containing $v$. Formally, the set system associated with $\Phi = (\cV, \cC)$ is $(\cU_\Phi, \cF_\Phi)$, where
\begin{align}
    \cU_\Phi & = \cC \text{, and} \label{eqn:formulatoset}\\
    \cF_\Phi &= \{ C_v \mid v \in \cV \}, \text{ where } C_v = \{c \in \cC \mid v \text{ or } \neg v \in c \}\label{eqn:formulatofamily}.
\end{align}

We define \textit{VC-dimension of a \cnf formula} analogous to VC-dimension of a set system.

\begin{definition}[VC-dimension of a \cnf formula]
\label{def: vc_dim_sat}
Given a set of clauses $C \subseteq \cC$, we define $\cF_C = \{ C \cap C_v \mid v \in \cV \}$. We say that a set of clauses $C$ is shattered by $\cV$ if $\cF_C = 2^{C}$. A \cnf formula $\Phi = (\cV, \cC)$ has VC-dimension $d$ if $d$ is the smallest integer such that no subset of $\cC$ of size $d+1$ can be shattered by $\cV$.   
\end{definition}

Observe that the VC dimension of a formula $\Phi$ is the same as that of its corresponding set system $(\cU_{\Phi}, \cF_{\Phi})$. We now state and prove a claim that shows that the VC dimension of the set system of any reduced \maxcov instance obtained by our reduction algorithm is bounded by that of the input \cmaxsat instance.

\begin{lemma}
\label{lem: vcdim_reduced_bound}
Let $\cI = (\Phi, k)$ be an instance of \cmaxsat and $\cM$ be a family of \maxcov instances that is output by \Cref{alg:framework-wsat}. Then, for any instance $\cJ = ((\cU, \cF), \wtilde{k})$ in $\cM$, VC-dimension of $(\cU, \cF)$ is at most VC-dimension of $\Phi$.
\end{lemma}

\begin{proof}
Let the VC-dimension of $\Phi$ be $d$. Fix an instance $\cJ \in \cM$ produced by some guessed assignment $\beta$ to $Z$, and identify each element of $\cJ$ with its corresponding labeled residual clause in
\[
\cC_{\wtilde{\Phi}}=\{c\setminus Z \mid c \in \cC \setminus (\cC' \cup \cC_Z)\}.
\]
There is a natural injective map $\pi:\cC_{\wtilde{\Phi}}\to \cC$ sending each labeled residual clause $c\setminus Z$ to its original clause $c$. For every variable $x \in \cV\setminus Z$ and every residual clause $\wtilde c \in \cC_{\wtilde{\Phi}}$, the variable $x$ occurs in $\wtilde c$ if and only if $x$ occurs in $\pi(\wtilde c)$ in the original formula.

Suppose, for a contradiction, that $(\cU,\cF)$ has VC-dimension larger than $d$. Then there is a set $D \subseteq \cC_{\wtilde{\Phi}}$ with $|D|>d$ that is shattered by the variables $\cV\setminus Z$ in the reduced set system. Let $\widehat D=\pi(D)$. Since clauses are kept with their original labels, $|\widehat D|=|D|$.

For every subset $D_0 \subseteq D$, shattering in the reduced instance gives a variable $x \in \cV\setminus Z$ whose incidence with $D$ is exactly $D_0$. By the incidence-preservation observation above, the same variable $x$ has incidence exactly $\pi(D_0)$ with $\widehat D$ in the original formula. Hence $\widehat D$ is shattered in $\Phi$, contradicting that $\Phi$ has VC-dimension $d$.
\end{proof}

Equipped with \Cref{thm: wtred_sat_to_cov}, we now present our \epas result for \cmaxsat for set systems of bounded VC-dimension. We use the \epas by~\cite{DBLP:conf/compgeom/BadanidiyuruKL12} stated in the proposition below as a subroutine.

\begin{proposition}[Section~$3$,~\cite{DBLP:conf/compgeom/BadanidiyuruKL12}, in the trace-enumerated form of \Cref{thm:bkl-trace-enumerated-runtime}]
\label{prop: bkl_satvcdim}
For any $\varepsilon > 0$, there is a deterministic algorithm that takes as input an instance $\cI = ((\cU, \cF), k)$ of \maxcov, and in time 
\[2^{\widetilde{\mathcal{O}}(kd/\varepsilon)}\,
N^{\mathcal{O}(1)}
\]
returns a subfamily of size at most $k$ such that the total number of elements covered by it is at least $(1-\varepsilon)\optc(\cI)$. Here, $d$ is the VC-dimension of the input set system $(\cU, \cF)$, and \( \optc(\mathcal{I}) \) denotes the maximum number of elements that can be covered by any subfamily of size at most \( k \).
\end{proposition}

\vcdimsatepas*
\begin{proof}
For an $\varepsilon > 0$, and an instance $\cI = (\Phi, k)$ of \cmaxsat with bounded VC-dimension $d$, we first run \Cref{alg:framework-wsat}\footnote[1]{We can use \Cref{alg:framework-wsat} on unweighted instances by setting the weight of every clause $c \in \cC$ to $1$.} to obtain a family of $\left(\frac{k \log k}{\varepsilon}\right)^{\mathcal{O}(k)}$ \maxcov instances. Each reduced \maxcov instance also  has VC-dimension no more than $d$ by~\Cref{lem: vcdim_reduced_bound}. By~\Cref{prop: bkl_satvcdim}, with $\wtilde{\varepsilon}$ set to $\frac{\varepsilon}{2}$, in time 
\[2^{\widetilde{\mathcal{O}}(kd/\varepsilon)}\,
N^{\mathcal{O}(1)},
\]
we get a $(1 - \wtilde{\varepsilon})$-approximate solution.
By \Cref{thm: wtred_sat_to_cov}, we get a $(1 - \varepsilon)$-approximate solution to $\cI$.
\end{proof}

%% file: applications.tex
\section{Applications}
\label{sec:applications}
Several well-studied coverage problems can be modeled as instances of
\wmrbdsfull{} with bounded \dintercomp{}, and hence admit an \epas{}. We first
apply our results to \prds. We then study an abstract partial-covering problem
on the structured set systems of Langerman and Morin~\cite{DBLP:journals/dcg/LangermanM05},
obtaining applications to points covered by hyperplanes, spheres, and
polynomial curves. Finally, we consider covering bounded-size sets by
bounded-size sets.

\input{partial_dominating_set.tex}

\subsection{An abstract partial-covering problem}
In this section, we introduce an abstract partial covering problem, inspired by the $\textsc{Dim-Set-Cover}$ problem defined in \cite{DBLP:journals/dcg/LangermanM05}. Let $U$ be a (possibly infinite) universe, and let $S \subseteq U$ be a finite ground set of size $n < \infty$. The universe $U$ is associated with a collection $R$ of its subsets, which together cover $U$. Given a set system $(U, S, R)$ that satisfies certain structural properties, a weight function $\w:\cU \to \mathbb{Q}^+$ and an integer $k$, the goal of the partial covering problem is to select $k$ sets $r_1, r_2, \dots, r_k \in R$ such that the total weight of elements of $S$ covered by their union, i.e., $\w(S \cap (\cup_{i=1}^{k} r_i))$, is maximized.

\paragraph{Properties of Set System $(U,S,R)$:}
The set family $R$ is partitioned into $d$ subsets: $R_0, \dots, R_{d-1}$. For convenience, we define $R_{-1} = \emptyset$ and $R_d = \{U\}$. Given a set $A \subseteq U$, we define the \emph{dimension} of $A$ as
\[
\dimension(A) = \min \left\{ i \colon \exists\, r \in R_i \text{ such that } A \subseteq r \right\},
\]
and the \emph{cover} of $A$ as
\[
\cover(A) = \left\{ r \in R_{\dimension(A)} \colon A \subseteq r \right\}.
\]
Since $R_d = \{U\}$, it follows that both $\dimension(A)$ and $\cover(A)$ are well-defined, and that $\cover(A)$ is non-empty for every $A \subseteq U$.

We further assume that the partition $R_0, \dots, R_{d-1}$ satisfies the following property:

\begin{property}[Intersection Reduces Dimension \cite{DBLP:journals/dcg/LangermanM05}]\label{property:one}
For any $r_1 \in R_i$ and $r_2 \in R_j$, $0\leq i,j \leq d$ such that $r_1 \nsubseteq r_2$ and $r_2 \nsubseteq r_1$, $\dimension(r_1 \cap r_2)<\min\{i,j\}$

\end{property}

Now we state some results related to the structure of the set system. 

\begin{proposition}[\cite{DBLP:journals/dcg/LangermanM05}]
    For any $A\subset U$ and any finite $S \subseteq U$ there is a set, $\set(A) \in \cover(A)$ such that $r\cap
    S \subseteq \set(A)\cap S$ for all $r \in \cover(A)$
\end{proposition}
\begin{proposition}[\cite{DBLP:journals/dcg/LangermanM05}]\label{prop:dimincr}
    For any $A \subseteq U$ and $p \in U\setminus \set(A)$, $\dimension( A \cup \{p\})>\dimension(A)$
\end{proposition}
\begin{proposition}[\cite{DBLP:journals/dcg/LangermanM05}]
    For any $r \in R$ such that $r=\set(A)$ for some $A \subseteq U$, there exist a basis $A' \subseteq A$ such that $r=\set(A')$ and $|A'|\leq \dimension(A)+1$
\end{proposition}

The sets $U$, $R$, and the partitions $R_0, \dots, R_{d-1}$ may be infinite and are typically represented implicitly. Any algorithm for \pardimsc{} accesses these sets through two operations:

\begin{enumerate}
    \item Given a set $r \in R_i$ and an element $p \in S$, the algorithm can query whether $p \in r$. Such a query takes $\mathcal{O}'(1)$ time.
    \item Given a set $A \subseteq U$, the algorithm can compute $\set(A)$ and $\dimension(A)$ in $\mathcal{O}'(|A|)$ time.
\end{enumerate}

It follows that the only sets in $R$ accessible to the algorithm are those of the form $r = \set(S')$ for some $S' \subseteq S$. We refer to such sets as \emph{accessible sets}. Hence, the goal of the partial covering problem is to select $k$ accessible sets $r_1, r_2, \dots, r_k \in R$ such that the total weight of elements of $S$ covered by their union, i.e., $\w(S \cap (\cup_{i=1}^{k} r_i))$, is maximized.
Here and throughout, we use the notation $\mathcal{O}'(\cdot)$ to denote asymptotic bounds that hide factors polynomial in $d$. That is, $\mathcal{O}'(f(n))$ denotes $\mathcal{O}(f(n) \cdot \text{poly}(d))$, where $d$ is the dimension parameter associated with the set system.

\defparopt{\pardimsc}{A set system $(U,S,R)$, where $R$ consists of only accessible sets, a partition $R_0, \dots, R_{d-1}$ that satisfies \Cref{property:one}, a weight function $\w: S \to \bQ^+$ and an integer $k$}{$k$}{ A set of $k$ sets $r_1, r_2, \dots, r_k \in R$ such that the total weight of elements of $S$ covered by their union, i.e., $\w(S \cap (\cup_{i=1}^{k} r_i))$, is maximized}

\paragraph{Associated Bipartite Graph:}  
Given a set system \( (U, S, R) \), weight function $\w$ and the partition \( \{R_0, \dots, R_{d-1}\} \), we construct the following incidence bipartite graph \( \graph \) with a weight function $ \w' : \cB \to \mathbb{Q}^+$ 
\begin{itemize}
  \item For every non-empty subset \( A \subseteq S \) of size at most \( d \), add a vertex \( a_{\set(A)} \) to \( \cR \), corresponding to \( \set(A) \in R \).
  \item For each \( u \in S \), add a vertex \( b_u \) to \( \cB \). Also, assign \( \w'(b_u) = \w(u) \).
  \item Add an edge between \( a_{\set(A)} \) and \( b_u \) if and only if \( u \in \set(A) \).
\end{itemize}

\noindent
Note that the size of the graph constructed is $n^{\cO(d)}$. Now, finding \( k \) accessible sets \( r_1, r_2, \dots, r_k \in R \) such that \( \w'(S \cap (\cup_{i=1}^{k} r_i)) \) is maximized reduces to finding a set of \( k \) vertices \( X \subseteq \cR \) such that \( \nbr(X) \) is maximized. In particular, solving the \wmrbds{} problem for the instance \( I = (\graph, \w', k) \) is sufficient. Now, all we need to show is that the \dintercomp{} of the incidence graph is bounded.
\paragraph{Bounded $\dintcmp{}$ of the Incidence Graph:}
\begin{lemma}
 \label{lem: contbipgraph}
  Given $(U,S,R)$, let \( \graph \) be the incidence graph. Then, we have  $\dintcmp(G) \leq d+3$.
\end{lemma}
 
\begin{proof}
    Let $d'$ denote the \smlindex of $\graph$. By definition we have two sequences, $a_1,\dots,a_{d'} \in \cR$ and $b_1,\dots,b_{d'} \in \cB$ that form a \sml of order $d'$ in $G$, that is, $(a_i,b_j) \in E(G)$ for all $i,j \in [d']$ with $i>j$, and $(a_i,b_i)\notin E(G)$ for all $i \in [d']$. For convenience, we will abuse notation and, for \( i \in [d'] \), denote by \( a_i \) both the vertex in \( \cR \) and the corresponding set in $R$. Similarly, we will use \( b_i \) to denote both the vertex in \( \cB \) and the corresponding element in \( S \).
Suppose we show that for $i \in [d']$, $\dimension(a_i)\geq i-2$ by applying induction on $i$. Then we have $\dimension(a_{d'})\geq d'-2$. Given that the dimension of any \( r \in R \) is bounded by \( d \), we obtain \( d' \leq d+2 \). Thus, the \smlindex of the incidence graph is bounded by \( d+2 \), which further implies  
\[
\dintcmp(G) \leq d+3
\]  
due to \Cref{lemma:boundingsmlindex}.

Now we will show that for $i \in [d']$, $\dimension(a_i)\geq i-2$ by applying induction on $i$. Since $\dimension(a_i)\geq \dimension(\{b_1,\dots b_{i-1}\})$, it is sufficient to show  $\dimension(\{b_1,\dots b_{i-1}\})\geq i-2$. By our assumption we already have $\dimension(a_1) \geq -1$, since $R_{-1}=\emptyset$ \\

  \noindent \textbf{Induction Hypothesis:} For all $i \in [d']$, $\dimension(\{b_1,\dots b_i\})\geq i-1$.
  
  \smallskip
  \noindent \textbf{Base Case:} $\dimension(\{b_1\})\geq 0$ holds since $\{b_1\}$ is non-empty and hence base case is true.
  
  \smallskip
  \noindent \textbf{Induction Assumption:} Let $i' \leq d'$ be an integer. For all $i< i'$ we have $\dimension(\{b_1,\dots b_i\})\geq i-1$.
  
  \smallskip
  \noindent \textbf{Inductive Step:} We will show that \( \dimension(\{b_1, \dots, b_{i'}\}) \geq i'-1 \). Since \( (a_{i'}, b_{i'}) \notin E(G) \), it follows that \( b_{i'} \notin a_{i'} \). Furthermore, since \( a_{i'} \supseteq \set(\{b_1, \dots, b_{i'-1}\}) \), we also have \( b_{i'} \notin \set(\{b_1, \dots, b_{i'-1}\}) \).  

Thus, by \Cref{prop:dimincr}, we obtain  
\[
\dimension(\set(\{b_1, \dots, b_{i'}\})) \geq \dimension(\set(\{b_1, \dots, b_{i'-1}\})) + 1 \geq i'-1.
\]  
The last inequality follows from the induction hypothesis, which states that \( \dimension(\{b_1, \dots, b_{i'-1}\}) \geq i' - 2 \). Thus by principle of mathematical induction we have for all $i \in [d']$, $\dimension(\{b_1,\dots b_i\})\geq i-1$. This completes the proof of the lemma. 
\end{proof}

\paragraph{Results:} Below, we state the results regarding \pardimsc we obtain after reducing to an instance of \wmrbds{}. Note that, for instance $\cI$ of \pardimsc we define \[  \wopt_k(\cI) =\max_{\{r_1,\dots r_k\}\subseteq R}{\w(S \cap (\cup_{i=1}^{k} r_i))} \]
\begin{theorem}
    Given an instance $\cI = (U,S,R,\w,k)$ of \pardimsc, a partition $
    R_0,\dots R_{d-1}$ of $R$ and an $\varepsilon>0$, there exists a randomized algorithm that runs in time \[\left(\frac{k}{\varepsilon} \right)^{\cO(dk)}\cdot n^{\mathcal{O}(d)} = \left( \frac{1}{\varepsilon}\right )^{\cO(kd \log k)} n^{\mathcal{O}(d)}\] and with probability at least $ 1- \frac{1}{e}$, outputs a subset $ \{r_1,\dots,r_k\} \subseteq R$ such that  \[\w(S \cap (\cup_{i=1}^{k} r_i)) \geq (1 - \varepsilon) \wopt_k(\cI),\]
\end{theorem}

Let \decpardimsc be the decision version of \pardimsc, where, along with the input, we are also given an integer \( W \). The goal is to find a $ k$-sized subset, say $\{r_1,\dots r_k\}$ of $R$, such that \(  \w(S \cap (\cup_{i=1}^{k} r_i)) \geq W \). By our previous arguments we obtain the following result. 
\begin{theorem}
    Given an instance $\cI = (U,S,R,\w,k,W)$ of \decpardimsc, a partition $
    R_0,\dots R_{d-1}$ of $R$ and an $\varepsilon>0$  there exists a randomized algorithm with running time 
\[
k^{\mathcal{O}(dk)} \cdot n^{\mathcal{O}(d)} =  2^{\cO(dk \log k)} n^{\mathcal{O}(d)}
\]
which, in the case of a \yes-instance, returns  \( \{r_1,\dots,r_{k+1}\} \subseteq R \)  such that 
\[
\w(S \cap (\cup_{i=1}^{k+1} r_i)) \geq W
\]
with probability at least \( 1 - \frac{1}{e} \). However, in the case of a \no-instance, the algorithm either correctly reports that the instance is a \no-instance, or returns  \( \{r_1,\dots,r_{k+1}\} \subseteq R \)  such that 
\[
\w(S \cap (\cup_{i=1}^{k+1} r_i)) \geq W
\]
\end{theorem}

We can derive a deterministic version of the above results using derandomization techniques and the results for \wmrbds in~\Cref{section: weighteddetfptas}.
\subsection{Further geometric applications}
Langerman and Morin~\cite{DBLP:journals/dcg/LangermanM05} show that several
covering problems are defined on set systems $(U,S,R)$ satisfying
\Cref{property:one}. Their partial versions can similarly be formulated as
instances of \pardimsc. Our algorithms also handle the weighted variants, in
which elements of $S$ carry weights.

\paragraph{Covering points with hyperplanes and vice versa.}
The \plc problem is a fundamental topic in geometry and has been extensively studied in the parameterized setting. The input consists of \( n \) points in the plane, and the objective is to cover these points using \( k \) lines. We focus on a partial variant of the problem, specifically the \pplc problem, where the goal is to cover the maximum number of points using \( k \) lines. To the best of our knowledge, the \pplc problem has not previously been shown to be \whard in the literature. We address this gap in~\Cref{section: pplc}, where we prove that \pplc is indeed \whard when parameterized by the solution size \( k \). 

We consider the more general version of the problem. Given a set $S$ of $n$ points in $\mathbb{R}^d$, where each point has a weight. The objective is to select a set of at most $k$ hyperplanes that maximize the total weight of points covered by these hyperplanes, that is, points that lie on at least one of the selected hyperplanes.

We model this as an instance of \pardimsc\ by setting \( U = \mathbb{R}^d \), and for each \( 0 \leq i \leq d - 1 \), the range set \( R_i \) consists of all \( i \)-flats embedded in \( \mathbb{R}^d \). This system satisfies Property 1 since the intersection of an \( i \)-flat and a \( j \)-flat, provided neither contains the other, is an \( \ell \)-flat for some \( \ell < \min\{i, j\} \). Therefore, this corresponds to an instance of \pardimsc where the parameter \( d \) matches the dimension of the Euclidean space.

In the dual setting, we are given a set \( S \) of \( n \) hyperplanes in \( \mathbb{R}^d \) and asked whether there exists a set of \( k \) points that maximizes the weight of hyperplanes in \( S \) contains at least one of the points. By standard point–hyperplane duality, this problem is equivalent to the original and can be solved with the same efficiency.

\paragraph{Covering points with spheres.}
In the partial covering version of the problem, we are given a set $S$ of $n$ points in $\mathbb{R}^d$ where each point has a weight, and the objective is to determine whether there exists a set of $k$ hyperspheres that maximizes the total weight of points from $S$ that lie on the surface of at least one of the hyperspheres.

To model this as an instance of \pardimsc, we set $U = \mathbb{R}^d$. For each $0 \leq i \leq d$, the range set $R_i$ consists of all $i$-spheres: $R_0$ consists of points, $R_1$ consists of pairs of points, $R_2$ consists of circles, and so on. Furthermore, the intersection of an $i$-sphere and a $j$-sphere (neither of which contains the other) is an $\ell$-sphere for some $\ell < \min\{i, j\}$, thereby satisfying \Cref{property:one}. This constructs an instance of \pardimsc whose dimension is one greater than that of the ambient Euclidean space.

\paragraph{Covering points with polynomials.}
Given a set $S = \{(x_1, y_1), \dots, (x_n, y_n)\}$ of $n$ points in $\mathbb{R}^2$, where each point has a weight, the objective in the partial covering version is to determine whether there exist $k$ polynomial functions $f_1, \dots, f_k$, each of degree at most $d$, that maximize the total weight of points from $S$ that lie on the graph of at least one of these polynomials. In other words, we aim to maximize the weight of points in $S$ covered by at least one polynomial in the set. This problem is a generalization of \pplc.

We say that two points in $\mathbb{R}^2$ are \textbf{x-distinct} if they have different $x$-coordinates. For this problem, $R_0$ denotes the set of all points, $R_1$ denotes the set of all pairs of x-distinct points, $R_2$ denotes the set of all triples of x-distinct points, and so on, up to $R_{d+1}$, which denotes the set of all degree $d$ polynomials. Because there exists a degree $d$ polynomial that can pass through any set of $d+1$ or fewer x-distinct points, each set in $R$ corresponds to points that can be covered by a degree $d$ polynomial (with coefficients possibly set to zero).

Furthermore, the intersection of a set of $i$ points and a (possibly infinite) set of $j \ge i$ points, neither of which contains the other, results in a set of at most $\ell < \min\{i, j\}$ points. The intersection of two non-identical degree $d$ polynomials comprises at most $d$ points. Thus, the above system of sets satisfies Property 1 and can be solved using our algorithms.

\input{point_line_cover_hardness}

\subsection{Covering small sets by small sets}
A set system \( (\mathcal{U}, \mathcal{F}) \) is defined as follows:  
\[
\mathcal{U} \coloneqq \{u_1, u_2, \dots, u_n\}, \quad \mathcal{F} \subseteq 2^{\mathcal{U}}, \quad \text{where } |\mathcal{F}| = m.
\]
Here, \( \mathcal{U} \) is a universe of \( n \) elements, and \( \mathcal{F} \) is a collection of \( m \) subsets of \( \mathcal{U} \). In this section, we consider set systems $(\cU,\cF)$ where the cardinality of sets in \( \cF \) is bounded by \( d \). We study the problem where the objective is to choose \( k \)  sets from \( \cF \) that maximizes the number of sets in \( \cF \) that are intersected by their union. 
 We begin by introducing a few key definitions that will be used throughout this section.

\paragraph{Definitions:}
\begin{description}
\setlength{\itemsep}{-2pt}
     \item[1.] For ${\cS} \subseteq \cF$, we define $\cov(\cS) = \cup_{f\in {\cS}} f$ to be the union of all the elements that appear in at least one set of $\cS$. When the family is a singleton set, say $f$, then we simply use $\cov(f)$ instead of $\cov(\{f\})$.
    \item[2.] For $S \subseteq \cU$, we define $$\hit_{\cF}(S) \coloneqq \{f \mid f \in \cF, f \cap S\neq \emptyset \}.$$
    In other words, $\hit_{\cF}(S)$ contains all the sets in $\cF$ that intersects with $S$.

    \item[3.]  Similarly, for $\cS \subseteq \cF$, we define $$\hit_{\cF}(\cS) \coloneqq \{f \mid f \in \cF \textnormal{~and~} \exists~ S \in \cS \textnormal{~such that~} f ~\cap~ S \neq \emptyset \} =  \{f \mid f \in \cF, f ~\cap ~\cov(\cS) \neq \emptyset \}.$$ 
    That is,  $\hit_{\cF}(\cS)$ includes all the sets in $\cF$ that intersect with at least one set in $\cS$.

 When the family $\cF$ is clear from the context we omit the subscript and simply write $\hit(S)$ and $\hit(\cS)$.

\end{description}
\paragraph{Problem.}
First, we define our problem formally as follows
\defparopt{\wphedsfull(\wpheds)}{A set system $(\cU,\cF)$, where $|f|\leq d$ for every $f \in \cF$, $\w:\cF \to \bQ^+$ and an integer $k$.}
{$k$}
{A set $S \subseteq \cF$ of size at most $k$ such that $\w(\hit(S))$ is maximum.}
Given an instance $ \mathcal{I} = (\mathcal{U}, \mathcal{F},\w, k) $, we define
\[
\wopt_k(\mathcal{I}) = \max_{\mathcal{F}' \subseteq \mathcal{F}, \ |\mathcal{F}'| \leq k} \w( \hit(\mathcal{F}')),
\]
\paragraph{Modeling as an instance of \wmrbds.}

Given any set system, $(\cU,\cF)$, we construct \intbip graph of $\cF$, say $\graph$, as follows 
\begin{itemize}
    \item For every $f\in \cF$ add a vertex $r_f$ to $\cR$ and a vertex $b_f$ to $\cB$.
    \item For $f,f' \in \cF$, we add $(r_f,b_{f'})$ and $(r_{f'},b_f)$ to $E(G)$ if and only if $f \cap f'\neq \emptyset$
\end{itemize}

The objective is to find a subset \( S \subseteq \cF \) of size at most \( k \) such that \( \w(\hit(S)) \) is maximized. This is equivalent to selecting a subset \( \cR' \subseteq \cR \) of \( k \) vertices in \( \graph \) such that \( \w(\nbr(\cR')) \) is maximized. In other words, we need to solve the \wmrbdsfull problem, where the input is the \intbip graph of \( (\cU, \cF) \). The parameter \( k \) and weight function \( \w \) remain unchanged. The weight of a vertex \( b_f \in \cB \) is defined as the weight of its corresponding set \( f \in \cF \).

\paragraph{Input size.}
We write $\sizesetsys$ for the encoding size of $(\cU,\cF,\w)$. The explicit
intersection bipartite graph has $2|\cF|$ vertices and at most $2|\cF|^2$
edges, and can be constructed in time polynomial in $\sizesetsys$.
\paragraph{Bounding Downward Intersection Complexity.}
Our objective now is to show that the intersection bipartite graph of a set system, where each set has a bounded size, has bounded \dintercomp. To this end, we want to use the ``skewed version'' of Bollobás’s theorem as stated below from the book~\cite{DBLP:series/txtcs/Jukna11}, originally proved by Frankl (1982)~\cite{DBLP:journals/ejc/Frankl82} and later proved in an equivalent form by Kalai (1984)~\cite{kalai1984intersection}.
\begin{proposition} \label{prop:bollobas}
	Let $A_1, \dots, A_m$ and $B_1, \dots, B_m$ be finite sets such that $A_i \cap B_i = \emptyset$ and $A_i \cap B_j  \neq \emptyset$ if $i<j$. Also suppose that $|A_i| \leq a$ and $|B_i| \leq b$. Then $m \leq  \binom{a+b}{b}$
\end{proposition}
Having equipped ourselves with the necessary tools we state the main lemma of this section. 
\begin{lemma}\label{lemma:boundusingbollobas}
    Given a set system \( (\mathcal{U}, \mathcal{F}) \) such that \( |f| \leq d \) for every \( f \in \mathcal{F} \), let $\graph$ be the \intbip graph of \( \mathcal{F} \). Then we have $\dintcmp(G)\leq \binom{2d}{d}+1$.
\end{lemma}
\begin{proof}
    Let $D=  \binom{2d}{d}+1$. Suppose for a contradiction $\dintcmp(G)\geq \binom{2d}{d}+2$ i.e, $\dintcmp(G)\geq D+1$ . Then due to~\Cref{defn:dic} there exists $X \subseteq \cB$ such that $\dintcmp(G,X)\geq D+1$ which implies that \smlindex of $G$ is at least $D$ due to ~\Cref{lemma:boundingsmlindex}. Let $r_1,\dots,r_{D} \in \cR$ and $b_1,\dots,b_{D} \in \cB$ be the two sequences that form a \sml of order $D$ in $G$. For ease of notation, let $r_i$ and $b_i$ for $i \in [D]$ also denote the corresponding sets in $\cF$. Due to~\Cref{defn:semiladder} we know that $(r_i,b_i) \notin E(G)$ and for all $i, j \in [D]$ with $i>j$ we have $(r_i,b_j) \in E(G)$. By the construction of \intbip graph this implies that $r_i \cap b_i =\emptyset$ for all $i \in [D]$ and for all $i, j \in [D]$ with $i>j$ we have $r_i \cap b_j \neq \emptyset$. Since $|f|\leq d$ for every $f \in \cF$,~\Cref{prop:bollobas} implies that we have $D\leq \binom{2d}{d}$ which is a contradiction since $D= \binom{2d}{d}+1$. Thus we have $\dintcmp(G)\leq \binom{2d}{d}+1$.
\end{proof}
Now we will show that the bound of \dintercomp shown in~\Cref{lemma:boundusingbollobas} is essentially tight. For this we give an explicit construction of a set system whose \intbip graph, say $\graph$ has $\dintcmp(G)=\binom{2d}{d}$.\\

\begin{lemma}
    Consider the set system $(\cU, \cF)$ where $\cU=\{1,\dots,2d\}$ and $\cF=\{f \subseteq \cU \mid |f| = d\}$. The \intbip graph of $\cF$, say $\graph$, of the set system has $\dintcmp(G)\geq \binom{2d}{d}$
\end{lemma}
\begin{proof}
Note that $|\cF|=\binom{2d}{d}$. For every $f\in\cF$, the complement
$f'=\cU\setminus f$ is the unique member of $\cF$ disjoint from $f$; every
$\widetilde f\in\cF\setminus\{f'\}$ intersects $f$. Now consider the \intbip
graph $\graph$ of $(\cU,\cF)$. Let
$r_1,\dots,r_{\binom{2d}{d}}\in\cR$ and
$b_1,\dots,b_{\binom{2d}{d}}\in\cB$ be ordered so that
$b_i=\cU\setminus r_i$. We use $r_i,b_i$ for both vertices and their
corresponding sets. Then $(r_i,b_i)\notin E(G)$ for every $i$, whereas
$(r_i,b_j)\in E(G)$ whenever $i\ne j$. Let
$X_i=\cB\setminus\{b_1,\dots,b_i\}$. Since $b_i \in \nbr(r_j)$ for all $j \neq i$ and $b_i \notin 
\nbr(r_i)$ we have $\incld(G,X_i)=\{r_1,\dots r_i\}$ and $X_i=\cap_{\iletter \in [i]}\nbr(r_\iletter)=\reex(G,X_i)$. Thus we have $X_i$ is \realizable for all $i \in [\binom{2d}{d}]$.
Hence, we get a sequence of \realizable sets,
\[
X_1\supsetneq \dots \supsetneq X_{\binom{2d}{d}} 
\]
which implies $\dintcmp(G) \geq \dintcmp(G,X_1)\geq \binom{2d}{d}$.
\end{proof}
\paragraph{Results.}
By \Cref{lemma:boundusingbollobas}, the \dintercomp of $G$ is at most
$\binom{2d}{d}+1$. We therefore obtain the following results.
\begin{theorem}
Let $\cI = (\cU,\cF, \w, k)$ be an instance of \wpheds and an $\varepsilon>0$. There exists a randomized algorithm that runs in time

    \[\left(\frac{k}{\varepsilon} \right)^{\cO \left (\binom{2d}{d}k \right )}\cdot \sizesetsys^{\cO(1)} = \left( \frac{1}{\varepsilon} \right)^{\cO \left( \binom{2d}{d} k \log k\right)} \sizesetsys^{\cO(1)}\] and with probability at least $ 1- \frac{1}{e}$, outputs a subfamily $\cS \subseteq \cF$ of size $k$ such that  \[\w(\hit(\cS)) \geq (1 - \varepsilon) \wopt_k(\cI).\]

Here  $d$ is the cardinality bound for sets in $\cF$.

\end{theorem}
Let \decwphedsfull(\decwpheds) be the decision version of \wpheds, where, along with the input, we are also given an integer \( W \). The goal is to find a set \( S \subseteq \cF \) of size at most \( k \) such that \( \w(\hit(S)) \geq W \). 

By our previous arguments, we obtain the following result.

\begin{theorem}
Given an instance $\cI=(\cU,\cF,\w,k,W)$ of \decwpheds, there is a
randomized algorithm running in time
\[
k^{\cO\left(\binom{2d}{d}k\right)}\sizesetsys^{\cO(1)}
=2^{\cO\left(\binom{2d}{d}k\log k\right)}\sizesetsys^{\cO(1)}.
\]
On a \yes-instance, with probability at least $1-1/e$ it outputs a subfamily
$\cS\subseteq\cF$ of size at most $k+1$ satisfying
$\w(\hit(\cS))\ge W$. On a \no-instance, it either correctly returns \no or
outputs such a subfamily. Here $d$ bounds the cardinality of every set in
$\cF$.
\end{theorem}
We can derive a deterministic version of the above results using derandomization techniques and the results for \wmrbds in~\Cref{section: weighteddetfptas}.

%% file: partial_dominating_set.tex
\subsection{\texorpdfstring{Partial $r$-dominating set}{Partial r-dominating set}}
In this section, we present EPASes for \prds on several graph classes. To this end, we first introduce the following quantity: the maximum number of vertices that can be covered by the $r$-neighborhood of any subset of size at most $k$ in the graph.

\[
\opt_k^r(G) = \max_{\substack{S \subseteq V(G) \\ |S| \leq k}} |\nbr_G^r(S)|.
\]

We now formally define the problem studied in this section.
\defparopt{\prds}{A graph $G$ on $n$ vertices and $m$ edges and an integer $k$.}
{$k$}
{A subset $D\subseteq V(G)$ of size at most $k$ such that $|\nbr_G^r(D)|=\opt_k^r(G)$.}

Golovach and Villanger~\cite{DBLP:conf/wg/GolovachV08} showed that 
\prds is W[1]-hard on $2$-degenerate graphs even for $r=1$ (called {\sc Partial Dominating Set}). \prds can be modeled as \prbds as follows. We construct a bipartite graph $H$ as follows.
We take both $\cR$ and $\cB$ to be copies of $V(G)$, and join
$u\in\cR$ to $v\in\cB$ if and only if $\dist_G(u,v)\le r$, where
$\dist_G$ denotes shortest-path distance in $G$. A set $D\subseteq V(G)$ of
size at most $k$ satisfies $|\nbr_G^r(D)|=\opt_k^r(G)$ if and only if the
corresponding set $\cD\subseteq\cR$ satisfies
$|\nbr_H(\cD)|=\opt_k^r(G)$. Thus, \Cref{thm:mainUnweighted} and the
bounded-semi-ladder classes discussed in the introduction give the following.

\begin{theorem}
Let \( s \in \mathbb{N} \), and let \( \mathscr{M} \) denote the class of map graphs and \( \mathscr{D} \) be a nowhere-dense class. Then, \prds admits the following algorithms on \( \mathscr{D}^s \) and $\mathscr{M}$:
\begin{description}
    \item[1.] There is an \epas that, given an \( \varepsilon > 0 \), outputs a set \( D \subseteq V(G) \) of size at most \( k \) such that \( |\nbr_G^r(D)| \geq (1 - \varepsilon) \opt_k^r(G) \).
    \item[2.] There is a parameterized algorithm (in fact, a \pas) that outputs a set \( D \subseteq V(G) \) of size at most \( k + 1 \) such that \( |\nbr_G^r(D)| \geq \opt_k^r(G) \).
\end{description}
\end{theorem}

%% file: point_line_cover_hardness.tex
\subsubsection{\texorpdfstring{\pplc\ is \whard\ parameterized by $k$}{\pplc\ is \whard\ parameterized by k}}
\label{section: pplc}

In this section, we show that the \pplc problem is \whard parameterized by the solution size $k$. A formal definition of the problem is as follows:

\defparopt{\pplc}
{A set $\cP$ of $n$ points in $\mathbb{R}^{2}$ and integers $k$ and $t$.}
{$k$}
{A set of at most $k$ lines in $\mathbb{R}^2$ that cover at least $t$ points of $\cP$.}

We present a reduction from \pvc, which is \whard parameterized by $k$~\cite{DBLP:journals/mst/GuoNW07}. The \pvc problem takes as input a graph $G$ along with integers $k$ and $t$, and the objective is to determine whether there exists a set of $k$ vertices that covers at least $t$ edges in $G$.

The reduction consists of three phases:
\begin{enumerate}
    \item First, we reduce \pvc in general graphs to \pvc in graphs that satisfy a specific requisite property.
    \item Next, we reduce \pvc in graphs with the requisite property to \plpc. 
    
    In the \plpc problem, we are given a set $\ccL$ of $n$ lines in $\mathbb{R}^2$ along with integers $k$ and $t$. The objective is to determine whether there exists a set of at most $k$ points in $\mathbb{R}^2$ that collectively intersect at least $t$ lines from $\ccL$.
    \item Finally, we reduce an instance of \plpc to an equivalent instance of \pplc.
\end{enumerate}

\paragraph{First Phase of the Reduction.}
Let $\cI = (G, k, t)$ be an instance of \pvc. We construct a new instance $\cI' = (G', k, t + 3k)$ as follows.

We modify the graph $G$ to obtain $G'$ by adding pendant vertices to each vertex in $G$:

\begin{itemize}
    \item For each vertex $v \in V(G)$, introduce three new pendant vertices and connect them to $v$.
    \item Formally, define:
    \[
    V(G') = V(G) \cup \{v_i \mid v \in V(G), \ i \in [3]\}
    \]
    \[
    E(G') = E(G) \cup \{(v, v_i) \mid v \in V(G), \ i \in [3]\}
    \]
\end{itemize}

In the resulting graph $G'$, if the original graph had $n$ vertices, then $G'$ contains $4n$ vertices: the $n$ original vertices and $3n$ additional pendant vertices. Among these, $n$ vertices have degree at least $3$, while the remaining $3n$ vertices are pendant.

By Claim~\ref{clm: pvctopvc}, this transformation preserves equivalence, meaning that the instance $(G', k, t + 3k)$ is a \yes instance of \pvc if and only if $(G, k, t)$ is an \yes instance for the same.

\begin{claim}\label{clm: pvctopvc}
     $\cI = (G, k, t)$ is an \yes instance of \pvc if and only if $\cI' = (G', k, t + 3k)$ is an \yes instance for the same.
\end{claim}
\begin{proof}
For the forward direction, assume that $\cI = (G, k, t)$ is a \yes instance of \pvc. This means there exists a set $S \subseteq V(G)$ of size at most $k$ that covers at least $t$ edges in $G$. If $S$ is strictly less than $k$ we can add arbitrary vertices to make $|S|=k$ without reducing the number of edges covered.

In the modified graph $G'$, each vertex $s \in S$ covers three additional edges, namely $(s, s_i)$ for $i \in [3]$. By construction, these edges are not covered by any other vertex of $S$. Consequently, the set $S$ in $G'$ covers at least $t + 3k$ edges. Therefore, $\cI' = (G', k, t + 3k)$ is also a \yes instance of \pvc.

For the reverse direction, assume that $\cI' = (G', k, t + 3k)$ is a \yes instance of \pvc. This means there exists a set $S' \subseteq V(G')$ of size at most $k$ that covers at least $t + 3k$ edges in $G'$.  

By construction, the vertex set of $G'$ can be partitioned as $V(G') = V_1 \uplus V_2$, where:
\begin{itemize}
    \item $V_1$ consists of the original (non-pendant) vertices from $V(G)$.
    \item $V_2$ consists of the newly added pendant vertices.
\end{itemize}  

If $S'$ contains any pendant vertex $s_1 \in V_2$, we replace it with its neighbor $s \in V_1$. There are two cases to consider:
\begin{itemize}
    \item If $s$ is already in $S'$, then $s_1$ was not covering any additional edges, so replacing it has no effect.
    \item If $s$ was not in $S'$, then replacing $s_1$ with $s$ ensures that the number of covered edges remains at least the same.
\end{itemize}  
Thus, we can assume that $S' \subseteq V_1$.  

In $G'$, each vertex in $S'$ covers exactly three additional edges due to the pendant vertices. Since $S'$ covers at least $t + 3k$ edges in $G'$, the number of edges it covers in $G$ is at least:
\[
t + 3k - 3k = t.
\]
Therefore, $\cI = (G, k, t)$ is also a \yes instance of \pvc.

\end{proof}
    
\paragraph{Second Phase of the Reduction.}
In the second phase of the reduction, we construct an instance $\cJ = (\cL, k, t+3k)$ of \plpc using the \pvc instance $\cI' = (G', k, t + 3k)$ obtained from the first phase.  

We define the point set $\cP$ as follows:
\begin{itemize}
    \item For each vertex $v \in V(G')$ with degree at least $3$, create a point $p_v$.
    \item For each pendant vertex $v_i \in V(G')$, where $v \in V(G)$ and $i \in [3]$, create a point $p_{v_i}$.
\end{itemize}  

Thus, we have $\cP = V_1 \uplus V_2$, where:
\begin{itemize}
    \item $V_1 = \{p_v \mid v \in V(G') \text{ and } \deg_{G'}(v) \geq 3\}$.
    \item $V_2 = \{p_{v_i} \mid v \in V(G), \ i \in [3], \text{ and } \deg_{G'}(v_i) = 1\}$.
\end{itemize}  

The points in $\cP$ are in general position satisfying the following geometric properties:
\begin{enumerate}
    \item No three points are collinear.
    \item No two lines passing through pairs of points in $\cP$ are parallel.
    \item No three lines passing through pairs of points in $\cP$ are concurrent at a common point that is not in $\cP$.
\end{enumerate}  

Next, we construct a set of lines $\ccL$ based on the edges of $G'$. Specifically, for each edge $(u,v) \in E(G')$, create a line $\ell_{uv}$ passing through $p_u$ and $p_v$:
\[
\ccL = \{\ell_{uv} = \overline{p_u p_v} \mid (u,v) \in E(G')\}.
\]  

By Claim~\ref{clm: pvctolpc}, the instance $(\cP, k, t+3k)$ is an equivalent instance of \plpc.
\begin{claim}\label{clm: pvctolpc}
     $\cI' = (G', k, t+3k)$ is an \yes instance of \pvc if and only if $\cJ = (\cL, k, t + 3k)$ is an \yes instance of \plpc.
\end{claim}
\begin{proof}

Suppose $\cI' = (G', k, t+3k)$ is a \yes instance of \pvc. This means there exists a set $S \subseteq V(G')$ of size at most $k$ that covers at least $t+3k$ edges in $G'$.  

By the construction of $\cJ$, for each edge $(u,v) \in E(G')$, we have a corresponding line $\ell_{uv}$ passing through the points $p_u$ and $p_v$. Since the vertices in $S$ cover at least $t+3k$ edges in $G'$, the corresponding lines in $\ccL$ are covered by at most $k$ points in the plane.  

Thus, the number of lines covered in $\cJ$ is at least $t+3k$, implying that $\cJ = (\cL, k, t+3k)$ is also a \yes instance of \plpc.

For the reverse direction, let $\cJ = (\cL, k, t + 3k)$ be a \yes instance of \plpc. That is, there exists a set $P$ of at most $k$ points that collectively cover at least $t+3k$ lines.

First, suppose $P$ contains a point from $V_2$, say $p_{s_1}$. By construction, we know that $p_{s_1}$ covers at most one line. Instead, we replace it with the corresponding point $p_s$ from $V_1$ that lies on the same line. Since $P \setminus \{p_{s_1}\} \cup \{p_s\}$ covers at least the same number of lines, we can assume without loss of generality that $P$ does not contain any points from $V_2$.

Next, suppose $P$ contains a point $v$ that is not in $\cP$. By construction, since the points in $\cP$ are in general position, the third property ensures that no three lines passing through pairs of points in $\cP$ are concurrent at a common point that is not in $\cP$. Consequently, $v$ can cover at most two lines, say $\ell_{ab}$ and $\ell_{cd}$. Let $p_a$ and $p_c$ be the corresponding points in $V_1$. If both $p_a$ and $p_c$ are already in $P$, then we can safely remove $v$. Otherwise, suppose $p_a \notin P$. Then, consider the modified set $P' = P \setminus \{v\} \cup \{p_a\}$. Since $P$ does not contain points from $V_2$, the inclusion of $p_a$ ensures that at least three additional lines are covered. Therefore, the total number of lines covered by $P'$ is at least as many as those covered by $P$, allowing us to assume that $P \subseteq V_1$.

Finally, consider the corresponding vertices in $G$. By construction, for each edge $(u,v) \in E(G')$, there exists a corresponding line $\ell_{uv}$ passing through the points $p_u$ and $p_v$. Since the points in $P$ cover at least $t + 3k$ lines in $\cJ$, the corresponding vertices in $G'$ must cover at least $t + 3k$ edges. Thus, $\cI' = (G', k, t+3k)$ is an \yes instance of \pvc.
\end{proof}
\paragraph{Third phase of the reduction.}
We reduce an instance $\cJ=(\ccL, k, t)$ of \plpc to an equivalent instance
of \pplc. Let $\ccL = \{\ell_1, \dots, \ell_m\}$ be the given set of lines,
and let $\cP_{\ccL} = \{p_1, \dots, p_q\}$, where $q = \binom{m}{2}$, be the
set of intersection points of all pairs of lines in $\ccL$.

We utilize the concept of \emph{point-line duality}~\cite{DBLP:books/lib/BergCKO08}, which states the following transformation rules:
\begin{itemize}
    \item The dual of a point $p = (p_x, p_y)$ is a line $\widetilde{\ell}$ given by the equation $y = p_x x - p_y$.
    \item The dual of a line $\ell$ with equation $y = mx + c$ is a point $\widetilde{p} = (m, -c)$.
    \item The duality transform preserves incidence; that is, a point $p$ lies on a line $\ell$ if and only if the dual point $\widetilde{p}$ lies on the dual line $\widetilde{\ell}$.
\end{itemize}

Using this duality transformation, we construct:
\begin{itemize}
    \item A set of points $\widetilde{\cP} = \{\widetilde{p}_1, \dots, \widetilde{p}_m\}$, where $\widetilde{p}_i$ is the dual of the line $\ell_i \in \ccL$, for each $i \in [m]$.
    \item A set of lines $\widetilde{\ccL} = \{\widetilde{\ell}_1, \dots, \widetilde{\ell}_q\}$, where $\widetilde{\ell}_j$ is the dual of the point $p_j \in \cP_{\ccL}$, for each $j \in [q]$.
\end{itemize}

By Claim~\ref{clm: lpctoplc}, the instance $\cM=(\widetilde{\cP}, k, t)$ is an equivalent instance of \pplc.

\begin{claim}\label{clm: lpctoplc}
    The instance $\cJ=(\ccL, k, t)$ is a \yes instance of \plpc if and only if the instance $\cM=(\widetilde{\cP}, k, t)$ is a \yes instance of \pplc.
\end{claim}

\begin{proof}
For the forward direction, assume that $\cJ=(\ccL, k, t)$ is a \yes instance of \plpc. This means there exists a set $S = \{p_{i_1}, \dots, p_{i_k}\} \subseteq \cP_{\ccL}$ of $k$ points that covers at least $t$ lines in $\ccL$. Consider the set of lines
\[
\widetilde \ccL_S = \{\widetilde \ell_a \in \widetilde \ccL \mid p_a \in S\}.
\]
We claim that $\widetilde \ccL_S$ is a solution to $(\widetilde \cP, k, t)$. 

Since each line $\ell_a \in \ccL$ is covered by some point $p_b \in S$, it follows from the point-line duality transform that the dual line $\widetilde \ell_b$ is in $\widetilde \ccL_S$. The incidence-preserving property of the transformation ensures that $\widetilde \ell_b$ passes through the point $\widetilde p_a$, which is the dual of $\ell_a$. Consequently, every point in $\widetilde \cP$ that corresponds to a covered line in $\ccL$ is also covered in the dual instance. Thus, $(\widetilde \cP, k, t)$ is a \yes instance of \pplc.

For the backward direction, assume that $(\widetilde \cP, k, t)$ is a \yes instance of \pplc. This implies that there exists a set of $k$ lines 
\[
\widetilde S = \{\widetilde \ell_{i_1}, \dots, \widetilde \ell_{i_k}\} \subseteq \widetilde \ccL
\]
that covers at least $t$ points in $\widetilde \cP$. Define the corresponding set of points in the primal space as
\[
\ccL_{\widetilde S} = \{p_a \in \cP_{\ccL} \mid \widetilde \ell_a \in \widetilde S\}.
\]
We claim that $\ccL_{\widetilde S}$ is a solution to $(\ccL, k, t)$ in \plpc. 

Each covered point $\widetilde p_a \in \widetilde \cP$ corresponds to a line $\ell_a \in \ccL$ in the primal space. Since the covering lines in $\widetilde S$ originate from points in $\ccL_{\widetilde S}$, the incidence-preserving property guarantees that the points in $\ccL_{\widetilde S}$ collectively cover at least $t$ lines in $\ccL$. Hence, $(\ccL, k, t)$ is a \yes instance of \plpc.
\end{proof}
We now state our main theorem.

\begin{theorem}
\label{thm: pplchard}
    \pplc is \whard parameterized by $k$.
\end{theorem}

\begin{proof}
We prove this by providing a polynomial-time parameter preserving reduction from \pvc, which is \whard, to \plpc, and then further reducing \plpc to \pplc. The explicit reduction steps have been described above, and the correctness of each reduction follows from Claims~\ref{clm: pvctopvc},~\ref{clm: pvctolpc}, and~\ref{clm: lpctoplc}.  
\end{proof}

%% file: matroid_partition_constraints.tex
\section{Covering with Matroid and Partition Constraints}\label{section:matroidpartition}
In this section, we consider the problem \wmrbdsfull{} (\wmrbds{}) under additional constraints, and design an \epas. The problem is defined as follows. The input consists of a bipartite graph $\graph$, a weight function $\w: \cB \rightarrow \bQ^+$, a surjective coloring function $\col: \cB \to [r]$, a tuple $t=\langle t_1, \dots, t_r \rangle$, an integer $k$, and a matroid $\matfull$ defined on $\cR$. The matroid is given via oracle access. The objective is to find a subset $\cS \subseteq \cR$ of size $k$ such that $\cS$ is independent in $\mat$ and, for each $i \in [r]$, 
\[
\w(\nbr(\cS) \cap \col^{-1}(i)) \geq t_i.
\]
 We refer to this problem as \wconprbds{}.
Given an instance $\cI  = \wconprbdsi $ of the \wconprbds problem, such a set $\cS$ is called a solution to $\cI$. All vertices in $\cB$, that have the same color, say $i$, will be referred to as vertices in color class $i$. 

For completeness, the problem is defined as follows. 

\defparopt{\wconprbds}{A bipartite graph $\graph$, a weight function $\w:\cB \rightarrow \bQ^{+}$, a surjective coloring function $\col: \cB \to [r]$, a tuple $t=\langle t_1, \dots, t_r \rangle$, an integer $k$, and a matroid $\matfull$ defined on $\cR$ (given via an independence oracle).}{$k$}{A subset of vertices $\cS \subseteq \cR$ such that $|\cS|\leq k$, $\cS$ is independent in $\mat$ and, for each $i \in [r]$, 
\[
\w(\nbr(\cS) \cap \col^{-1}(i)) \geq t_i.
\]}

\subsection{A brief description of the algorithm}

The algorithm for \wconprbds{} follows the same conceptual idea as the algorithm for \prbds{}. 
However, in this case, we must ensure that the solution also satisfies the additional constraints. 
Let $\bO = \langle o_1, \dots, o_k \rangle$ be a solution. 
The algorithm first fixes an $\iletter \in [k]$ and then focuses on $\pot_\iletter$. 
Observe that, by definition (analogous to the definition in the algorithm for \prbds{}), 
we have $o_\iletter \in \pot_\iletter$. 
These definitions are restated formally in~\Cref{def:matpartup}. 

Next, we partition $\pot_\iletter$ into \emph{bags} such that the total weight of the neighbors of every vertex within a bag is approximately the same (differing by at most a factor of $(1 + \frac{\varepsilon}{2})$) for each color class. Furthermore, if any vertex in a bag has neighbors of total weight exceeding the coverage requirement for a particular color class, then all vertices in that bag have neighbors whose total weight also exceeds the corresponding coverage requirement. Similarly, if a vertex in a bag has neighbors whose total weight is at most an $\frac{\varepsilon}{4k}$-fraction of the coverage requirement, then all vertices in that bag satisfy this condition as well. This is because, in the former case, any vertex in the bag can individually satisfy the coverage requirement of that color class, whereas in the latter case, even all $k$ vertices together can cover at most a $\frac{\varepsilon}{4}$-fraction of the required coverage. Since the number of bags is bounded as shown in~\Cref{obs:bucketing} 
we guess the bag that contains $o_\iletter$. 
However, since the size of a bag can be large, direct branching is infeasible. 
To address this, we compute a $(k-1)$-representative set of the bag, which is of size $k$ as shown in~\Cref{lem:oraclerepset}, 
which guarantees the existence of an element $h$ such that 
$\bO[\downarrow \alpha][\uparrow \alpha, h]$ is independent in $\mat$. 
Since $h$ and $o_\iletter$ lie in the same bag, the weights of their neighborhood are nearly identical 
in every color class. Using this property, we show that either there exists a color class $i$ such that 
$\nbr^i(h) = \nbr(h) \cap \col^{-1}(i)$ has a large intersection with $\nbr^i(\bO)$ in which case 
we can choose an element covered by $\bO$ with high probability or, if no such color class exists, 
then $h$ itself is a good choice. 
The remaining part of the algorithm proceeds similarly to the algorithm for \wmrbds{}. A complete description of the algorithm with pseudocode is provided in Algorithm~\ref{alg:framework-wconprbds}.

In the following, we state some relevant definitions followed by a detailed description of the \buck procedure, which partitions a set into bags of vertices with almost similar degree(weight of neighbors) in every color class. Then we provide a detailed description of the algorithm and analysis.

 \begin{definition}[Potential, Freely Covered and Non-Free neighbors]\label{def:matpartup}
Let 
\[
\cJ=(\wconprbdsi,\mi = \langle \mcon_1, \dots, \mcon_{k} \rangle,\free ),
\]  
be an instance for \Cref{alg:framework-wconprbds}.  Then, the potential solution family, the set of freely covered, newly-free and non-free neighbors of any subset of $\cB$ are defined as follows. 
\begin{description}
    \item[1.]
    ${\sf PotentialSolutionFamily}_\iletter =\pot_\iletter= \{ v \in \cR \mid \mcon_\iletter \subseteq \nbr(v) \}$ for all $\iletter \in [k]$,  and $\pot=\cup_{\iletter \in [k]}\pot_\iletter $
    \item[2.] 
$\fcov_\iletter =\bigcap_{v \in \pot_\iletter} \nbr(v)$ for all $\iletter \in [k]$ and $\fcov = \cup_{\iletter \in [k]}\fcov_\iletter$
\item[3.] $\nfree_\iletter  =   \fcov_\iletter \setminus \free$ for all $\iletter \in [k]$ and 
 $\nfree = \fcov \setminus \free$\;
 \item[4.] $\nbrst(\cS, \cJ) \coloneq \nbr(\cS) \setminus \free$,
where $\cS \subseteq \cR $. The definition of $\nbrst(,)$ captures the idea of newly covered vertices in $\cB$.
\end{description} 
For $j \in [r]$ we use $\fcov^j,\fcov_\iletter^j,\nfree^j,\nbr^{\star,j}(\dot)$ to denote the intersection of the sets with color class $j$.
\end{definition}
Finally, we define the notion of a consistent tuple, which helps us compare objects across different recursive calls of our algorithm.

 \begin{definition}[Consistent Tuple]\label{def:matparcontup}
Given an instance 
\[
\cJ = (\wconprbdsi,\, \mi = \langle \mcon_1, \dots, \mcon_{k} \rangle,\, \free),
\]
we say that a tuple $\cS = \langle s_1, \dots, s_k \rangle$ is \emph{consistent for} $\cJ$ if
\[
s_i \in \cR, \quad 
\fcov_i = \bigcap_{v \in \pot_i} \nbr(v) \subseteq \nbr(s_i), 
\quad \forall i \in [k],
\]
and $\cS$ is independent in $\mat$.  
We use $\cons{\cJ}$ to denote the set of all consistent tuples for $\cJ$.
\end{definition}

\begin{definition}[$(\alpha_i,\beta_i)$-approximate solution]
Let $$\cJ = (\wconprbdsi,\, \mi = \langle \mcon_1, \dots, \mcon_{k} \rangle,\, \free)$$ be an instance.  
A consistent tuple $\cS$ is said to be a \emph{$(\alpha_i,\beta_i)$-approximate solution} for $\cJ$ if, for every $i \in [r]$, it holds that
\[
\w\!\big( \nbr^{\star,i}(\cS,\cJ)\big) 
\geq \alpha_i t_i - \beta_i.
\] If $\alpha_i=1$ and $\beta_i=0$ for all $i \in [r]$ we have a \emph{solution tuple}. Also if $\alpha_i=\alpha$ for all $i\in [r]$ and $\beta_i=0$ for all $i \in [r]$ then it is referred to as $\alpha$-approximate solution.
\end{definition}

\subsection{Bucketing procedure}

In this section, we design a subroutine, called \buck, which (or slight variations thereof) will be crucially used to design our \epas . We divide the set of vertices of $\cR$ into a bounded number of equivalence classes depending on the weight of their neighborhoods in color class $j$ ($j$-neighborhoods) for a color $j$. 

At a high level, two vertices $u$ and $v$ belonging to a particular equivalence class will have $j$-degrees (weight of $ j$-neighborhoods) that are approximately equal. There are two exceptions to this:
\begin{enumerate}
    \item For a color $j$, all vertices of degree at least $t_j$ are treated as equivalent as far as color $j$ is concerned (i.e., we still classify such vertices based on their $j'$-degrees for other colors $j'$), since any single vertex from such a class is sufficient to entirely take care of color $j$.
    \item All vertices of $j$-degree less than $\varepsilon t_j / 2k$ are treated as equivalent as far as color $j$ is concerned. The reason is as follows: the difference between the $j$-degrees of such two vertices is at most $\varepsilon t_j / 2k$. Thus, even if we make a bad choice at most $k$ times, we only lose at most $\varepsilon t_j$ coverage for color $j$.
\end{enumerate}

Thus, the interesting range of $j$-degrees is between $\left[ \varepsilon t_j / 2k,\, t_j \right]$, which we sub-divide into intervals of range $(1 + \varepsilon)$. It is easy to see that the vertices are partitioned into $\mathcal{O}(\log_{1+\varepsilon} k)$ classes for a particular color. We proceed similarly for each color $j \in [r]$, and then obtain our final set of equivalence classes, such that all vertices in a particular class are equivalent with respect to each color in the sense described above.

\paragraph{Bucketing.} 
Given an instance  $\cI=\wconprbdsi$ of \wconprbds, we define the \buck procedure on an induced subgraph $G'=(R'\cup B')$ of $\graph$ and a given $\varepsilon'$ (different from $\varepsilon$ of the instance)  as follows. Let 
\[
\lambda := \left\lceil \log_{1+\varepsilon'} \frac{2k}{\varepsilon'} \right\rceil.
\]
Fix a color $j \in [r]$. For every $1 \le \alpha \le \lambda$, define
\[
G'_{\varepsilon'}(j, \alpha) := \left\{ v \in R' : \frac{t_j}{(1 + \varepsilon')^{\alpha}} \le \w(\nbr_{G'}^{j}(v)) < \frac{t_j}{(1 + \varepsilon')^{\alpha-1}} \right\}.
\]
The functions $\w,\col$ and the tuple $t$ are omitted from the notation as they are the same as those of the instance and will be clear in the context. We also define
\[
G'_{\varepsilon'}(j, 0) := \{ v \in R' :\w(\nbr_{G'}^{j}(v)) \ge t_j \}, \quad \text{and} \quad G'_{\varepsilon'}(j, \lambda + 1) := \{ v \in R': \w(\nbr_{G'}^{j}(v)) < \frac{t_j}{(1+\varepsilon')^{\lambda}} \}.
\]
Let $\cV = \{0, 1, \ldots, \lambda, \lambda+1\}^r$, and consider an arbitrary vector $\mathbf{v} = (\alpha_1, \alpha_2, \ldots, \alpha_r) \in \cV$. We define
\[
G'_{\varepsilon'}(\mathbf{v}) := \bigcap_{j=1}^{r} G'_{\varepsilon'}(j, \alpha_j).
\]
We refer to any such $G'_{\varepsilon'}(\mathbf{v})$ as a \bag. The \buck procedure first computes the set of bags as defined above, and returns only the set of non-empty bags, which form a partition of $R'$. Thus, we define the following notation.
\[
\bucket(G',\varepsilon')=\{G'_{\varepsilon'}(\mathbf{v}) \mid v \in \cV, G'_{\varepsilon'}(\mathbf{v})\neq \emptyset \}
\]It is easy to see that this procedure can be implemented in polynomial time.
\begin{observation}\label{obs:bucketing} The following properties hold for any color $j \in [r]$.
\begin{enumerate}
    \item For any $v \in G'_{\varepsilon'}(j, \lambda + 1)$, it holds that $\w(\nbr_{G'}^{j}(v)) < \varepsilon' t_j / (2k)$.
    \item For any $v \in G'_{\varepsilon'}(j, 0)$, it holds that $\w(\nbr_{G'}^{j}(v)) \ge t_j$.
    \item For any $1 \le \alpha \le \lambda$, and for any $u, v \in G'_{\varepsilon'}(j, \alpha)$, it holds that
    \[
        \frac{\w(\nbr_{G'}^{j}(u))}{1 + \varepsilon'} \le \w(\nbr_{G'}^{j}(v)) \le (1 + \varepsilon') \cdot \w(\nbr_{G'}^{j}(u)).
    \]
\end{enumerate}
Furthermore, the number of bags returned by the \buck procedure is bounded by
\[
     (\lambda + 2)^r \;\le\; \left\lceil \frac{6 \ln k}{(\varepsilon')^2} \right\rceil^{r} \;=:\; L.
\]
\end{observation}

\begin{algorithm}[t]
\caption{\algwconprbds}
\label{alg:framework-wconprbds}
\KwInput{$\cJ = (\wconprbdsi, \mi=\langle \mcon_1,\dots,\mcon_k\rangle, \free)$}
\KwOutput{A $(1-\varepsilon,\frac{(1-\varepsilon)\varepsilon t_i}{4}\ln{k})$-approximate solution for $\cJ$}

\ForEach{$\iletter\in[k]$}{
  $\pot_\iletter=\{v\in\cR\mid\mcon_\iletter\subseteq\nbr(v)\}$\;
  $\fcov_\iletter=\bigcap_{v\in\pot_\iletter}\nbr(v)$
}
$\nfree=\left(\bigcup_{\iletter\in[k]}\fcov_\iletter\right)\setminus\free$\;

\If{$\mi$ is a $0$-tuple and $t=0$}{\Return $0$-tuple$ $}

\If{$\exists\,\iletter\in[k]:\bigcup_{v\in\pot_\iletter}\nbr(v)\subseteq(\free\cup\nfree)$ \label{algwconprbds:basetwo}}{
  Let $\pot'_\iletter\subseteq^{k-1}_{\sf rep}\pot_\iletter$\;
  \ForEach{$v\in\pot'_\iletter$}{
    $\cS\gets\algwconprbds((\graph,\w,\col,\,t',\,k-1,\,\mat/\{v\},\,\varepsilon),\,\mi[\downarrow\iletter],\,\free\cup\nfree\cup\nbr(v))$\;
    where $t'=\left\langle t_i-\w(\nfree^i)-
    \w\bigl(\nbr^i(v)\setminus(\free^i\cup\nfree^i)\bigr)
    \right\rangle_{i\in[r]}$\;
    \If{$\cS[\uparrow\iletter,v]$ is $(1-\varepsilon, \frac{(1-\varepsilon)\varepsilon t_i}{4}\ln{k})$-approximate solution}{\Return $\cS[\uparrow\iletter,v]$ \label{algwconprbds:return_red_rule}}
  }
}

Fix $\iletter\in[k]$ \label{algwconprbds:fixedpot}, $\cH=\bucket((\pot_\iletter,\cB\setminus (\free \cup\nfree)),\,\varepsilon'=\frac{\varepsilon}{2})$\;
\ForEach{bag $G'_{\varepsilon'}(\mathbf{v})\in\cH$, $\mathbf{v}\in\{0,1,\dots,\lambda+1\}^r$ \label{algwconprbds:choosebag}}{
  $\widetilde{G'_{\varepsilon'}(\mathbf{v})}\subseteq^{k-1}_{\sf rep}G'_{\varepsilon'}(\mathbf{v})$\;
  \ForEach{$h\in\widetilde{G'_{\varepsilon'}(\mathbf{v})}$}{
    \ForEach{$i\in[r]$\label{algwconprbds:bagcolorguessed}}{
      Let $(\nbr^i(h)\setminus(\free^i \cup \nfree^i))=\{u_1,\dots,u_\ell\}$, $\omega=\sum_{j=1}^\ell\w(u_j)$\;
      Choose $\clubsuit\in\{h,u_1,\dots,u_\ell\}$ with $\Pr[h]=\frac12$, $\Pr[u_j]=\frac{\w(u_j)}{2\omega}$ \label{algwconprbds:prob_dist}

      \uIf{$\clubsuit=h\in\pot_\iletter$}{
        $\cS\gets\algwconprbds((\graph,\w,\col,\,t',\,k-1,\,\mat/\{h\},\,\varepsilon),\,\mi[\downarrow\iletter],\,\free\cup\nfree\cup\nbr(h))$ where 
      $$ t' =\begin{cases}
  t_i - \w(\nfree^i) - (1+\varepsilon)\,\w\big(\nbr^i(h)\setminus(\free^i\cup\nfree^i)\big), & \text{(Case 1)}\\[4pt]
  t_i - \w(\nfree^i) - \frac{\varepsilon t_i}{4k}-\frac{\varepsilon}{2}\w\big(\nbr^i(h)\setminus(\free^i\cup\nfree^i)\big), & \text{(Case 2)}.
\end{cases}
$$
\text{// Case 1: } $\displaystyle \frac{\varepsilon t_i}{4k} < \w(\nbr^i(h)\setminus(\free^i\cup\nfree^i))$\; 
\smallskip
\text{ // Case 2: } $\displaystyle \frac{\varepsilon t_i}{4k} \ge \w(\nbr^i(h)\setminus(\free^i\cup\nfree^i))$ \;
        \If{$\cS[\uparrow\iletter,h]$ is $(1-\varepsilon, \frac{(1-\varepsilon)\varepsilon t_i}{4}\ln{k})$-approximate solution}{\Return $\cS[\uparrow\iletter,h]$ \label{algwconprbds:return_high_deg}}
      }
      \Else{
        $\clubsuit=u\in\{u_1,\dots,u_\ell\}$\; choose $\beta\in[k]$ uniformly\;
        $\mi\gets\langle\mcon_1,\dots,\mcon_\beta\cup\{u\},\dots,\mcon_k\rangle$\;
        $t'=\left\langle t_i-\w(\nfree^i)
        -\w\bigl(\{u\}\cap\col^{-1}(i)\bigr)\right\rangle_{i\in[r]}$\;
        $\cS\gets\algwconprbds((\graph,\w,\col,t',k,\mat,\varepsilon),
        \,\mi,\,\free\cup\nfree\cup\{u\})$\;
        \If{$\cS$ is $(1-\varepsilon, \frac{(1-\varepsilon)\varepsilon t_i}{4}\ln{k})$-approximate solution}{\Return $\cS$ \label{algwconprbds:return_opt_elem}}
      }
    }
  }
}
\end{algorithm}

\subsection{Algorithm analysis}
In this section, we show that our algorithm indeed outputs a solution with the desired approximation ratio, that is \epas.
Let \( \mathbb{O}=\langle o_1,\dots, o_k\rangle\) be a solution tuple.
 In the following, we present a sequence of lemmata that capture relationships between two structurally similar instances arising from consecutive recursive calls. To avoid ambiguity caused by identically named objects across different calls, we annotate each object with its corresponding instance in brackets and omit those when the context is clear.

\begin{lemma}
\label{lem:wconpropertyonealgoone}
 Let $\cJ$ denote the input instance 
 \[(\wconprbdsi,\mi = \langle \mcon_1, \dots, \mcon_{k} \rangle,\free).\] 
 Consider $v\in \pot_{\iletter}(\cJ)$, for some $\iletter \in [k]$ and let $\cJ'$ denote the instance
  \[ ((\graph, \w,\col,t',k-1,\mat / \{v\}, \varepsilon), \mi[\downarrow \iletter],\free(\cJ) \cup \nfree(\cJ) \cup \nbr(v)).\]   for some $t'$.
Then, for any consistent tuple $\cS$ for $\cJ'$, we have the following.
\begin{description}[leftmargin=1.2em, labelsep=0.5em]
    \item[\textbf{1) Consistent:}] 
    $\cS[\uparrow \iletter, v]$ is a consistent tuple for $\cJ$.
    
    \item[\textbf{2) Weight Equality:}]
    For every color class $i \in [r]$, we have
    \begin{equation}\label{eqn:wconfirstpartcovstarsum}
    \begin{aligned}
        \w\big(\nbr^{\star,i}(\cS[\uparrow \iletter, v], \cJ) \big)
        &= \w\big(\nbr^{\star,i}(\cS, \cJ') \big) + \w\big(\nfree^i(\cJ)\big) \\
        &\quad + \w\big((\nbr^i(v) \setminus (\free^i(\cJ) \cup \nfree^i(\cJ))) \big).
    \end{aligned}
    \end{equation}
\end{description}

\end{lemma}
\begin{proof}

 {\rm \bf 1) Consistent:} We first show that \( \cS[\uparrow \iletter, v] \) is a consistent tuple for \( \cJ \). First by definition of $\mat / \{v\}$ it follows that \( \cS[\uparrow \iletter, v] \) is independent in  $\mat$ since $\cS$ is independent in  $\mat / \{v\}$.  Now, observe that since \( \cS \) is a consistent tuple for \( \cJ' \), we have
\[
\fcov_i(\cJ') \subseteq \nbr(s_i) \quad \text{for all } i \in [k] \setminus \{\iletter\}.
\]
Moreover, for all \( i \in [k] \setminus \{\iletter\} \), we have \( \mcon_i(\cJ) = \mcon_i(\cJ') \), and thus \( \fcov_i(\cJ) = \fcov_i(\cJ') \subseteq \nbr(s_i) \). Now consider index \( \iletter \). Since \( v \in \pot_\iletter(\cJ) \), we know that
\[
\mcon_\iletter(\cJ) \subseteq \fcov_\iletter(\cJ) \subseteq \nbr(v).
\]
Therefore, for all \( i \in [k] \), the neighborhood of the \( i^\text{th} \) vertex in the tuple \( \cS[\uparrow \iletter, v] \) contains \( \fcov_i(\cJ) \). This proves that \( \cS[\uparrow \iletter, v] \) is a consistent tuple for \( \cJ \).
\medskip

\noindent 
 {\rm \bf 2) Weight Equality:} Due to the  {\bf Pairwise Disjoint Property} of \Cref{lem:propertyonealgoone}, we know that in this case, the following holds:
\begin{equation}
\label{eqn: wconteq}
   \nbrst(\cS[\uparrow \iletter,v], \cJ)= \nbrst(\cS,\cJ') \uplus \nfree(\cJ) \uplus (\nbr(v) \setminus (\free(\cJ) \cup \nfree(\cJ))).
\end{equation}
This implies for every $i \in [r]$
\begin{equation}
\begin{aligned}
   \nbr^{\star,i}(\cS[\uparrow \iletter, v], \cJ) 
    &= \nbr^{\star,i}(\cS, \cJ')\uplus\,\nfree^i(\cJ) \\
    &\quad \uplus\, \big(\nbr^i(v) \setminus (\free^i(\cJ) \cup \nfree^i(\cJ))\big).
\end{aligned}
\end{equation}
Hence, \Cref{eqn:wconfirstpartcovstarsum} holds.
\end{proof}

\begin{lemma}
\label{lem:wconpropertytwoalgoone}
Let \( \cJ \) be the input instance  
\[(\wconprbdsi,\mi = \langle \mcon_1, \dots, \mcon_{k} \rangle,\free).\]  
Consider \( u \in \cB \setminus (\free(\cJ)\cup \nfree(\cJ)) \). For any \( \beta \in [k] \),  let $$\mi' \gets \langle \mcon_1, \dots, \mcon_{\beta} \cup \{u\},\dots, \mcon_{k} \rangle,$$ and  let \( \cJ' \) denote the instance  
\[
((\graph, \w,\col,t',k,\mat,\varepsilon) \mi', \free(\cJ)\cup \nfree(\cJ) \cup \{u\}).
\]for some $t'$. 
 
Then, for any consistent tuple $\cS$ for $\cJ'$, we have the following.
\begin{description}
    \item {\rm \bf 1) Consistent:}  $\cS$ is a consistent tuple for $\cJ$. 
    \item {\rm \bf 2) Weight Equality:} For every $i\in [r]$
\begin{equation}\label{eqn:wcontequal2}
\begin{aligned}
\w(\nbr^{\star,i}(\cS, \cJ)
&= \w(\nbr^{\star,i}(\cS, \cJ'))  + \w(\nfree^i(\cJ))
+ \w\big(u \cap \col^{-1}(i)\big).
\end{aligned}
\end{equation}

\item  {\rm \bf 3) Intersection Complexity:} $\dintcmp(G,\mcon_\beta \cup \{u\})\geq \dintcmp(G,\mcon_\beta)+1$
\end{description}

\end{lemma}
\begin{proof}
Let $\cS = \langle s_1, \dots, s_k \rangle$ be a consistent tuple for $\cJ'$. Observe that the matroid remains unchanged in both instances and hence $\cS$ remains independent. Since \( \cS \) is a consistent tuple for \( \cJ' \), for any \( i \in [k] \setminus \{\beta\} \), we have that $\fcov_{i}(\cJ') \subseteq \nbr(s_i)$(by definition). Further by property of $\cJ$ and $\cJ'$, for every $i \in [k] \setminus \beta$, we have that $\mcon_{i}(\cJ) = \mcon_{i}(\cJ') \subseteq \nbr(s_i)$ which implies that $\fcov_{i}(\cJ) = \fcov_{i}(\cJ') \subseteq \nbr(s_i)$.

Now, consider the coordinate \( \beta \). Since $\mcon_{\beta}(\cJ') = \mcon_{\beta}(\cJ) \cup \{u\}$, this implies that $\mcon_{\beta}(\cJ) \subseteq \mcon_{\beta}(\cJ')$.
Since \( \cS \) is a consistent tuple for \( \cJ' \), it follows that $\mcon_{\beta}(\cJ') \subseteq \fcov_{\beta}(\cJ') \subseteq \nbr(s_{\beta})$. 
Therefore,  
$\mcon_{\beta}(\cJ) \subseteq \nbr(s_{\beta})$.  
This implies that \( s_{\beta} \in \pot_{\beta}(\cJ) \). Consequently, we also have that  $\fcov_{\beta}(\cJ) \subseteq \nbr(s_{\beta})$. Thus, \( \cS \) is a consistent tuple for the instance \( \cJ \).

The remaining part of proof for \textbf{Intersection Complexity} follow identically from the corresponding argument in \Cref{lem:propertytwoalgoone}, and is therefore omitted.

\smallskip
\noindent
{\rm \bf 2) Weight Equality:} 
Due to the \textbf{Pairwise Disjoint} property in \Cref{lem:propertytwoalgoone}, we know that in this case,
\begin{equation}
\label{eqn:wconteqn2}
\nbrst(\cS, \cJ) = \nbrst(\cS,\cJ') \uplus \nfree(\cJ) \uplus \{u\}.
\end{equation}
This implies for every $i\in [r]$ we have 
\begin{equation}
\nbr^{\star,i}(\cS, \cJ)
= \big(\nbr^{\star,i}(\cS, \cJ') \big)
\uplus \big(\nfree^i(\cJ) \big)
\uplus \big(u \cap \col^{-1}(i)\big).
\end{equation}

Hence, \Cref{eqn:wcontequal2} holds.
 \end{proof}

 Let \(\intcmp = \dintcmp(\graph)\). We define a measure associated with \(\cJ\), denoted by \(\mu(\cJ)\), as  
\[
\mu(\cJ) = k + k\intcmp - \sum_{\iletter \in [k]} \dintcmp( G, \mcon_{\iletter})
\]  
We use this measure for induction in our proof for both approximation as well as probability analysis. We now present the main technical lemma of this section, along with its proof.
\begin{lemma} \label{lem:framework-wconprbds}
For an input instance
\[
\cJ=(\wconprbdsi,\mi=\langle\mcon_1,\dots,\mcon_k\rangle,\free),
\]
Algorithm~\ref{alg:framework-wconprbds} runs in time
\[
\left(\frac{kr\ln k}{\varepsilon}\right)^{\cO(r\mu(\cJ))}\sizeg
\]
and, provided that a solution tuple exists, outputs a
$\bigl(1-\varepsilon,\frac{(1-\varepsilon)\varepsilon t_i}{4}\ln k\bigr)$-approximate
solution $\cS=\langle s_1,\dots,s_k\rangle$ with probability at least
$\left(\frac{\varepsilon}{4k}\right)^{\mu(\cJ)}$, where
\[
\mu(\cJ)=k+k\intcmp-\sum_{\iletter\in[k]}\dintcmp(G,\mcon_{\iletter}).
\]
\end{lemma}

\begin{proof}
    We divide the proof into two parts. First, we establish the approximation ratio of our algorithm, and then we analyze its success probability. Our proof proceeds by induction, using the measure defined above: 
\[
\mu(\cJ) = k + k\intcmp - \sum_{\iletter \in [k]} \dintcmp( G, \mcon_{\iletter})
\]
In particular, we prove the desired statement by induction on \(\mu(\cJ)\).

\paragraph{\bf Non-Negativity of Measure:} 
Note that \(\mu(\cJ) = k + k\intcmp - \sum_{\iletter \in [k]} \dintcmp( G, \mcon_{\iletter}(\cJ))\) is always non-negative, since for all \(\iletter \in [k]\), we have \( \dintcmp( G,\mcon_{\iletter}(\cJ)) \leq \intcmp \), which implies \( \sum_{\iletter =1}^k \dintcmp (G,\mcon_{\iletter}(\cJ)) \leq k \intcmp \). Moreover, since any subset cannot have a negative size, we know that \( k \geq 0 \). Hence, it follows that  
\[
\mu(\cJ) = k + k\intcmp - \sum_{\iletter \in [k]} \dintcmp( G, \mcon_{\iletter}) \geq 0.
\]

\paragraph{\bf Analysis of the algorithm.} 
Consider a solution tuple \(\bo = \langle o_1, \dots, o_k \rangle\) for the instance \(\cJ\). We prove the statement of the lemma using induction on \(\mu(\cJ)\). That is, the statement of the lemma serves as our induction hypothesis.

\smallskip

\noindent 
	\textbf{Base Case:} When $\mu(\cJ) = 0$, then we have that $k=0$ since $\intcmp \geq \dintcmp(G,\mcon_{\iletter})$ for $\iletter \in [k]$.  This implies that $\mi$ is a $0$-tuple. Thus, $0$-tuple is the only consistent tuple in this case, and hence $t_i$ must be $0$ for all $i \in [r]$, for there to exist a solution tuple. Thus \Cref{alg:framework-wconprbds} correctly returns a $0$-tuple with {\color{blue}\texttt{probability $1$}}.  

    \smallskip

\noindent 
\textbf{Induction Assumption:} Let $\mu' >0$. Assume that the statement of Lemma~\ref{lem:framework-wconprbds} holds for all instances, $\cJ$,  for which  $\mu(\cJ) < \mu'$. That is, for an  input instance $$ \cJ =(\wconprbdsi, \mi = \langle \mcon_1, \dots ,\mcon_{k} \rangle,\free) $$ such that $\mu(\cJ) <\mu'$, \Cref{alg:framework-wconprbds} $\textnormal{(\algwconprbds)} $ outputs a $(1-\varepsilon,\frac{(1-\varepsilon)\varepsilon t_i}{4}\ln{k})$-approximate solution $\cS = \langle s_1, s_2, \dots, s_k \rangle$ 
with probability at least $\left( \frac{\varepsilon}{2k}\right)^{\mu(\cJ)}$.

\smallskip

\noindent 
\textbf{Inductive Step:} We prove the statement of Lemma \ref{lem:framework-wconprbds} for an arbitrary instance, $\cJ$, for which $\mu(\cJ) = \mu'$.

\medskip

\noindent 
\underline{\textbf{Approximation Analysis.}}   We first perform the approximation analysis \textbf{assuming} that when the algorithm recursively calls an instance \( \cJ' \) such that \( \mu(\cJ') < \mu(\cJ) \), it indeed returns a $(1-\varepsilon,\frac{(1-\varepsilon)\varepsilon t_i}{4}\ln{k})$-approximate solution \( \cS' \) for \( \cJ' \).
In reality, this outcome occurs with some probability, which we will analyze after establishing the desired approximation factor. In simple terms, we condition on the recursive subroutines ``correctly'' returning the expected approximate solution and then proceed to analyze the resulting approximation factor.

We divide the approximation analysis in two parts. In the first part, we analyse the case when the condition of ~\Cref{algwconprbds:basetwo} holds. Then, we analyse the case when the algorithm executes~\Cref{algwconprbds:bagcolorguessed}. 

\smallskip
\noindent 
\textbf{Case A: Analysis for the case when condition of ~\Cref{algwconprbds:basetwo} holds:}
 Let $\cJ$ denote the input instance 
 \[(\wconprbdsi, \mi = \langle \mcon_1, \dots, \mcon_{k} \rangle,\free)\] 
 and $\cJ'$ denote the instance
  \[ ((\graph,\w,\col,\,t',\,k-1,\,\mat/\{v\},\,\varepsilon),\,\mi[\downarrow\iletter],\,\free\cup\nfree\cup\nbr(v))\] where 
  \[
  t'=\left\langle t_i-\w\bigl(\nfree^i(\cJ)\bigr)
  -\w\bigl(\nbr^i(v)\setminus(\free^i(\cJ)\cup\nfree^i(\cJ))\bigr)
  \right\rangle_{i\in[r]}.
  \]

Since $\pot_\iletter'$ is a $(k-1)$-representative set of $\pot_\iletter$,
there exists $v\in\pot_\iletter'$ such that
$\bO[\downarrow\iletter][\uparrow\iletter,v]$ is independent in $\mat$.
Since $\bigcup_{v\in\pot_\iletter}\nbr(v)\subseteq(\free\cup\nfree)$ there does not exist $u_1,u_2 \in \pot_\iletter$ such that there is a vertex $u_b \in\cB\setminus \free$ and $u_b\in \nbr(u_1)$ but $u_b \notin \nbr(u_2)$. Thus we have $\nbr(v)\setminus \free=\nbr(o_\iletter)\setminus \free$. Hence we get 
 \[
 t'=\left\langle t_i-\w\bigl(\nfree^i(\cJ)\bigr)
 -\w\bigl(\nbr^i(o_\iletter)\setminus
 (\free^i(\cJ)\cup\nfree^i(\cJ))\bigr)\right\rangle_{i\in[r]}.
 \]
 Thus $ \bO[\downarrow \iletter]$ is solution tuple for $\cJ'$.
 As a step towards our proof, we first establish that \( \mu(\cJ') < \mu(\cJ) \), allowing us to apply the induction hypothesis to \( \cJ' \). Observe that 
\begin{align}
	&\mu(\cJ) - \mu(\cJ') \nonumber \\ 
	 &=\big(k + k\intcmp - \sum_{\iletter \in [k]} \dintcmp( G, \mcon_{\iletter}) \big)- \big( (k-1) + (k-1)\intcmp - \sum_{\substack{\iletter \in [k]\\ \iletter\neq j}} \dintcmp(G,  \mcon_{\iletter})\big) \nonumber \\
	&=(1+\intcmp-\dintcmp(G,\mcon_{j})) \stackrel{(\P)}{>}0 \label{measure:wconcaseA}
\end{align}
where $(\P)$ follows from the fact that $\intcmp\geq \dintcmp(G,\mcon_j)$.

 Hence, by the induction hypothesis we can get, \( \cS \), a $(1-\varepsilon,\frac{(1-\varepsilon)\varepsilon t_i}{4}\ln{k})$-approximate tuple for \( \cJ' \) returned in \Cref{algwconprbds:return_red_rule} of the algorithm. Then, by Lemma~\ref{lem:wconpropertyonealgoone}, we know that the returned solution $\cS[\uparrow \iletter, v]$ is a consistent tuple for $\cJ$. Next, we show that it also a $(1-\varepsilon, \frac{(1-\varepsilon)\varepsilon t_i}{4}\ln{k})$-approximate solution.
 
  By induction hypothesis, we have for every $i \in [r]$
\begin{equation}\label{eqn:wconinducone}
\begin{aligned}
\w(\nbr^{\star,i}(\cS,\cJ'))
&\ge (1-\varepsilon)\Big(t_i
-\w(\nfree^i(\cJ))\\
&\quad
-\w(\nbr^i(v)\!\setminus\!(\free^i(\cJ)\!\cup\!\nfree^i(\cJ)))\Big)-\frac{(1-\varepsilon)\varepsilon t_i'}{4}\ln{(k-1)}.
\end{aligned}
\end{equation}

where $t_i'=t_i- \w\big(\nfree^i(\cJ) \big) - \w\big((\nbr^i(v) \setminus (\free^i(\cJ) \cup \nfree^i(\cJ))$. Again for every $i\in [r]$ we get the following due to~\Cref{eqn:wconfirstpartcovstarsum}
  \begin{equation}\label{eqn:wconwteq}
    \begin{aligned}
        \w\big(\nbr^{\star,i}(\cS[\uparrow \iletter, v], \cJ) \big)
        &= \w\big(\nbr^{\star,i}(\cS, \cJ')\big) + \w\big(\nfree^i(\cJ) \big) \\
        &\quad + \w\big(\nbr^i(v) \setminus (\free^i(\cJ) \cup \nfree^i(\cJ))\big).
    \end{aligned}
    \end{equation}

Let \(N := \w(\nfree^i(\cJ)\)) and 
\(W_v :\w=( \big(\nbr^i(v)\setminus(\free^i(\cJ)\cup\nfree^i(\cJ))\big)\)).
Thus by combining equations~\Cref{eqn:wconinducone} and~\Cref{eqn:wconwteq}, we get for every $i \in [r]$
\begin{equation}\label{eqn:wcon-detailed}
\begin{aligned}
\w\big(\nbr^{\star,i}(\cS[\uparrow \iletter, v], \cJ) \big)
&= (1-\varepsilon)\big(t_i - N - W_v\big) + N + W_v-\frac{(1-\varepsilon)\varepsilon t_i'}{4}\ln{(k-1)}\\
&= (1-\varepsilon)t_i - (1-\varepsilon)N - (1-\varepsilon)W_v + N + W_v-\frac{(1-\varepsilon)\varepsilon t_i'}{4}\ln{(k-1)}\\
&= (1-\varepsilon)t_i + \varepsilon\,N + \varepsilon\,W_v-\frac{(1-\varepsilon)\varepsilon t_i'}{4}\ln{(k-1)}\\
&\ge (1-\varepsilon)\,t_i - \frac{(1-\varepsilon)\varepsilon t_i}{4}\ln{k}
\end{aligned}
\end{equation}
where the last inequality follows from the fact that $\frac{(1-\varepsilon)\varepsilon t_i}{4}\ln{k}\geq \frac{(1-\varepsilon)\varepsilon t_i'}{4}\ln{(k-1)}$ since $t_i\geq t_i'$ and $\ln{k}\geq \ln{(k-1)}$

\noindent  \textbf{Case B: Analysis for the case when~\Cref{algwconprbds:bagcolorguessed} is executed:} Consider the $\iletter$ fixed in~\Cref{algwconprbds:fixedpot}. We know that $\bO$ is a solution tuple. There exists a \bag in $\cH = \bucket((\pot_\iletter, \cB \setminus \free),\, \varepsilon' = \frac{\varepsilon}{2})$, say $G'_{\varepsilon'}(\mathbf{v})$, that contains $o_\iletter$, since $\cH$ forms a partition of $\pot_\iletter$. Consider the branch that chooses $G'_{\varepsilon'}(\mathbf{v})$ in~\Cref{algwconprbds:choosebag}. Since $\widetilde{G'_{\varepsilon'}(\mathbf{v})}$ is a $(k-1)$-representative set of $G'_{\varepsilon'}(\mathbf{v})$, there exists $h \in G'_{\varepsilon'}(\mathbf{v})$ such that $\bO[\downarrow \iletter][\uparrow \iletter, h]$ is independent in $\mat$. Consider the branch that chooses $h$. We now divide the analysis into two possible cases.

\medskip
\noindent
\textbf{Case $\textnormal{\textbf{B}}_{\textbf{1}}$:} For every $i \in [r]$, we have 
\[
    \w\big((\nbr^i(\bO) \cap \nbr^i(h)) \setminus (\nfree^i \cup \free^i)\big)
    < \frac{\varepsilon}{2} \,
    \w\big(\nbr^i(h) \setminus (\nfree^i \cup \free^i)\big).
\]

In this case, consider the branch where the algorithm returns 
$\cS[\uparrow \iletter, h]$ in~\Cref{algwconprbds:return_high_deg}, 
where $\cS$ is the tuple returned by
\[
    \algwconprbds\big(
        (\graph, \w, \col,\, t',\, k-1,\, \mat/\{h\},\, \varepsilon),\,
        \mi[\downarrow \iletter],\,
        \free \cup \nfree \cup \nbr(h)
    \big),
\]
where
\[
t' =
\begin{cases}
  t_i - \w(\nfree^i) - (1+\varepsilon)\,\w\big(\nbr^i(h)\setminus(\free^i\cup\nfree^i)\big), & \text{(Case 1)}\\[4pt]
   t_i - \w(\nfree^i) - \frac{\varepsilon t_i}{4k}-\frac{\varepsilon}{2}\w\big(\nbr^i(h)\setminus(\free^i\cup\nfree^i)\big), & \text{(Case 2)}.
\end{cases}
\]
\smallskip
\text{// Case 1: } $\displaystyle \frac{\varepsilon t_i}{4k} < \w(\nbr^i(h)\setminus(\free^i\cup\nfree^i))$\\

\noindent\text{ // Case 2: } $\displaystyle \frac{\varepsilon t_i}{4k} \ge \w(\nbr^i(h)\setminus(\free^i\cup\nfree^i))$ \\

Let $\cJ$ denote the input instance
\[
    (\wconprbdsi,\, 
      \mi = \langle \mcon_1, \dots, \mcon_k \rangle,\, 
      \free),
\]
and let $\cJ'$ denote the instance
\[
    ((\graph, \w, \col,\, t',\, k-1,\, \mat/\{h\},\, \varepsilon),\,
     \mi[\downarrow \iletter],\,
     \free \cup \nfree \cup \nbr(h)),
\]
which the algorithm recursively solves in this branch.
As a step toward our proof, we first establish that \( \mu(\cJ') < \mu(\cJ) \), allowing us to apply the induction hypothesis on \( \cJ' \). Observe that 
\begin{align}
	&\mu(\cJ) - \mu(\cJ')\nonumber\\
	 &=\big(k + k\intcmp - \sum_{\iletter \in [k]} \dintcmp( G, \mcon_{\iletter}) \big)- \big( (k-1) + (k-1)\intcmp - \sum_{\substack{\iletter \in [k]\\ \iletter\neq j}} \dintcmp( G, \mcon_{\iletter})\big) \nonumber \\
	&=(1+\intcmp-\dintcmp(G,\mcon_{j}))\stackrel{(\P)}{>}0 \label{measure:wconcaseB}
\end{align}
where $(\P)$ follows from the fact that $\intcmp = \dintcmp(G) \geq \dintcmp(G,\mcon_j)$.

Now we divide the analysis based on whether $i\in [r]$ satisfies Case 1 or Case 2.
\paragraph{For $i \in [r]$ such that Case 1 is satisfied:}
Since we are in the branch where $o_\iletter$ and $h$ are in the same bag, 
for the color classes where 
$\w\big(\nbr^i(h) \setminus (\free^i \cup \nfree^i)\big) \ge t_i$, 
the tuple $\cS[\uparrow \iletter, h]$ already satisfies the covering constraints 
of those color classes. Hence, we analyze only the remaining color classes. 
Throughout this paragraph, whenever we say for every $i \in [r]$, we refer to those color classes
that satisfies Case 1, and for which the condition 
$\w\big(\nbr^i(h) \setminus (\free^i \cup \nfree^i)\big) \ge t_i$ 
is not satisfied.

Due to~\Cref{obs:bucketing}, we have for every  color $i \in [r]$,
\begin{equation}\label{eqn:optnear}
    \w\big(\nbr^i(o_\iletter) \setminus (\free^i \cup \nfree^i)\big)
    \leq
    \Big(1 + \frac{\varepsilon}{2}\Big)
    \w\big(\nbr^i(h) \setminus (\free^i \cup \nfree^i)\big).
\end{equation}

Since for every color $i \in [r]$,
\[
\begin{aligned}
    \w\big(\nbr^{\star,i}(\bO[\downarrow \iletter]), \cJ'\big)
    &\geq 
      \w\big(\nbr^{\star,i}(\bO), \cJ\big)
      - \w\big(\nfree^i\big)
      - \w\big(\nbr^i(o_\iletter) \setminus (\free^i \cup \nfree^i)\big) \\
    &\quad
      - \w\big((\nbr^i(\bO) \cap \nbr^i(h)) 
      \setminus (\nfree^i \cup \free^i)\big),
\end{aligned}
\]
by~\Cref{eqn:optnear} and our case assumption, we get for every color $i \in [r]$,
\[
\begin{aligned}
    \w\big(\nbr^{\star,i}(\bO[\downarrow \iletter]), \cJ'\big)
    &\geq 
      t_i - \w\big(\nfree^i\big)
      - \Big(1 + \frac{\varepsilon}{2}\Big)
        \w\big(\nbr^i(h) \setminus (\free^i \cup \nfree^i)\big) \\
    &\quad
      - \frac{\varepsilon}{2}
        \w\big(\nbr^i(h) \setminus (\free^i \cup \nfree^i)\big) \\
    &= 
      t_i - \w\big(\nfree^i\big)-(1 + \varepsilon)
      \w\big(\nbr^i(h) \setminus (\free^i \cup \nfree^i)\big).
\end{aligned}
\]

Hence, $\bO[\downarrow \iletter]$ is a solution tuple for the instance $\cJ'$.

Thus, by the induction hypothesis, for every $i \in [r]$ we have
\begin{equation}\label{eqn:indhyp}
\begin{aligned}
    \w\big(\nbr^{\star,i}(\cS,\cJ')\big)
    &\;\ge\;
    \Big(1 - \varepsilon\Big)
    \Big(t_i'
    \Big)  - \frac{(1-\varepsilon)\varepsilon t_i'}{4}\ln{(k-1)}  \\[6pt]
\end{aligned}
\end{equation}
where $$t_i'= t_i
        - \w\big(\nfree^i(\cJ)\big)
        - (1+\varepsilon)\,
          \w\big(\nbr^i(v)\setminus(\free^i(\cJ)\cup\nfree^i(\cJ))\big)
          $$

Next, by~\Cref{lem:wconpropertyonealgoone}, for every color $i\in [r]$ we have
\begin{equation}\label{eqn:wconprop}
\begin{aligned}
    \w\big(\nbr^{\star,i}(\cS[\uparrow \iletter, h], \cJ)\big)
    = \w\big(\nbr^{\star,i}(\cS, \cJ')\big)
      + \w\big(\nfree^i(\cJ)\big)
      + \w\big(\nbr^i(h)\setminus(\free^i(\cJ)\cup\nfree^i(\cJ))\big).
\end{aligned}
\end{equation}

Substituting the lower bound from~\eqref{eqn:indhyp} into~\eqref{eqn:wconprop}, we obtain
\begin{align*}
    \w\big(\nbr^{\star,i}(\cS[\uparrow \iletter, h], \cJ)\big)
    &\;\ge\;
    \left[
    (1 - \varepsilon) t_i' - \frac{(1-\varepsilon)\varepsilon t_i'}{4}\ln(k-1)
    \right] \\
    &\qquad
    + \w\big(\nfree^i(\cJ)\big)
    + \w\big(\nbr^i(h)\setminus(\free^i(\cJ)\cup\nfree^i(\cJ))\big)
    \stepcounter{equation}\tag{\theequation}\label{eqn:subst}
\end{align*}
Let $N = \w\big(\nfree^i(\cJ)\big)$ and $W_h = \w\big(\nbr^i(h)\setminus(\free^i(\cJ)\cup\nfree^i(\cJ))\big)$.
Recall that $t_i' = t_i - N - (1+\varepsilon)W_h$.
We substitute this definition of $t_i'$ into the first term of~\eqref{eqn:subst}:
\[
\begin{aligned}
    \w\big(\nbr^{\star,i}(\cS[\uparrow \iletter, h], \cJ)\big)
    &\;\ge\;
    (1 - \varepsilon) \Big( t_i - N - (1+\varepsilon)W_h \Big)
    + N + W_h
    - \frac{(1-\varepsilon)\varepsilon t_i'}{4}\ln(k-1) \\
    &\;=\;
    (1 - \varepsilon)t_i - (1-\varepsilon)N - (1-\varepsilon^2)W_h
    + N + W_h
    - \frac{(1-\varepsilon)\varepsilon t_i'}{4}\ln(k-1) \\
    &\;=\;
    (1 - \varepsilon)t_i + \big(N - (1-\varepsilon)N\big)
    + \big(W_h - (1-\varepsilon^2)W_h\big)
    - \frac{(1-\varepsilon)\varepsilon t_i'}{4}\ln(k-1) \\
    &\;=\;
    (1 - \varepsilon)t_i + \varepsilon N + \varepsilon^2 W_h
    - \frac{(1-\varepsilon)\varepsilon t_i'}{4}\ln(k-1).
\end{aligned}
\]
Since weights are non-negative and $\varepsilon > 0$, we have $\varepsilon N \ge 0$ and $\varepsilon^2 W_h \ge 0$.
Thus, we can lower bound the expression by dropping these terms:
\[
    \ge (1 - \varepsilon)t_i
    - \frac{(1-\varepsilon)\varepsilon t_i'}{4}\ln(k-1).
\]
From the definition of $t_i'$, since $N \ge 0$ and $W_h \ge 0$, we have $t_i' \le t_i$.
Furthermore, for the induction to hold, $k \ge 2$, which implies $\ln(k-1) < \ln(k)$.
Since $t_i', t_i, \ln(k-1), \ln(k)$ are all non-negative, we have
$t_i' \ln(k-1) \le t_i \ln(k-1) < t_i \ln(k)$.
Multiplying by the negative constant $-\frac{(1-\varepsilon)\varepsilon}{4}$ reverses the inequality:
\[
    - \frac{(1-\varepsilon)\varepsilon t_i'}{4}\ln(k-1)
    \ge
    - \frac{(1-\varepsilon)\varepsilon t_i}{4}\ln(k).
\]
Substituting this back, we get the desired result:
\[
\begin{aligned}
    \w\big(\nbr^{\star,i}(\cS[\uparrow \iletter, v], \cJ)\big)
    &\;\ge\;
    (1 - \varepsilon)t_i
    - \frac{(1-\varepsilon)\varepsilon t_i'}{4}\ln(k-1) \\
    &\;\ge\;
    (1 - \varepsilon)t_i
    - \frac{(1-\varepsilon)\varepsilon t_i}{4}\ln(k).
\end{aligned}
\]

\paragraph{For $i \in [r]$ such that Case 2 is satisfied:}
In this case, both
$\w\big(\nbr^i(h)\setminus(\free^i\cup\nfree^i)\big) \leq \frac{\varepsilon t_i}{4k}$
and
$\w\big(\nbr^i(o_\iletter)\setminus(\free^i\cup\nfree^i)\big) \leq \frac{\varepsilon t_i}{4k}$.
Throughout this paragraph, for every $i \in [r]$ refers to those color classes that satisfy the condition of Case~2.

Since for every color $i \in [r]$ we have
\[
\begin{aligned}
    \w\big(\nbr^{\star,i}(\bO[\downarrow \iletter]), \cJ'\big)
    &\ge
      \w\big(\nbr^{\star,i}(\bO), \cJ\big)
      - \w\big(\nfree^i\big)
      - \w\big(\nbr^i(o_\iletter)\setminus(\free^i\cup\nfree^i)\big) \\
    &\quad
      - \w\big((\nbr^i(\bO)\cap\nbr^i(h))\setminus(\nfree^i\cup\free^i)\big),
\end{aligned}
\]
by the Case~2 assumption on $\nbr^i(o_\iletter)$ and the bound on the intersecting neighborhood, we obtain
\[
\begin{aligned}
    \w\big(\nbr^{\star,i}(\bO[\downarrow \iletter]), \cJ'\big)
    &\ge
      t_i - \w(\nfree^i)
      - \frac{\varepsilon t_i}{4k}
      - \frac{\varepsilon}{2}\,
        \w\big(\nbr^i(h)\setminus(\free^i\cup\nfree^i)\big).
\end{aligned}
\]
Hence, $\bO[\downarrow \iletter]$ is a feasible solution tuple for the instance~$\cJ'$ with threshold $t_i'$, where
\[
t_i' := t_i - \w(\nfree^i)
            - \frac{\varepsilon t_i}{4k}
            - \frac{\varepsilon}{2}\,
              \w\big(\nbr^i(h)\setminus(\free^i\cup\nfree^i)\big).
\]
Let $N:=\w(\nfree^i(\cJ))$ and
$W_h:=\w\big(\nbr^i(h)\setminus(\free^i(\cJ)\cup\nfree^i(\cJ))\big)$, so $t_i' = t_i - N - \frac{\varepsilon t_i}{4k} - \frac{\varepsilon}{2}W_h$.

\smallskip
\noindent
By the induction hypothesis, for every $i\in[r]$,
\begin{equation}\label{eq:indhyp_case2_final}
\w\big(\nbr^{\star,i}(\cS,\cJ')\big)
\ge
(1-\varepsilon)t_i'
-\frac{(1-\varepsilon)\varepsilon t_i'}{4}\ln(k-1).
\end{equation}
By \Cref{lem:wconpropertyonealgoone}, the combined solution satisfies
$\w\big(\nbr^{\star,i}(\cS[\uparrow \iletter,h],\cJ)\big)
=
\w\big(\nbr^{\star,i}(\cS,\cJ')\big)
+N+W_h$.
Substituting the lower bound from~\eqref{eq:indhyp_case2_final} gives
\[
\begin{aligned}
\w\big(\nbr^{\star,i}(\cS[\uparrow \iletter,h],\cJ)\big)
&\ge
(1-\varepsilon)\Big(t_i - N - \frac{\varepsilon t_i}{4k} - \frac{\varepsilon}{2}W_h\Big)
-\frac{(1-\varepsilon)\varepsilon t_i'}{4}\ln(k-1)
+N+W_h\\[4pt]
&=
(1-\varepsilon)t_i
+\varepsilon N
+\Big(1 - \frac{\varepsilon(1-\varepsilon)}{2}\Big)W_h
-(1-\varepsilon)\frac{\varepsilon t_i}{4k}
-\frac{(1-\varepsilon)\varepsilon t_i'}{4}\ln(k-1).
\end{aligned}
\]
Since $N \ge 0$, $W_h \ge 0$, and $\big(1 - \frac{\varepsilon(1-\varepsilon)}{2}\big) > 0$, we may drop the non-negative terms involving $N$ and $W_h$ to obtain the lower bound:
\begin{equation}\label{eq:case2_main_bound}
\w\big(\nbr^{\star,i}(\cS[\uparrow \iletter,h],\cJ)\big)
\ge
(1-\varepsilon)t_i
-(1-\varepsilon)\frac{\varepsilon t_i}{4k}
-\frac{(1-\varepsilon)\varepsilon t_i'}{4}\ln(k-1).
\end{equation}
We now prove that the error terms can be absorbed. We show:
\begin{equation}\label{eq:combined-goal}
-(1-\varepsilon)\frac{\varepsilon t_i}{4k}
-\frac{(1-\varepsilon)\varepsilon t_i'}{4}\ln(k-1)
\;\ge\;
-\frac{(1-\varepsilon)\varepsilon t_i}{4}\ln k.
\end{equation}
Dividing both sides of \eqref{eq:combined-goal} by the positive scalar $\frac{(1-\varepsilon)\varepsilon}{4}$ and rearranging, it is equivalent to prove:
\begin{equation}
\frac{t_i}{k} \;\le\; t_i\ln k - t_i'\ln(k-1).
\label{eq:divided}
\end{equation}
Let $D := t_i - t_i' \ge 0$. The right-hand side (RHS) of \eqref{eq:divided} expands as:
\begin{align}
t_i\ln k - t_i'\ln(k-1)
&= t_i\ln k - (t_i - D)\ln(k-1) \nonumber \\
&= t_i\big(\ln k - \ln(k-1)\big) + D\ln(k-1). \label{eq:rhs-expand}
\end{align}
Since $D\ln(k-1)\ge0$ (for $k \ge 2$), we obtain the lower bound:
\begin{equation}
t_i\ln k - t_i'\ln(k-1) \;\ge\; t_i\big(\ln k - \ln(k-1)\big)
= t_i\ln\!\frac{k}{\,k-1\,}.
\label{eq:rhs-lower}
\end{equation}
Hence, to prove \eqref{eq:divided}, it suffices to show the stronger inequality:
\begin{equation}
\frac{t_i}{k} \;\le\; t_i\ln\!\frac{k}{k-1},
\label{eq:reduce}
\end{equation}
which, after cancelling the positive factor $t_i$, reduces to verifying:
\begin{equation}
\frac{1}{k} \;\le\; \ln\!\frac{k}{k-1}.
\label{eq:final-check}
\end{equation}
To prove \eqref{eq:final-check}, set $x:=\frac{1}{k-1}$ (so $x>0$ for $k\ge2$). Then
$\ln\!\frac{k}{k-1}=\ln(1+x)$.
We use the standard inequality $\ln(1+x) \ge \frac{x}{1+x}$ for $x \ge 0$.
Substituting $x=\frac{1}{k-1}$ into this inequality yields:
\[
\ln\!\frac{k}{k-1}=\ln(1+x)\ge\frac{x}{1+x}=\frac{1/(k-1)}{1+1/(k-1)}=\frac{1}{k-1+1}=\frac{1}{k},
\]
which is precisely \eqref{eq:final-check}.
Thus, \eqref{eq:combined-goal} is valid. Substituting this result into \eqref{eq:case2_main_bound} completes the proof for this case:
\[
\w\big(\nbr^{\star,i}(\cS[\uparrow \iletter, h], \cJ)\big)
\;\ge\;
(1-\varepsilon)t_i- \frac{(1-\varepsilon)\varepsilon t_i}{4}\ln(k).
\]

\noindent \textbf{Case $\textnormal{\textbf{B}}_{\textbf{2}}$:} There exists $i \in [r]$, such that  
\[
    \w\big((\nbr^i(\bO) \cap \nbr^i(h)) \setminus (\nfree^i \cup \free^i)\big)
    \geq \frac{\varepsilon}{2} \,
    \w\big(\nbr^i(h) \setminus (\nfree^i \cup \free^i)\big).
\]\\ 
In this case, consider the branch where the algorithm executes 
\Cref{algwconprbds:return_opt_elem} and returns $\cS$, 
where $\cS$ is the tuple returned by
\[
    \algwconprbds\big(
        (\graph, \w, \col, t', k, \mat, \varepsilon),\,
        \mi',\,
        \free \cup \nfree \cup \{u\}
    \big),
\]
where
\[
    t'_i = 
    \Big\{
        t_i - 
        \big(
            \w\big(\nfree^i(\cJ)(i)\big)
            + \w\big(u \cap \col^{-1}(i)\big)
        \big)
    \Big\}_{i \in [r]}.
\]

Also,
\[
    \mi' \gets 
    \langle 
        \mcon_1, \dots, 
        \mcon_{\beta} \cup \{u\}, \dots, 
        \mcon_k 
    \rangle,
\]
and the selected element $u$ belongs to $\nbr(o_{\beta})$ for some $\beta \in [k]$.

Let $\cJ$ denote the input instance
\[
    (\graph, \w, \col, t, k, \mat, \varepsilon),\, 
    \mi,\, 
    \free),
\]
and let $\cJ'$ denote the instance
\[
    (\graph, \w, \col, t', k, \mat, \varepsilon),\, 
    \mi',\, 
    \free \cup \nfree \cup \{u\}),
\]

which the algorithm recursively solves in this branch.
Note that in this branch,  $\bO= \langle o_1,\dots, o_k\rangle$ is a consistent tuple of $\cJ'$ since the selected element $u$ belongs to $\nbr(o_{\beta})$. We must show that $\w(\nbr^{\star,i}(\bO, \cJ')) \ge t'_i$ for all $i \in [r]$.
Let $N_i = \w\big(\nfree^i(\cJ)(i)\big)$ and $W_{u,i} = \w\big(u \cap \col^{-1}(i)\big)$. The threshold for $\cJ'$ is $t'_i = t_i - N_i - W_{u,i}$. By \Cref{lem:wconpropertytwoalgoone}, the coverage of $\bO$ in $\cJ$ and $\cJ'$ is related by:
\[
\w(\nbr^{\star,i}(\bO, \cJ)) = \w(\nbr^{\star,i}(\bO, \cJ')) + N_i + W_{u,i}.
\]
Since $\bO$ is a solution for $\cJ$, we have $\w(\nbr^{\star,i}(\bO, \cJ)) \ge t_i$. Rearranging the equation, we get:
\[
\w(\nbr^{\star,i}(\bO, \cJ')) = \w(\nbr^{\star,i}(\bO, \cJ)) - N_i - W_{u,i} \ge t_i - N_i - W_{u,i} = t'_i.
\]
Thus, $\bO$ satisfies the coverage requirement for $\cJ'$, and $\cJ'$ is a feasible instance. 

As a step towards our proof, we establish that \( \mu(\cJ') < \mu(\cJ) \), allowing us to apply the induction hypothesis on \( \cJ' \). Observe that $\mu(\cJ) - \mu(\cJ')$ is at least 
\begin{align}
	 & \left(k + k\intcmp - \sum_{\iletter \in [k]} \dintcmp(  G, \mcon_{\iletter}) \right) - \left(k + k\intcmp - \big(\sum_{\iletter \in [k]} \dintcmp(  G, \mcon_{\iletter})+1\big)\right)\nonumber\\
        & \textnormal{ (since }\dintcmp(G,\mcon_\iletter \cup \{u\})\geq \dintcmp(G, \mcon_\iletter)+1 \textnormal{ due to~\Cref{lem:wpropertytwoalgoone})} \nonumber\\
	&=1 >0  \label{measure:wconcaseC}
\end{align}

Let $\cS$ be a $(1-\varepsilon,\frac{(1-\varepsilon)\varepsilon t_i}{4}\ln{k})$-approximate tuple for $\cJ'$, then due to~\Cref{lem:wconpropertytwoalgoone}, $\cS$ is also a consistent tuple for $\cJ$.  Next, we show that it also satisfies the approximation guarantee. That is, we need to show that for every $i \in [r]$
\[
\w\big(\nbr^{\star,i}(\cS, \cJ)\big)
\;\ge\;
(1-\varepsilon)t_i- \frac{(1-\varepsilon)\varepsilon t_i}{4}\ln(k).
\]

Now since $\mu(\cJ') < \mu(\cJ)$, and $\bO$ is solution tuple for $cJ'$ by the induction hypothesis on $\cJ'$, we have:
\begin{equation}\label{eq:caseB2_indhyp}
    \w(\nbr^{\star,i}(\cS,\cJ')) \geq (1-\varepsilon)t_i'- \frac{(1-\varepsilon)\varepsilon t_i'}{4}\ln(k).
\end{equation}
By \Cref{lem:wconpropertytwoalgoone}, we have the following coverage relation:
\begin{equation}\label{eq:caseB2_cover}
\w(\nbr^{\star,i}(\cS, \cJ)) = \w(\nbr^{\star,i}(\cS, \cJ')) + N_i + W_{u,i}.
\end{equation}
Substitute the inductive bound \eqref{eq:caseB2_indhyp} into \eqref{eq:caseB2_cover}:
\[
\begin{aligned}
    \w(\nbr^{\star,i}(\cS, \cJ))
    &\;\ge\;
    \left[ (1-\varepsilon)t_i' - \frac{(1-\varepsilon)\varepsilon t_i'}{4}\ln(k) \right]
    + N_i + W_{u,i} \\
    &\;=\;
    (1-\varepsilon)(t_i - N_i - W_{u,i})
    - \frac{(1-\varepsilon)\varepsilon t_i'}{4}\ln(k)
    + N_i + W_{u,i} \\
    &\;=\;
    (1-\varepsilon)t_i - (1-\varepsilon)N_i - (1-\varepsilon)W_{u,i}
    + N_i + W_{u,i}
    - \frac{(1-\varepsilon)\varepsilon t_i'}{4}\ln(k) \\
    &\;=\;
    (1-\varepsilon)t_i + \varepsilon N_i + \varepsilon W_{u,i}
    - \frac{(1-\varepsilon)\varepsilon t_i'}{4}\ln(k).
\end{aligned}
\]
Since $N_i \ge 0$, $W_{u,i} \ge 0$, and $\varepsilon > 0$, the term $\varepsilon N_i + \varepsilon W_{u,i} \ge 0$. We can drop this non-negative term to find a lower bound:
\[
\w(\nbr^{\star,i}(\cS, \cJ))
    \;\ge\;
    (1-\varepsilon)t_i - \frac{(1-\varepsilon)\varepsilon t_i'}{4}\ln(k).
\]
From the definition $t_i' = t_i - N_i - W_{u,i}$, we have $t_i' \le t_i$. Since the loss term is negative, replacing $t_i'$ with $t_i$ provides a valid (weaker) lower bound:
\[
    - \frac{(1-\varepsilon)\varepsilon t_i'}{4}\ln(k)
    \;\ge\;
    - \frac{(1-\varepsilon)\varepsilon t_i}{4}\ln(k).
\]
Substituting this back, we obtain the final desired result:
\[
\begin{aligned}
    \w(\nbr^{\star,i}(\cS, \cJ))
    &\;\ge\;
    (1-\varepsilon)t_i - \frac{(1-\varepsilon)\varepsilon t_i'}{4}\ln(k) \\
    &\;\ge\;
    (1-\varepsilon)t_i - \frac{(1-\varepsilon)\varepsilon t_i}{4}\ln(k).
\end{aligned}
\]

This concludes the approximation analysis of the algorithm. 

\medskip

\noindent 
\underline{\textbf{Probability  Analysis:}}  We next prove the success probability of our algorithm. Just as we do in approximation analysis, we go through each case and analyze the desired probability of success.  For an input instance $$\cJ=(\wconprbdsi, \mi = \langle \mcon_1, \dots \mcon_{k} \rangle,\free)$$, we say that a tuple $\cQ$ is {\em good for $\cJ$} if $\w(\nbrst(\cQ,\cJ))$ is a $(1-\varepsilon,\frac{(1-\varepsilon)\varepsilon t_i}{4}\ln{k})$-approximate solution otherwise it is {\em bad for $\cJ$}.

\smallskip
\noindent 
\textbf{Case A: Analysis for the case when condition of ~\Cref{algwconprbds:basetwo} holds:}
 Let $\cJ$ denote the input instance 
 \[(\wconprbdsi, \mi = \langle \mcon_1, \dots, \mcon_{k} \rangle,\free)\] 
 and $\cJ'$ denote the instance
  \[ ((\graph,\w,\col,\,t',\,k-1,\,\mat/\{v\},\,\varepsilon),\,\mi[\downarrow\iletter],\,\free\cup\nfree\cup\nbr(v))\] where 
  \[
  t'=\left\langle t_i-\w\bigl(\nfree^i(\cJ)\bigr)
  -\w\bigl(\nbr^i(v)\setminus(\free^i(\cJ)\cup\nfree^i(\cJ))\bigr)
  \right\rangle_{i\in[r]}.
  \]
Let \( \cS \) be a consistent tuple for \( \cJ' \) returned in \Cref{algwprbds:return_red_rule} of the algorithm. By Equation \ref{measure:wconcaseA}, we know that $\mu(\cJ')< \mu(\cJ)$. We have also shown  $ \bO[\downarrow \iletter]$ is solution tuple for $\cJ'$. Thus, we can claim the following by induction hypothesis. Let $\mathbf{P}=\prob{\cS[\uparrow \iletter,v] \mbox{ is good for }  \cJ}$. Then
\begin{eqnarray*}
\mathbf{P} & = &  \prob{\cS[\uparrow \iletter,v] \mbox{ is good for }  \cJ \mid \cS \mbox{ is good for }  \cJ'}  
 \prob{\cS \mbox{ is good for }  \cJ'} \\
 && +   \prob{\cS[\uparrow \iletter,v] \mbox{ is good for }  \cJ \mid \cS \mbox{ is bad for }  \cJ'}  
 \prob{\cS \mbox{ is bad for }  \cJ'} \\
 &\geq & \prob{\cS[\uparrow \iletter,v] \mbox{ is good for }  \cJ \mid \cS \mbox{ is good for }  \cJ'}  
 \prob{\cS \mbox{ is good for }  \cJ'} \\
 & =& 1 \times  \prob{\cS \mbox{ is good for }  \cJ'} \\
& \geq  &  \left( \frac{\varepsilon}{4k} \right)^{\mu(\cJ')} (\textnormal{by induction hypothesis})\\
& \geq &  \left( \frac{\varepsilon}{4k} \right)^{\mu(\cJ)}
 \end{eqnarray*}

\smallskip
\noindent 
\textbf{Case B: When \Cref{algwconprbds:bagcolorguessed} is executed:}
Next, we will consider the case when we select a vertex  $h \in G'_{\varepsilon'}(\mathbf{v})$ such that $G'_{\varepsilon'}(\mathbf{v})$ contains $o_\iletter$ and $\bO[\downarrow \iletter][\uparrow \iletter, h]$ is independent in $\mat$. In this case the algorithm randomly selects the vertex $h$ with probability $\frac{1}{2}$, or a non-free neighbor of $h$, say $u$, with probability $\frac{\w(u)}{2\omega}$  (\Cref{algwconprbds:prob_dist}).  

Recall that \(\bo = \langle o_1, \dots, o_k \rangle\) is  an optimal tuple for the instance \(\cJ\). 
Now we consider two cases based on whether \\
($a$)~For every $i \in [r]$, we have 
\[
    \w\big((\nbr^i(\bO) \cap \nbr^i(h)) \setminus (\nfree^i \cup \free^i)\big)
    < \frac{\varepsilon}{2} \,
    \w\big(\nbr^i(h) \setminus (\nfree^i \cup \free^i)\big).
\]
($b$)~There exists $i \in [r]$, such that  
\[
    \w\big((\nbr^i(\bO) \cap \nbr^i(h)) \setminus (\nfree^i \cup \free^i)\big)
    \geq \frac{\varepsilon}{2} \,
    \w\big(\nbr^i(h) \setminus (\nfree^i \cup \free^i)\big).
\]

This corresponds to executing either \Cref{algwconprbds:return_high_deg} or
\Cref{algwconprbds:return_opt_elem}. Let $\cQ$ be the returned tuple, and let
$E_a$ and $E_b$ denote the events that $\cQ$ is good for $\cJ$ and that case
$(a)$ or $(b)$, respectively, holds. Since the two cases are exhaustive,
\[
\prob{\cQ\text{ is good for }\cJ}
=\prob{E_a}+\prob{E_b}
\ge \min\{\prob{E_a},\prob{E_b}\}.
\]
We will show that in either case  $\prob{\cQ \mbox{ is good for }  \cJ} \geq \left( \frac{\varepsilon}{4k} \right)^{\mu(\cJ)}.$

\medskip

\noindent 
\textbf{Case $\textnormal{\textbf{B}}_{\textbf{1}}$:} When ($a$) is satisfied \\
 In this case, consider the branch where the algorithm returns $\cS[\uparrow \iletter,h]$ in \Cref{algwconprbds:return_high_deg}, where  $\cS$ is the tuple returned by \[
    \algwconprbds\big(
        (\graph, \w, \col,\, t',\, k-1,\, \mat/\{h\},\, \varepsilon),\,
        \mi[\downarrow \iletter],\,
        \free \cup \nfree \cup \nbr(h)
    \big),
\]
where
\[
t' =
\begin{cases}
  t_i - \w(\nfree^i) - (1+\varepsilon)\,\w\big(\nbr^i(h)\setminus(\free^i\cup\nfree^i)\big), & \text{(Case 1)}\\[4pt]
   t_i - \w(\nfree^i) - \frac{\varepsilon t_i}{4k}-\frac{\varepsilon}{2}\w\big(\nbr^i(h)\setminus(\free^i\cup\nfree^i)\big), & \text{(Case 2)}.
\end{cases}
\]
Let $\cJ$ denote the input instance
\[
    (\wconprbdsi,\, 
      \mi = \langle \mcon_1, \dots, \mcon_k \rangle,\, 
      \free),
\]
and let $\cJ'$ denote the instance
\[
    ((\graph, \w, \col,\, t',\, k-1,\, \mat/\{h\},\, \varepsilon),\,
     \mi[\downarrow \iletter],\,
     \free \cup \nfree \cup \nbr(h)),
\]
which the algorithm recursively solves in this branch.
Let \( \cS \) be a consistent tuple for \( \cJ' \) returned in \Cref{algwconprbds:return_high_deg} of the algorithm. By Equation \ref{measure:wconcaseB}, we know that 
$\mu(\cJ')< \mu(\cJ)$. We have also shown  $ \bO[\downarrow \iletter]$ is a solution tuple for $\cJ'$. Thus, we can claim the following by induction hypothesis.  Let $\mathcal{E}$ denote the event that ($a$) is satisfied. Let $\mathbf{P}=\prob{\cS[\uparrow \iletter,h] \mbox{ is good for }   \cJ  \wedge \mathcal{E}}$

\begin{eqnarray*}
 \mathbf{P} & \geq & \prob{\cS[\uparrow \iletter,h]  \mbox{ is good for }  \cJ \wedge \mathcal{E} \mid \cS \mbox{ is good for }  \cJ'} \times\\ 
 & & \prob{\cS \mbox{ is good for }  \cJ'} \\
 & = & \prob{ \cS[\uparrow \iletter,h] \mbox{ is a } (1-\varepsilon, \frac{(1-\varepsilon)\varepsilon t_i}{4}\ln{k})\mbox{-approx sol} \wedge \mathcal{E} \mid \cS \mbox{ is good for }  \cJ'}\times \\ 
 & & \prob{\cS \mbox{ is good for }  \cJ'}\\
 & \geq & \prob{\clubsuit=h^\star }\times  \prob{\cS \mbox{ is good for }  \cJ'} \\
& \geq  &  \frac{1}{2}\left( \frac{\varepsilon}{4k} \right)^{\mu(\cJ')} (\textnormal{by induction hypothesis})\\
& \geq &  \left( \frac{\varepsilon}{4k} \right)^{\mu(\cJ)}
 \end{eqnarray*}

\medskip

\noindent 
\textbf{Case $\textnormal{\textbf{B}}_{\textbf{2}}$:} There exists $i \in [r]$, such that  
\[
    \w\big((\nbr^i(\bO) \cap \nbr^i(h)) \setminus (\nfree^i \cup \free^i)\big)
    \geq \frac{\varepsilon}{2} \,
    \w\big(\nbr^i(h) \setminus (\nfree^i \cup \free^i)\big).
\]\\ 
In this case, consider the branch where the algorithm chooses the $i$ that satisifes the above condition and executes 
\Cref{algwconprbds:return_opt_elem} where it returns $\cS$, 
where $\cS$ is the tuple returned by
\[
    \algwconprbds\big(
        (\graph, \w, \col, t', k, \mat, \varepsilon),\,
        \mi',\,
        \free \cup \nfree \cup \{u\}
    \big),
\]
where
\[
    t'_i = 
    \Big\{
        t_i - 
        \big(
            \w\big(\nfree^i(\cJ)(i)\big)
            + \w\big(u \cap \col^{-1}(i)\big)
        \big)
    \Big\}_{i \in [r]}.
\]

Also,
\[
    \mi' \gets 
    \langle 
        \mcon_1, \dots, 
        \mcon_{\beta} \cup \{u\}, \dots, 
        \mcon_k 
    \rangle,
\]
and the selected element $u$ belongs to $\nbr(o_{\beta})$ for some $\beta \in [k]$.

Let $\cJ$ denote the input instance
\[
    (\graph, \w, \col, t, k, \mat, \varepsilon),\, 
    \mi,\, 
    \free),
\]
and let $\cJ'$ denote the instance
\[
    (\graph, \w, \col, t', k, \mat, \varepsilon),\, 
    \mi',\, 
    \free \cup \nfree \cup \{u\}),
\]

which the algorithm recursively solves in this branch.

Let \( \cS \) be a consistent tuple for \( \cJ' \) returned in \Cref{algwconprbds:return_opt_elem} of the algorithm. By \Cref{measure:wconcaseC} we know that 
$\mu(\cJ')< \mu(\cJ)$. We have also shown  $ \bO$ is a solution tuple for $\cJ'$ in this case. Thus, we can claim the following by induction hypothesis.  Let $\mathcal{E}$ denote the event that ($b$) is satisfied. Let $\mathbf{P}=\prob{\cS \mbox{ is good for }   \cJ  \wedge \mathcal{E}}$
\begin{eqnarray*}
\mathbf{P} & \geq & \prob{\cS \mbox{ is good for }  \cJ \wedge \mathcal{E} \mid \cS \mbox{ is good for }  \cJ'} \times\\ 
 & & \prob{\cS \mbox{ is good for }  \cJ'} \\
 & = & \prob{\cS  \mbox{ is a } (1-\varepsilon, \frac{(1-\varepsilon)\varepsilon t_i}{4}\ln{k})\mbox{-approx sol} \wedge \mathcal{E} | \cS \mbox{ is good for }  \cJ'}\times \\ 
 & & \prob{\cS \mbox{ is good for }  \cJ'}\\
   &\geq &   \prob{\clubsuit= u \in \{u_1, \dots, u_\ell\} \cap(\nbr^i(\bO)\setminus (\free^i \cup \nfree^i ))\wedge u\in \nbr(o_\alpha)}\times \\ & & \prob{\cS \mbox{ is good for }  \cJ'} \\
& \geq  &    \prob{\clubsuit= u \in \{u_1, \dots, u_\ell\} \cap(\nbr^i(\bO)\setminus (\free^i \cup \nfree^i ))\wedge u\in \nbr(o_\alpha)}\times \\ & & \left( \frac{\varepsilon}{2k} \right)^{\mu(\cJ')} \\ & & (\textnormal{by induction hypothesis})\\
& \geq &  \frac{\varepsilon \omega}{2.2 \omega} \frac{1}{k}\left( \frac{\varepsilon}{4k} \right)^{\mu(\cJ')}\\
& \geq & \left( \frac{\varepsilon}{4k} \right)^{\mu(\cJ)}
 \end{eqnarray*}
 Thus, we have shown that in either case 
\[
\Pr\bigl[\cQ \mbox{ is good for } \cJ\bigr] \ge \left(\frac{\varepsilon}{4k}\right)^{\mu(\cJ)}.
\]
This concludes the probability analysis.

\noindent\underline{\textbf{Running Time Analysis:}} 
Let $T(\mu(\cJ))$ denote the running time of \algwconprbds for an instance with measure $\mu(\cJ)$. 
The algorithm branches on the number of bags 
$
L=\left\lceil \tfrac{6 \ln k}{(\varepsilon/2)^2} \right\rceil^{r},
$
the number of vertices in the $(k-1)$ representative sets of each bag (which is $k$, by \Cref{lem:oraclerepset}), 
and the number of colors $r$. 
The running time of the \buck procedure is linear in the size of the input, 
since it is proportional to the sum of degrees of the vertices in $\cR$. 
Similarly, the $(k-1)$ representative sets can be computed in time linear in the number of vertices, 
as shown in \Cref{lem:oraclerepset}. 
Hence, the total running time of the non-recursive part of each call is $\cO(\sizeg)$, 
following the linear dependence on input size established in the proof of \Cref{thm:mainUnweighted}. 
Thus, we obtain the recurrence
\[
T(\mu(\cJ)) \;=\; Lkr\,T(\mu(\cJ)-1) + \cO(\sizeg),
\qquad T(0)=1.
\]
Substituting the value of $L$ and solving the recurrence, we get
\[
\begin{aligned}
T(\mu(\cJ))
&= \cO\!\left(\sizeg\,(Lkr)^{\mu(\cJ)}\right)
= \cO\!\left(\sizeg\,\Big(k r \Big\lceil \tfrac{6\ln k}{(\varepsilon/2)^2}\Big\rceil^{r}\Big)^{\mu(\cJ)}\right) \\[4pt]
&=  \left(\frac{kr\ln(k)}{\varepsilon}\right)^{\,\cO(r\mu(\cJ))}\sizeg,
\end{aligned}
\]
This completes the proof of the lemma.
\end{proof}

To boost the success probability, we run \algwconprbds (\Cref{alg:framework-wconprbds}) independently \( \left( \frac{4k}{\varepsilon} \right)^{\mu(\cJ)} \) times and return the tuple \( \cS \) with the maximum value of \( \w(\nbr(\cS)) \). The probability that all runs fail is at most
\[
\left(1 - \left( \frac{\varepsilon}{4k} \right)^{\mu(\cJ)} \right)^{\left( \frac{4k}{\varepsilon} \right)^{\mu(\cJ)}} \leq \frac{1}{e},
\]
so the algorithm succeeds with probability at least \( 1 - \frac{1}{e} \). This gives us the following lemma. Note that we refer to the new algorithm with boosted success probability by \algwconprbds also.

\begin{lemma} \label{lem:framework-wconprbdsth}
For a given input instance $$\cJ = (\wconprbdsi, \mi = \langle \mcon_1, \dots, \\ \mcon_{k} \rangle, \free)$$ \algwconprbds runs in time 
\[ \left( \frac{4k}{\varepsilon} \right)^{\mu(\cJ)} \left(\frac{kr\ln(k)}{\varepsilon}\right)^{\,\cO(r\mu(\cJ))}\sizeg=  \left(\frac{kr\ln(k)}{\varepsilon}\right)^{\,\cO(r\mu(\cJ))}\sizeg \] 
and outputs a $(1-\varepsilon, \frac{(1-\varepsilon)\varepsilon t_i}{4}\ln{k})$-approximate solution  $\cS = \langle s_1, s_2, \dots, s_k \rangle$
with probability at least  \( 1 - \frac{1}{e} \). Here 
\[\mu(\cJ) = k + k\intcmp - \sum_{\iletter \in [k]} \dintcmp( G, \mcon_{\iletter})\]
\end{lemma}

We now state our main theorem.
\mainWeightedcon*
\begin{proof}
For $k=1$, the claim follows by checking every vertex of $\cR$, so assume
$k>1$. Given an instance $\cI$ of \wconprbds, set
\[
\alpha=\frac{\varepsilon}{1+\frac14\ln k}.
\]
We initialize the $k$-tuple $\mi$ by setting
$\mcon_{\iletter}=\emptyset$ for every $\iletter\in[k]$, and set
$\free=\emptyset$. We then run \algwconprbds on
\[
\cJ=\bigl((\graph,\w,\col,t,k,\mat,\alpha),
\langle\emptyset,\dots,\emptyset\rangle,\emptyset\bigr).
\]
A solution tuple for $\cI$ is also a solution tuple for $\cJ$. Hence, with
probability at least $1-1/e$, \Cref{lem:framework-wconprbdsth} returns a
$\bigl(1-\alpha,\frac{(1-\alpha)\alpha t_i}{4}\ln k\bigr)$-approximate
tuple $\cS$.

\begin{claim}
For $k>1$, $t_i>0$, and $\varepsilon\in[0,1]$, let
\[
\alpha=\frac{\varepsilon}{1+\frac14\ln k}.
\]
Then the inequality
\[
(1-\varepsilon)t_i \;\ge\; (1-\alpha)t_i \;-\; \frac{(1-\alpha)\alpha\, t_i}{4}\ln k
\]
holds.
\end{claim}

\begin{proof}
Divide by $t_i>0$ and set $c=\frac14\ln k$. Since
$(1+c)\alpha=\varepsilon$, we have
\[
(1-\varepsilon)-(1-\alpha)(1-c\alpha)
=c\alpha^2\ge0,
\]
which is precisely the desired inequality.
\end{proof}

We have the upper bound on the measure as follows
\[
\mu(\cJ) = k + k \cdot \intcmp - \sum_{\iletter \in [k]} \dintcmp(G, \mcon_{\iletter}) \leq k + k \Gamma,
\]
where \( \Gamma = \dintcmp(G) \). 
Since $1/\alpha=(1+\frac14\ln k)/\varepsilon$, the additional logarithmic
factor is absorbed in the exponent. Thus, by
\Cref{lem:framework-wconprbdsth}, the running time is at most
\[ \left(\frac{kr\ln(k)}{\varepsilon}\right)^{\,\cO(r\Gamma k)}\sizeg\]

\medskip

This completes the proof.
\end{proof}

%% file: constrained_maxsat.tex
\subsection{\wmscfull}
\label{sec: weightedconstrainedsat}

This section presents an \epas for \wmscfull (or \wmsc) for generic \cnf formulae. 
This is achieved by reducing the problem to \wmccfull (or \wmcc).  

\begin{tcolorbox}[float=!htp,colback=white!5!white,colframe=gray!75!black]
	\wmsc
	\\\textbf{Input:} An instance $\cI = (\Phi, \mat, r, f, \w, k, t)$, where
    \begin{itemize}
		\item $\Phi = (\cV, \cC)$ is a {\sf CNF-SAT} formula on $n$ variables and $m$ clauses,
        \item $\mat = (\cV, I)$ is a matroid defined on the set of variables $\cV$,
		\item $f: \cC \to [r]$ is a surjective function, and for each $j \in [r]$, we say that $\cC_j \coloneqq \{ c \in \cC: f(c) = j \}$ is the set of clauses of \emph{color} $j$,
        \item $\w: \cC \rightarrow \bQ^{+}$ is a weight function on the clauses,
		\item $k$ is a non-negative integer, and
        \item for each color $j \in [r]$, $t_j \coloneqq t(j) \in \bQ^{+}$ is a 
        \emph{coverage requirement}. 
    \end{itemize}
	\textbf{Question:} Does there exist an assignment $\beta$ of weight at most $k$ to $\cV$, such that
    \begin{itemize}
        \item $\beta^{-1}(1) \in I$, and
        \item for each $j \in [r]$, $\w(\sat{\Phi_j}{\beta}) \ge t_j$?
    \end{itemize}
         
	Here, $\sat{\Phi_j}{\beta} \subseteq \cC_j$ is the set of clauses of color $j \in [r]$ that are satisfied by $\beta$ in the subformula $\Phi_j = (\cV, \cC_j)$. Recall that the weight of an assignment $\beta$ is the number of variables it sets to $1$, i.e. $|\beta^{-1}(1)|$.
\end{tcolorbox}

\medskip
\noindent 
We give a procedure that reduces an instance of \wmsc to a family of \wmcc instances. This reduction is deterministic and for any $\varepsilon > 0$, runs in time $q(k, r, \varepsilon)$ for some computable function $q$. Furthermore, the reduction is robust. It extends naturally to the weighted setting and preserves approximation guarantees up to a multiplicative factor of $(1 - \varepsilon)$. Furthermore, this reduction is robust enough to accommodate fairness constraints on the clauses and matroid constraints on the variables.

This establishes the equivalence of \wmsc and \wmcc in the context of parameterized approximation schemes, which enables us to utilize the results obtained for \wmcc to solve \wmsc. We state the definition of \wmcc below for the reader's convenience.

\begin{tcolorbox}[float=!htp,colback=white!5!white,colframe=gray!75!black]
	\wmcc
	\\\textbf{Input:} An instance $\cJ = ((\cU, \cF), \mat, r, f, \w, k, t)$, where
    \begin{itemize}
		\item  $(\cU, \cF)$ is a set system on $n$ elements and $m$ sets,
        \item $\mat = (\cF, I)$ is a matroid defined on the family $\cF$,
		\item $f: \cU \to [r]$ is a surjective function, and for each $j \in [r]$, we say that $\cU_j \coloneqq \{ e \in \cU: f(e) = j \}$ is the set of elements of \emph{color} $j$,
        \item $\w: \cU \rightarrow \bQ^{+}$ is a weight function on the elements,
		\item $k$ is a non-negative integer, and
        \item for each color $j \in [r]$, $t_j \coloneqq t(j) \in \bQ^{+}$ is a \emph{coverage requirement}. 
    \end{itemize}
	\textbf{Question:} Does there exist a subfamily $\cF' \subseteq \cF$ of size at most $k$, such that
    \begin{itemize}
        \item $\cF' \in I$, and
        \item elements covered by $\cF'$ in $\cU_j$ has weight at least $t_j$?
    \end{itemize}
\end{tcolorbox}

\paragraph{Notations.} We recall our notations for {\sf SAT} here for convenience.
Let $\Phi$ be a \cnf formula on $n$ variables and $m$ clauses. Variables and clauses in $\Phi$ are denoted by $\cV_{\Phi}$ and $\cC_{\Phi}$. When the context is clear, the subscripts are omitted. A variable is said to be positive if it doesn't occur as a negative literal in the formula $\Phi$, otherwise it is a negative variable. We denote $\cV_{neg} \subseteq \cV$ as the set of negative variables in the formula. Let $\cC_{neg} \subseteq \cC$ be the set of clauses with at least one negative literal. Then, $\cC_{pos} = \cC \setminus \cC_{neg}$ is the set of clauses that do not contain any negative literal.

For a clause $c \in \cC_{\Phi}$, let $\n{c}$ denote the set of variables that appear as negative literals in $c$. For a set of clauses $C \subseteq \cC_{\Phi}$, we define
\[
\n{C} = \bigcup_{c \in C} \n{c}.
\]

Given a weight function \(\w: \cC_{\Phi} \to \bQ^+\), the weight of a set of clauses \(C \subseteq \cC_{\Phi}\) is defined as
\[
\w(C) = \sum_{c \in C} \w(c).
\]

An assignment $\psi$ is a binary function on $\cV_{\Phi}$, i.e., $\psi: \cV_{\Phi} \rightarrow \{0, 1\}$. The \emph{weight} of an assignment $\psi$, denoted by $\wt{\Phi}{\psi}$, is the number of variables set to $1$ by $\psi$. For an assignment $\psi$, we denote by $\sat{\Phi}{\psi}$ the set of clauses of $\Phi$ satisfied by $\psi$. Similary, for $S \subseteq \cV$, we have $\sat{\Phi}{S}=\sat{\Phi}{\psi}$ where the assignment $\psi$ assigns $1$ only to the variables in $S$.
The incidence graph of $\Phi$, denoted by $G_{\Phi} = (P, Q, E)$, is the bipartite graph that has a vertex in $P$ for each variable in $\cV_{\Phi}$, a vertex in $Q$ for each clause in $\cC_{\Phi}$ and an edge $(u, v)$ if the variable corresponding to $u \in P$ occurs as a positive or negative literal in the clause corresponding to $v \in Q$.
\begin{tcolorbox}
[colback=red!5!white,colframe=gray!75!black]
For a formula \( \Phi = (\cV, \cC) \), we use \( n = |\cV| \), \( m = |\cC| \), and \( m^\star = \sum_{c \in \cC} |C| \) to denote the size of the variables, the number of clauses, and the total number of variable-clause incidences, respectively. Let $\w$ be the weight function defined on the set of clauses $\cC$ and $W_{\text{rep}} = \sum_{c \in \cC} \ln \w(c)$ be the number of bits required to represent the weights on the clauses. We use $\sizeinst$ to denote the number of bits required to store the instance, i.e., $m+n+m^\star+ W_{\text{rep}}$.
\end{tcolorbox}

\subsubsection{A Brief description of the reduction algorithm.} 

We briefly explain our algorithm for the unweighted case. The arguments follow similarly in the presence of a weight function on the set of clauses, which we prove in detail in our analysis. 

The algorithm takes a \wmsc instance as input with a constant $\varepsilon > 0$. It retains negative variables with {\em high} negative degree with respect to at least one color $j \in [r]$. Here, we refer to the negative degree of a variable with respect to color $j$ as the number of clauses in color class $j$ in which it is present as a negative literal, with the precise definition coming up later in the section. We show that for every $j \in [r]$, either this subset of variables is bounded by a function of $r$, $k$ and $\varepsilon$ or for any assignment of weight at most $k$, the coverage requirement of the color class is achieved. In the latter case, we can forget about such a  color class by removing the clauses belonging to it, as their coverage requirement is always met, and clause deletion is safe for matroid constraint. The remaining negative variables (ones with {\em low} negative degree with respect to every color $j$) are transformed into positive variables by removing clauses that contain them as negative literals. This procedure is safe because we show that any assignment of weight at most $k$ to these low negative degree variables can unsatisfy only a small fraction of these deleted clauses, in fact, even a small fraction of the clauses satisfied by any assignment of weight at most $k$ that satisfies the coverage requirement for all colors. In fact, this procedure is also safe in regards to the matroid constraint as only clauses get deleted and the matroid constraint is over the variables. The only negative variables at this point in the formula are the ones that have high negative degree with respect to at least one color $j \in [r]$ and are bounded in size. To deal with the high negative degree variables, since they are of bounded size, we guess all possible assignments of weight at most $k$ to this set. For each such assignment to the high negative degree variables such that set of variables assigned $1$, say $Q$, form an independent set in the input matroid, we remove clauses that are satisfied by this assignment and the resulting formula is essentially a \wmcc instance. The matroid defined for the reduced instance is compatible with $Q$, i.e., a subfamily is independent in the matroid of the reduced instance only if the corresponding set of variables forms an independent set along with $Q$ in the matroid of the input instance. Since we create a \wmcc instance for each assignment to the high negative degree variables, we get a family of \wmcc instances of size bounded by $g(r, k, \varepsilon)$ for some computable function $g$.
Furthermore, we show that there is at least one instance in this family of reduced instances such that an approximate solution to it can be used to construct an approximate solution to our input \wmsc instance. 

We use the concept of {\em weighted normalized negative degree with respect to colors} defined below in our reduction algorithm.
This has been inspired by a similar concept in \cite{DBLP:journals/tcs/Manurangsi25}

\begin{definition}[Weighted normalized negative degree with respect to colors]
Let $\Phi = (\cV, \cC)$ be a \cnf formula, where $\cC = \biguplus_{j \in [r]} \cC_j$ and $\w: \cC \rightarrow \bQ^{+}$ is a weight function on clauses. Let $\wtilde{\cV}$ be a variable subset in $\cV$. The weighted normalized negative degree of a variable $x \in \wtilde{\cV}$ with respect to the subset $\wtilde{\cV}$ and a color $j \in [r]$, denoted by $\wndc_{\wtilde{\cV}, j}(x)$, is defined as follows.
\[
\wndc_{\wtilde{\cV}, j}(x) = \sum_{\substack{c \in \cC_{j} \\ x \in \n{c}}} \frac{\w(c)}{|\n{c} \cap \wtilde{\cV}|}
\] 
\end{definition}

We begin by applying the following reduction rule to bound the number of negative literals in a negative clause. 

\begin{redrule}
\label{rr: clargeneglit}
Let $\cI = (\Phi, \mat, r, f, \w, k, t)$ be an instance of \wmsc. If there exists a clause $c \in \cC_j$, for some $j \in [r]$, such that $|\n{c}| \geq k+1$, then remove clause $c$ from $\Phi$. Return the instance $\cI' = (\Phi' = (\cV, \cC'), \mat, r, f, \w', k, t')$, where 
\begin{itemize}
    \item $\cC' = \cC \setminus \{c\}$,
    \item $\w'$ is $\w$ restricted to $\cC'$, and
    \item $t'_j = \max\{0,t_j-\w(c)\}$, and $t'_i=t_i$ for every
    $i\in[r]\setminus\{j\}$.
\end{itemize}  
\end{redrule}

Reduction Rule~\ref{rr: clargeneglit} is safe: every assignment of weight at
most $k$ sets at least one variable in $\n{c}$ to $0$, and hence satisfies
$c$. Its weight can therefore be subtracted from the target for its color.

A complete description of the algorithm with pseudocode is provided in Algorithm~\ref{alg:framework-wcsat}.
        
\medskip
\begin{algorithm}[!htp]

\caption{\algwcsat}\label{alg:framework-wcsat}
\KwInput{An instance $\cI=(\Phi,\mat,r,f,\w,k,t)$ of \wmsc and
$\varepsilon\in(0,1)$.}
\KwOutput{A family $\cR$ of \wmcc instances, one of whose
$(1-\varepsilon/2)$-approximate solutions lifts to a
$(1-\varepsilon)$-approximate solution to $\cI$.}
Apply~\Cref{rr: clargeneglit} exhaustively.\label{redwms: crrapply}\\
Set $Y_{\text{high}} = \emptyset$ and $\cL = \emptyset$.\\
\For{$j = 1$ to $r$\label{redwms: cstartfor}}
{
    Set $\tau_j = \frac{\varepsilon}{2k} \cdot t_j$.\\
    Let $Y_0^j = \cV$ and $i = 1$.\\
    \While{there exists a variable $x_{i} \in Y_{i-1}^j$ such that $\wndc_{Y_{i-1}^j, j}(x_{i}) > \tau_j$\label{redwms: cgreedybegin}} 
    {
        Set $Y_i^j = Y_{i-1}^j \setminus \{x_i\}$ and $i = i+1$.
    }
    Set $Y_{\text{high}}^j = \cV \setminus Y_i^j$ and $Y_{\text{low}}^j = Y_i^j$ \label{redwms: cvarpart}.\\
    \If{$|Y_{\text{high}}^j| < \frac{2k^2}{\varepsilon} + k$}
    {
        $Y_{\text{high}} = Y_{\text{high}} \cup Y_{\text{high}}^j$
    }
    \Else 
    {
        $\cL = \cL \cup \{j\}$ 
    }       
}\label{redwms: cendfor}
Set $Y_{\text{low}} = \bigcap_{j \in [r] \setminus \cL} Y_{\text{low}}^j$. \\
Let $\cC'=\{c\in\cC\mid\n{c}\cap Y_{\text{low}}\ne\emptyset\}$ and
$\cC_{\text{del}}=\cC'\cup\bigcup_{j\in\cL}\cC_j$.
\label{redwms: cdelclause}\\
Let $Z = \cV_{neg} \setminus Y_{\text{low}}$. \label{redwms: cz_bound}\\
Set $\cR \gets \emptyset$.\\
\For{each assignment $\beta$ to $Z$ of weight at most $k$ such that $\beta^{-1}(1) \in I$\label{redwms: cguess_assgn}}
{
Let $\cC_Z = \sat{\Phi'}{\beta}$, where $\Phi' = (Z, \cC \setminus \cC_{\text{del}})$.\\
Let $\wtilde{\Phi} = (\cV_{\wtilde{\Phi}}, \cC_{\wtilde{\Phi}})$, where $\cV_{\wtilde{\Phi}} = \cV \setminus Z$ and $\cC_{\wtilde{\Phi}} = \{ c \setminus Z \mid c \in \cC \setminus (\cC_{\text{del}} \cup \cC_Z)\}.$\\
{Create an instance of \wmcc $((\cU, \cF), \wtilde{\mat}, \wtilde{r}, \wtilde{f}, \wtilde{\w}, \wtilde{k}, \wtilde{t})$ as follows:\label{redwms: creducetowmc}
\begin{description}
\setlength{\itemsep}{-1pt}
      \item[$\star$] For each clause $c \in \cC_{\wtilde{\Phi}}$, we have an element $e_c \in \cU$, set $\wtilde{\w}(e_c) = \w(c),$ and $\wtilde{f}(e_c) = f(c),$\\
      \item[$\star$] For each variable $x \in \cV_{\wtilde{\Phi}}$, we have a set $h_x \in \cF$, where $h_x = \{e_c \in \cU \mid \text{clause } c \text{ contains variable } x\},$\\
      \item[$\star$] Let $Q=\beta^{-1}(1)$. Define
      $\wtilde{\mat}=(\cF,\wtilde I)$ by declaring $\cA\subseteq\cF$
      independent if
      $Q\cup\{x\in\cV\setminus Z\mid h_x\in\cA\}\in I$.
      Equivalently, $\wtilde{\mat}$ is the restriction of $\mat/Q$ to
      $\cV\setminus Z$.\\
      \item[$\star$] Set $\wtilde{k}=k-|Q|$ and $\wtilde{r}=r-|\cL|$.\\
      \item[$\star$] Restrict the colors to $[r]\setminus\cL$ and relabel them
      as $[\wtilde r]$. For each retained color $j$, set
      $\wtilde{t}_j=\max\!\left\{0,t_j-
      \frac{\w(\sat{\Phi'_j}{\beta})}{1-\varepsilon/2}\right\}$.
\end{description}}
$\cJ_\beta\gets((\cU,\cF),\wtilde{\mat},\wtilde r,\wtilde f,
\wtilde{\w},\wtilde k,\wtilde t)$\\
$\cR\gets\cR\cup\{\cJ_\beta\}$\\
}
\Return $\cR$\\
\end{algorithm}

\subsubsection{Algorithm Analysis}

We begin by proving some claims and lemmas that will help us in establishing our desired result.

\begin{claim}
\label{clm: wcnndeg_lowerbound}
    For any variable $y \in Y^j_{\text{high}}$, $\wndc_{Y^{j}_{0}, j}(y) \geq \frac{\tau_j}{k}$.
\end{claim}

\begin{proof}
    Let $x_{i} \in \cV$ be the variable that is selected in Step~\ref{redwms: cgreedybegin} of Algorithm~\ref{alg:framework-wcsat} during the $i^{\text{th}}$ iteration of the while loop when the color considered is $j \in [r]$. Then, its normalized negative degree with respect to the subset $Y^j_{i-1}$ and color $j$ is at least $\tau_j$, i.e., $\wndc_{Y^{j}_{i-1}, j}(x_{i}) > \tau_j$. 
    
    Due to Reduction Rule~\ref{rr: clargeneglit}, every clause in the formula contains no more than $k$ negative literals. Hence, $\frac{\w(c)}{|\n{c} \cap Y|} \in [\frac{\w(c)}{k}, \w(c)]$ for every clause $c$ that contains $x_i$ as a negative literal. Moreover, the normalized negative degree of a variable $x_i$ is non-decreasing. This is because the while loop progresses as a negative variable is removed from the subset $Y^j_{\ell}$ during the $\ell^{\text{th}}$ iteration of the while loop, where $\ell \in [i]$. Thus, if the normalized negative degree of a variable $x_i$ with respect to color $j$ at the beginning of the greedy procedure, i.e., with respect to subset $Y^j_0$, is $a_0$, then its normalized negative degree with respect to color $j$ at the time of being chosen, i.e., with respect to subset $Y^j_{i-1}$, is at most $a_0 \cdot k$. This implies that $a_0 \geq \frac{\tau_j}{k}$.
\end{proof}

\begin{claim}
\label{clm: csuffnegvar}
If, for some $j\in[r]$,
$|Y^j_{\text{high}}|\ge \frac{2k^2}{\varepsilon}+k$, then every assignment of
weight at most $k$ satisfies clauses in $\cC_j$ of total weight at least $t_j$.
\end{claim}

\begin{proof}
    Any assignment of weight at most $k$ to $\cV$ can set at most $k$ variables to $1$, implying that at least $\frac{2k^2}{\varepsilon}$ variables in $Y^j_{\text{high}}$ get assigned $0$. We now show that the total weight of clauses satisfied by any $\frac{2k^2}{\varepsilon}$ variables from $Y^j_{\text{high}}$, when set to $0$, is at least $t_j$. By Claim~\ref{clm: wcnndeg_lowerbound}, we know that for a variable $x \in Y^j_{\text{high}}$, its weighted normalized negative degree with respect to color $j$ at the beginning of the greedy procedure with respect to color $j$ is at least $\frac{\tau_j}{k}$. Let $\wtilde{Y}$ be a subset of any $\frac{2k^2}{\varepsilon}$ variables from $Y^j_{\text{high}}$.
    \[
    \sum_{x \in \wtilde{Y}} \wndc_{Y^j_0, j}(x) \geq \frac{2k^2}{\varepsilon} \cdot \frac{\tau_j}{k} = t_j
    \]
    Therefore, the total weight of clauses satisfied when the variables in
    $\wtilde Y$ are set to $0$ is
    \begin{align*}
    \w\!\left(\bigcup_{x \in \wtilde{Y}}
    \{c \in \cC_j \mid c \text{ is satisfied because }x=0\}\right)
    & = \sum_{x \in \wtilde{Y}} \sum_{\substack{c \in \cC_j\\x \in \n{c}}} \frac{\w(c)}{|\n{c} \cap \wtilde{Y}|} \\
    & \stackrel{(\diamond)}{\geq} \sum_{x \in \wtilde{Y}} \sum_{\substack{c \in \cC_j\\x \in \n{c}}} \frac{\w(c)}{|\n{c} \cap Y^j_0|} \\
    & = \sum_{x \in \wtilde{Y}} \wndc_{Y^j_0, j}(x) \\
    & \geq t_j
    \end{align*}
    
$(\diamond)$ holds due to the fact that $\wtilde{Y} \subseteq Y^j_0 = \cV$.
\end{proof}

\begin{lemma}
\label{lem: cunsatclausebound}
Let $\cC_{\text{del}}$ be the set of clauses in Step~\ref{redwms: cdelclause} of Algorithm~\ref{alg:framework-wcsat} that is deleted and let $\beta$ be any assignment of weight at most $k$ to $\cV$. Then, for each $j \in [r] \setminus \mathcal{L}$, the sum of weights of clauses among $\cC_{\text{del}, j}$ that is unsatisfied by $\beta$ is at most $\frac{\varepsilon}{2} \cdot t_j$. Here, $\cC_{\text{del}, j}$ is the subset of $\cC_{\text{del}}$ restricted to clauses of color class $j$.
\end{lemma}

\begin{proof}
    Consider the color $j \in [r] \setminus \mathcal{L}$. 
    
    Let $\cC_{\text{del}, j}^{\text{unsat}(\beta)}$ be the set of clauses among $\cC_{\text{del}, j}$, the clauses deleted by the algorithm from $\cC_j$, that is unsatisfied by $\beta$. 
    Notice that any clause $c \in \cC_{\text{del}, j}$ is unsatisfied because $\n{c} \subseteq \beta^{-1}(1)$. Moreover, any clause $c$ is in $\cC_{\text{del}, j}^{\text{unsat}(\beta)}$ because $\n{c} \cap Y^{j}_{\text{low}} \neq \emptyset$. Let $Y^{j}_{\text{low}}$ be the set of variables that are retained after the while loop ends in Step~\ref{redwms: cvarpart} of Algorithm~\ref{alg:framework-wcsat}. Then, $Y^{j}_{\text{low}}$ is the set of variables that have low normalized negative degree with respect to itself and color $j$.

    \begin{align*}
        \w(\cC_{\text{del}, j}^{\text{unsat}(\beta)}) & = \sum_{c \in \cC_{\text{del}, j}^{\text{unsat}(\beta)}} \w(c) \\
        & = \sum_{c \in \cC_{\text{del}, j}^{\text{unsat}(\beta)}} \sum_{x \in (\n{c} \cap Y^{j}_{\text{low}})} \frac{\w(c)}{|\n{c} \cap Y^{j}_{\text{low}}|} \\
        & = \sum_{x \in (\beta^{-1}(1) \cap Y^{j}_{\text{low}})} \sum_{\substack{c \in \cC_{\text{del}, j}^{\text{unsat}(\beta)}\\ x \in \n{c}}}  \frac{\w(c)}{|\n{c} \cap Y^{j}_{\text{low}}|} \\
        & = \sum_{x \in (\beta^{-1}(1) \cap Y^{j}_{\text{low}})} \wndc_{Y^{j}_{\text{low}}, j}(x) \\
        & \leq \sum_{x \in (\beta^{-1}(1) \cap Y^{j}_{\text{low}})} \tau_j \\
        & \leq k \tau_j = \frac{\varepsilon}{2}t_j \tag{as $\beta$ has weight at most $k$}
    \end{align*}
\end{proof}

\begin{observation}
\label{obs: chighnegvarbound}      
The set of negative variables with high normalized negative degree with respect to at least one color $j \in [r] \setminus \mathcal{L}$, denoted by $Y_{\text{high}}$, contains at most $\mathcal{O}(\frac{rk^2}{\varepsilon})$ variables.

The negative variables in the formula of the reduced instance are only the variables in $Y_{\text{high}}$ since all the clauses with negative occurrences of variables in $Y_{\text{low}}$ are deleted.
\end{observation}

\begin{lemma}
\label{lem: capprox_preserve}
    Let $\cI = (\Phi, \mat, r, f, \w, k, t)$ be an instance of \wmsc and $\cR$ be the set of \wmcc instances returned by \Cref{alg:framework-wcsat} on input $\cI$. Then, there exists at least one \wmcc instance $\cJ = ((\cU, \cF), \wtilde{\mat}, \wtilde{r}, \wtilde{f}, \wtilde{\w}, \wtilde{k}, \wtilde{t})$ in $\cR$ such that given a $\left( 1-\frac{\varepsilon}{2} \right)$-approximate solution to $\cJ$, we can compute a $(1-\varepsilon)$-approximate solution to $\cI$ of \wmsc.
\end{lemma}

\begin{proof}
Let $\beta^{\star}$ be an assignment of weight at most $k$ to $\cI$ such that
$(\beta^{\star})^{-1}(1)\in I$ and
$\w(\sat{\Phi_j}{\beta^{\star}})\ge t_j$ for each $j\in[r]$.

Let $Z$ be the set of remaining negative variables in $\Phi$ that we get in \Cref{redwms: cz_bound} of the reduction algorithm. Clearly, $\beta^{\star}\restriction_Z$, i.e. the assignment $\beta^{\star}$ restricted to the variables in $Z$, is one of the assignments guessed in \Cref{redwms: cguess_assgn} of the algorithm. Let $\cJ_{\beta^{\star}\restriction_Z} = ((\cU, \cF), \wtilde{\mat}, \wtilde{r}, \wtilde{f}, \wtilde{\w}, \wtilde{k}, \wtilde{t})$ be an instance of \wmcc corresponding to $\beta^{\star}\restriction_Z$ that is constructed by the algorithm. Thus, we know that $\cJ_{\beta^{\star}\restriction_Z} \in \cR$. 

Let $h_\cS$ be a $\left(1-\frac{\varepsilon}{2} \right)$-approximate solution to $\cJ_{\beta^{\star}\restriction_Z}$. Then, $h_\cS$ has size at most $\wtilde{k}$, is independent in $\wtilde{\mat}$, and covers elements of each retained color $j$ of total weight at least $\left( 1-\frac{\varepsilon}{2} \right)\wtilde{t}_j$.

Let $\cV_{h_\cS}=\{x\in\cV\setminus Z\mid h_x\in h_\cS\}$, and define
the assignment $\sigma$ to $\cV$ as follows.
\[
\sigma(x) = \begin{cases} 
      \beta^{\star}\restriction_Z(x) & \text{if } x \in Z \\
      1 & \text{if } x \in \cV \setminus Z \text{ and the corresponding set } h_x \in h_\cS \\
      0 & \text{otherwise}
   \end{cases}
\]
 We claim that $\sigma$ is an assignment of weight at most $k$, $\sigma^{-1}(1) \in I$, and $\w(\sat{\Phi_j}{\sigma}) \geq (1 - \varepsilon)t_j$, for each $j \in [r]$. 
 
 Clearly, $\wt{\Phi}{\sigma} \leq \wtilde{k} + \wt{}{\beta^{\star}\restriction_Z} \leq k$. Moreover, as $h_\cS$ is an independent set in $\wtilde{\mat}$, this implies that $(\beta^{\star}\restriction_Z)^{-1}(1)  \cup \cS \in I$, where $\cS \subseteq \cV$ is the set of variables corresponding to the subfamily $h_\cS$. Thus, we have $\sigma^{-1}(1) \in I$.

We next show the approximation guarantee. Recall that $\cC'$ contains the
clauses with a negative occurrence of a variable in $Y_{\text{low}}$, and that
$\cC_{\text{del}}=\cC'\cup\bigcup_{j\in\cL}\cC_j$.
For any color $j \in \cL$, by \Cref{clm: csuffnegvar}, $\w(\sat{\Phi_j}{\sigma}) \geq t_j$. 
Now, consider the colors in $j \in [r] \setminus \cL$. Let $\sat{\Phi'_j}{\beta^{\star}\restriction_Z}$, where $\Phi'_j = (Z, \cC_j \setminus \cC_{\text{del}})$, be the set of clauses in $\cC_j \setminus \cC_{\text{del}}$ that is satisfied by the assignment $\beta^{\star}\restriction_Z$ to the variables in $Z$. Recall that $\wtilde{\Phi}_j = (\cV_{\wtilde{\Phi}_j}, \cC_{\wtilde{\Phi}_j})$, where 
\[\cV_{\wtilde{\Phi}_j} = \cV \setminus Z \text{ and } \cC_{\wtilde{\Phi}_j} = \cC_j \setminus (\cC' \cup \sat{\Phi'_j}{\beta^{\star}\restriction_Z}).\]
For any assignment $\psi$ to $\cV$ that agrees with $\beta^{\star}\restriction_Z$ and is of weight at most $k$, we have 
\begin{align*}
\w(\sat{\wtilde{\Phi}_j}{\psi}) + \w(\sat{\Phi'_j}{\beta^{\star}\restriction_Z}) + \w(\cC'_j) - \w(\cC'_{\text{unsat}(\psi)}) = \w(\sat{\Phi_j}{\psi}).
\end{align*}
This is because $\cC_{\wtilde{\Phi}_j}$, $\cC_{\Phi'_j}$, and $\cC'_j$
are pairwise disjoint families of clauses.

In particular, for any assignment $\psi^{\star}$ to $\cV$ that agrees with $\beta^{\star}\restriction_Z$, is of weight at most $k$, and satisfies $\w(\sat{\Phi_j}{\psi^{\star}}) \geq t_j$, for each $j \in [r]$, we have
\begin{equation}\label{eqn: ccon}
    \w(\sat{\wtilde{\Phi}_j}{\psi^{\star}}) + \w(\sat{\Phi'_j}{\beta^{\star}\restriction_Z}) + \w(\cC'_j) - \w(\cC'_{\text{unsat}(\psi^{\star})})  = \w(\sat{\Phi_j}{\psi^{\star}}) 
\end{equation}

Let $\cJ_{\beta^{\star}\restriction_Z}$ correspond to the \wmcc instance
created from $\wtilde{\Phi}$ in \Cref{redwms: creducetowmc}. Fix a retained
color $j$. If
$\w(\sat{\Phi'_j}{\beta^{\star}\restriction_Z})
\ge(1-\varepsilon/2)t_j$, then $\sigma$ already satisfies clauses of color $j$
of weight at least $(1-\varepsilon)t_j$. Otherwise $\wtilde t_j>0$, and the
$\left(1-\varepsilon/2\right)$-approximation $h_\cS$ ensures that $\sigma$
satisfies clauses in
$\cC_j\setminus(\cC_{\text{del}}\cup\cC_Z)$ of weight at least
$\left(1-\frac{\varepsilon}{2}\right)\wtilde t_j$, where
$\cC_Z=\sat{\Phi'}{\beta^{\star}\restriction_Z}$.

Thus, the sum of weights of clauses in $\cC_j$ satisfied by the assignment $\sigma$ is at least 
\begin{align*}
\w(\sat{\Phi_j}{\sigma}) & \stackrel{(\blacklozenge)}{=}
\w(\sat{\wtilde{\Phi}_j}{\sigma}) + \w(\sat{\Phi'_j}{\beta^{\star}\restriction_Z}) + \w(\cC'_j) - \w(\cC'_{\text{unsat}(\sigma)})\\
& \stackrel{(\diamondsuit)}{\geq} \left(1 - \frac{\varepsilon}{2} \right)\wtilde{t}_j + \w(\sat{\Phi'_j}{\beta^{\star}\restriction_Z}) + \w(\cC'_j) - \frac{\varepsilon}{2}t_j  \\
& = \left(1 - \frac{\varepsilon}{2} \right)\left(t_j - \frac{\w(\sat{\Phi'_j}{\beta^{\star}\restriction_Z})}{(1 - \frac{\varepsilon}{2})}\right) + \w(\sat{\Phi'_j}{\beta^{\star}\restriction_Z}) + \w(\cC'_j) - \frac{\varepsilon}{2}t_j \\
& \geq \left(1 - \frac{\varepsilon}{2} \right)t_j - \frac{\varepsilon}{2}t_j \\
& = (1 - \varepsilon)t_j
\end{align*}

Note that $(\blacklozenge)$ holds due to Equation~\ref{eqn: ccon} and $(\diamondsuit)$ holds because \Cref{lem: cunsatclausebound} implies that weight of clauses unsatisfied by any assignment to $Y_{\text{low}}$ of weight at most $k$ is at most $\frac{\varepsilon}{2}t_j$ in $\cC'_j$, and hence, the weight of satisfied clauses in $\cC'_j$ is at least $\w(\cC'_j) - \frac{\varepsilon}{2}t_j$. Moreover, as $\sigma$ agrees with $\beta^{\star}\restriction_Z$ on $Z$, it satisfies all clauses in $\sat{\Phi'_j}{\beta^{\star}\restriction_Z}$. This concludes our proof.
\end{proof}

Thus, we get our main result below.

\begin{theorem}
\label{thm: cwtred_sat_to_cov}
Given an instance $\cI = (\Phi, \mat, r, f, \w, k, t)$ of \wmsc, there exists a deterministic algorithm that, in time
\[
\left(\frac{rk^2}{\varepsilon}\right)^{\mathcal{O}(k)} \cdot \sizeinst,
\]
produces a family \( \mathcal{R} \) consisting of at most \( \left(\frac{rk^2}{\varepsilon}\right)^{\mathcal{O}(k)} \) instances of \wmcc. Moreover, there exists an instance \( \mathcal{J} \in \mathcal{R} \) such that, given a \( \left(1 - \frac{\varepsilon}{2} \right) \)-approximate solution for \( \mathcal{J} \), we can compute a \( (1 - \varepsilon) \)-approximate solution for \( \mathcal{I} \) in polynomial time.
\end{theorem}

\begin{proof}
The algorithm first applies \Cref{rr: clargeneglit} to remove all clauses with at least $k+1$ negative literals. Then, it greedily reduces the set of negative variables in the input formula in Steps \ref{redwms: cstartfor} to \ref{redwms: cendfor}. At the end of this greedy procedure, the set of negative variables with high weighted normalized negative degree with respect to at least one color either has bounded size as seen in \Cref{obs: chighnegvarbound}, or the coverage requirement is always satisfied with respect to that color, letting us remove that color class from further consideration. Whereas, the set of negative variables with low weighted normalized negative degree with respect to all colors are made positive by removing clauses that contain them as negative literals.

It then constructs a family $\cR$ of \wmcc instances. For each assignment $\eta$ of weight at most $k$ to $Z$ such that set of variables assigned $1$, say $Q$, form an independent set in the input matroid, it creates a \wmcc instance $\cJ_{\eta}$ as follows. It removes all clauses satisfied by $\eta$ among the remaining clauses. The resulting formula does not contain negative variables or clauses with negative literals in them, thus making it equivalent to a \wmcc instance. We use matroid contraction operation with $Q$ to define the matroid for the reduced instance to ensure that the solution to the reduced instance forms an independent set with $Q$ in the input matroid.

By \Cref{lem: capprox_preserve}, we know that there is at least one instance $\cJ \in \cR$ such that given a $\left( 1 - \frac{\varepsilon}{2}\right)$-approximate solution to $\cJ$, we can get a $(1 - \varepsilon)$-approximate solution for our input instance $\cI$. The correctness of the algorithm follows from the correctness of \Cref{rr: clargeneglit} and \Cref{lem: capprox_preserve}.

\medskip
\noindent 
\textbf{Running time:} Most of the steps of \Cref{alg:framework-wcsat} can be executed in polynomial time and the only overhead comes from \Cref{redwms: cguess_assgn} of the algorithm. In fact, we can compute the set $Z$ in time \( \cO\left(\left(\frac{rk^2}{\varepsilon}\right)^{\cO(1)} \sizeinst \right) \) using appropriate data structures and bookkeeping, as done in the proof of \Cref{thm:mainUnweighted} to achieve linear dependence on the input size. 
By \Cref{obs: chighnegvarbound}, $|Z| \leq \mathcal{O} \left(\frac{rk^2}{\varepsilon} \right)$, so there are at most
\[
\sum_{i=0}^k\binom{\cO(rk^2/\varepsilon)}{i}
\leq \left(\frac{rk^2}{\varepsilon}\right)^{\mathcal{O}(k)}
\]
assignments of weight at most $k$ to the variables in $Z$. For each assignment of weight at most $k$, a \wmcc instance is constructed in $\cO(\sizeinst)$ time. Hence, the total running time of the algorithm is $\left(\frac{rk^2}{\varepsilon}\right)^{\mathcal{O}(k)}\sizeinst$.
\end{proof}

Before proceeding further, we state and prove a claim that helps in bounding the downward intersection complexity of the incidence graph of any reduced \wmcc instance obtained by the reduction algorithm.
The incidence graph of a \wmcc instance $\cJ = ((\cU, \cF), \w, k)$ is a bipartite graph $G_{\cJ} = (P, Q)$ where $P$ contains a vertex for each set in $\cF$, $Q$ contains a vertex for each element in $\cU$ and an edge joins a set vertex and an element vertex if that element is present in the set.

\begin{claim}
\label{clm: cdi_reduced_bound}
Let $\cI = (\Phi, \mat, r, f, \w, k, t)$ be an instance of \wmsc and $\cR$ be a family of \wmcc instances that is output by \Cref{alg:framework-wcsat}. Then, for any instance $\cJ$ in $\cR$, $\dintcmp(G_{\cJ}) \leq \dintcmp(G_{\cI}) + 1$.
\end{claim}

\begin{proof}
Let $\cI = (\Phi, \mat, r, f, \w, k, t)$ be an instance of \wmsc, where $G_{\cI}$ or $G_{\Phi} = (P_{\cV}, Q_{\cC})$ is the incidence graph of $\Phi$.
Let $\cJ$ in $\cR$ be a reduced \wmcc instance, where $G_{\cJ}$ or $G_{(\cU, \cF)} = (P_{\cU}, Q_{\cF})$ is the incidence graph of $(\cU, \cF)$.   

Notice that every \wmcc instance $\cJ$ in $\cR$ corresponds to a subformula $\wtilde{\Phi} = (\wtilde{\cV}, \wtilde{\cC})$, which is obtained by removing variables and clauses from $\Phi$. Thus, $G_{\cJ} = G_{\wtilde{\Phi}}$, where $P_{\cU} = Q_{\wtilde{\cC}}$ and 
$Q_{\cF} = P_{\wtilde{\cV}}$.
Let the $\smlindex$ of $G_{\Phi}$ be $\lambda$ and the $\smlindex$ of $G_{\cJ}$ be $\lambda'$. 
It is easy to see that no step in the reduction algorithm involves removal of edges without removing both its end points in the incidence graph $G_{\Phi}$, i.e., an edge is removed only when one of its endpoints is also removed. Thus, $G_{\wtilde{\Phi}}$ is an induced subgraph of $G_{\Phi}$,
and hence $G_{\wtilde{\Phi}}$ cannot have a larger induced semi-ladder than $G_{\Phi}$. Thus, we have $\lambda' \leq \lambda$.
Due to \Cref{thm: smli_dic_rel}, we know that $\lambda \leq \dintcmp(G) \leq \lambda + 1$, for any graph $G$. 
Thus, we have $\dintcmp(G_{\cJ}) \leq \lambda' + 1 \leq \lambda + 1 \leq \dintcmp(G_{\Phi}) + 1$.
\end{proof}

Equipped with \Cref{thm: cwtred_sat_to_cov}, we now present our randomized \epas result for \wmsc.

\applicationwtconccsat*

\medskip

For an $\varepsilon > 0$, and an instance $\cI = (\Phi, \mat, r, f, \w, k, t)$ of \wmsc, we first run \Cref{alg:framework-wcsat} to obtain a family of $\left(\frac{rk^2}{\varepsilon}\right)^{\mathcal{O}(k)}$ \wmcc instances. The incidence graph of each instance in the family is a \wconprbds instance that is given as input to
\Cref{alg:framework-wconprbds} with $\wtilde{\varepsilon}$ set to $\frac{\varepsilon}{2}$.  
By \Cref{thm:mainWeightedcon},
this algorithm runs in time 
\[ \left(\frac{kr\ln k}{\wtilde{\varepsilon}}\right)^{\,\cO(r\Gamma k)}\sizeinst\]
and outputs a $(1 - \wtilde{\varepsilon})$-approximate solution which is also an independent set in the input matroid with probability at least $1 - \frac{1}{e}$. By \Cref{thm: cwtred_sat_to_cov}, we get a $(1 - \varepsilon)$-approximate solution to $\cI$.

%% file: conclusion.tex
\section{Conclusion}
\label{section: conclusion}

We studied how structural restrictions within the bounded-VC-dimension regime affect
the parameterized approximability of coverage, uncovering a sharp asymmetry between its
two formulations.  Bounded VC-dimension already gives an \epas\ for the fixed-budget
problem \maxcov, yet it does not suffice for the target-covering problem \psc: already
at VC-dimension seven, \psc\ admits no \fpt\ approximation to any factor below two
(assuming \(\mathrm{FPT}\neq\mathrm{W[1]}\)) and no approximation scheme at all
(assuming ETH, even in time \(F(k)\,N^{k/\log^{C}k}\)).  Tractability returns on both
sides under the stronger hypothesis of bounded semi-ladder index \(\Gamma\), which
still strictly generalizes the \(K_{d,d}\)-free setting: we gave an \epas\ for \wmc\
and an additive \pas\ for \wpsc\ that finds \(k+1\) sets whenever \(k\) sets suffice,
both running in time linear in the input size.  A structure-preserving reduction from
\cmaxsat\ to \maxcov\ then extends the \epas\ to \cmaxsat, on bounded-semi-ladder
instances and, through a faster deterministic implementation of the
Badanidiyuru--Kleinberg--Lee coverage scheme, on bounded-VC-dimension instances.
The same branching framework handles simultaneous matroid-independence and
per-color coverage constraints, and its combination with the satisfiability
reduction yields a constrained cardinality-MaxSAT algorithm. Applications to
partial dominating set, geometric partial covering, and bounded-size set
systems illustrate the reach of the structural condition.

We conclude with the following open questions.
\begin{enumerate}
\item Can the running time of the \epas\ for \maxcov\ on bounded-VC-dimension instances
  be improved further?
   \item Does \psc\ admit a lossy kernel on instances of bounded semi-ladder index, or
  even on those in which every element has frequency at most~$3$?
\end{enumerate}

\paragraph{Acknowledgments.} We thank the anonymous reviewers of an earlier version for their valuable comments and suggestions, especially for directing us to the notion of \wvpn.

%% file: appendix.tex
\section{Appendix}
\label{sec: appendix}

\subsection{Related Work}
\label{sec: related_work}

A polynomial-time algorithm with the ratio $1 - \frac{1}{e}$ exists for \maxcov and \cmaxsat~\cite{DBLP:journals/mp/NemhauserWF78, Sviridenko01}. Feige~\cite{DBLP:journals/jacm/Feige98} showed that this achieved ratio is optimal for {\sc Maximum Coverage} in polynomial time, unless some well known complexity theory assumption fails. A special case of \cmaxsat, called, \textsc{CC-Max-2-SAT}, where the input formula is $2$-CNF has been studied extensively~\cite{BlaserM02,Hofmeister03,RaghavendraT12,Sviridenko01}. The best algorithm up to date achieves an approximation factor of roughly $0.929$~\cite{RaghavendraT12}, which improved over the series of works in~\cite{BlaserM02,Hofmeister03,Sviridenko01}.  Austrin and Stankovi\'{c}~\cite{AustrinS19} showed that it is hard to approximate  within $0.929+\varepsilon$, for any $\varepsilon>0$, assuming Unique Games Conjecture (UGC). \textsc{Partial Vertex Cover}, which is a special case of \textsc{CC-Max-2-SAT} where negative literals are not allowed is known to be approximable within $0.929$ and is also hard to approximate  within $0.929$, assuming UGC~\cite{AustrinS19, DBLP:journals/corr/abs-1810-03792}. 

As \maxcov is known to be {\sf W}$[2]$-hard, Cohen-Addad et al.~\cite{DBLP:conf/icalp/Cohen-AddadG0LL19} initiated its study from the perspective of FPT-approximation during their work on {\sc $k$-Median} and {\sc $k$-Mean} and showed that there is no \epas that improves over the polynomial time greedy approximation algorithm. Later, this was also studied by Manurangsi~\cite{Manurangsi20}, who obtained the following strengthening over~\cite{DBLP:conf/icalp/Cohen-AddadG0LL19}. Manurangsi ruled out any possibility of having an approximation algorithm for \cmaxsat and \maxcov with factor 
$(1-\frac{1}{e} +\varepsilon)$ and running time of the form $f(k,\varepsilon)n^{\OO(1)}$. Skowron and  Faliszewski~\cite{DBLP:journals/jair/SkowronF17}  showed that on set families where each element in $U$ appears in at most $p$ sets, there exists an algorithm, that given an $\varepsilon >0$, runs in time $(\frac{p}{\varepsilon})^{\OO(k)}n^{\OO(1)}$ and returns a $(1-\varepsilon)$-approximate solution.

The decision version of \maxcov, \dmaxcov, and its special case on graphs called {\sc Partial Vertex Cover},  
 have been studied from the approximation perspective under the name of ``partial covering''~\cite{Bar-Yehuda01,BshoutyB98,Hochbaum98a,GandhiKS04}. {\sc Partial Vertex Cover}  is known to admit a factor $2$ approximation, which cannot be improved further, assuming UGC~\cite{Bar-Yehuda01,BshoutyB98,Hochbaum98a,GandhiKS04,Khot02}. It is also known to be \whard with respect to $k$~\cite{GuoNW07}. Furthermore, \dmaxcov generalizes the classical {\sc Set Cover} problem, and hence, it is unlikely to admit any FPT-approximation algorithms~\cite{SLM19}. 

Studies have been undertaken for {\sc Set Cover} and {\sc Backdoor Sets} that show that these problems admit \epas~\cite{LokshtanovP022,DBLP:journals/talg/PhilipRS12,DBLP:journals/tcs/TelleV19}. It is well known that {\sc Set Cover} is  $W[2]$-hard but admits an \fpt algorithm when the incidence graph excludes some $K_{d,d}$ as a subgraph~\cite{DBLP:journals/talg/PhilipRS12,DBLP:journals/tcs/TelleV19}.

The decision version of the {\sc Maximum Hitting Set} problem,  {\sc Dec-MHS} (where, additionally we are given an integer $t$, and we seek a set $S$ of size $k$, that intersects at least $t$ sets), is \whard, even when each set in the family has size at most $2$~\cite{GuoNW07}. When each set in the family has size at most $2$, then  {\sc Maximum Hitting Set} is popularly called {\sc Maximum Vertex Cover}, we will stick with the latter. Marx gave the first FPT-AS for the {\sc Maximum Vertex Cover} problem which was followed by several algorithms with improved running time~\cite{DBLP:journals/cj/Marx08}. The current fastest algorithm was independently given by  Manurangsi~\cite{DBLP:journals/corr/abs-1810-03792} and Skowron and  Faliszewski~\cite{DBLP:journals/jair/SkowronF17}, and runs in time $(\frac{1}{\varepsilon})^{\OO(k)}n^{\OO(1)}$. It is known to be W[1]-hard~\cite{GuoNW07}, parameterized by $k$, but admits \fpt algorithms on planar graphs, graphs of bounded degeneracy,  $K_{d,d}$-free graphs, and bipartite graphs, parameterized by $k$~\cite{AminiFS11,FominLRS11,KoanaKNS22}. 
Sellier gave an FPT-AS and an approximate kernel of {\sc Maximum Coverage} problem with instances having bounded frequency under matroid constraints \cite{DBLP:conf/esa/Sellier23}.
Indeed, it is among the first problems to admit FPT-AS~\cite{DBLP:journals/cj/Marx08,DBLP:journals/corr/abs-1810-03792,DBLP:journals/jair/SkowronF17}. It is also known to have ``lossy kernels''~\cite{DBLP:journals/cj/Marx08,LokshtanovPRS17}, a lossy version of classical kernelization. Panolan et al.~\cite{DBLP:journals/tcs/PanolanY25} studied \textsc{CC-Max-2-SAT} and gave a lossy kernel parameterized by $k$. Moreover, they give an \fpt algorithm parameterized by $k$ for the weighted version of \cmaxsat and also give a polynomial-time approximation scheme for the same. 

\Cref{tab:comparisons} provides a consolidated snapshot of results for the different problem variants across the various set systems considered, summarizing both previously known results and those established in this work.
\Cref{tab:comparisonpoly} presents the polynomial-time approximation landscape of the considered problems, organized according to structural properties of the underlying set systems.

\begin{table}[ht!]
\centering
\renewcommand{\arraystretch}{2.2} 
\setlength{\tabcolsep}{9pt} 
\small
\begin{adjustbox}{max width=\textwidth}
\begin{tabular}{|>{\columncolor{firstcolgray}\centering\arraybackslash}m{4.3cm}|
                >{\centering\arraybackslash}m{5.5cm}|
                >{\centering\arraybackslash}m{5.5cm}|
                >{\centering\arraybackslash}m{5.5cm}|
                >{\centering\arraybackslash}m{5.5cm}|}
\hline
\rowcolor{headergray}
\diagbox[width=4.3cm]{\textbf{Problem}}{\textbf{Set System}} &
\textbf{Frequency - $d$} &
\textbf{$K_{d,d}$-free} &
\textbf{$d$-\slf} &
\textbf{VC dimension - $d$} \\ 
\hline

\makecell[{{p{4.3cm}}}]{\centering\rule{0pt}{2.2em}\maxcov
\rule[-0.4em]{0pt}{1.2em}} &
\makecell[{{p{5.5cm}}}]{\centering\rule{0pt}{2.2em}
$\Bigl(\tfrac{1}{\varepsilon}\Bigr)^{\!\mathcal{O}(k\ln d)}{\sizesetsys}^{\mathcal{O}(1)}$\\[4pt]
\scriptsize\cite{DBLP:journals/jair/SkowronF17}
\rule[-0.4em]{0pt}{1.2em}} &
\makecell[{{p{5.5cm}}}]{\centering\rule{0pt}{2.2em}
$\Bigl(\tfrac{1}{\varepsilon}\Bigr)^{\!\mathcal{O}(dk\ln(dk))}{\sizesetsys}^{\mathcal{O}(1)}$\\[4pt]
\scriptsize\cite{DBLP:conf/soda/0001KPSS0U23}\rule[-0.4em]{0pt}{1.2em}} &
\makecell[{{p{5.5cm}}}]{\centering\rule{0pt}{2.2em}
$\textcolor{red}{\Bigl(\tfrac{1}{\varepsilon}\Bigr)^{\!\mathcal{O}(dk\ln k)}\mathcal{O}(\sizesetsys)}$
\rule[-0.4em]{0pt}{1.2em}} &
\makecell[{{p{5.5cm}}}]{\centering\rule{0pt}{2.2em}
$2^{\widetilde{\mathcal{O}}(kd/\varepsilon)}{\sizesetsys}^{\mathcal{O}(1)}$\\[4pt]
\scriptsize\cite{DBLP:conf/compgeom/BadanidiyuruKL12} + Sec.~\ref{sec:bkl-trace-enumeration}\rule[-0.4em]{0pt}{1.2em}} \\ \hline

\makecell[{{p{4.3cm}}}]{\centering\rule{0pt}{2.2em}{\sc Weighted}\\{\sc Maximum Coverage}\rule[-0.4em]{0pt}{1.2em}} &
\makecell{\centering\rule{0pt}{2.2em}$\diamond$\rule[-0.4em]{0pt}{1.2em}} &
\makecell[{{p{5.5cm}}}]{\centering\rule{0pt}{2.2em}
$\Bigl(\tfrac{1}{\varepsilon}\Bigr)^{\!\mathcal{O}(d^{2}k\ln(dk))}{\sizesetsys}^{\mathcal{O}(1)}$\\[4pt]
\scriptsize\cite{DBLP:conf/ijcai/Gupta000U25}\rule[-0.4em]{0pt}{1.2em}} &
\makecell[{{p{5.5cm}}}]{\centering\rule{0pt}{2.2em}
$\textcolor{red}{\Bigl(\tfrac{1}{\varepsilon}\Bigr)^{\!\mathcal{O}(dk\ln k)}\mathcal{O}(\sizesetsys)}$
\rule[-0.4em]{0pt}{1.2em}} &
\makecell[{{p{5.5cm}}}]{\centering\rule{0pt}{2.2em}
$2^{\widetilde{\mathcal{O}}(kd/\varepsilon)}{\sizesetsys}^{\mathcal{O}(1)}$\\[4pt]
\scriptsize\cite{DBLP:conf/compgeom/BadanidiyuruKL12} + Sec.~\ref{sec:bkl-trace-enumeration}\rule[-0.4em]{0pt}{1.2em}} \\ \hline

\rowcolor{lightgray}
\makecell[{{p{4.3cm}}}]{\centering\rule{0pt}{2.2em}{\sc Constrained}\\\maxcov\rule[-0.4em]{0pt}{1.2em}} &
\makecell[{{p{5.5cm}}}]{\centering\rule{0pt}{2.2em}
$\Bigl(\tfrac{1}{\varepsilon}\Bigr)^{\!\mathcal{O}(kr\ln(d\ln k))}{\sizesetsys}^{\mathcal{O}(1)}$\\[4pt]
\scriptsize\cite{DBLP:conf/icalp/00020LS0U24}\rule[-0.4em]{0pt}{1.2em}} &
\makecell[{{p{5.5cm}}}]{\centering\rule{0pt}{2.2em}
$2^{\mathcal{O}(\tfrac{k^{3}rd\ln r}{\varepsilon})}{\sizesetsys}^{\mathcal{O}(1)}$\\[4pt]
\scriptsize\cite{DBLP:conf/icalp/00020LS0U24}\rule[-0.4em]{0pt}{1.2em}} &
\makecell{\centering\rule{0pt}{2.2em}$\diamond$\rule[-0.4em]{0pt}{1.2em}} &
\makecell{\centering\rule{0pt}{2.2em}Not Known\rule[-0.4em]{0pt}{1.2em}} \\ \hline

\makecell[{{p{4.3cm}}}]{\centering\rule{0pt}{2.2em}{\sc Constrained}\\{\sc Weighted}\\{\sc Max Coverage}\rule[-0.4em]{0pt}{1.2em}} &
\makecell{\centering\rule{0pt}{2.2em}$\diamond$\rule[-0.4em]{0pt}{1.2em}} &
\makecell{\centering\rule{0pt}{2.2em}$\diamond$\rule[-0.4em]{0pt}{1.2em}} &
\makecell[{{p{5.5cm}}}]{\centering\rule{0pt}{2.2em}
$\textcolor{red}{\Bigl(\tfrac{1}{\varepsilon}\Bigr)^{\!\mathcal{O}(rdk\ln(kr))}\mathcal{O}(\sizesetsys)}$
\rule[-0.4em]{0pt}{1.2em}} &
\makecell{\centering\rule{0pt}{2.2em}Not Known\rule[-0.4em]{0pt}{1.2em}} \\ \hline

\rowcolor{lightgray}
\makecell[{{p{4.3cm}}}]{\centering\rule{0pt}{2.2em}{\sc Partial}\\{\sc Set Cover}}&
\makecell[{{p{5.5cm}}}]{\centering\rule{0pt}{2.2em}
$\diamond$
\rule[-0.4em]{0pt}{1.2em}} &
\makecell[{{p{5.5cm}}}]{\centering\rule{0pt}{2.2em}
$2^{\mathcal{O}(dk\ln(dk))}{\sizesetsys}^{\mathcal{O}(1)}$\\[4pt]
\scriptsize\cite{DBLP:conf/soda/0001KPSS0U23}\rule[-0.4em]{0pt}{1.2em}} &
\makecell{\centering\rule{0pt}{2.2em}$\diamond$\rule[-0.4em]{0pt}{1.2em}} &
\makecell[{{p{5.5cm}}}]{\centering\rule{0pt}{2.2em}
{\color{red}No \pas}\\[2pt]
{\scriptsize\color{red}even VC-dim.\ at most $7$; Thm.~\ref{thm:vc-no-aptas}}\rule[-0.4em]{0pt}{1.2em}} \\ \hline

\makecell[{{p{4.3cm}}}]{\centering\rule{0pt}{2.2em}{\sc Weighted}\\{\sc Partial}\\{\sc Set Cover}\rule[-0.4em]{0pt}{1.2em}} &
\makecell{\centering\rule{0pt}{2.2em}$\diamond$\rule[-0.4em]{0pt}{1.2em}} &
\makecell[{{p{5.5cm}}}]{\centering\rule{0pt}{2.2em}
$2^{\mathcal{O}(d^{2}k\ln(dk))}{\sizesetsys}^{\mathcal{O}(1)}$\\[4pt]
\scriptsize\cite{DBLP:conf/ijcai/Gupta000U25}\rule[-0.4em]{0pt}{1.2em}} &
\makecell[{{p{5.5cm}}}]{\centering\rule{0pt}{2.2em}
$\textcolor{red}{2^{\mathcal{O}(dk\ln(dk))}\mathcal{O}(\sizesetsys)}$
\rule[-0.4em]{0pt}{1.2em}} &
\makecell[{{p{5.5cm}}}]{\centering\rule{0pt}{2.2em}
{\color{red}No \pas}\\[2pt]
{\scriptsize\color{red}already impossible unweighted; Thm.~\ref{thm:vc-no-aptas}}\rule[-0.4em]{0pt}{1.2em}} \\ \hline

\rowcolor{lightgray}
\makecell[{{p{4.3cm}}}]{\centering\rule{0pt}{2.2em}\cmaxsat\rule[-0.4em]{0pt}{1.2em}} &
\makecell[{{p{5.5cm}}}]{\centering\rule{0pt}{2.2em}
$\diamond$ \rule[-0.4em]{0pt}{1.2em}} &
\makecell[{{p{5.5cm}}}]{\centering\rule{0pt}{2.2em}
$\Bigl(\tfrac{1}{\varepsilon}\Bigr)^{\!\mathcal{O}(dk\ln(dk))}{\sizephi}^{\mathcal{O}(1)}$\\[4pt]
\scriptsize\cite{DBLP:conf/icalp/00020LS0U24,DBLP:journals/tcs/Manurangsi25}\rule[-0.4em]{0pt}{1.2em}} &
\makecell{\centering\rule{0pt}{2.2em}$\diamond$\rule[-0.4em]{0pt}{1.2em}} &
\makecell[{{p{5.5cm}}}]{\centering\rule{0pt}{2.2em}
\textcolor{red}{$2^{\widetilde{\mathcal{O}}(kd/\varepsilon)}{\sizephi}^{\mathcal{O}(1)}$}\\[4pt]
\scriptsize\textcolor{red}{This paper}\rule[-0.4em]{0pt}{1.2em}} \\ \hline

\makecell[{{p{4.3cm}}}]{\centering\rule{0pt}{2.2em}{\sc Weighted}\\\cmaxsat\rule[-0.4em]{0pt}{1.2em}} &
\makecell{\centering\rule{0pt}{2.2em}$\diamond$\rule[-0.4em]{0pt}{1.2em}} &
\makecell{\centering\rule{0pt}{2.2em}$\diamond$\rule[-0.4em]{0pt}{1.2em}} &
\makecell[{{p{5.5cm}}}]{\centering\rule{0pt}{2.2em}
$\textcolor{red}{\Bigl(\tfrac{1}{\varepsilon}\Bigr)^{\!\mathcal{O}(dk\ln k)}\mathcal{O}(\sizephi)}$
\rule[-0.4em]{0pt}{1.2em}} &
\makecell[{{p{5.5cm}}}]{\centering\rule{0pt}{2.2em}
Not Known
\rule[-0.4em]{0pt}{1.2em}} \\ \hline

\rowcolor{lightgray}
\makecell[{{p{4.3cm}}}]{\centering\rule{0pt}{2.2em}{\sc Constrained}\\\cmaxsat\rule[-0.4em]{0pt}{1.2em}} &
\makecell[{{p{5.5cm}}}]{\centering\rule{0pt}{2.2em}
$\Bigl(\tfrac{1}{\varepsilon}\Bigr)^{\!\mathcal{O}(kr\ln(d\ln k))}{\sizephi}^{\mathcal{O}(1)}$\\[4pt]
\scriptsize\cite{DBLP:conf/icalp/00020LS0U24}\rule[-0.4em]{0pt}{1.2em}} &
\makecell[{{p{5.5cm}}}]{\centering\rule{0pt}{2.2em}
$2^{\mathcal{O}(\tfrac{k^{3}rd\ln r}{\varepsilon})}{\sizephi}^{\mathcal{O}(1)}$\\[4pt]
\scriptsize\cite{DBLP:conf/icalp/00020LS0U24}\rule[-0.4em]{0pt}{1.2em}} &
\makecell{\centering\rule{0pt}{2.2em}$\diamond$\rule[-0.4em]{0pt}{1.2em}} &
\makecell{\centering\rule{0pt}{2.2em}Not Known\rule[-0.4em]{0pt}{1.2em}} \\ \hline

\makecell[{{p{4.3cm}}}]{\centering\rule{0pt}{2.2em}{\sc Weighted}\\{\sc Constrained}\\\cmaxsat\rule[-0.4em]{0pt}{1.2em}} &
\makecell{\centering\rule{0pt}{2.2em}$\diamond$\rule[-0.4em]{0pt}{1.2em}} &
\makecell{\centering\rule{0pt}{2.2em}$\diamond$\rule[-0.4em]{0pt}{1.2em}} &
\makecell[{{p{5.5cm}}}]{\centering\rule{0pt}{2.2em}
$\textcolor{red}{\Bigl(\tfrac{1}{\varepsilon}\Bigr)^{\!\mathcal{O}(rdk\ln(kr))}\mathcal{O}(\sizephi)}$
\rule[-0.4em]{0pt}{1.2em}} &
\makecell{\centering\rule{0pt}{2.2em}Not Known\rule[-0.4em]{0pt}{1.2em}} \\ \hline

\end{tabular}
\end{adjustbox}
\caption{\noindent Summary of results for different problem variants across various set systems.
Entries marked with $\diamond$ indicate cases that are subsumed by a more general result, and for which no separate algorithm for the particular set system is known. 
\textcolor{red}{Results in red} are obtained in this paper
for the specified set systems.}
\label{tab:comparisons}
\end{table}

\begin{table}[ht!]
\centering
\renewcommand{\arraystretch}{2.2}
\setlength{\tabcolsep}{9pt}
\small
\begin{adjustbox}{max width=\textwidth}
\begin{tabular}{|
>{\columncolor{firstcolgray}\centering\arraybackslash}m{3.2cm}|
>{\centering\arraybackslash}m{5.8cm}|
>{\centering\arraybackslash}m{5.8cm}|}
\hline
\rowcolor{headergray}
\diagbox[width=2.7cm,height=1.9cm,dir=NW,outerleftsep=0pt,
  innerrightsep=0pt]{\textbf{Problem}}{ \textbf{Objective}} &
\textbf{Maximum Version} &
\textbf{Partial Version} \\
\hline

\makecell[{{p{2.7cm}}}]{\centering \textbf{Hitting Set}} & \vspace{2pt}
\makecell[{{p{5.8cm}}}]{
\textbf{\textsc{Hardness of Approximation}}\\[2pt]
\textit{Set size $2$:} $0.929+\varepsilon$ (UGC)\\
\cite{AustrinS19, DBLP:journals/corr/abs-1810-03792}\\[2pt]
\textit{General set system:} $(1-\tfrac{1}{e}+\varepsilon)$ ($P \neq NP$)\\
\cite{DBLP:journals/jacm/Feige98}\\[4pt]
\textbf{\textsc{Best Known Approximations}}\\[2pt]
\textit{Set size $2$:} $0.929$-approximation\\
\cite{DBLP:journals/corr/abs-1810-03792, RaghavendraT12}\\
\textit{General case:} $(1-\tfrac{1}{e})$\\
\cite{DBLP:journals/mp/NemhauserWF78}
} & \vspace{2pt}
\makecell[{{p{5.8cm}}}]{
\textbf{\textsc{Hardness of Approximation}}\\[2pt]
\textit{Set size $2$:} $(2-\varepsilon)$ (UGC)\\
\cite{DBLP:journals/jcss/KhotR08, DBLP:conf/focs/BansalK09}\\
\textit{Set size $d \ge 2$:} $(d-\varepsilon)$ (UGC)\\
\textit{General set system:} $(1-o(1))\ln n$ ($P \neq NP$)\\
\cite{DBLP:journals/jacm/Feige98, DBLP:journals/corr/abs-1305-1979}\\[4pt]
\textbf{\textsc{Best Known Approximations}}\\[2pt]
\textit{Set size $d$:} $d$-approximation\\
\cite{DBLP:conf/wg/Bar-YehudaE83, GandhiKS04}
}\\
\hline

\rowcolor{lightgray}
\makecell[{{p{2.7cm}}}]{\centering \textbf{Coverage}} &
\makecell[{{p{5.8cm}}}]{
\textbf{\textsc{Hardness of Approximation}}\\[2pt]
\textit{Frequency $2$:} $0.929+\varepsilon$ (UGC)\\
\cite{AustrinS19, DBLP:journals/corr/abs-1810-03792}\\
\textit{General set system:} $(1-\tfrac{1}{e}+\varepsilon)$ ($P \neq NP$)\\
\cite{DBLP:journals/jacm/Feige98}\\[4pt]
\textbf{\textsc{Best Known Approximations}}\\[2pt]
\textit{Frequency $2$:} $0.929$-approximation\\
\cite{DBLP:journals/corr/abs-1810-03792, RaghavendraT12}\\
\textit{General case:} $(1-\tfrac{1}{e})$\\
\cite{DBLP:journals/mp/NemhauserWF78}
} & 
\makecell[{{p{5.8cm}}}]{ 
\textbf{\textsc{Hardness of Approximation}}\\[2pt]
\textit{Frequency $2$:} $(2-\varepsilon)$ (UGC)\\
\cite{DBLP:journals/jcss/KhotR08, DBLP:conf/focs/BansalK09}\\
\textit{Frequency $d \ge 2$:} $(d-\varepsilon)$ (UGC)\\
\textit{Intersection $1$ ($K_{2,2}$-free):} $o(\log n)$ 
\\
\cite{DBLP:conf/icalp/KumarAR00}\\
\textit{General case:} $(1-o(1))\ln n$ ($P \neq NP$)\\
\cite{DBLP:journals/jacm/Feige98, DBLP:journals/corr/abs-1305-1979}\\[4pt]
\textbf{\textsc{Best Known Approximations}}\\[2pt]
\textit{Set size $d$:} \\$(1+\tfrac{1}{2}+\dots+\tfrac{1}{d})$-approximation\\
\cite{DBLP:journals/ipl/Slavik97, DBLP:journals/combinatorica/Wolsey82}\\
\textit{Frequency $d$:} $d$-approximation\\
\cite{DBLP:conf/wg/Bar-YehudaE83, GandhiKS04}
}\\
\hline

\makecell[{{p{2.7cm}}}]{\centering \textbf{Satisfiability}} &
\makecell[{{p{5.8cm}}}]{
\textbf{\textsc{Hardness of Approximation}}\\[2pt]
\textit{Clause size $2$:} $0.929+\varepsilon$ (UGC)\\
\cite{AustrinS19}\\
\textit{General formula:} $(1-\tfrac{1}{e}+\varepsilon)$ ($P \neq NP$)\\
\cite{DBLP:journals/jacm/Feige98}\\[4pt]
\textbf{\textsc{Best Known Approximations}}\\[2pt]
\textit{Clause size $2$:} $0.929$\\
\cite{RaghavendraT12}\\
\textit{General formula:} $(1-\tfrac{1}{e})$\\
\cite{Sviridenko01}
} &
\makecell[{{p{5.8cm}}}]{
\textbf{\textsc{Hardness of Approximation}}\\[2pt]
\textit{Clause size $2$:} $2-\varepsilon$ (UGC)\\
\cite{DBLP:journals/sigact/Hochba97}
}\\
\hline

\end{tabular}
\end{adjustbox}
\caption{
Polynomial-time approximation thresholds under various structural restrictions on the underlying set system. 
Entries under \textsc{Hardness of Approximation} list known lower bounds, annotated by set system constraints such as bounded size, frequency, or intersection. 
Entries under \textsc{Best Known Approximations} list the best known polynomial-time approximation ratios for the same settings.
}
\label{tab:comparisonpoly}
\end{table}

%% file: main.bbl
\newcommand{\etalchar}[1]{$^{#1}$}
\begin{thebibliography}{dBCvKO08}

\bibitem[ABB{\etalchar{+}}23]{DBLP:conf/focs/AbbasiBBCGKMSS23}
Fateme Abbasi, Sandip Banerjee, Jaroslaw Byrka, Parinya Chalermsook, Ameet
  Gadekar, Kamyar Khodamoradi, D{\'{a}}niel Marx, Roohani Sharma, and Joachim
  Spoerhase.
\newblock Parameterized approximation schemes for clustering with general norm
  objectives.
\newblock In {\em 64th {IEEE} Annual Symposium on Foundations of Computer
  Science, {FOCS} 2023, Santa Cruz, CA, USA, November 6-9, 2023}, pages
  1377--1399. {IEEE}, 2023.

\bibitem[AFS11]{AminiFS11}
Omid Amini, Fedor~V. Fomin, and Saket Saurabh.
\newblock Implicit branching and parameterized partial cover problems.
\newblock {\em J. Comput. Syst. Sci.}, 77(6):1159--1171, 2011.

\bibitem[AS19]{AustrinS19}
Per Austrin and Aleksa Stankovic.
\newblock Global cardinality constraints make approximating some max-2-csps
  harder.
\newblock In Dimitris Achlioptas and L{\'{a}}szl{\'{o}}~A. V{\'{e}}gh, editors,
  {\em Approximation, Randomization, and Combinatorial Optimization. Algorithms
  and Techniques, {APPROX/RANDOM} 2019, September 20-22, 2019, Massachusetts
  Institute of Technology, Cambridge, MA, {USA}}, volume 145 of {\em LIPIcs},
  pages 24:1--24:17. Schloss Dagstuhl - Leibniz-Zentrum f{\"{u}}r Informatik,
  2019.

\bibitem[Bar01]{Bar-Yehuda01}
Reuven Bar{-}Yehuda.
\newblock Using homogeneous weights for approximating the partial cover
  problem.
\newblock {\em J. Algorithms}, 39(2):137--144, 2001.

\bibitem[BB98]{BshoutyB98}
Nader~H. Bshouty and Lynn Burroughs.
\newblock Massaging a linear programming solution to give a 2-approximation for
  a generalization of the vertex cover problem.
\newblock In Michel Morvan, Christoph Meinel, and Daniel Krob, editors, {\em
  {STACS} 98, 15th Annual Symposium on Theoretical Aspects of Computer Science,
  Paris, France, February 25-27, 1998, Proceedings}, volume 1373 of {\em
  Lecture Notes in Computer Science}, pages 298--308. Springer, 1998.

\bibitem[BCM99]{DBLP:journals/siamcomp/BronnimannCM99}
Herv{\'{e}} Br{\"{o}}nnimann, Bernard Chazelle, and Jir{\'{\i}} Matousek.
\newblock Product range spaces, sensitive sampling, and derandomization.
\newblock {\em {SIAM} J. Comput.}, 28(5):1552--1575, 1999.

\bibitem[BE83]{DBLP:conf/wg/Bar-YehudaE83}
Reuven Bar{-}Yehuda and Shimon Even.
\newblock A local-ratio theorem for approximating the weighted vertex cover
  problem.
\newblock In Manfred Nagl and J{\"{u}}rgen Perl, editors, {\em Proceedings of
  the {WG} '83, International Workshop on Graphtheoretic Concepts in Computer
  Science, June 16-18, 1983, Haus Ohrbeck, near Osnabr{\"{u}}ck, Germany},
  pages 17--28. Universit{\"{a}}tsverlag Rudolf Trauner, Linz, 1983.

\bibitem[BEHW86]{DBLP:conf/stoc/BlumerEHW86}
Anselm Blumer, Andrzej Ehrenfeucht, David Haussler, and Manfred~K. Warmuth.
\newblock Classifying learnable geometric concepts with the vapnik-chervonenkis
  dimension (extended abstract).
\newblock In Juris Hartmanis, editor, {\em Proceedings of the 18th Annual {ACM}
  Symposium on Theory of Computing, May 28-30, 1986, Berkeley, California,
  {USA}}, pages 273--282. {ACM}, 1986.

\bibitem[BG95]{DBLP:journals/dcg/BronnimannG95}
Herv{\'{e}} Br{\"{o}}nnimann and Michael~T. Goodrich.
\newblock Almost optimal set covers in finite vc-dimension.
\newblock {\em Discret. Comput. Geom.}, 14(4):463--479, 1995.

\bibitem[BK09]{DBLP:conf/focs/BansalK09}
Nikhil Bansal and Subhash Khot.
\newblock Optimal long code test with one free bit.
\newblock In {\em 50th Annual {IEEE} Symposium on Foundations of Computer
  Science, {FOCS} 2009, Atlanta, Georgia, USA, October 25-27, 2009}, pages
  453--462. {IEEE} Computer Society, 2009.

\bibitem[BKL12]{DBLP:conf/compgeom/BadanidiyuruKL12}
Ashwinkumar Badanidiyuru, Robert Kleinberg, and Hooyeon Lee.
\newblock Approximating low-dimensional coverage problems.
\newblock In Tamal~K. Dey and Sue Whitesides, editors, {\em Proceedings of the
  28th {ACM} Symposium on Computational Geometry, Chapel Hill, NC, USA, June
  17-20, 2012}, pages 161--170. {ACM}, 2012.

\bibitem[BM02]{BlaserM02}
Markus Bl{\"{a}}ser and Bodo Manthey.
\newblock Improved approximation algorithms for max-2sat with cardinality
  constraint.
\newblock In Prosenjit Bose and Pat Morin, editors, {\em Algorithms and
  Computation, 13th International Symposium, {ISAAC} 2002 Vancouver, BC,
  Canada, November 21-23, 2002, Proceedings}, volume 2518 of {\em Lecture Notes
  in Computer Science}, pages 187--198. Springer, 2002.

\bibitem[BP25]{DBLP:conf/soda/BourneufP25}
Romain Bourneuf and Marcin Pilipczuk.
\newblock Bounding {$\varepsilon$}-scatter dimension via metric sparsity.
\newblock In Yossi Azar and Debmalya Panigrahi, editors, {\em Proceedings of
  the 2025 Annual {ACM-SIAM} Symposium on Discrete Algorithms, {SODA} 2025, New
  Orleans, LA, USA, January 12-15, 2025}, pages 3155--3171. {SIAM}, 2025.

\bibitem[BSM25]{BafnaKarthikMinzer2025}
Mitali Bafna, Karthik~C. S., and Dor Minzer.
\newblock Near optimal constant inapproximability under {ETH} for fundamental
  problems in parameterized complexity.
\newblock In {\em Proceedings of the 57th Annual {ACM} Symposium on Theory of
  Computing}, pages 2118--2129, 2025.

\bibitem[CGK{\etalchar{+}}19]{DBLP:conf/icalp/Cohen-AddadG0LL19}
Vincent Cohen{-}Addad, Anupam Gupta, Amit Kumar, Euiwoong Lee, and Jason Li.
\newblock Tight {FPT} approximations for k-median and k-means.
\newblock In Christel Baier, Ioannis Chatzigiannakis, Paola Flocchini, and
  Stefano Leonardi, editors, {\em 46th International Colloquium on Automata,
  Languages, and Programming, {ICALP} 2019, July 9-12, 2019, Patras, Greece},
  volume 132 of {\em LIPIcs}, pages 42:1--42:14. Schloss Dagstuhl -
  Leibniz-Zentrum f{\"{u}}r Informatik, 2019.

\bibitem[dBCvKO08]{DBLP:books/lib/BergCKO08}
Mark de~Berg, Otfried Cheong, Marc~J. van Kreveld, and Mark~H. Overmars.
\newblock {\em Computational geometry: algorithms and applications, 3rd
  Edition}.
\newblock Springer, 2008.

\bibitem[DS13]{DBLP:journals/corr/abs-1305-1979}
Irit Dinur and David Steurer.
\newblock Analytical approach to parallel repetition.
\newblock {\em CoRR}, abs/1305.1979, 2013.

\bibitem[Fei98]{DBLP:journals/jacm/Feige98}
Uriel Feige.
\newblock A threshold of ln \emph{n} for approximating set cover.
\newblock {\em J. {ACM}}, 45(4):634--652, 1998.

\bibitem[FKLM20]{DBLP:journals/algorithms/FeldmannSLM20}
Andreas~Emil Feldmann, {Karthik {C. S.}}, Euiwoong Lee, and Pasin Manurangsi.
\newblock A survey on approximation in parameterized complexity: Hardness and
  algorithms.
\newblock {\em Algorithms}, 13(6):146, 2020.

\bibitem[FLRS11]{FominLRS11}
Fedor~V. Fomin, Daniel Lokshtanov, Venkatesh Raman, and Saket Saurabh.
\newblock Subexponential algorithms for partial cover problems.
\newblock {\em Inf. Process. Lett.}, 111(16):814--818, 2011.

\bibitem[FPST18]{DBLP:journals/corr/abs-1811-06799}
Grzegorz Fabianski, Michal Pilipczuk, Sebastian Siebertz, and Szymon Torunczyk.
\newblock Progressive algorithms for domination and independence.
\newblock {\em CoRR}, abs/1811.06799, 2018.

\bibitem[FPST19]{DBLP:conf/stacs/FabianskiPST19}
Grzegorz Fabianski, Michal Pilipczuk, Sebastian Siebertz, and Szymon Torunczyk.
\newblock Progressive algorithms for domination and independence.
\newblock In Rolf Niedermeier and Christophe Paul, editors, {\em 36th
  International Symposium on Theoretical Aspects of Computer Science, {STACS}
  2019, March 13-16, 2019, Berlin, Germany}, volume 126 of {\em LIPIcs}, pages
  27:1--27:16. Schloss Dagstuhl - Leibniz-Zentrum f{\"{u}}r Informatik, 2019.

\bibitem[Fra82]{DBLP:journals/ejc/Frankl82}
Peter Frankl.
\newblock An extremal problem for two families of sets.
\newblock {\em Eur. J. Comb.}, 3(2):125--127, 1982.

\bibitem[GJS{\etalchar{+}}25]{DBLP:conf/ijcai/Gupta000U25}
Sushmita Gupta, Pallavi Jain, Souvik Saha, Saket Saurabh, and Anannya Upasana.
\newblock More efforts towards fixed-parameter approximability of multiwinner
  rules.
\newblock In {\em Proceedings of the Thirty-Fourth International Joint
  Conference on Artificial Intelligence, {IJCAI} 2025, Montreal, Canada, August
  16-22, 2025}, pages 3891--3899. ijcai.org, 2025.

\bibitem[GKS04]{GandhiKS04}
Rajiv Gandhi, Samir Khuller, and Aravind Srinivasan.
\newblock Approximation algorithms for partial covering problems.
\newblock {\em J. Algorithms}, 53(1):55--84, 2004.

\bibitem[GNW07a]{DBLP:journals/mst/GuoNW07}
Jiong Guo, Rolf Niedermeier, and Sebastian Wernicke.
\newblock Parameterized complexity of vertex cover variants.
\newblock {\em Theory Comput. Syst.}, 41(3):501--520, 2007.

\bibitem[GNW07b]{GuoNW07}
Jiong Guo, Rolf Niedermeier, and Sebastian Wernicke.
\newblock Parameterized complexity of vertex cover variants.
\newblock {\em Theory Comput. Syst.}, 41(3):501--520, 2007.

\bibitem[Gui25]{DBLP:journals/toct/G25}
Sylvain Guillemot.
\newblock Parameterized covering in semi-ladder-free hypergraphs.
\newblock {\em {ACM} Trans. Comput. Theory}, 2025.

\bibitem[GV08]{DBLP:conf/wg/GolovachV08}
Petr~A. Golovach and Yngve Villanger.
\newblock Parameterized complexity for domination problems on degenerate
  graphs.
\newblock In Hajo Broersma, Thomas Erlebach, Tom Friedetzky, and Dani{\"{e}}l
  Paulusma, editors, {\em Graph-Theoretic Concepts in Computer Science, 34th
  International Workshop, {WG} 2008, Durham, UK, June 30 - July 2, 2008.
  Revised Papers}, volume 5344 of {\em Lecture Notes in Computer Science},
  pages 195--205, 2008.

\bibitem[Hoc97]{DBLP:journals/sigact/Hochba97}
Dorit~S. Hochbaum.
\newblock Approximation algorithms for np-hard problems.
\newblock {\em {SIGACT} News}, 28(2):40--52, 1997.

\bibitem[Hoc98]{Hochbaum98a}
Dorit~S. Hochbaum.
\newblock The {$t$}-vertex cover problem: Extending the half integrality
  framework with budget constraints.
\newblock In Klaus Jansen and Dorit~S. Hochbaum, editors, {\em Approximation
  Algorithms for Combinatorial Optimization, International Workshop APPROX'98,
  Aalborg, Denmark, July 18-19, 1998, Proceedings}, volume 1444 of {\em Lecture
  Notes in Computer Science}, pages 111--122. Springer, 1998.

\bibitem[Hof03]{Hofmeister03}
Thomas Hofmeister.
\newblock An approximation algorithm for {MAX-2-SAT} with cardinality
  constraint.
\newblock In Giuseppe~Di Battista and Uri Zwick, editors, {\em Algorithms -
  {ESA} 2003, 11th Annual European Symposium, Budapest, Hungary, September
  16-19, 2003, Proceedings}, volume 2832 of {\em Lecture Notes in Computer
  Science}, pages 301--312. Springer, 2003.

\bibitem[Hp11]{10.5555/2031416}
Sariel Har-peled.
\newblock {\em Geometric Approximation Algorithms}.
\newblock American Mathematical Society, USA, 2011.

\bibitem[HW86]{DBLP:conf/compgeom/HausslerW86}
David Haussler and Emo Welzl.
\newblock Epsilon-nets and simplex range queries.
\newblock In Alok Aggarwal, editor, {\em Proceedings of the Second Annual {ACM}
  {SIGACT/SIGGRAPH} Symposium on Computational Geometry, Yorktown Heights, NY,
  USA, June 2-4, 1986}, pages 61--71. {ACM}, 1986.

\bibitem[IJL{\etalchar{+}}24]{DBLP:conf/icalp/00020LS0U24}
Tanmay Inamdar, Pallavi Jain, Daniel Lokshtanov, Abhishek Sahu, Saket Saurabh,
  and Anannya Upasana.
\newblock Satisfiability to coverage in presence of fairness, matroid, and
  global constraints.
\newblock In Karl Bringmann, Martin Grohe, Gabriele Puppis, and Ola Svensson,
  editors, {\em 51st International Colloquium on Automata, Languages, and
  Programming, {ICALP} 2024, July 8-12, 2024, Tallinn, Estonia}, volume 297 of
  {\em LIPIcs}, pages 88:1--88:18. Schloss Dagstuhl - Leibniz-Zentrum f{\"{u}}r
  Informatik, 2024.

\bibitem[JKP{\etalchar{+}}23]{DBLP:conf/soda/0001KPSS0U23}
Pallavi Jain, Lawqueen Kanesh, Fahad Panolan, Souvik Saha, Abhishek Sahu, Saket
  Saurabh, and Anannya Upasana.
\newblock Parameterized approximation scheme for biclique-free max
  \emph{k}-weight {SAT} and max coverage.
\newblock In Nikhil Bansal and Viswanath Nagarajan, editors, {\em Proceedings
  of the 2023 {ACM-SIAM} Symposium on Discrete Algorithms, {SODA} 2023,
  Florence, Italy, January 22-25, 2023}, pages 3713--3733. {SIAM}, 2023.

\bibitem[Juk11]{DBLP:series/txtcs/Jukna11}
Stasys Jukna.
\newblock {\em Extremal Combinatorics - With Applications in Computer Science}.
\newblock Texts in Theoretical Computer Science. An {EATCS} Series. Springer,
  2011.

\bibitem[Kal84]{kalai1984intersection}
Gil Kalai.
\newblock Intersection patterns of convex sets.
\newblock {\em Israel Journal of Mathematics}, 48:161--174, 1984.

\bibitem[KAR00]{DBLP:conf/icalp/KumarAR00}
V.~S.~Anil Kumar, Sunil Arya, and H.~Ramesh.
\newblock Hardness of set cover with intersection 1.
\newblock In Ugo Montanari, Jos{\'{e}} D.~P. Rolim, and Emo Welzl, editors,
  {\em Automata, Languages and Programming, 27th International Colloquium,
  {ICALP} 2000, Geneva, Switzerland, July 9-15, 2000, Proceedings}, volume 1853
  of {\em Lecture Notes in Computer Science}, pages 624--635. Springer, 2000.

\bibitem[Kho02]{Khot02}
Subhash Khot.
\newblock On the power of unique 2-prover 1-round games.
\newblock In {\em Proceedings of the 17th Annual {IEEE} Conference on
  Computational Complexity, Montr{\'{e}}al, Qu{\'{e}}bec, Canada, May 21-24,
  2002}, page~25. {IEEE} Computer Society, 2002.

\bibitem[KKNS22]{KoanaKNS22}
Tomohiro Koana, Christian Komusiewicz, Andr{\'{e}} Nichterlein, and Frank
  Sommer.
\newblock Covering many (or few) edges with k vertices in sparse graphs.
\newblock In Petra Berenbrink and Benjamin Monmege, editors, {\em 39th
  International Symposium on Theoretical Aspects of Computer Science, {STACS}
  2022, March 15-18, 2022, Marseille, France (Virtual Conference)}, volume 219
  of {\em LIPIcs}, pages 42:1--42:18. Schloss Dagstuhl - Leibniz-Zentrum
  f{\"{u}}r Informatik, 2022.

\bibitem[KLM19]{SLM19}
{Karthik {C. S.}}, Bundit Laekhanukit, and Pasin Manurangsi.
\newblock On the parameterized complexity of approximating dominating set.
\newblock {\em J. {ACM}}, 66(5):33:1--33:38, 2019.

\bibitem[KPW92]{komlos1992almost}
J{\'a}nos Koml{\'o}s, J{\'a}nos Pach, and Gerhard Woeginger.
\newblock Almost tight bounds for $\varepsilon$-nets.
\newblock {\em Discrete \& Computational Geometry}, 7(2):163--173, 1992.

\bibitem[KR08]{DBLP:journals/jcss/KhotR08}
Subhash Khot and Oded Regev.
\newblock Vertex cover might be hard to approximate to within 2-epsilon.
\newblock {\em J. Comput. Syst. Sci.}, 74(3):335--349, 2008.

\bibitem[Lin21]{DBLP:conf/stoc/Lin21}
Bingkai Lin.
\newblock Constant approximating k-clique is w[1]-hard.
\newblock In Samir Khuller and Virginia~Vassilevska Williams, editors, {\em
  {STOC} '21: 53rd Annual {ACM} {SIGACT} Symposium on Theory of Computing,
  Virtual Event, Italy, June 21-25, 2021}, pages 1749--1756. {ACM}, 2021.

\bibitem[LM05]{DBLP:journals/dcg/LangermanM05}
Stefan Langerman and Pat Morin.
\newblock Covering things with things.
\newblock {\em Discret. Comput. Geom.}, 33(4):717--729, 2005.

\bibitem[LPR22]{LokshtanovP022}
Daniel Lokshtanov, Fahad Panolan, and M.~S. Ramanujan.
\newblock Backdoor sets on nowhere dense {SAT}.
\newblock In Mikolaj Bojanczyk, Emanuela Merelli, and David~P. Woodruff,
  editors, {\em 49th International Colloquium on Automata, Languages, and
  Programming, {ICALP} 2022, July 4-8, 2022, Paris, France}, volume 229 of {\em
  LIPIcs}, pages 91:1--91:20. Schloss Dagstuhl - Leibniz-Zentrum f{\"{u}}r
  Informatik, 2022.

\bibitem[LPRS17a]{DBLP:conf/stoc/LokshtanovPRS17}
Daniel Lokshtanov, Fahad Panolan, M.~S. Ramanujan, and Saket Saurabh.
\newblock Lossy kernelization.
\newblock In Hamed Hatami, Pierre McKenzie, and Valerie King, editors, {\em
  Proceedings of the 49th Annual {ACM} {SIGACT} Symposium on Theory of
  Computing, {STOC} 2017, Montreal, QC, Canada, June 19-23, 2017}, pages
  224--237. {ACM}, 2017.

\bibitem[LPRS17b]{LokshtanovPRS17}
Daniel Lokshtanov, Fahad Panolan, M.~S. Ramanujan, and Saket Saurabh.
\newblock Lossy kernelization.
\newblock In Hamed Hatami, Pierre McKenzie, and Valerie King, editors, {\em
  Proceedings of the 49th Annual {ACM} {SIGACT} Symposium on Theory of
  Computing, {STOC} 2017, Montreal, QC, Canada, June 19-23, 2017}, pages
  224--237. {ACM}, 2017.

\bibitem[Man19]{DBLP:journals/corr/abs-1810-03792}
Pasin Manurangsi.
\newblock A note on max k-vertex cover: Faster fpt-as, smaller approximate
  kernel and improved approximation.
\newblock In Jeremy~T. Fineman and Michael Mitzenmacher, editors, {\em 2nd
  Symposium on Simplicity in Algorithms, {SOSA} 2019, January 8-9, 2019, San
  Diego, CA, {USA}}, volume~69 of {\em OASIcs}, pages 15:1--15:21. Schloss
  Dagstuhl - Leibniz-Zentrum f{\"{u}}r Informatik, 2019.

\bibitem[Man20]{Manurangsi20}
Pasin Manurangsi.
\newblock Tight running time lower bounds for strong inapproximability of
  maximum {$k$}-coverage, unique set cover and related problems (via {$t$}-wise
  agreement testing theorem).
\newblock In Shuchi Chawla, editor, {\em Proceedings of the 2020 {ACM-SIAM}
  Symposium on Discrete Algorithms, {SODA} 2020, Salt Lake City, UT, USA,
  January 5-8, 2020}, pages 62--81. {SIAM}, 2020.

\bibitem[Man25]{DBLP:journals/tcs/Manurangsi25}
Pasin Manurangsi.
\newblock Improved {FPT} approximation scheme and approximate kernel for
  biclique-free max k-weight {SAT:} greedy strikes back.
\newblock {\em Theor. Comput. Sci.}, 1028:115033, 2025.

\bibitem[Mar08]{DBLP:journals/cj/Marx08}
D{\'{a}}niel Marx.
\newblock Parameterized complexity and approximation algorithms.
\newblock {\em Comput. J.}, 51(1):60--78, 2008.

\bibitem[Mat90]{DBLP:conf/compgeom/Matousek90}
Jir{\'{\i}} Matousek.
\newblock Cutting hyperplane arrangements.
\newblock In Raimund Seidel, editor, {\em Proceedings of the Sixth Annual
  Symposium on Computational Geometry, Berkeley, CA, USA, June 6-8, 1990},
  pages 1--9. {ACM}, 1990.

\bibitem[NWF78]{DBLP:journals/mp/NemhauserWF78}
George~L. Nemhauser, Laurence~A. Wolsey, and Marshall~L. Fisher.
\newblock An analysis of approximations for maximizing submodular set functions
  - {I}.
\newblock {\em Math. Program.}, 14(1):265--294, 1978.

\bibitem[PRS12]{DBLP:journals/talg/PhilipRS12}
Geevarghese Philip, Venkatesh Raman, and Somnath Sikdar.
\newblock Polynomial kernels for dominating set in graphs of bounded degeneracy
  and beyond.
\newblock {\em {ACM} Trans. Algorithms}, 9(1):11:1--11:23, 2012.

\bibitem[PY25]{DBLP:journals/tcs/PanolanY25}
Fahad Panolan and Hannane Yaghoubizade.
\newblock On {MAX-SAT} with cardinality constraint.
\newblock {\em Theor. Comput. Sci.}, 1025:114971, 2025.

\bibitem[RT12]{RaghavendraT12}
Prasad Raghavendra and Ning Tan.
\newblock Approximating csps with global cardinality constraints using {SDP}
  hierarchies.
\newblock In Yuval Rabani, editor, {\em Proceedings of the Twenty-Third Annual
  {ACM-SIAM} Symposium on Discrete Algorithms, {SODA} 2012, Kyoto, Japan,
  January 17-19, 2012}, pages 373--387. {SIAM}, 2012.

\bibitem[Sau72]{DBLP:journals/jct/Sauer72}
Norbert Sauer.
\newblock On the density of families of sets.
\newblock {\em J. Comb. Theory {A}}, 13(1):145--147, 1972.

\bibitem[Sch03]{schrijver2003combinatorial}
Alexander Schrijver.
\newblock {\em Combinatorial optimization: polyhedra and efficiency},
  volume~24.
\newblock Springer Science \& Business Media, 2003.

\bibitem[Sel23]{DBLP:conf/esa/Sellier23}
Fran{\c{c}}ois Sellier.
\newblock Parameterized matroid-constrained maximum coverage.
\newblock In Inge~Li G{\o}rtz, Martin Farach{-}Colton, Simon~J. Puglisi, and
  Grzegorz Herman, editors, {\em 31st Annual European Symposium on Algorithms,
  {ESA} 2023, September 4-6, 2023, Amsterdam, The Netherlands}, volume 274 of
  {\em LIPIcs}, pages 94:1--94:16. Schloss Dagstuhl - Leibniz-Zentrum f{\"{u}}r
  Informatik, 2023.

\bibitem[SF17]{DBLP:journals/jair/SkowronF17}
Piotr Skowron and Piotr Faliszewski.
\newblock Chamberlin-courant rule with approval ballots: Approximating the
  maxcover problem with bounded frequencies in {FPT} time.
\newblock {\em J. Artif. Intell. Res.}, 60:687--716, 2017.

\bibitem[She72]{Shelah1972}
Saharon Shelah.
\newblock A combinatorial problem; stability and order for models and theories
  in infinitary languages.
\newblock {\em Pacific Journal of Mathematics}, 41(1):247--261, 1972.

\bibitem[Sla97]{DBLP:journals/ipl/Slavik97}
Petr Slav{\'{\i}}k.
\newblock Improved performance of the greedy algorithm for partial cover.
\newblock {\em Inf. Process. Lett.}, 64(5):251--254, 1997.

\bibitem[Svi01]{Sviridenko01}
Maxim Sviridenko.
\newblock Best possible approximation algorithm for {MAX} {SAT} with
  cardinality constraint.
\newblock {\em Algorithmica}, 30(3):398--405, 2001.

\bibitem[TV19]{DBLP:journals/tcs/TelleV19}
Jan~Arne Telle and Yngve Villanger.
\newblock {FPT} algorithms for domination in sparse graphs and beyond.
\newblock {\em Theor. Comput. Sci.}, 770:62--68, 2019.

\bibitem[Vaz01]{DBLP:books/daglib/0004338}
Vijay~V. Vazirani.
\newblock {\em Approximation algorithms}.
\newblock Springer, 2001.

\bibitem[VC71]{doi:10.1137/1116025}
V.~N. Vapnik and A.~Ya. Chervonenkis.
\newblock On the uniform convergence of relative frequencies of events to their
  probabilities.
\newblock {\em Theory of Probability \& Its Applications}, 16(2):264--280,
  1971.

\bibitem[Wol82]{DBLP:journals/combinatorica/Wolsey82}
Laurence~A. Wolsey.
\newblock An analysis of the greedy algorithm for the submodular set covering
  problem.
\newblock {\em Comb.}, 2(4):385--393, 1982.

\end{thebibliography}
